\documentclass[rmp,aps,amssymb,twocolumn,nofootinbib,floatfix]{revtex4}

\usepackage{amsmath,amssymb,bm}
\usepackage{graphics}
\usepackage{epsfig}
\usepackage{graphicx}

\def\inbar{\,\vrule height1.5ex width.4pt depth0pt}
\def\IR{\relax{\rm I\kern-.18em R}}
\def\IC{\relax\hbox{$\inbar\kern-.3em{\rm C}$}}

\begin{document}
\title{Critical phenomena in complex networks}

\author{S. N. Dorogovtsev}
\email{sdorogov@fis.ua.pt}
\affiliation{Departamento de F{\'\i}sica da Universidade de Aveiro, 3810-193 Aveiro, Portugal}
\affiliation{A.F. Ioffe Physico-Technical Institute, 194021 St. Petersburg, Russia}
\author{A.~V.~Goltsev}
\email{goltsev@fis.ua.pt}
\affiliation{Departamento de F{\'\i}sica da Universidade de Aveiro, 3810-193 Aveiro, Portugal}
\affiliation{A.F. Ioffe Physico-Technical Institute, 194021 St. Petersburg, Russia}
\author{J. F. F. Mendes}
\email{jfmendes@fis.ua.pt}
\affiliation{Departamento de F{\'\i}sica da Universidade de Aveiro, 3810-193 Aveiro, Portugal}

\begin{abstract}
The combination of the compactness of networks, featuring small diameters, and their complex
architectures results in a variety of critical effects dramatically
different from those in cooperative systems on lattices. In the last
few years, researchers have made important steps toward
understanding the qualitatively new critical phenomena in complex
networks. We review the results, concepts, and methods of this
rapidly developing field. Here we mostly consider two closely
related classes of these critical phenomena, namely structural phase
transitions in the network architectures and transitions in
cooperative models on networks as substrates. We also discuss
systems where a network and
interacting agents on it influence each other. We overview a wide
range of critical phenomena in equilibrium and growing networks
including the birth of the giant connected component, percolation,
$k$-core percolation, phenomena near epidemic thresholds,
condensation transitions, critical phenomena in spin models placed
on networks, synchronization, and self-organized criticality
effects in interacting systems on networks. We also discuss strong
finite size effects in these
systems and highlight open problems and perspectives.
\end{abstract}

\maketitle \tableofcontents





\section{INTRODUCTION}
\label{sec:introduction}

By definition, complex networks are networks with more complex architectures than classical random graphs with their ``simple'' Poissonian distributions of connections. 
The great majority of real-world networks, including the World Wide
Web, the Internet, basic cellular networks, and many others, are
complex ones.
The complex organization of these nets typically implies
a skewed distribution of connections with many hubs, strong
inhomogeneity and high clustering, as well as non-trivial temporal
evolution. These architectures are quite compact (with a small degree of separation between vertices), infinitely
dimensional, which is a fundamental property of various
networks---small worlds.

Physicists intensively studied
structural properties of complex networks since the end of 90's, but
the current focus is essentially on cooperative systems defined on
networks and on
dynamic processes taking place on networks. In
recent years it was revealed that the extreme compactness of
networks together with their complex organization result in a wide
spectrum of non-traditional critical effects and intriguing
singularities. This paper reviews the progress in the understanding
of the unusual critical phenomena in networked systems.

One should note that the tremendous current interest in critical
effects in networks is explained not only by numerous important
applications. Critical phenomena in disordered systems were among
the hottest fundamental topics of condensed matter theory and
statistical physics in the end of XX century.
Complex networks imply a new,
practically unknown in condensed matter, type of strong disorder,
where fluctuations of structural characteristics of vertices (e.g.,
the number of neighbors) may greatly exceed their mean values.
One should add to this large-scale inhomogeneity which is
significant in many complex networks---statistical properties of
vertices may strongly differ in different parts of a network.

The first studies of a critical phenomenon in a network were made by
\textcite{Solomonoff:sr51} and \textcite{Erdos:er59} who introduced
classical random graphs and described the structural phase
transition of the birth of the giant connected component.
These simplest random graphs were widely used by physicists as
substrates for various cooperative models.

Another
basic small-world substrate in statistical mechanics and condensed
matter theory is the Bethe lattice---an infinite regular tree---and
its diluted variations. The Bethe lattice usually allows exact
analytical treatment, and, typically, each new cooperative model is
inspected on this network (as well as on the infinite fully
connected graph).

Studies of critical phenomena in complex networks essentially use
approaches developed for these two fundamental, related classes of
networks---classical random graphs and the Bethe lattices. In these
graphs and many others, small and finite loops (cycles) are rare and
not essential, the architectures are locally tree-like, which is a
great simplifying feature extensively exploited. One may say,
the existing analytical and algorithmic approaches already allow
one to exhaustively analyse any locally tree-like network and to
describe cooperative models on it. Moreover, the tree ansatz works
well even in numerous important situations for loopy and clustered
networks. We will discuss in detail various techniques based on
this
standard approximation. It is these techniques, including, in
particular, the Bethe-Peierls approximation, that are main
instruments for study the critical effects in networks.

Critical phenomena in networks include a wide range of issues:
structural changes in networks,
the emergence of critical---scale-free---network architectures,
various percolation phenomena, epidemic thresholds, phase
transitions in cooperative models defined on networks, critical
points of diverse optimization problems, transitions in co-evolving
couples---a cooperative model and its network substrate, transitions
between different regimes in processes taking place on networks, and
many others. We will show that many of these critical effects are
closely related and universal for different models and may be
described and explained in the framework of a unified approach.


The outline of this review is as follows.
In Sec.~\ref{sec:models} we briefly describe basic models of complex
networks.
Section~\ref{sec:birth} contains a discussion of structural phase
transitions in networks: the birth of the giant connected component
of a complex random network and various related percolation
problems.
In Sec.~\ref{sec:condensation} we describe condensation phenomena,
where a finite fraction of edges, triangles, etc. are attached to a
single vertex.
Section~\ref{sec:diseases} overviews main critical effects in the disease spreading.
Sections~\ref{sec:Ising}, \ref{sec:Potts} and
\ref{sec:xy model} discuss the Ising, Potts  and $XY$ models on
networks. We use the Ising model to
introduce main techniques of analysis of interacting systems in
networks. We place a comprehensive description of
this analytical apparatus, more useful for theoretical physicists,
in the Appendix. Section~\ref{sec:phenomenology} contains a general
phenomenological approach to critical phenomena in networks. In
Secs.~\ref{sec:synchronization} and \ref{sec:self-organized} we
discuss specifics of synchronization and self-organized criticality
on networks. Section~\ref{sec:other} briefly describes a number of
other critical effects in networks. In Sec.~\ref{sec:summary} we
indicate open problems and perspectives of this field. Note that for
a few interesting problems, as yet uninvestigated for complex
networks, we
discuss only the classical random graph case.




\section{MODELS OF COMPLEX NETWORKS}
\label{sec:models}

In this section we briefly introduce basic networks, which are used
as substrates for models, and basic terms. For more detail see books
and reviews of \textcite{Albert:ab02},
\textcite{Dorogovtsev:dmbook03,Dorogovtsev:dm02},
\textcite{Newman:n03a}, \textcite{Bollobas:br03},
\textcite{Pastor-Satorras:pvbook04}, \textcite{Boccaletti:blm06},
\textcite{Durrett:dbook06}, and \textcite{Caldarelli:cbook07}.



\subsection{Structural characteristics of networks}
\label{ssec:characteristics}

A random network is a statistical ensemble, where each member---a
particular configuration of vertices and edges---is realized with
some prescribed probability (statistical weights). Each graph of
$N$ vertices may be described by its {\em adjacency} $N {\times}
N$ {\em matrix} $(a_{ij})$, where $a_{ij}=0$ if edges between
vertices $i$ and $j$ are absent, and $a_{ij}>0$ otherwise. In
simple graphs, $a_{ij}=0,1$. In weighted networks, the adjacency
matrix elements are non-negative numbers which may be
non-integer---weights of edges. The simplest characteristic of a
vertex in a graph is its {\em degree} $q$, that is the number of
its nearest neighbors. In physics this is often called
connectivity. In directed graphs, at least some of edges are
directed, and one should introduce in- and out-degrees. For random
networks, a vertex degree distribution $P(q)$ is the first
statistical measure.

The presence of connections between the nearest neighbors of a
vertex $i$ is described by its {\em clustering coefficient}
$C(q_i)\equiv t_i/[q_i(q_i-1)/2]$, where $t_i$ is the number of
triangles (loops of length $3$) attached to this vertex, and
$q_i(q_i-1)/2$ is the maximum possible number of such triangles.
Note that in general, the mean clustering $\langle C \rangle \equiv
\sum_q P(q)C(q)$ should not coincide with the clustering coefficient
(transitivity) $C \equiv \langle t_i \rangle/\langle q_i(q_i-1)/2
\rangle$ which is three times the ratio of the total number of
triangles in the network and the total number of connected triples
of vertices.
A connected triple here is a vertex with its two nearest neighbors.
A triangle can be treated as a three connected triples, which explains the coefficient $3$. 

{\em A loop} (simple cycle) is a closed path
visiting each its vertex only once. By definition, {\em trees} are
graphs without loops.


For each pair of vertices $i$ and $j$ connected by at least one
path, one can introduce the shortest path length, the so-called {\em
intervertex distance} $\ell_{ij}$, the corresponding number of edges
in the shortest path. The distribution of intervertex distances
$\cal{P}(\ell)$ describes the global structure of a random network,
and the mean intervertex distance $\overline{\ell}(N)$ characterizes
the ``compactness'' of a network. In finite-dimensional systems,
$\overline{\ell}(N) \sim N^{1/d}$.
We, however, mostly discuss networks with {\em the small-world
phenomenon}---the so called {\em small worlds}, where
$\overline{\ell}$ increases with the total number of vertices $N$
slower than any positive power, i.e., $d=\infty$
\cite{Watts:wbook99}. Typically in networks,
$\overline{\ell}(N) \sim \ln N$.

Another important characteristic of a vertex (or edge) is its {\em
betweenness centrality} (or, which is the same, {\em load}):
the
number of the shortest paths between other vertices which run
through this vertex (or edge).
In more strict terms, the betweenness centrality $b(v)$ of vertex $v$ is defined as follows. Let $s(i,j){>}0$ be the number of the shor\-test paths between vertices $i$ and $j$. Let $s(i,v,j)$ be the number of these paths, passing through vertex $v$. Then $b(v) \equiv \sum_{i\neq v \neq j} s(i,v,j)/s(i,j)$.
A betweenness centrality distribution
is introduced for a random network.

A basic notion is {\em a giant connected component} analogous to
the percolation cluster in condensed matter. This is a set of
mutually reachable vertices and their interconnections, containing a
finite fraction of vertices of an infinite network. Note that in
physics the infinite network limit, $N\to\infty$, is also called
the thermodynamic limit.
The relative size of the giant component (the relative number of its
vertices) and the size distribution of finite connected components
describe the topology of a random network.




\subsection{Cayley tree versus Bethe lattice}
\label{ssec:cayley_versus_bethe}

\begin{figure}[t]
\begin{center}
\scalebox{0.26}{\includegraphics[angle=0]{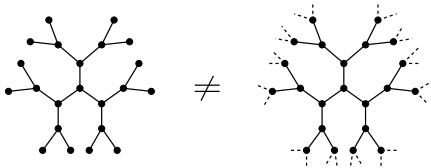}}
\end{center}
\caption{ The Cayley tree (on the left) versus the Bethe lattice (on
the right). }
\label{f1}%
\end{figure}

Two very different regular graphs
are extensively used as substrates for cooperative models. Both are
small worlds if the degree of their vertices exceeds $2$. 
In the (regular) {\em Cayley tree}, explained on Fig.~\ref{f1}, a
finite fraction of vertices are dead ends. These vertices form a
sharp border of this tree. There is a central vertex, equidistant
from the boundary vertices. The presence of the border essentially
determines the physics of interacting systems on the Cayley tree.

{\em The Bethe lattice} is an infinite regular graph
(see Fig.~\ref{f1}).
All vertices in a Bethe lattice are topologically equivalent, and
boundaries are absent.
Note that in the thermodynamic limit, the so called {\em random
regular graphs} asymptotically approach the Bethe lattices
\cite{Johnston:jp98}. The random regular graph is a maximally random
network of vertices of equal degree. It is constructed of vertices
with the same number (degree) of stubs by connecting pairs of the
stubs in all possible ways.




\subsection{Equilibrium random trees versus growing ones}
\label{ssec:equilibrium_trees}

\begin{figure}[t]
\begin{center}
\scalebox{0.30}{\includegraphics[angle=0]{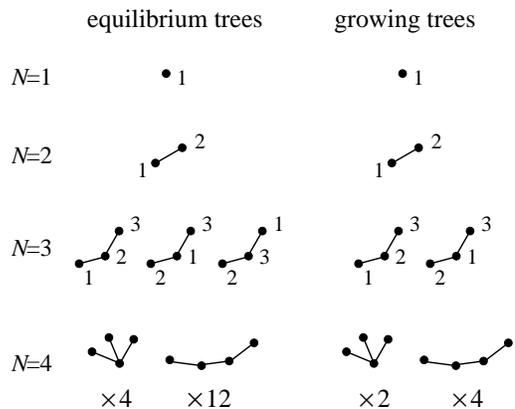}}
\end{center}
\caption{ Statistical ensembles of equilibrium random connected
trees (left-hand side) and of growing connected trees (right-hand
side) for $N=1,2,3,4$. The ensemble of equilibrium trees consists of
all possible connected trees of $N$ labelled vertices, where each
tree is taken with the same weight. The ensemble of growing (causal)
trees is the following construction.
Its members are the all possible connected trees of size $N$
that can be made by sequential attachment of new labelled vertices.
Each of these trees of $N$ vertices is taken with the same weight.
Notice that at $N=3$,
one of the labelled graphs of the equilibrium ensemble is absent in
the ensemble of growing trees. At $N=4$, we indicate the numbers of
isomorphic graphs in both ensembles.
(By definition, isomorphic graphs differ from each other only by vertex labels.)
Already at $N=4$,
the equilibrium random tree is less compact, since the probability
of realization of the chain is higher in this case.
}
\label{f1001}%
\end{figure}

Remarkably, random connected trees (i.e., consisting of a single
connected component) may or may not be small worlds
\cite{Burda:bck01,Bialas:bbj02}.
{\em The equilibrium random connected trees} have extremely extended
architectures characterising by the fractal (Hausdorff) dimension
$d_h=2$, i.e., $\overline{\ell}(N) \sim N^{1/2}$. These random trees
are the statistical ensembles
that consist of all possible connected trees with $N$ labelled
vertices, taken with equal probability---Fig.~\ref{f1001}, left
side. The degree distributions of these networks are
rapidly decreasing, $P(q) = e^{-1}/(q-1)!$. However one may arrive at scale-free degree
distributions $P(q) \sim q^{-\gamma}$ by, for example, introducing
special degree dependent
statistical weights of different members of these ensembles. In this
case, if $\gamma\geq 3$, then $d_h=2$, and if $2<\gamma<3$, then the
fractal dimension is $d_h=(\gamma-1)/(\gamma-2)>2$.

In contrast to this, {\em the growing (causal, recursive) random
connected trees} are small worlds. These trees are constructed by
sequential attachment of new (labelled) vertices---Fig.~\ref{f1001},
right side. The rule of this attachment or, alternatively, specially
introduced degree dependent weights for different realizations,
determine the resulting degree distributions. The mean intervertex
distance in these graphs $\overline{\ell} \sim \ln N$. Thus, even
with identical degree distributions, the equilibrium random trees
and growing ones have quite different geometries.



\subsection{Classical random graphs}
\label{ssec:classical}

Two simplest models of random networks are so close (one may say,
asymptotically coincident in the thermodynamic limit) that they are
together called classical random graphs. The
{\em Gilbert model}, or the $G_{np}$ model,
\cite{Solomonoff:sr51,Gilbert:g59} is a random graph where an edge
between each pair of $N$ vertices is present with a fixed
probability $p$.

The slightly more difficult for analytical treatment {\em Erd\H
os-R\'enyi model} \cite{Erdos:er59}, which is also called the
$G_{nm}$ model, is a statistical ensemble where all members---all
possible graphs with a given numbers of vertices, $N$, and edges,
$M$,---have equal probability of realization. The relationship
between the Erd\H os-R\'enyi model and the Gilbert one is given by
the following equalities for the mean degree: $\langle q \rangle =
2M/N = pN$. If $\langle q \rangle/N \to 0$ as $N \to \infty$, a
network is {\em sparse}, i.e., it is far more sparse than a fully
connected graph. So, the Gilbert model is sparse when $p(N \to
\infty)\to 0$.

The classical random graphs are maximally random graphs under a
single constraint---a fixed mean degree $\langle q \rangle$. Their
degree distribution is Poissonian,
$P(q) = e^{-\langle q \rangle}\langle q \rangle^q/q!$.



\subsection{Uncorrelated networks with arbitrary degree distributions}
\label{ssec:uncorrelated}


\begin{figure}[b]
\begin{center}
\scalebox{0.34}{\includegraphics[angle=0]{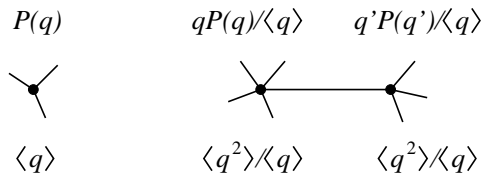}}
\end{center}
\caption{ The distribution of connections and the mean degree of a
randomly chosen vertex (on the left) differ sharply from those of
end vertices of a randomly chosen edge (on the right). }
\label{f2}%
\end{figure}

One should emphasize that in a random network, the degree
distribution of the nearest neighbor $P_{\text{nn}}(q)$ (or, which
is the same, the degree distribution of an end vertex of a randomly
chosen edge) does not coincide with the vertex degree distribution
$P(q)$. In general random networks, 
\begin{equation}
P_{\text{nn}}(q) = \frac{qP(q)}{\langle q \rangle}, \ \ \ \ \langle
q \rangle_{\text{nn}} = \frac{\langle q^2 \rangle}{\langle q
\rangle}>\langle q \rangle , \label{e1.1}
\end{equation}
see Fig.~\ref{f2}. These simple relations play a key role in the
theory of complex networks.

By definition, in uncorrelated networks correlations are absent, in
particular, there are no correlations between degrees of the nearest
neighbors. That is, the joint distribution of degrees of the nearest
neighbor vertices factors into the product: 
\begin{equation}
P(q,q') = \frac{qP(q)q'P(q')}{\langle q \rangle^2} . \label{e1.2}
\end{equation}

Thus, the architectures of uncorrelated networks are determined by
their degree distributions. The Erd\H os-R\'enyi and Gilbert models
are simple uncorrelated networks. Below we list the models of
complex uncorrelated networks, which are actually very close to each
other in the thermodynamic limit. In this limit all these networks
are locally tree-like (if they are sparse, of course), with only
infinite loops.


\subsubsection{Configuration model}
\label{sssec:configuration_model}

The direct generalization of the Erd\H os-R\'enyi graphs is the
famous configuration model formulated by \textcite{Bollobas:b80},
see also the work of \textcite{Bender:bc78}. In graph theory, these
networks are also called random labelled graphs with a given degree
sequence. The configuration model is the statistical ensemble,
whose members are realized with equal probability. These members are
all possible graphs with a given set $\{N_q=NP(q)\}$,
$q=0,1,2,3,\ldots$, where $N_q$ is the number of vertices of degree
$q$. In simple terms,
the configuration model provides maximally random graphs with a
given degree distribution $P(q)$.

This construction may be also portrayed in more graphic terms:
(i) Attach stubs---edge-halves---to $N$ vertices according to a
given sequence of numbers $\{N_q\}$. (ii) Pair randomly chosen stubs
together into edges. Since stubs of the same vertex may be paired
together, the configuration model, in principle, allows a number of
loops of length one as well as multiple connections. Fortunately,
these may be neglected in many problems.

Using relation (\ref{e1.1}) gives the formula $z_2 = \langle q^2
\rangle - \langle q \rangle$ for the mean number of the second
nearest neighbors of a vertex. That is, the mean branching
coefficient of the configuration model and, generally, of an
uncorrelated network is 
\begin{equation}
B = \frac{z_2}{z_1} = \frac{\langle q^2 \rangle - \langle q
\rangle}{\langle q \rangle} , \label{e1.2a}
\end{equation}
where $z_1 = \langle q \rangle$. Consequently, the mean number of
the $\ell$th nearest neighbors of a vertex is $z_\ell = z_1
(z_2/z_1)^{\ell-1}$. So the mean intervertex distance is
$\overline{\ell}(N) \cong \ln N/\ln(z_2/z_1)$ \cite{Newman:nsw01}.

The distribution of the intervertex distances in the configuration
model is quite narrow. Its relative width approaches zero in the
thermodynamic limit. In other words, in this limit, almost all
vertices of the configuration model are mutually equidistant
\cite{Dorogovtsev:dms03a}. We emphasise that this remarkable
property is valid for a very wide class of networks with the small-world phenomenon.

The configuration model was generalized to bipartite networks
\cite{Newman:nsw01}. By definition, {\em a bipartite graph} contains
two kinds of vertices, and only vertices of different kinds may be
interlinked.
In short, the configuration model of a bipartite network is a
maximally random bipartite graph with two given degree distributions
for two types of vertices.


\subsubsection{Static model}
\label{sssec:static_model}


The direct generalization of the Gilbert model is the static one
\cite{Goh:gkk01}, see also works of \textcite{Chung:cl02},
\textcite{Soderberg:s02}, and \textcite{Caldarelli:ccd02}. These are graphs with a given sequence of
desired degrees. These desired degrees $\{d_i\}$ play role of
``hidden variables'' defined on vertices $i=1,2,\ldots,N$.
Pairs of vertices $(ij)$ are connected with probabilities
$p_{ij}=1-\exp(-d_id_j/N\langle d\rangle)$.
The degree distribution of resulting
network $P(q)$ tends to a given distribution of desired degrees at
sufficiently large $q$.
It is important that at small enough $d_i$, the probability $p_{ij} \cong d_id_j/(N\langle d\rangle)$. The exponential function keeps the probability below $1$ even if $d_id_j>N\langle d\rangle$ which is possible if the desired degree distribution is heavy tailed.


\subsubsection{Statistical mechanics of uncorrelated networks}
\label{sssec:statistical_ensembles}

It is also easy to generate random networks by using a standard
thermodynamic approach, see \textcite{Burda:bck01},
\textcite{Bauer:bb02}, and \textcite{Dorogovtsev:dms03b}. In
particular, assuming that the number of vertices is constant, one
may introduce ``thermal'' hopping of edges or their rewiring. These
processes lead to relaxational dynamics in the system of edges
connecting vertices. The final state of this relaxation process---an
equilibrium statistical ensemble---may be treated as an
``equilibrium random network''. This network is uncorrelated if the
rate/probability of rewiring depends only on degrees of host
vertices and on degrees of targets, and, in addition, if rewirings
are independent.
The resulting diverse degree distributions are determined by two
factors: a specific degree dependent rewiring and the mean vertex
degree in the network. Note that if multiple connections are
allowed, this construction is essentially equivalent to the simple
{\em balls-in-boxes (or backgammon) model}
\cite{Bialas:bbj97,Bialas:bbb00}, where ends of edges---balls---are
statistically distributed among vertices---boxes.



\subsubsection{Cutoffs of degree distributions}
\label{sssec:cutoffs}

Heavy tailed degree distributions $P(q)=\langle N(q)\rangle/N$ in
finite networks,
inevitably end by a rapid drop at large degrees---cutoff. Here,
$\langle N(q)\rangle$ is the number of vertices of degree $q$ in a
random network, averaged over all members of the corresponding
statistical ensemble. The knowledge of the size dependence of the
cutoff position, $q_{\text{cut}}(N)$ is critically important for the
estimation of various size effects in complex networks. The
difficulty is that the form of $q_{\text{cut}}(N)$ is highly model
dependent.

We here present estimates of $q_{\text{cut}}(N)$ in uncorrelated
scale-free networks, where $P(q)\sim q^{-\gamma}$. The results
essentially depend on (i) whether exponent $\gamma$ is above or
below $3$, and (ii) whether multiple connections are allowed in the
network or not.

In the range $\gamma\geq 3$, the resulting estimates are the same in
networks with multiple connections \cite{Burda:bck01} and without
them \cite{Dorogovtsev:dmp05}. In this range, strict calculation of
a degree distribution taking into account all members of a
statistical network ensemble leads to $q_{\text{cut}}(N) \sim
N^{1/2}$.
The total number of the members of an equilibrium network ensemble (e.g., for the configuration model)
is huge, say, of the order of $N!$.
However, in empirical research or simulations, ensembles under
investigation have rather small number $n$ of members---a whole
ensemble may consist of a single empirically studied map or of a few
runs in a simulation. Often,
only a single network configuration is used as a substrate in
simulations of a cooperative model.
In these measurements, {\em a natural cutoff} of an observed degree distribution arises \cite{Dorogovtsev:dms01c,Cohen:ceb00}. Its degree, much lower than $N^{1/2}$,
is estimated from the following condition.   
In the $n$ studied ensemble members, a vertex degree exceeding $q_{\text{cut}}$ should occur one time: 
$nN\int^\infty_{q_{\text{cut}}(N)}dq P(q) \sim 1$. 
This gives the
really observable cutoff: 
\begin{equation}
q_{\text{cut}}(N,\gamma\geq 3) \sim (nN)^{1/(\gamma-1)}
\label{e1.2aa}
\end{equation}
if $n \ll N^{(\gamma-3)/2}$, which is a typical situation, and
$q_{\text{cut}}(N,\gamma\geq 3) \sim N^{1/2}$ otherwise.


In the interesting range $2<\gamma<3$, the cutoff
essentially depends on the kind of an uncorrelated network. If in an
uncorrelated network, multiple connections are allowed, then
$q_{\text{cut}}(N,2{<}\gamma{<} 3) \sim N^{1/(\gamma-1)}$.
In uncorrelated networks without multiple connections,
$q_{\text{cut}}(N,2{<}\gamma{<} 3) \sim N^{1/2}\ll
N^{1/(\gamma-1)}$ \cite{Burda:bk03}, although see
\textcite{Dorogovtsev:dmp05} for a different estimate for a specific
model without multiple connections.
For discussion of the cutoff problem in the static model in this range of exponent $\gamma$, see \textcite{Lee:lgk06}.

\textcite{Seyed-allaei:sbm06} found that in scale-free uncorrelated networks with exponent $\gamma<2$, the cutoff is $q_{\text{cut}}(N,1{<}\gamma{<}2) \sim N^{1/\gamma}$. They showed that the mean degree of these networks increases with $N$: namely, $\langle q \rangle \sim N^{(2-\gamma)/\gamma}$.

For the sake of completeness, we here mention that in growing scale-free recursive networks, $q_{\text{cut}}(N,\gamma{>}2) \sim N^{1/(\gamma-1)}$ \cite{Dorogovtsev:dms01c,Krapivsky:kr02,Waclaw:ws07}. Note that the growing networks are surely correlated.



\subsection{Equilibrium correlated networks}
\label{ssec:correlated}


The simplest kind of correlations in a network are correlations
between degrees of the nearest neighbor vertices. These correlations
are described by the joint degree--degree distribution $P(q,q')$. If
$P(q,q')$ is not factorized, unlike equality (\ref{e1.2}), the
network is correlated \cite{Newman:n02b,Maslov:ms02}.

The natural generalization of uncorrelated networks, which is still
sometimes analytically treatable, are networks maximally random
under the constraint that their joint degree-degree distributions
$P(q,q')$ are fixed. That is, only these correlations are present.
In the hierarchy of equilibrium network models,
this is the next, higher, level, after the classical random graphs
and uncorrelated networks with an arbitrary degree distribution.
Note that networks with
this kind of correlations are still locally tree-like in the sparse
network regime. In this sense they may be treated as random Bethe
lattices.

These networks may be constructed in the spirit of the configuration
model.
An alternative construction---{\em networks with hidden
variables}---directly generalizes the static model. These are
networks, where (i) a random hidden variable $h_i$ with distribution
$P_h(h)$ is assigned to each vertex, and (ii) each pair of vertices
$(ij)$ is connected by an edge with probability $p(h_i,h_j)$
\cite{Soderberg:s02,Caldarelli:ccd02,Boguna:bp03}. The resulting
joint degree--degree distribution is determined by $P_h(h)$ and
$p(h,h')$ functions.



\subsection{Loops in networks}
\label{ssec:loops}

The above-described equilibrium network models share the convenient
locally tree-like structure in the sparse network regime. The number
of loops  ${\cal N}_L$ of length $L$ in a network allows us to
quantify this important property. We stress that the total number of
loops in these networks is in fact very large. Indeed, the typical
intervertex distance $\sim \ln N$, so that the number of loops with
lengths $\gtrsim  \ln N$
should be huge. On the other hand, there is few loops of smaller
lengths.
In simple terms, if the second moment of the degree distribution is
finite in the thermodynamic limit, then
the number of loops of any given finite length is finite even in an infinite network. Consequently, the 
probability that a finite loop passes through a
vertex is quite small, which explains the tree-likeness.

In more precise terms, the number of loops in uncorrelated undirected networks
is given by the following expression
\cite{Bianconi:bm05a,Bianconi:bc03}: 
\begin{equation}
{\cal N}_L \sim \frac{1}{2L}
\left(\frac{\langle q^2 \rangle-\langle q\rangle}{\langle q\rangle}\right)^L
,
\label{e1.3}
\end{equation}
which is valid for sufficiently short (at least, for finite) loops,
so that the clustering coefficient $C(k)=C=\langle C\rangle=(\langle
q^2 \rangle{-}\langle q \rangle)^2/(N\langle q \rangle^3)$
\cite{Newman:n03b}. In addition, there are exponentially many, $\ln
{\cal N}_L \propto N$, loops of essentially longer lengths (roughly speaking, longer than the network diameter).
These ``infinite loops'',
as they are longer than a correlation length for a cooperative system, do not violate the validity of the tree approximation.
Moreover, without these loops---in perfect trees---phase transitions are often impossible, as, e.g.,  in the Ising model on a tree. 
The mean number of loops of length $L$ passing through a vertex of degree $k$ is 
${\cal N}_L(k)  \approx [k(k{-}1)/(\langle q \rangle N)]\,[(L{-}1)/L]\,{\cal N}_{L-1}$. 
With
degree distribution cutoffs represented in Sec.~\ref{sssec:cutoffs},
formula (\ref{e1.3}) leads to finite ${\cal N}_L$ in uncorrelated
networks with $\gamma>3$, and to a large number of loops 
\begin{equation}
{\cal N}_L \sim \frac{1}{2L}
\left(a/\langle q \rangle\right)^L N^{L(3-\gamma)/2} , \label{e1.4}
\end{equation}
for $2<\gamma<3$ and $\langle q^2 \rangle \cong a N^{(3-\gamma)/2}$,
 where $a$ is a constant.
For the statistics of loops in directed networks, see \textcite{Bianconi:bgm07}. Note that formulas (\ref{e1.3}) and (\ref{e1.4}) indicate that even the sparse uncorrelated networks are actually loopy if $\gamma<3$. Nonetheless, we suppose that the tree ansatz still works even in this situation (see discussion in following sections).




\subsection{Evolving networks}
\label{ssec:evolving_networks}


Self-organization of non-equilibrium networks during their evolution
(usually growth) is one of traditional explanations of network
architectures with a great role of highly connected hubs. One should
also stress that non-equilibrium networks inevitably have a wide
spectrum of correlations.

The simplest random growing network is {\em a random recursive tree}
defined as follows. The evolution starts from a single vertex. At
each time step, a new vertex is attached to a random existing one by
an edge. The resulting random tree has an exponential degree
distribution.


\subsubsection{Preferential attachment}
\label{sssec:preferential}


To arrive at a heavy-tailed degree distribution, one may use
preferential attachment---vertices for linking are chosen with
probability proportional to a special function $f(q)$ of their
degrees (preference function). In particular, the scale-free
networks are generated with a linear preference function.

A recursive network growing by the following way is rather
representative. The growth starts with some initial configuration,
and at each time step, a new vertex is attached to preferentially
chosen $m\geq 1$ existing vertices by $m$ edges. Each vertex for
attachment is chosen with probability, proportional to a linear
function of its degree, $q+A$, where the constant $A>-m$. In
particular, if $A{=}0$---the proportional preference,---this is the
Barab\'asi-Albert model \cite{Barabasi:ba99}, where the $\gamma$
exponent of the degree distribution is equal to $3$. In general, for
a linear preferential attachment, the degree distribution exponent
is $\gamma=3+A/m$ \cite{Dorogovtsev:dms00,Krapivsky:krl00}.

Among these recursive networks, the Barab\'asi-Albert model is a
very special case: it has anomalously weak degree--degree
correlations for the nearest neighbors, and so it is frequently
treated as ``almost uncorrelated''.

The idea of preferential attachment providing
complex network architectures was well explored.
The smooth variations of these diverse structures with various model parameters were extensively studied.
For example, \textcite{Szabo:sak03} described the variations of the degree-dependent clustering in simple generalizations of the Barab\'asi-Albert model.



\subsubsection{Deterministic graphs}
\label{sssec:deterministic}

\begin{figure}[t]
\begin{center}
\scalebox{0.178}{\includegraphics[angle=0]{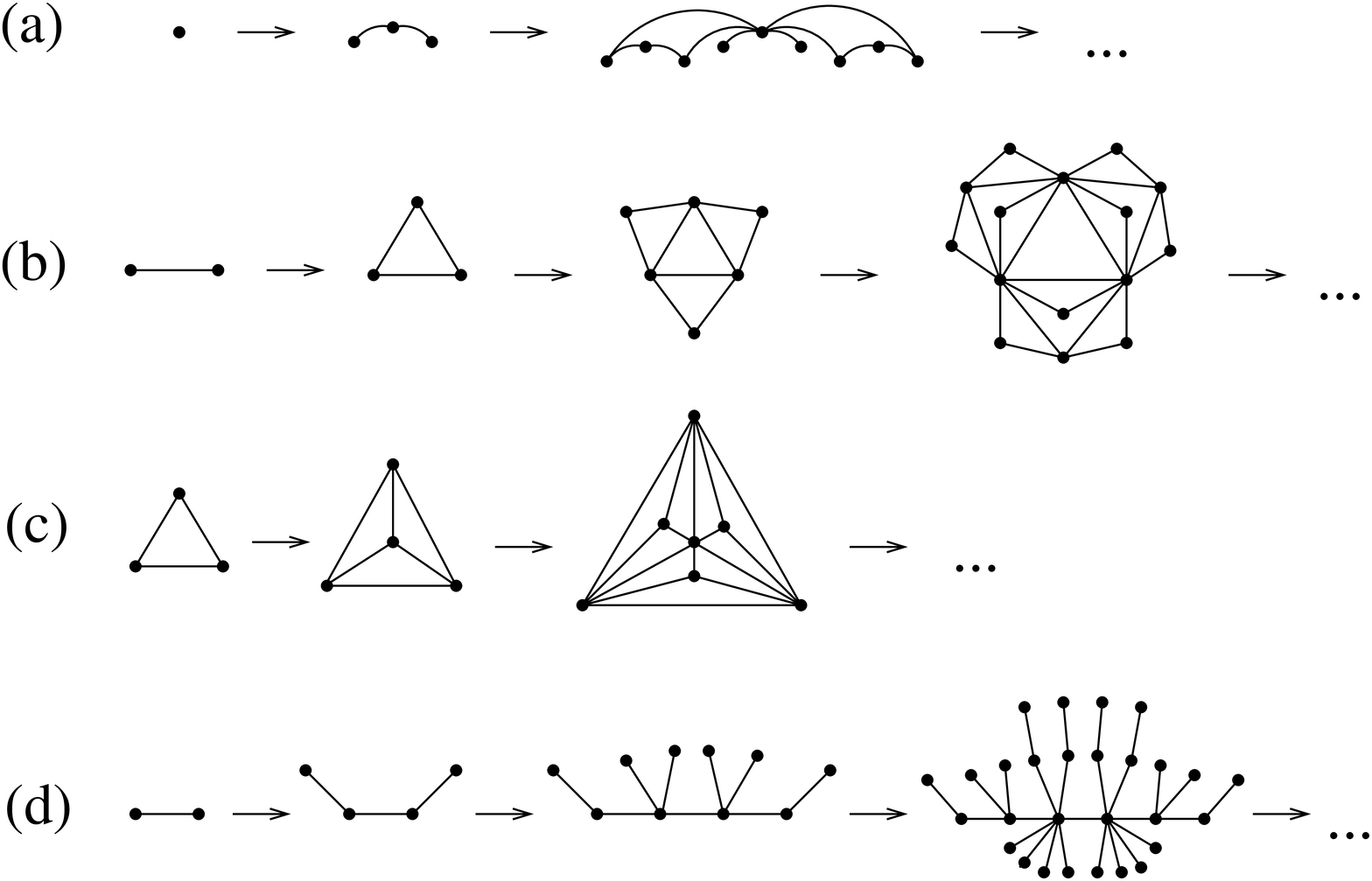}}
\end{center}
\caption{Examples of deterministic small worlds: (a) of
\textcite{Barabasi:brv01}, (b) of \textcite{Dorogovtsev:dm02} and
\textcite{Dorogovtsev:dgm02a}, (c) of \textcite{Andrade:aha05} and
\textcite{Doye:dm05}, (d) of \textcite{Jung:jkk02}. The $\gamma$
exponent for each of these four deterministic graphs equals
$1+\ln3/\ln2 = 2.585\ldots$. } \label{f3}
\end{figure}

Deterministic graphs often provide the only possibility for
analytical treatment of difficult problems. Moreover, by using these
graphs, one may mimic complex random networks surprisingly well.
Fig.~\ref{f3} demonstrates a few simple
``scale-free''
deterministic graphs, which show the small-world phenomenon and
whose
discrete degree distribution have a power-law envelope.



\subsection{Small-world networks}
\label{ssec:small-world_networks}

\begin{figure}[t]
\begin{center}
\scalebox{0.17}{\includegraphics[angle=0]{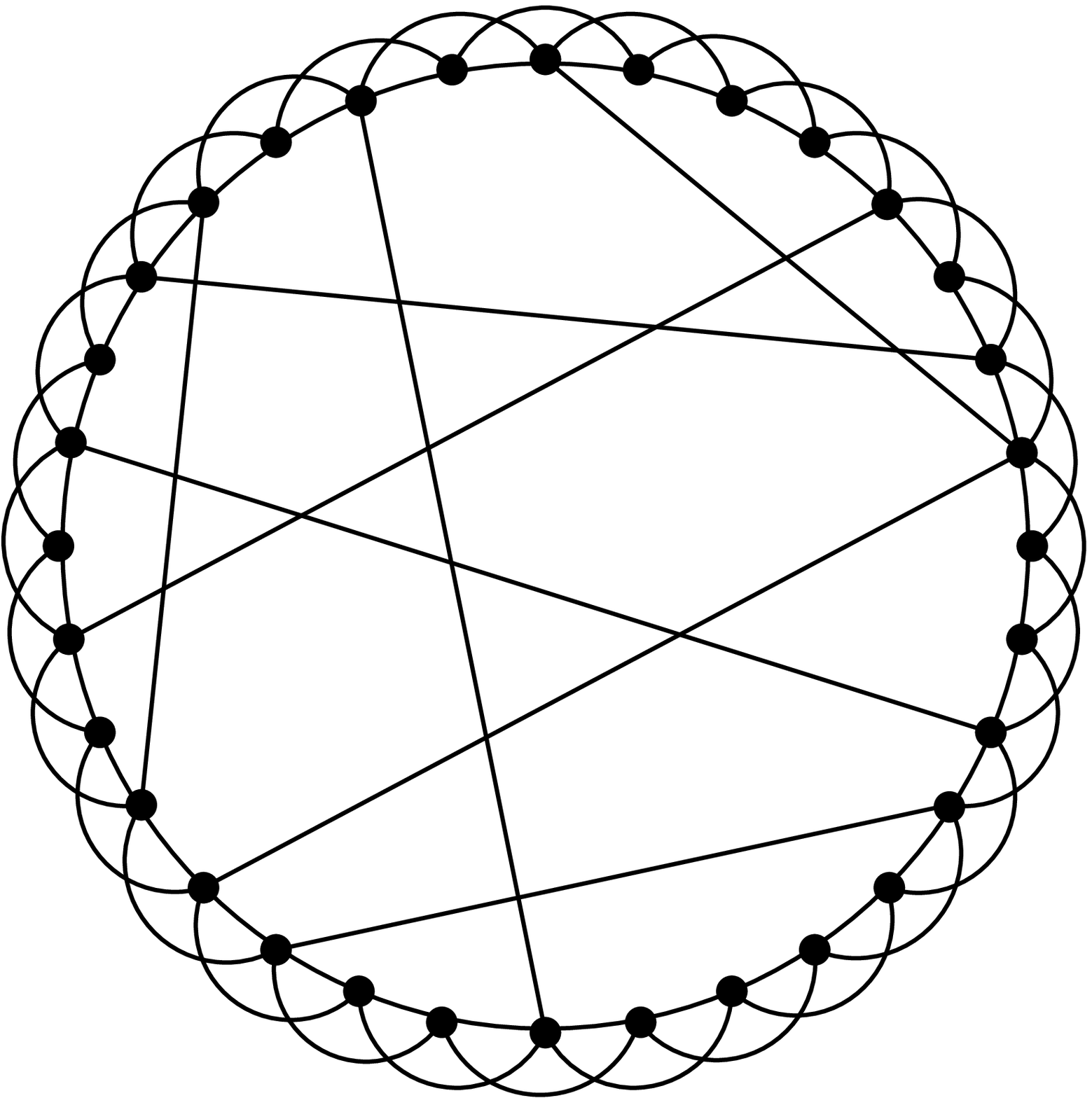}}
\end{center}
\caption{ A simple variation of the Watts-Strogatz model
\cite{Watts:ws98,Watts:wbook99}. Adapted from
\protect\textcite{Newman:n00}. } \label{f4}
\end{figure}

The small-world networks introduced by \textcite{Watts:ws98} are
superpositions of finite dimensional lattices and classical random
graphs, thus combining their properties. One of variations of the
Watts-Strogatz model is explained in Fig.~\ref{f4}: randomly chosen
pairs of vertices in a one-dimensional lattice are connected by
shortcuts. There is a smooth crossover from a lattice to a
small-world geometry with an increasing number of shortcuts.
Remarkably, even with extremely low relative numbers of shortcuts,
these networks demonstrate the small-world phenomenon.


\textcite{Kleinberg:k99,Kleinberg:k00}
used an important generalization of the Watts-Strogatz model. In the
Kleinberg network (``the grid-based model with exponent $\alpha$''),
the probability that a shortcut connects a pair of vertices
separated by Euclidean distance $r$ decreases as $r^{-\alpha}$.
The resulting network geometry critically depends on the value of
exponent $\alpha$.

We end this section with a short remark. In solid state physics,
boundary conditions play an important role. We stress that as a
rule, the networks under discussion have no borders. So the question
of boundary conditions is meaningless here. There are very few
exceptions, e.g., the Cayley tree.




\section{THE BIRTH OF A GIANT COMPONENT}
\label{sec:birth}

This is a basic structural transition in the network architecture.
Numerous critical phenomena in cooperative models on networks can be
explained by taking into account the specifics of this transition in
complex networks.
The birth of a giant connected component corresponds to the percolation threshold notion in condensed matter.
The study of random graphs was started with the discovery and description of this transition \cite{Solomonoff:sr51,Erdos:er59}. Remarkably, it takes place in sparse networks, at $\langle q \rangle \sim \text{const}$, which makes this range of mean degrees most interesting.



\subsection{Tree ansatz}
\label{ssec:tree_ansatz}

The great majority of analytical results for cooperative models on
complex networks were obtained in the framework of the tree
approximation. This ansatz assumes the absence of finite loops in a
network in the thermodynamic limit
and allows only infinite loops. The allowance of the infinite loops
is of primary importance since they greatly influence the critical
behavior.
Indeed, without loops, that is on perfect trees, the ferromagnetic
order, say, in the Ising model
occurs only at zero temperature. Also, the removal of even a
vanishingly small fraction of vertices or edges from a perfect tree
eliminates the giant connected component.

The tree ansatz allows one to use the convenient techniques of the
theory of random branching processes. On the other hand, in the
framework of this ansatz, equilibrium networks
are actually equivalent to random Bethe lattices.



\subsection{Organization of uncorrelated networks}
\label{ssec:organization_uncorrelated}

The mathematical solution of the problem of organization of
arbitrary uncorrelated networks as a system of connected components
was proposed by \textcite{Molloy:mr95,Molloy:mr98}. In the works of
\textcite{Newman:nsw01} and of \textcite{Callaway:cns00} these ideas
were represented and developed using the apparatus and language of
physics. Here we describe
these fundamental results and ideas in simple terms. The reader may
refer to the papers of \textcite{Newman:nsw01} and
\textcite{Newman:n03b} for the details of this theory based on the
generating function technique.



\subsubsection{Evolution of the giant connected component}
\label{sssec:giant_connected_component}

The theory of uncorrelated networks (we mostly discuss the
configuration model, which is completely described by the degree
distribution $P(q)$ and
size $N$) is based on their following
simplifying features:

\begin{figure}[t]
\begin{center}
\scalebox{0.23}{\includegraphics[angle=0]{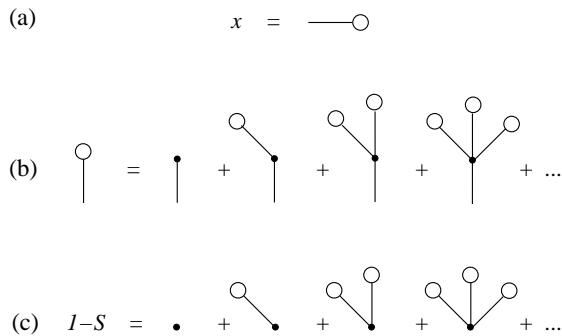}}
\end{center}
\caption{ (a) The graphic notation for the probability $x$ that,
following a randomly chosen edge to one of its end vertices, we arrive at a finite
connected component.
(b) Equation (\protect\ref{e3.1}) or, equivalently,
Eq.~(\protect\ref{e3.3}) in graphic form. (c) The graphic
representation of formula~(\protect\ref{e3.2}) and of equivalent
relation~(\protect\ref{e3.4}) for the relative size $S$ of the giant
connected component. } \label{f5}
\end{figure}

\begin{itemize}

\item[(i)] The sole characteristic of a vertex in these networks is its degree, in any other respect, the vertices are statistically equivalent---there are no borders or centers, or older or younger vertices in these models. The same is valid for edges.

\item[(ii)] The tree ansatz is supposed to be valid.

\item[(iii)]
Formulas (\ref{e1.1}) and (\ref{e1.2}) are valid (see Fig.~\ref{f2}).

\end{itemize}

Feature (i) allows one to introduce the probability $x$ that,
following a randomly chosen edge to one of its end vertices, he or
she
arrives at a finite connected component.
In more strict terms, choose a random edge; choose its random end;
then $x$ is the probability that after removing this edge, the
chosen end vertex will belong to a finite connected component. A
graphic representation of $x$ is introduced in Fig.~\ref{f5}(a). The
probability that an edge belongs to one of finite components
is, graphically, 
\begin{equation}
\bigcirc \!\!-\!\!\!-\!\!\!-\!\!\!-\! \bigcirc = x^2 .
\label{e3.10000}
\end{equation}
This is the probability that following an edge in any direction, we arrive at finite trees.
Thus $1-x^2$ is a fraction of edges which are in the giant connected
component. This simple relation enables us to measure $x$. Using
features (i), (ii), and (iii) immediately leads to the following
self-consistent equation for $x$ and expression for the probability
$1-S$ that a vertex belongs to a finite connected component: 
\begin{eqnarray}
&& x = \sum_{q}\frac{qP(q)}{\langle q \rangle} x^{q-1} ,
\label{e3.1}
\\[5pt]
&& 1-S = \sum_{q} P(q) x^{q} . \label{e3.2}
\end{eqnarray}
In particular, relation~(\ref{e3.2}) is explained as follows.
A vertex belongs to a finite connected component if and only if following every its edge in direction from this vertex we arrive at a finite tree. The probability of this event is $x^q$ for a vertex of degree $q$. For a randomly chosen vertex, we must sum over $q$ the products of $x^q$ and the probability $P(q)$.
One can see that $S$ is the relative size of the giant connected
component. Figures~\ref{f5}(b) and (c) present these formulas in
graphic form and explain them. Note that if $P(q=0,1)=0$, then
Eq.~(\ref{e3.1}) has the only solution $x=1$, and so $S=1$, i.e., the
giant connected component coincides with the network.
Using the generating function of the degree distribution, $\phi(z)
\equiv \sum_q P(k)z^q$ and the notation $\phi_1(z) \equiv
\phi'(z)/\phi'(1) = \phi'(z)/\langle q \rangle$ gives
\begin{eqnarray}
&& x = \phi_1(x) , \label{e3.3}
\\[5pt]
&& S = 1 - \phi(x) . \label{e3.4}
\end{eqnarray}
These relations demonstrate the usefulness of the generating
function technique in network theory. The deviation
$1-x$ plays the role of the order parameter. If
Eq.~(\ref{e3.1}) has a non-trivial solution $x<1$, then
the network has the giant connected component. The size of
this component can be found
by substituting the solution of Eqs.~(\ref{e3.1}) or (\ref{e3.3})
into formulas (\ref{e3.2}) or (\ref{e3.4}).
Remarkably, the resulting $S$ is obtained by only considering finite
connected components [which are (almost) surely trees in these
networks], see Fig.~\ref{f5}.
Knowing the size of the giant connected component and the total number of finite components, one can find the number of loops in the giant component. For the calculation of this number, see \textcite{Lee:lgk04c}.
Applying generating function
techniques in a similar way one may also describe the organization
of connected components in
the bipartite uncorrelated networks, see, e.g.,
\textcite{Soderberg:s02}.

The analysis of Eq.~(\ref{e3.1}) shows that an uncorrelated network
has a giant connected component when the mean number of
second nearest neighbors of a randomly chosen vertex $z_2 = \langle
q^2 \rangle - \langle q \rangle$ exceeds the mean number of
nearest neighbors: $z_2 > z_1$. This is {\em the Molloy-Reed
criterion}: 
\begin{equation}
\langle q^2 \rangle - 2\langle q \rangle > 0
\label{e3.5}
\end{equation}
\cite{Molloy:mr95}. For the Poisson degree dis\-tri\-bu\-tion, i.e.,
for the classical random graphs, $z_2=
\langle q \rangle^2$, and so the birth point of the giant connected
component is $z_1=1$. In the Gilbert model, this corresponds to the
critical probability $p_c(N\to\infty)\cong 1/N$ that a pair vertices
is connected. These relations explain the importance of the sparse
network regime, where this transition takes place. The Molloy-Reed
criterion shows that the divergence of the second moment of the
degree distribution guarantees the presence of the giant connected
component.

Exactly at the birth point of the giant connected component, the
mean size of a finite component to which a randomly chosen vertex
belongs diverges as follows:
\begin{equation}
\langle s \rangle = \frac{\langle q \rangle^2}{2\langle q
\rangle-\langle q^2 \rangle} + 1
\label{e3.5p}
\end{equation}
\textcite{Newman:nsw01}. This formula is given for the phase without
the giant connected component. In this problem, $\langle s \rangle$
plays the role of susceptibility.
Usually, it is convenient to express
the variation of the giant component near the critical point and
other critical properties
in terms of the deviation of one parameter,
e.g., the mean degree $\langle q \rangle$, from its critical value,
$\langle q \rangle_c$.
Usually, the resulting singularities in terms of $\langle q \rangle-\langle q
\rangle_c$ are the same as in terms of $p-p_c$ in the percolation
problem on complex networks ($p$ is the concentration of undeleted
vertices, see below).
Note that in scale-free networks with fixed exponent $\gamma$ one
may vary the mean degree by changing the low degree part of a degree
distribution.


\subsubsection{Percolation on uncorrelated networks}
\label{sssec:percolation_uncorrelated}


What
happens with a network if a random fraction $1-p$ of its
vertices (or edges) are removed? In this site (or bond) percolation
problem, the giant connected component plays the role of the
percolation cluster which may be destroyed by decreasing $p$. Two
equivalent approaches to this problem are possible. The first way
\cite{Cohen:ceb00} uses the following idea. (i) Find the degree
distribution of the damaged network, which is $\tilde{P}(q) =
\sum_{r=q}^\infty P(r)C_q^r p^q (1-p)^{r-q}$ both for the site and
bond percolation. (ii) Since the damaged network is obviously still
uncorrelated, Eqs.~(\ref{e3.1}) and (\ref{e3.2}) with this
$\tilde{P}(q)$ describe the percolation.

The second way is technically more convenient:
derive direct generalizations of Eqs.~(\ref{e3.1}) and (\ref{e3.2})
with the parameter $p$ and the degree distribution $P(q)$ of the
original, undamaged network
\cite{Callaway:cns00}. Simple arguments, similar to those
illustrated by Fig.~\ref{f5}, immediately lead to 
\begin{eqnarray}
&& x = 1-p+p\sum_{q}\frac{qP(q)}{\langle q \rangle} x^{q-1} ,
\label{e3.6}
\\[5pt]
&& 1-S = 1-p+p\sum_{q} P(q) x^{q} . \label{e3.7}
\end{eqnarray}
Although Eq.~(\ref{e3.6}) is valid for both the site and bond
percolation, relation~(\ref{e3.7}) is valid only for site
percolation. For the bond percolation problem, use Eq.~(\ref{e3.2}).
One can see that the giant connected component is present when 
\begin{equation}
p z_2 > z_1 , \label{e3.8}
\end{equation}
that is, the percolation threshold is at 
\begin{equation}
p_c = \frac{z_1}{z_2} = \frac{\langle q \rangle}{\langle q^2
\rangle-\langle q \rangle} , \label{e3.9}
\end{equation}
So, in particular, $p_c=1/\langle q \rangle$ for classical random
graphs, and $p_c=1/(q-1)$ for random regular graphs.
Relations~(\ref{e3.8}) and (\ref{e3.9}) show that it is practically
impossible to eliminate the giant connected component in an infinite
uncorrelated network if the second moment of its degree distribution
diverges---{\em the network is ultraresilient against random damage
or failures} \cite{Albert:ajb00,Cohen:ceb00}. In scale-free
networks, this takes place if $\gamma\leq 3$.
\textcite{Callaway:cns00} considered a more general problem, where
the probability $p(q)$ that a vertex is removed depends on its
degree. As is natural, the removal of highly connected hubs from a
scale-free network---{\em an intentional damage}---effectively
destroys its giant connected component
\cite{Albert:ajb00,Cohen:ceb01}.

\begin{figure}[t]
\begin{center}
\scalebox{0.31}{\includegraphics[angle=0]{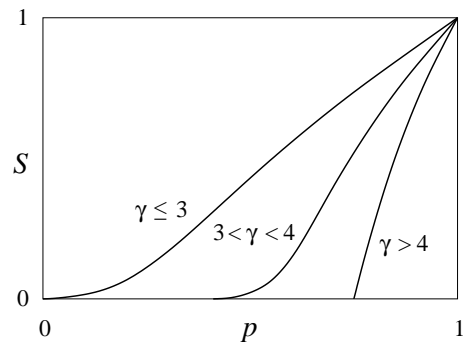}}
\end{center}
\caption{ The effect of the heavy-tailed architecture of a network
on the variation of its giant connected component under random
damage. The relative size of the giant connected component, $S$, is
shown as a function of the concentration $p$ of the retained
vertices in the infinite network. }
\label{f6}
\end{figure}

Near the critical point, the right hand side of Eq.~(\ref{e3.6}) for
the order parameter $1-x$ becomes non-analytic if the higher moments
of the degree distribution diverge. This leads to unusual critical
singularities in these percolation problems and, more generally, to
unusual critical phenomena at the birth point of the giant connected
component in networks with heavy-tailed degree distributions
\cite{Cohen:cbh02,Cohen:chb03a}. For the sake of convenience, let
the infinite uncorrelated network be scale-free. In this case,
the critical behavior of the size $S$ of the giant connected
component is as follows \cite{Cohen:cbh02}:

\begin{itemize}

\item[(i)] if $\gamma>4$, i.e., $\langle q^3 \rangle < \infty$, then $S \propto p-p_c$, which is the standard mean-field result, also valid for classical random graphs;

\item[(ii)] if $3<\gamma<4$, then $S \propto (p-p_c)^{1/(\gamma-3)}$, i.e., the $\beta$ exponent equals $1/(\gamma-3)$;

\item[(iii)] if $\gamma=3$, then $p_c=0$ and $S \propto p\,\exp[-2/(p\langle q \rangle)]$;

\item[(iv)] if $2<\gamma<3$, then $p_c=0$ and $S \propto p^{1/(3-\gamma)}$.

\end{itemize}

\noindent These results are schematically shown in Fig.~\ref{f6}. We
stress that the unusual critical exponents here are only the
consequence of a fat-tailed degree distribution, and the theory is
essentially of mean-field nature. Note that we discuss only
unweighted networks, where edges have unit weights.
For percolation
on weighted networks, see \textcite{Braunstein:bbc03,Braunstein:bbs04,Li:lbb07} and references therein. In weighted networks one can naturally introduce a mean length of the path along edges with the minimum sum of weights, $\overline{\ell}_{\text{opt}}$.
Based on the percolation theory Braunstein {\em et al.} showed that in the Erd\H{os}-R\'enyi graphs with a wide weight distribution, the optimal path length $\overline{\ell}_{\text{opt}} \sim N^{1/3}$.

Numerous variations of percolation on networks may be considered. In particular, one may remove vertices from a network with a degree-dependent probability \cite{Albert:ajb00,Callaway:cns00,Gallos:gca05}.

The probability that a vertex of degree $q$ belongs to the giant
connected component is $1-x^q$ [compare with Eq~(\ref{e3.2})], so
that it is high for highly connected vertices.
Here $x$ is the physical root of Eq.~(\ref{e3.1}) for the order
parameter.
The degree distribution of vertices
in the giant connected component (GCC) is
\begin{equation}
P_{\text{GCC}}(q) = \frac{P(q) (1-x^q)}{1-\sum_{q}P(q) x^q}
. \label{e3.10}
\end{equation}
Therefore at the birth point ($x{\to}1$) of the giant connected
component, the degree distribution of its vertices is proportional
to $qP(q)$. Thus, in networks with slowly decreasing degree
distributions, the giant connected component near its birth point
mostly consists of vertices with high degrees.



\textcite{Cohen:ceb01,Cohen:chb03a} found that at the birth point,
the giant connected component does
not have a small-world geometry (that is, with a diameter growing with the number of vertices $N$ slower than any positive power of $N$) but a fractal one. Its fractal
dimension---a chemical dimension $d_l$ in their notations---equals
$d_l(\gamma>4)=2$ and $d_l(3<\gamma<4)=(\gamma-2)/(\gamma-3)$. That
is, the mean intervertex distance in the giant connected component
(of size $n$) at the point of its disappearance is quite large,
$\overline{\ell} \sim n^{d_l}$. To be clear, suppose that we are
destroying a small world by deleting its vertices. Then
precisely at the moment of destruction, a tiny
remnant of the network
has a much
greater diameter than the original compact network.
It is important that this remnant is an equilibrium tree with a
degree distribution characterized by exponent $\gamma-1$. Indeed,
recall that in Sec.~\ref{ssec:equilibrium_trees} we indicated that
equilibrium connected trees have a fractal structure. So
substituting $\gamma-1$ for $\gamma$ in the expression for the
fractal dimension of equilibrium connected trees
[\textcite{Burda:bck01}, see Sec.~\ref{ssec:equilibrium_trees}], we
readily explain the form of $d_l(\gamma)$.









\subsubsection{Statistics of finite connected components}
\label{sssec:statistics_finite}


The
sizes of largest connected components $s^{(i)}$
depend on the number of vertices in a network, $N$. Here the index
$i=1$ is for the largest component, $i=2$ is for the second largest
component, and so on. In the classical random graphs, $s^{(i)}(N)$
with a fixed $i$ and $N \to \infty$ are as follows (for more detail
see the graph theory papers of \textcite{Borgs:bck01} and of
\textcite{Bollobas:br03}):

\begin{itemize}

\item[(i)]   for
$p < p_c(1-CN^{-1/3})$, \ $s^{(i\geq 1)}(N) \sim \ln N$;

\item[(ii)]   within the so called scaling window
$|p-p_c| < CN^{-1/3}$, \ $s^{(i\geq 1)}(N) {\sim} N^{2/3}$;

\item[(iii)] for
$p > p_c(1+CN^{-1/3})$, \
$s^{(1)}(N) {\sim} N$, $s^{(i>1)}(N) \sim \ln N$
\cite{Bollobas:b84}.

\end{itemize}
Here $C$ denotes corresponding constants and $p=\langle q \rangle/N$.

In Sec.~\ref{ssec:finite_size_scaling} we will present a general
phenomenological approach to finite-size scaling in complex
networks. The application of this approach to scale-free networks
with degree distribution exponent $\gamma$
allows one to describe the sizes of the largest connected
components:

\begin{itemize}

\item[(i)] if $\gamma>4$, the same formulas hold, as for the classical random graphs;

\item[(ii)] if $3<\gamma<4$, then $s^{(i\geq 1)}(N) \sim N^{(\gamma-2)/(\gamma-1)}$ within  the scaling window
$|p-p_c|< CN^{-(\gamma-3)/(\gamma-1)}$ \cite{Kalisky:kc05}, and the
classical results, represented above, hold outside of the scaling
window.

\end{itemize}

\noindent Similarly, one can write 
\begin{equation}
p_c(N=\infty) - p_c(N) \sim N^{-(\gamma-3)/(\gamma-1)}
\label{e3.11}
\end{equation}
for the deviation of the percolation threshold in the range
$3<\gamma<4$.
(Note that rigorously speaking, $p_c$ is well defined only in the $N \to \infty$ limit.)
We will discuss the size effect in networks with
$2<\gamma<3$ in Sec.~\ref{sssec:size}.




Let us compare these results with the corresponding formulas for the
standard percolation on lattices. If the dimension of a lattice is
below the upper critical dimension for the percolation problem,
$d<d_u=6$, then
\begin{equation}
s^{(i\geq 1)}(N) \sim N^{d_f/d}
\label{e3.011}
\end{equation}
within the scaling window $|p-p_c| < \text{const}\,N^{-1/(\nu d)}$.
Here $d_f=(d+2-\eta)/2=\beta/\nu+2-\eta$ is the fractal dimension of
the percolation cluster in the critical point measured in the
$d$-dimensional space by using a
box counting procedure,
$\nu$ is the correlation length exponent, and $\eta$ is the Fisher
exponent.
(The boxes in this box counting procedure are based on an original, undamaged network.)
Above the upper critical dimension, which is the case for
the small worlds, one must replace, as is usual, $d$ in these
formulas (and in scaling relations) by $d_u$ and substitute the
mean-field values of the critical exponents $\nu$, $\eta$, and
$\beta$. Namely, use $\nu=1/2$ and $\eta=0$. For networks, the
mean-field exponent $\beta=\beta(\gamma)$, and so, similarly to
\textcite{Hong:hhp07}, we may formally introduce the upper critical
dimension $d_u(\gamma)=2\beta/\nu+2-\eta=4\beta(\gamma)+2$ and the
fractal dimension $d_f(\gamma)= \beta/\nu+2-\eta= 2\beta(\gamma)+2$.


With the known order parameter exponent $\beta(\gamma)$ from
Sec.~\ref{sssec:percolation_uncorrelated} this heuristic approach
gives
the fractal dimension
\begin{equation}
d_f(\gamma\geq 4) = 4 \ \
\text{and} \ \
d_f(3<\gamma<4) = 2\frac{\gamma-2}{\gamma-3}
\label{e3.011a}
\end{equation}
\cite{Cohen:chb03a}. Note that this fractal dimension $d_f$ does not
coincide with the ``chemical dimension'' $d_l$ discussed above but
rather $d_f=2d_l$.
Similarly,
\begin{equation}
d_u(\gamma\geq 4) = 6  \ \
\text{and} \ \
d_u(3<\gamma<4) = 2\frac{\gamma-1}{\gamma-3}
\label{e3.011b}
\end{equation}
\cite{Cohen:chb03a,Hong:hhp07,Wu:wlb07}.
With these $d_u(\gamma)$ and $d_f(\gamma)$, we reproduce the above
formulas for finite-size networks.


The distribution of sizes of connected components in the
configuration model was derived by using the generating function
technique \cite{Newman:nsw01,Newman:n07}.
Let ${\cal P}(s)$ be the size distribution of a finite component to which a
randomly chosen vertex belongs and ${\cal Q}(s)$ be the distribution of the total
number of vertices reachable following a randomly chosen edge. $h(z) \equiv \sum_s {\cal P}(s) z^s$ and $h_1(z) \equiv \sum_s {\cal Q}(s) z^s$ are the corresponding generating functions. Then
\begin{eqnarray}
&&
h(z) = z \phi(h_1(z)),
\label{e3.011c}
\\[5pt]
&&
h_1(z) = z \phi_1(h_1(z))
\label{e3.011d}
\end{eqnarray}
\cite{Newman:nsw01}.
To get $h(z)$ and its inverse transformation ${\cal P}(s)$, one should substitute the solution of Eq.~(\ref{e3.011d}) into relation~(\ref{e3.011c}).

Equations~(\ref{e3.011c}), (\ref{e3.011d}) have an interesting consequence for scale-free networks without a giant connected component. If the degree distribution exponent is $\gamma>3$, then in this situation the size distribution ${\cal P}(s)$ is also asymptotically power-law, ${\cal P}(s) \sim s^{-(\gamma-1)}$ \cite{Newman:n07}.
To arrive at this result, one must recall that if a function
is power-law, $P(k) \sim k^{-\gamma}$, then its generating function near $z=1$ is
$\phi(z) = a(z)+C(1-z)^{\gamma-1}$, where $a(z)$ is some function, analytic at $z=1$ and $C$ is a constant. Substituting this $\phi(z)$ into Eqs.~(\ref{e3.011c}) and (\ref{e3.011d}) immediately results in the nonanalytic contribution $\sim (1-z)^{\gamma-2}$ to $h(z)$.
[One must also take into account that 
$h(1)=h_1(1)=1$ when a giant component is absent.]
This corresponds to the power-law asymptotics of ${\cal P}(s)$.
Remarkably, there is a qualitative difference in the component size distribution between undamaged networks and networks with randomly removed vertices or edges.
In percolation problems for arbitrary uncorrelated networks, the power law for the distribution ${\cal P}(s)$ fails everywhere except a percolation threshold (see below).

In uncorrelated scale-free networks without a giant connected component, the largest connected component contains $\sim N^{1/(\gamma-1)}$ vertices \cite{Durrett:dbook06,Janson:j07}, where we assume $\gamma>3$. As is natural, this size coincides with the cutoff $k_{\text{cut}}(N)$ in these networks.   

Near the critical point in uncorrelated scale-free networks with a giant connected component, the size distribution of
finite connected components to which a randomly chosen vertex
belongs is
\begin{equation}
{\cal P}(s)\sim s^{-\tau+1}e^{-s/s^*(p)} , \label{e3.012}
\end{equation}
where $s^*(p_c) {\to} \infty$: $s^*(p) \sim (p-p_c)^{-1/\sigma}$
near $p_c$ \cite{Newman:nsw01}. The distribution of the sizes of finite connected
components is  ${\cal P}_s(s) \sim {\cal P}(s)/s
$. 
In uncorrelated networks with rapidly decreasing degree distributions, relation (\ref{e3.012}) is also valid in the absence of a giant connected component. 
Note that this situation, in particular, includes randomly damaged scale-free networks---percolation. 
The distribution ${\cal P}(s)$ near critical point in undamaged scale-free networks without a giant component, in simple terms, looks as follows: ${\cal P}(s) \sim s^{-\tau+1}$ at sufficiently small $s$, and ${\cal P}(s) \sim s^{-\gamma+1}$ at sufficiently large $s$ ($\gamma>3$, in this region the inequality $\gamma>\tau$ is valid).  
Exponents $\tau$, $\sigma$, and $\beta$ satisfy
the
scaling relations
$\tau-1=\sigma\beta+1 = \sigma d_u/2 = d_u/d_f$.
We stress that the mean size of a finite connected component,
i.e., the first moment of the distribution ${\cal P}_s(s)$, is
finite at the critical point.
A divergent quantity (and an
analogue of susceptibility) is the mean size of a finite connected
component to which a randomly chosen vertex belongs, 
\begin{equation}
\langle s \rangle = \sum_s s{\cal P}(s) \sim
|p-p_c|^{-\tilde{\gamma}} , \label{e3.013}
\end{equation}
where $\tilde{\gamma}$ is the ``susceptibility'' critical exponent.
This exponent does not depend on the form of the degree
distribution. Indeed, the well-known scaling relation
$\tilde{\gamma}/\nu = 2 - \eta$ with $\nu=1/2$ and $\eta=0$
substituted leads to $\tilde{\gamma}=1$ within the entire region
$\gamma>3$.


The resulting exponents for finite connected components in the
scale-free configuration model
are as follows:

\begin{itemize}

\item[(i)] for $\gamma>4$, \  the exponents are $\tau=5/2$, $\sigma=1/2$,
$\tilde{\gamma}=1$,
which is also valid for classical random graphs;

\item[(ii)] for $3<\gamma<4$, $\tau=2+1/(\gamma-2)$, $\sigma=(\gamma-3)/(\gamma-2)$,
$\tilde{\gamma}=1$ \cite{Cohen:chb03a}.

\end{itemize}

\noindent The situation in the range $2<\gamma<3$ is not so clear.
The difficulty is that in this interesting region, the giant
connected component disappears at
$p=0$, i.e., only with disappearance of the network itself.
Consequently, one cannot separate ``critical'' and non-critical
contributions, and so scaling relations fail.
In this range,

\begin{itemize}

\item[(iii)]
i.e., for $2<\gamma<3$, $\langle s \rangle \propto p$, $\tau=3$,
$\sigma=3-\gamma$.

\end{itemize}

\noindent Note that the last two values imply a specific cutoff of
the degree distribution, namely $q_{\text{cut}} \sim N^{1/2}$.

In principle, the statistics of connected components in the bond
percolation problem for a network
may be obtained by analysing the solution of the $p$-state Potts
model
(Sec.~\ref{sec:Potts})
with $p{=}1$ placed on this net. \textcite{Lee:lgk04c}
realized
this approach
for the static model.

{\em The correlation volume of a vertex} is defined as 
\begin{equation}
V_i \equiv \sum_{\ell=0}z_\ell(i) b^\ell , \label{e3.014}
\end{equation}
where $z_\ell$ is the number of the $\ell$-th nearest neighbors of
vertex $i$, and $b$ is a parameter characterizing the decay of
correlations.
The parameter $b$ may be calculated for specific cooperative models
and depends on their control parameters,
Sec.~\ref{sssec:correlation_volume}. In particular, if $b{=}1$, the
correlation volume is reduced to the size of a connected component.
Let us estimate the mean correlation volume in uncorrelated network
with the mean branching coefficient $B=z_2/z_1$: $\overline{V} \sim
\sum_{\ell}(bB)^\ell$ (we assume that the network has the giant
connected component). So $\overline{V}(N\to\infty)$ diverges at and
above the critical value of the parameter, $b_c=1/B$. At the
critical point, $\overline{V}(b_c)=\sum_{\ell}1 \sim \ln N$. Since
$B^{\overline{\ell}(N)} \sim N$, we obtain $\overline{V} \sim
N^{\ln(bB)/\ln B}$ for $b>b_c$. Thus, as $b$ increases from $b_c$ to $1$,
the exponent of the correlation volume grows from $0$ to $1$.

The correlation volume takes into account remote neighbors with
exponentially decreasing (if $b<1$) weights. A somewhat related
quantity---the mean number vertices at a distance less than
$a\overline{\ell}(N)$ from a vertex,
where $a \geq 1$,---was analysed by
\textcite{Lopez:lpc07} in their study of ``limited path
percolation''.
This number is
of the order of $N^\delta$, where exponent $\delta=\delta(a,B)\leq 1$.


\subsubsection{Finite size effects}
\label{sssec:size}

Practically all real-world networks are small, which makes the
factor of finite size of paramount importance. For example, empirically studied
metabolic networks contain about $10^3$ vertices.
Even the largest artificial net---the World Wide Web, whose size
will soon approach $10^{11}$
Web pages, show qualitatively strong finite size effects
\cite{May:ml01,Dorogovtsev:dm02,Boguna:bpv04}. To understand the
strong effect of finite size in real scale-free networks one must
recall that exponent $\gamma \leq 3$ in most of them, that is the
second moment of a degree distribution diverges in the infinite
network limit.

Note that the tree ansatz may be used even in this region ($\gamma
\leq 3$), where the uncorrelated networks are loopy. The same is
true for at least the great majority of interacting systems on these
networks. The reason for this surprising applicability is not clear
up to now.

Let us demonstrate a poor-man's approach to percolation on a finite
size (uncorrelated) network with $\gamma \leq 3$, where
$p_c(N\to\infty)\to 0$. To be specific, let us, for example, find
the size dependence of the percolation threshold, $p_c(N)$. The idea
of this estimate is quite simple. Use Eq.~(\ref{e3.9}), which was
derived for an infinite network, but with the finite network's
degree distribution substituted. Then, if the cutoff of the degree
distribution is $q_{\text{cut}} \sim N^{1/2}$, we readily arrive at
the following results: 
\begin{equation}
 p_c(N,2{<}\gamma{<}3) \sim N^{-(3-\gamma)/2} \!, \  p_c(N,\gamma{=}3) {\sim} 1/\ln N
. \label{e3.110}
\end{equation}
These relations suggest the emergence of the noticeable percolation
thresholds even in surprisingly large networks. In other words, the
ultraresilience against random failures is effectively broken
in finite networks.

Calculations of other quantities for percolation (and for a wide
circle of cooperative models) on finite nets are analogous.
Physicists, unlike mathematicians, routinely apply estimates of this
sort to various problems defined on networks. Usually, these
intuitive estimates work but evidently demand
thorough verification. Unfortunately, a strict statistical mechanics
theory of finite size effects for networks is technically hard and
was developed only for very special models (see
Sec.~\ref{ssec:condensation_equilibrium}). For a phenomenological
approach to this problem, see Sec.~\ref{ssec:finite_size_scaling}.




\subsubsection{$k$-core architecture of networks}
\label{sssec:k-core_architecture}

\begin{figure}[t]
\begin{center}
\scalebox{0.31}{\includegraphics[angle=270]{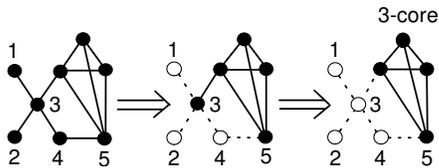}}
\end{center}
\caption{
Construction of the $3$-core of a given graph.
First we remove vertices 1, 2 and 4 together with their links because they have degrees smaller than 3. In the
obtained graph, vertex 3 has degree 1. Removing it, we
get the 3-core of the graph.
}
\label{f6p}
\end{figure}

The $k$-core of a network is its largest subgraph whose vertices
have degree at least $k$ \cite{Chalupa:clr79,Bollobas:b84}. In other
words, each of vertices in the $k$-core has at least $k$ nearest
neighbors within this subgraph. The notion of the $k$-core naturally
generalizes the giant connected component and offers a more
comprehensive view of the network organization. The $k$-core of a
graph may be obtained by
the ``pruning algorithm'' which looks as follows (see Fig.~\ref{f6p}). Remove from the
graph all vertices of degrees less than $k$. Some of remaining
vertices may now have less than $k$ edges.
Prune these vertices, and so on until no further pruning is
possible. The result, if it exists, is the $k$-core. Thus, a network
is hierarchically organized as a set of successfully enclosed
$k$-cores, similarly to a Russian nesting doll---``matrioshka''.
\textcite{Alvarez-Hamelin:adb05a} used this $k$-core architecture to
produce a set of beautiful visualizations of diverse networks.

{\em The $k$-core (bootstrap) percolation} implies the breakdown of
the giant $k$-core at a threshold concentration of vertices or edges
removed at random from an infinite network. \textcite{Pittel:p96}
found the way to analytically describe the $k$-core architecture of
classical random graphs. More recently, \textcite{Fernholz:fr04}
mathematically proved that the $k$-core organization of the
configuration model is
asymptotically
exactly described in the framework of a simple tree ansatz.

Let us discuss the $k$-core percolation in the configuration model
with degree distribution $P(q)$ by using intuitive arguments based
on the tree ansatz
\cite{Dorogovtsev:dgm06a,Goltsev:gdm06,Dorogovtsev:dgm06b}.
The validity of the tree ansatz here is non-trivial since in this theory it is applied to a giant $k$-core which has loops.
Note
that in tree-like networks, $(k{\geq}3)$-cores (if they exist) are
giant---finite $(k{\geq}3)$-cores are impossible.
In contrast to the giant connected component problem, the tree ansatz in application to higher $k$-cores fails far from the $k$-core birth points.
We assume that a
vertex in the network is present with probability $p=1-Q$. In this
locally tree-like network, the giant $k$-core coincides with the
infinite $(k{-}1)$-ary subtree. By definition, the $m$-ary tree is a
tree where all vertices have branching at least $m$.

\begin{figure}[t]
\begin{center}
\scalebox{0.255}{\includegraphics[angle=0]{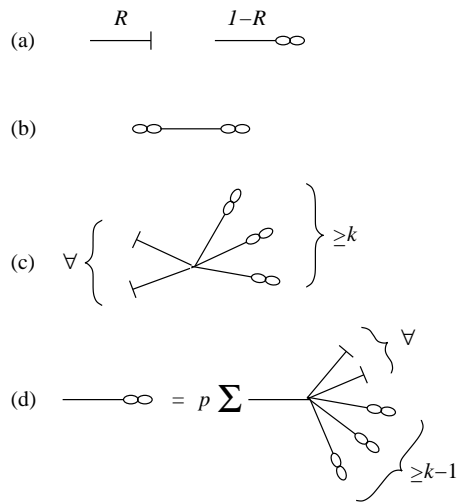}}
\end{center}
\caption{ Diagrammatic representation of
Eqs.~(\protect\ref{e3.111})--(\protect\ref{e3.113}). (a) Graphic
notations for the order parameter $R$ and for $1-R$.
(b) The probability that both ends of an edge are in the $k$-core,
Eq.~(\protect\ref{e3.111}). (c) Configurations contributing to
$M_k$, which is the probability that a vertex is in the $k$-core,
Eq.~(\protect\ref{e3.112}). The symbol $\forall$ here indicates that
there may be any number of the nearest neighbors which are not trees
of infinite $(k-1)$-ary subtrees. (d) A graphic representation of
Eq.~(\protect\ref{e3.113}) for the order parameter. Adapted from
\protect\textcite{Goltsev:gdm06}. }
\label{f8}
\end{figure}

Let the order parameter in the problem, $R$, be the probability that
a given end of an edge of a network is not the root of an infinite
$(k{-}1)$-ary subtree.
(Of course, $R$ depends on $k$.) An edge is in the $k$-core if both
ends of this edge are roots of infinite $(k{-}1)$-ary subtrees,
which happens with the probability $(1-R)^2$.
In other words, 
\begin{equation}
(1-R)^2 = \frac{\text{number of edges in the $k$-core}}{\text{number
of edges in the network}} ,
\label{e3.111}
\end{equation}
which expresses the order parameter $R$ in terms of observables.
Figure~\ref{f8} graphically explains this and the following two
relations.
A vertex is in the $k$-core if at least $k$ of its neighbors are
roots of infinite $(k{-}1)$-ary trees. So, the probability $M_k$
that a random vertex belongs to the $k$-core (the relative size of
the $k$-core) is given by the equation: 
\begin{equation}
M_k = p\sum_{n\geq k}\sum_{q\geq n}P(q)C_n^q R^{q-n}(1-R)^{n} ,
\label{e3.112}
\end{equation}
where $C_n^q=q!/[(q-n)!n!]$. To obtain the relative size of the
$k$-core, one must substitute the physical solution of the equation
for the order parameter into Eq.~(\ref{e3.112}). We write the
equation for the order parameter, noticing that a given end of an
edge is a root of an infinite $(k{-}1)$-ary subtree if it has at
least $k-1$ children which are roots of infinite $(k{-}1)$-ary
subtrees. Therefore, 
\begin{equation}
1{-}R = p\!\!\sum_{n=k-1}^{\infty}
\,\sum_{i=n}^{\infty }\frac{(i{+}1)P(i{+}1)}{z_{1}}\,C_{n}^{i}R^{i-n}(1{-}R)^{n}
. \label{e3.113}
\end{equation}

\begin{figure}[t]
\begin{center}
\scalebox{0.28}{\includegraphics[angle=270]{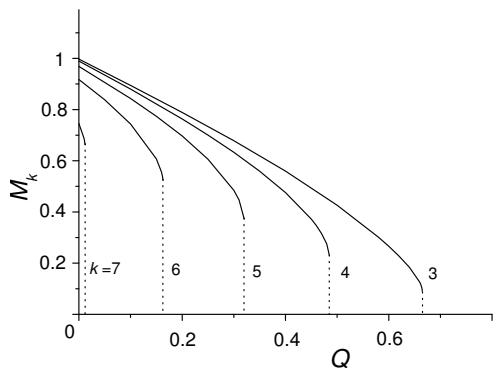}}
\end{center}
\caption{ Relative sizes of the $k$-cores, $M_k$, in classical
random graphs with the mean degree $z_1=10$ versus the concentration
$Q=1-p$ of randomly removed vertices.Adapted from
\protect\textcite{Dorogovtsev:dgm06a}. } \label{f9}
\end{figure}

This equation strongly differs from that for the order parameter in
the ordinary percolation, compare with Eq.~(\ref{e3.6}). The
solution of Eq.~(\ref{e3.114}) at $k{\geq} 3$ indicates a quite
unusual critical phenomenon. The order parameter (and also the size
of the $k$-core) has a jump at the critical point like a first order
phase transition. On the other hand, it has a square root critical
singularity: 
\begin{equation}
R_c-R \propto [p-p_c(k)]^{1/2} \propto M_k-M_{kc} , \label{e3.114}
\end{equation}
see Fig.~\ref{f9}. This intriguing critical phenomenon is often
called {\em a hybrid phase transition}
\cite{Schwartz:slc06,Parisi:pr06}. Relations (\ref{e3.114}) are
valid if the second moment of the degree distribution is finite.
Otherwise, the picture is very similar to what we observed for
ordinary percolation. In this range, the $k$-cores, even of high
order, practically cannot be destroyed by the random removal of
vertices from an infinite network.

The $2$-core of a graph can be obtained from the giant connected component of this graph by pruning dangling branches. 
At $k=2$, 
Eq.~(\ref{e3.113}) for the order parameter is identical to
Eq.~(\ref{e3.6}) for the ordinary percolation. Therefore the birth point of the $2$-core coincides with that of the giant connected component, and the phase
transition is continuous. According to Eq. (\ref{e3.112}) 
the size $M_2$ of the $2$-core is proportional to $(1-R)^2$ near the critical point, and so it is proportional to the square of the size of the giant connected component. This gives $M_2 \propto (p-p_c)^2$ if the degree distribution decays rapidly. 

In stark contrast to ordinary percolation,
the birth of $(k{>}2)$-cores is not related to the divergence of
corresponding finite components
which are absent in tree-like networks. Then, is there any
divergence associated with this hybrid transition?
The answer is yes. To unravel the nature of this divergence, let us
introduce a new notion. {\em The $k$-core's corona} is a subset of
vertices in the $k$-core (with their edges) which have exactly $k$
nearest neighbors in the $k$-core, i.e., the minimum possible number
of connections. One may see that the corona itself is a set of
disconnected clusters. Let $N_{\text{crn}}$ be the mean total size
of corona clusters attached to a vertex in the $k$-core. It turns
out that it is $N_{\text{crn}}(p)$ which diverges at the birth point
of the $k$-core, 
\begin{equation}
N_{\text{crn}}(p) \propto [p-p_c(k)]^{-1/2}
\label{e3.115}
\end{equation}
\cite{Schwartz:slc06,Goltsev:gdm06}. Moreover, the mean intervertex
distance in the corona clusters diverges by the same law as
$N_{\text{crn}}(p)$ \cite{Goltsev:gdm06}. It looks like the corona
clusters ``merge together'' exactly at the $k$-core percolation
threshold and simultaneously disappear together with the $k$-core,
which, of course, does not exist at $p>p_c(k)$.

\begin{figure}[t]
\begin{center}
\scalebox{0.22}{\includegraphics[angle=270]{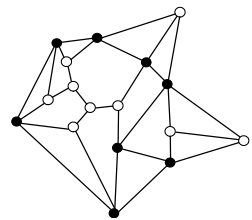}}
\end{center}
\caption{3-core of a graph and its corona (removed vertices and
links are not shown).
The corona consists of
a set of clusters with vertices (open circles) having
exactly 3 nearest
neighbors in this 3-core.
}
\label{f-new3}%
\end{figure}

Similarly to the mean size of a cluster to which a vertex belongs in
ordinary percolation, $N_{\text{crn}}$ plays the role of
susceptibility in this problem, see \textcite{Schwartz:slc06} for more detail. The exponent of the singularity in
Eq.~(\ref{e3.115}), $1/2$, dramatically differs from the standard
mean-field value of exponent $\tilde{\gamma}=1$ (see
Sec.~\ref{sssec:statistics_finite}). At this point, it is
appropriate to mention a useful association.
Recall the temperature dependence of the order parameter $m(T)$ in a
first order phase transition. In normal thermodynamics, metastable
states cannot be realized. Nonetheless, consider the metastable
branch of $m(T)$. One may easily find that near the end ($T_0$) of
this branch, $m(T) = m(T_0) + \text{const}[T_0-T]^{1/2}$, and the
susceptibility $\chi(T) \propto [T_0-T]^{-1/2}$. Compare these
singularities with those of Eqs.~(\ref{e3.114}) and (\ref{e3.115}).
The only essential difference is that, in contrast to the $k$-core
percolation, in the ordinary thermodynamics this region is not
approachable. Parallels of this kind were discussed already by
\textcite{Aizenman:al88}.

By using Eqs.~(\ref{e3.112}) and (\ref{e3.113}), we can easily find
the $k$-core sizes, $M_k$ in the important range $2<\gamma<3$: 
\begin{equation}
M_k = p^{1/(3-\gamma)} (q_0/k)^{(\gamma-1)/(3-\gamma)} ,
\label{e3.116}
\end{equation}
where $q_0$ is the minimal degree in the scale-free degree
distribution \cite{Dorogovtsev:dgm06a}. The exponent of this power
law agrees with the observed one in a real-world network---the
Internet at the Autonomous System level and the map of routers
\cite{Kirkpatrick:k05,Carmi:chk06,Alvarez-Hamelin:adb05b}. In the
infinite scale-free networks of this kind, there is an infinite
sequence of the $k$-cores (\ref{e3.116}). All these cores have a
practically identical architecture---their degree distributions
asymptotically coincide with the degree distribution of the network
in the range of high degrees.

\begin{figure}[t]
\begin{center}
\scalebox{0.35}{\includegraphics[angle=0]{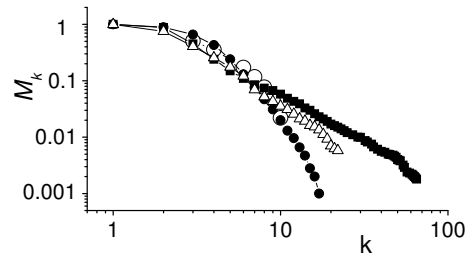}}
\end{center}
\caption{ Relative size of the $k$-cores vs. $k$ in several
networks. $\bigcirc$,
$M_k$ calculated neglecting correlations,
by using the degree distribution of Internet router network,
$N\approx 190\,000$, adapted from
\protect\textcite{Dorogovtsev:dgm06a}. $\triangle$, measurements for
the Autonomous System network (CAIDA map), $N=8542$, adapted from
\protect\textcite{Alvarez-Hamelin:adb05b}. $\bullet$, results for a
maximally random scale-free ($\gamma{=}2.5$) network of $10^6$
vertices, and $\blacksquare$, for a similar network but with a given
strong clustering, $C{=}0.71$, adapted from
\protect\textcite{Serrano:sb06a}.
} \label{f10}
\end{figure}

The finiteness of networks restricts the $k$-core sequence by some
maximum number $k_h$ for the highest $k$-core.
\textcite{Goltsev:gdm06} and
\textcite{Dorogovtsev:dgm06a,Dorogovtsev:dgm06b} estimated $k_h$
substituting empirical degree distributions into the equations for
uncorrelated networks. Unfortunately, the resulting $k_h$ turned out
to be several ($3$) times smaller than the observed values
\cite{Carmi:chk07,Alvarez-Hamelin:adb05b}. Later
\textcite{Serrano:sb06a,Serrano:sb06c} arrived at much more
realistic $k_h$, taking into account high clustering (see
Fig.~\ref{f10}). (They simulated a maximally random network with a
given degree distribution and a given clustering.) There is also
another way to diminish $k_h$: random damaging first destroys the
highest $k$-core, then the second highest, and so on.





\subsection{Percolation on degree-degree correlated networks}
\label{ssec:percolation_on_correlated}

Let in a random network only pair correlations between nearest
neighbor degrees be present. Then this network has a locally
tree-like structure, and so one can easily analyse the organization
of connected components
\cite{Newman:n02b,Vazquez:vm03,Boguna:bpv03}. The network is
completely described by the joint degree-degree distribution
$P(q,q')$, see Sec.~\ref{ssec:correlated} (and, of course, by $N$).
It is convenient to use a conditional probability $P(q'|q)$ that if
an end vertex of an edge has degree $q$, then the second end has
degree $q'$. In uncorrelated networks, $P(q'|q) = q'P(q')/\langle q
\rangle$ is independent of $q$. Obviously, $P(q'|q)=\langle q
\rangle P(q,q')/[qP(q)]$. The important quantity in this problem is
the probability $x_q$ that if an edge is attached to a vertex of
degree $q$, then, following this edge to its second end, we will not
appear in the giant connected component.
For the sake of brevity, let us discuss only the site percolation
problem, where $p$ is the probability that a vertex is retained. For
this problem, equations for $x_q$ and an expression for the relative
size of the giant connected component take the following form: 
\begin{eqnarray}
&& x_q = 1-p+p\sum_{q'}P(q'|q) (x_{q'})^{q'-1} , \label{e3.120}
\\[5pt]
&& 1-S = 1-p+p\sum_{q} P(q) (x_q)^{q}
\label{e3.121}
\end{eqnarray}
\cite{Vazquez:vm03}, which naturally generalizes
Eqs.~(\ref{e3.6}) and (\ref{e3.7}). Solving the system of equations~(\ref{e3.120})
gives the full set $\{x_q\}$. Substituting
$\{x_q\}$ into Eq.~(\ref{e3.121}) provides $S$.
\textcite{Newman:n02b} originally derived these equations in a more
formal way, using generating functions, and numerically solved them
for various networks.
The resulting curve $S(p)$
was found to significantly depend on the type of correlations---whether the degree-degree correlations were assortative or disassortative. 
Compared to an uncorrelated network with the same degree distribution, 
the assortative correlations increase the resilience of a network against random damage, while the disassortative correlations diminish this resilience.  
See \textcite{Noh:n07} for a similar observation in another network model with correlations.

Equation~(\ref{e3.120}) shows that the birth of the giant connected
component is a continuous phase transition. The percolation
threshold is found by linearizing Eq.~(\ref{e3.120}) for small
$y_q=1-x_q$, which results in the condition: $\sum_{q'}
C_{qq'} y_{q'} = 0$, where the matrix elements $C_{qq'} =
-\delta_{qq'}+p(q'-1)P(q'|q)$. With this matrix, the generalization
of the Molloy Reed criterion to the correlated networks is the
following condition: {\em if the largest eigenvalue of the matrix
$C_{qq'}$ is positive, then the correlated network has a giant
connected component.}  The percolation threshold may be obtained by
equating the largest eigenvalue of this matrix to zero.
In uncorrelated networks this reduces to criterion~(\ref{e3.9}).

Interestingly, the condition of ultra-resilience against random
damage does not depend on correlations. As in uncorrelated networks,
if the second moment $\langle q^2 \rangle$ diverges in an infinite
network, the giant connected component cannot be eliminated by
random removal of vertices \cite{Vazquez:vm03,Boguna:bpv03}. Very
simple calculations show that the mean number $z_2$ of the second
nearest neighbors of a vertex in a degree-degree correlated network
diverges simultaneously with $\langle q^2 \rangle$. It is this
divergence of $z_2$ that guarantees the ultra-resilience.

Percolation and optimal shortest path problems were also
studied for weighted networks with correlated weights \cite{Wu:wlb07}.




\subsection{The role of clustering}
\label{ssec:role_of_clustering}

The statistics of connected components in highly clus\-tered
networks, with numerous triangles (i.e., the clustering coefficient
$C$ does not approach zero
as $N {\to} \infty$), is a difficult and poorly
studied problem.
An important step to the resolution of this problem has been made by
\textcite{Serrano:sb06a,Serrano:sb06b,Serrano:sb06c}. These authors
studied constructions of networks with given degree distributions
and given mean clusterings of vertices of degree $q$, $C(q)$.
It turns out that only if $C(q)<1/(q-1)$, it is possible to build an
uncorrelated network with a given pair of characteristics: $P(q)$
and $C(q)$. Since clustering of this kind does not induce
degree--degree correlations, the regime $C(q)<1/(q-1)$ was
conventionally called ``weak clustering''.
(When $C(q)<1/(q-1)$, then the number of triangles based on an edge in the network is one or zero.)
On the other hand, if
$C(q)$ is higher than $1/(q-1)$ at least at some
degrees---``strong clustering'',---then the constructed networks
necessarily have at least correlations between the degrees of the
nearest neighbors.

\begin{figure}[t]
\begin{center}
\scalebox{0.31}{\includegraphics[angle=0]{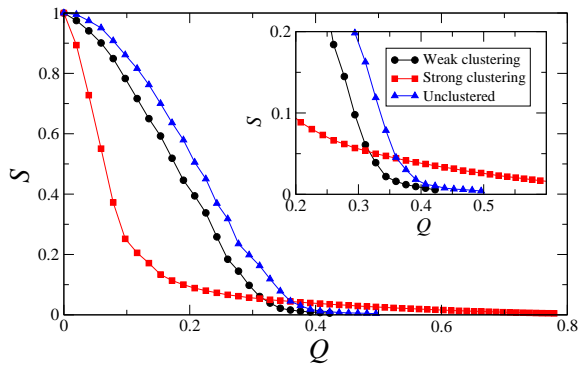}}
\end{center}
\caption{ Bond percolation on unclustered, and ``strongly'' and
``weakly'' clustered scale-free networks. Exponent $\gamma=3.5$. The
relative size $S$ of the giant connected component is shown as a
function of the concentration $Q=1-p$ of removed edges. From
\protect\textcite{Serrano:sb06c}. } \label{f11}
\end{figure}

\textcite{Serrano:sb06a,Serrano:sb06b,Serrano:sb06c} made a helpful
simplifying assumption that the triangles in a network cannot have
joint edges and neglected long loops. This assumption allowed them
to effectively use a variation of the ``tree ansatz''. In particular
they studied the bond percolation problem for these networks. The
conclusions of this work are as follows:

\begin{itemize}

\item[(i)] If the second moment of the degree distribution is finite, the ``weak clustering'' makes the
network less resilient to random damage---the percolation threshold (in terms of $Q=1-p$, where $Q$ is the fraction of removed edges) decreases, see Fig.~\ref{f11}. Contrastingly, the ``strong clustering'' moves the percolation threshold in the opposite direction, although small damage (low $Q$) noticeably 
diminishes the giant connected component.

\item[(ii)] If the second moment of the degree distribution diverges, neither ``weak'' nor ``strong'' clustering can
destroy the giant connected component in an infinite network.

\end{itemize}


\textcite{Newman:n03b} proposed a different approach to highly
clustered networks. He used the fact that a one-mode projection of a
bipartite uncorrelated network
has high clustering, while the original bipartite network has a
locally tree-like structure. (In this projection, two vertices of,
say, type $1$, are the nearest neighbors if they have at least one
joint vertex of type $2$.)
This convenient feature allows one to describe properties of the
clustered one-partite network with a tunable clustering and a
tunable degree distribution by applying the tree ansatz to the
bipartite network. For details---applications to percolation and
epidemic processes,---see \textcite{Newman:n03b}.


\subsection{Giant component in directed networks}
\label{ssec:directed_networks}

The structure of the giant connected component in uncorrelated
directed networks was studied by \textcite{Dorogovtsev:dms01a}. By
definition, edges of directed networks are directed, so that the
configuration model is described by the joint in-, out-degree
distribution $P(q_i,q_o)$. Directed networks have a far more complex
organization and topology of the giant connected components than
undirected ones. This organization may include specifically
interconnected giant subcomponents with different birth points.
Applying the tree ansatz, these authors found the birth points of
various giant components and obtained their sizes for an arbitrary
$P(q_i,q_o)$,
see also \textcite{Schwartz:scb02}. For more detailed description of the giant components in directed networks see \textcite{Serrano:sd07}.

\textcite{Boguna:bs05} generalized this theory to uncorrelated
networks which contain both directed and undirected connections.
These networks are characterized by a distribution $P(q,q_i,q_o)$,
where $q$, $q_i$, and $q_o$ are the numbers of undirected,
in-directed, and out-directed connections of a vertex, respectively.

The exponents of the critical singularities for the transitions of
the birth of various giant connected components in directed networks
were calculated by \textcite{Schwartz:scb02}. Note that although the
in-, out-degrees of different vertices in these networks are
uncorrelated, there may be arbitrary correlations between in- and
out-degrees of the same vertex. The critical exponents, as well as
the critical points, essentially dependent on these in-, out-degree
correlations.



\subsection{Giant component in growing networks}
\label{ssec:giant_in_growing}

The intrinsic large-scale inhomogeneity of nonequilibrium (e.g.,
growing)  networks may produce a surprising critical phenomenon. The
large-scale inhomogeneity here means the difference between
properties of vertices according to their age.
This difference usually makes the ``old'' part of a growing network
more ``dense'' than the ``young'' one.

\textcite{Callaway:chk01} found an unexpected effect in the birth of
the giant connected component already in a very simple model of the
growing network. In their model, the network grows due to two
parallel processes: (i) there is an inflow of new vertices
with the unit rate, and, in addition, (ii) there is an inflow of
edges with rate $b$, which interconnect randomly chosen vertex
pairs. The rate $b$ plays the role of the control parameter. As one
could expect, the resulting degree distribution is very
simple---exponential.
The inspection of this network when it is already infinite shows
that it has a giant connected component for $b>b_c$, where $b_c$ is
some critical value, unimportant for us. Remarkably, the birth of
the giant connected component in this net strongly resembles the
famous {\em Berezinskii-Kosterlitz-Thouless (BKT) phase transition}
in condensed matter \cite{Berezinskii:b70,Kosterlitz:kt73}. Near the
critical point, the relative size of the giant connected component
has the specific BKT singularity: 
\begin{equation}
S \propto \exp (-\text{const}/\sqrt{b-b_c}) . \label{e3.12}
\end{equation}
Note that in an equilibrium network with the same degree
distribution, $S$ would be proportional to the small deviation
$b-b_c$. The singularity (\ref{e3.12}), with all derivatives
vanishing at the critical point, implies an infinite order phase
transition.

Normally, the BKT transition occurs at the lower critical dimension
of an interacting system, where critical fluctuations are strong,
e.g., dimension $2$ for the $XY$ model. Most of known models with this
transition have a continuous symmetry of the order parameter. So
that the discovery of the BKT singularity in infinite dimensional
small worlds, that is in the mean-field regime, was somewhat
surprising. The mean size of a finite connected component to which a
vertex belongs in this network was also found to be nontraditional.
This characteristic---an analogy of susceptibility---has a finite
jump at this transition and not a divergence generic for equilibrium
networks and disordered lattices.

\textcite{Dorogovtsev:dms01b} analytically studied a much wider
class of growing networks with an arbitrary linear preferential
attachment (which may be scale-free or exponential) and arrived at
very similar results. In particular, they found that the constant
and $b_c$ in Eq.~(\ref{e3.12}) depend on the rules of the growth.
Looking for clues and parallels with the canonical BKT transition,
they calculated the size distribution of connected components,
${\cal P}_s(s)$, characterizing correlations. The resulting picture
looks as follows.

\begin{itemize}

\item
The distribution ${\cal P}_s(s)$ slowly (in a power-law fashion)
decays in the whole phase without the giant connected component, and
this distribution rapidly decreases in the phase with the giant
connected component.

\end{itemize}

\noindent This picture is in stark contrast to the equilibrium
networks, where

\begin{itemize}

\item
the distribution ${\cal P}_s(s)$ slowly decays only at the birth
point of the giant connected component (if a network is non-scale-free, see Sec.~\ref{sssec:statistics_finite}).

\end{itemize}

\noindent In this respect, the observed transition in growing
networks strongly resembles the canonical BKT transitions, where the
critical point separates a phase with rapidly decreasing
correlations and ``a critical phase'' with correlations decaying in
a power-law fashion. 
(Note, however the inverted order of the phases with a power-law
decay and with a rapid drop in these transitions.)

This phase transition was later observed in many other growing
networks with exponential and scale-free degree distributions (only
for some of these networks, see \textcite{Lancaster:l02},
\textcite{Coulomb:cb03}, \textcite{Krapivsky:kd04},
\textcite{Bollobas:br05}, and \textcite{Durrett:dbook06}). Moreover,
even ordinary, ``equilibrium'' bond percolation considered on
special networks
has the same critical phenomenon. For example, (i) grow up an
infinite random recursive graph (at each time step, add a new vertex
and attach it to $m$ randomly chosen vertices of the graph), (ii)
consider the bond percolation problem on this infinite network. We
emphasize that the attachment must be only random here. It is easy
to see that the resulting network may be equivalently prepared by
using a stochastic growth process which just leads to the BKT-like
transition. Similar effects were observed on the Ising and Potts
models placed on growing networks (see
Sec.~\ref{sssec:deterministic_bkt}).
A more realistic model of a growing protein interaction network where a giant connected component is born with the BKT-type singularity was described by \textcite{Kim:kkk02}.

Various percolation problems on deterministic (growing) graphs may
be solved exactly. Surprisingly, percolation properties of
deterministic graphs are rather similar to those of their random
analogs. For detailed discussion of these problems, see, e.g.,
\textcite{Dorogovtsev:dgm02a}, \textcite{Dorogovtsev:d03}, and
\textcite{Rozenfeld:rb07}.





\subsection{Percolation on small-world networks}
\label{ssec:percolation_small-world}


Let us consider a small-world network based on a $d$-dimensional
hypercubic lattice ($N\cong L^d$) with random shortcuts added with
probability $\phi$ per lattice edge. Note that in this network, in
the infinite network limit, there are no finite loops including
shortcuts. All finite loops are only of lattice edges. This fact
allows one to apply the usual tree ansatz to this actually loopy
network. In this way \textcite{Newman:njz02} obtained the statistics
of connected components in the bond percolation problem for
two-dimensional small-world networks. Their qualitative conclusions
are also valid for bond and site percolation on one-dimensional
\cite{Newman:nw99a,Newman:nw99b,Moore:mn00a,Moore:mn00b} and
arbitrary-dimensional small-world networks.

In the spirit of classical random graphs, at the percolation
threshold point, $p_c$, there must be one end of a retained shortcut
per connected component in the lattice substrate. In more strict
terms, this condition is $2d\phi p_c = 1/\langle n_0 \rangle(p_c)$,
i.e., the mean density of the ends of shortcuts on the lattice
substrate must be equal to the mean size $\langle n_0 \rangle$ of a
connected component (on a lattice) to which a vertex belongs. In the
standard percolation problem on a lattice, $\langle n_0 \rangle(p)
\propto (p_{c0}-p)^{-\tilde{\gamma}}$, where $p_{c0}$ and
$\tilde{\gamma}$ are the percolation threshold and the ``
susceptibility'' critical exponent in the standard percolation. So,
the percolation threshold is displaced by 
\begin{equation}
p_{c0}-p_c \propto \phi^{1/\tilde{\gamma}}
\label{e3.31}
\end{equation}
if $\phi$ is small \cite{Warren:wss01}. For example, for the bond
percolation on the two-dimensional small-world network, $p_{c0}=1/2$
and $\tilde{\gamma}=43/18=2.39\ldots$. The mean size of a connected
component to which a random vertex belongs is also easily
calculated: 
\begin{equation}
\langle n \rangle = \langle n_0 \rangle/(1-2d\phi p\langle n_0
\rangle) \propto (p_c-p)^{-1} . \label{e3.32}
\end{equation}
So that its critical exponent equals $1$, as in the classical random
graphs. The other percolation exponents also coincide with their
values for classical graphs. In general, this claim is equally valid
for
other cooperative models on small-world networks in a close
environment of a critical point.

\textcite{Ozana:o01} described the entire crossover from the lattice
regime to the small-world one and finite size effects by using
scaling functions with dimensionless combinations of the three
characteristic lengths:
(i) $L$, (ii) the mean Euclidean distance between the neighboring
shortcut ends, $\xi_{\text{sw}} \equiv 1/(2d\phi p)^{1/d}$, and
(iii) the usual correlation length $\xi_\text{l}$ for percolation on
the lattice. For an arbitrary physical quantity, $X(L) = L^x
f(\xi_{\text{sw}}/L, \xi_\text{l}/L)$, where $x$ and $f(\ ,\ )$ are
scaling exponent and function. In the case of $L \to \infty$, this
gives $X=\xi_{\text{sw}}^y \xi_\text{l}^z
g(\xi_{\text{sw}}/\xi_\text{l})$, where $y$, $z$, and $g(\ )$ are
other scaling exponents and function. This scaling is equally
applicable to many other cooperative models on small-world networks.






\subsection{$k$-clique percolation}
\label{ssec:k-clique}

A possible generalization of percolation was put forward by
\textcite{Derenyi:dpv05}. They considered percolation on the
complete set of the $k$-cliques of a network. The $k$-clique is a
fully connected subgraph of $k$ vertices. Two $k$-cliques are
adjacent if they share $k-1$ vertices. For example, the smallest
non-trivial clique, the $3$-clique, is a triangle, and so that two
triangles must have a common edge to allow the ``$3$-percolation''.

In fact, \textcite{Derenyi:dpv05} described the birth of the giant
connected component in the set of the $k$-cliques of a classical
random graph---the Gilbert model. The $k$-clique graph has
vertices---$k$-cliques---and edges---connections between adjacent
$k$-cliques. The total number of $k$-cliques approximately equals
$N^k p^{k(k-1)/2}/k!$. The degree distribution of this graph is
Poissonian, and the mean degree is $\langle q \rangle \cong
Nkp^{k-1}$, which may be much less than the mean degree in the
Gilbert model, $Np$.

Since the sparse classical random graphs have few
$(k{\geq}3)$-cliques, this kind of percolation obviously implies the
dense networks with a divergent mean degree. The application of the
Molloy-Reed criterion to the $k$-clique graph gives the birth point
of the $k$-clique giant connected component 
\begin{equation}
p_c(k) N = \frac{1}{k-1}\,N^{(k-2)/(k-1)} \ \ \text{as} \ N\to\infty
\label{e3.101}
\end{equation}
(for more detail, see \textcite{Palla:pdv06}).

The birth of the giant connected component in the $k$-clique graph
looks quite standard and so that its relative size is proportional
to the deviation $[p-p_c(k)]$ near the critical point.
On the other hand, the relative size $S_k$ of the $(k\geq 3)$-clique
giant connected component in the original graph (namely, the
relative number of vertices in this component) evolves with $p$ in a
quite different manner.
This component emerges abruptly, and for any $p$ above the threshold $p_c(k)$ it 
contains almost all vertices of the network: $S_k(p{<}p_c(k))=0$ and
$S_k(p{>}p_c(k))=1$.

\subsection{$e$-core}
\label{ssec:e-core}

\begin{figure}
\begin{center}
\scalebox{0.31}{\includegraphics[angle=270]{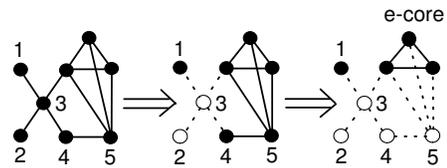}}
\end{center}
\caption{
Construction of the $e$-core of a given graph. Consequently removing the leaf
[23] and a new leaf [45] we obtain one isolated vertex and the $e$-core of the
graph. Removing at first the leaf [13] and then the leaf [45] leads to the
same $e$-core and the same number of isolated vertices. Compare with the
$3$-core of the same graph in Fig.~\protect\ref{f6p}.
}
\label{f11p}
\end{figure}

Let us define {\em a leaf} as the triple: a dead end vertex, its
sole nearest neighbor vertex, and the edge between them.
A more traditional definition does not include the neighbor, but
here for the sake of convenience we modify it. A number of
algorithms for networks are based on successive removal of these
leaves from a graph. In particular, algorithms of this kind are used
in the matching problem and in minimal vertex covers.
\textcite{Bauer:bg01a} described the final result of the recursive
removal of all leaves from the Erd\H os-R\'enyi graph. They found
that if the mean degree $\langle q \rangle >e=2.718\ldots$, the
resulting network contains a giant connected component---we call
it {\em the $e$-core} to distinct from resembling terms.
The $e$-core is explained in Fig.~\ref{f11p}.
$e$-cores
in other networks were not studied yet.

For $\langle q \rangle \leq e$, the removal procedure destroys the
graph---only $O(N)$ isolated vertices and small connected components
consisting in sum of $o(N)$ vertices remain. At  $\langle q
\rangle=e$, a second order phase transition of the birth of the
$e$-core takes place. For $\langle q \rangle>e$, a finite fraction
$S_e$ of $N$ vertices are in the $e$-core, a fraction $I$ are
isolated vertices, and negligible fraction of vertices are in finite
components. In the critical region, 
\begin{equation}
S_e \cong 12(\langle q \rangle-e)/e
, \label{e3.131}
\end{equation}
and the mean degree $\langle q \rangle_e$ of the vertices in the
$e$-core at the moment of its birth is exactly $2$ which corresponds
to a tree graph. In the critical region, 
\begin{equation}
\langle q \rangle_e \cong 2 + \sqrt{8/3}\,\sqrt{\langle q \rangle-e}
. \label{e3.132}
\end{equation}
This singularity is in sharp contrast to the analytic behavior of
the mean degree of the usual giant connected component of this graph
at the point of its birth. The relative number of isolated vertices
has a jump only in the second derivative: 
\begin{equation}
I \cong \frac{3{-}e}{e} - \frac{1}{e}(\langle q \rangle - e) +
\frac{1+3\theta(\langle q \rangle{-}e)}{2e}
(\langle q \rangle - e)^2 , \label{e3.133}
\end{equation}
where $\theta(x<0) = 0$ and $\theta(x>0)=1$. Interestingly, the leaf
removal algorithm slows down as $\langle q \rangle$ approaches the
critical point, which is a direct analog of the well-known
critical slowing down for usual continuous phase transitions.

The same threshold $\langle q \rangle=e$ is present in several
combinatorial optimization problems on the classical random graphs.
In simple terms, in each of these problems, a solution may be found
``rapidly'' only if $\langle q \rangle<e$. Above $e$, any algorithm
applied needs a very long time. Note that this statement is valid
both for the efficiently solvable in polynomial time, P
(deterministic polynomial-time) problems and for the NP
(non-deterministic polynomial time) problems. In particular, the $e$
threshold takes place in {\em the matching (P) problem}---find in a
graph the maximum set of edges without common vertices
\cite{Karp:ks81}, and, also, in {\em the minimum vertex cover (NP)
problem}---if a guard sitting at a vertex controls the incident
edges, find the minimum set of guards needed to watch over all the
edges of a graph \cite{Weigt:wh00}. The matching problem, belonging
to the P class \cite{Aronson:afp98}, is actually equivalent to the
model of dimers with repulsion. We will discuss the minimum vertex
cover in Sec.~\ref{ssec:vertex_cover}.
Here we only mention that in the combinatorial optimization
problems, the $e$ threshold separates the phase $\langle q
\rangle<e$ with a ``simple'' structure of the ``ground state'',
where the replica symmetry solution is stable from the phase with
huge degeneracy of the ``ground state'', where the replica symmetry
breaks. In particular, this degeneracy implies a huge number of
minimum covers.

Note another class of problems, where leaf removal is essential.
The adjacency matrix spectrum
is relevant to the localization/delocalization of a quantum particle
on a graph, see Sec.~\ref{ssec:localization}. It turns out that that
leaf removal does not change the degeneracy of the zero eigenvalue
of the adjacency matrix, and so the $e$-core notion is closely
related to the structure of this spectrum and to localization
phenomena. \textcite{Bauer:bg01a} showed that the number of
eigenvectors with zero eigenvalue equals the product $IN$, see
Eq.~(\ref{e3.133}), and thus has a jump in the second derivative at
$\langle q \rangle=e$. See Sec.~\ref{ssec:localization} for more
detail.

\section{CONDENSATION TRANSITION}
\label{sec:condensation}

Numerous models of complex networks show the following phenomenon. A
finite fraction of typical structural elements in a network
(motifs)---edges, triangles, etc.---turn out to be aggregated into
an ultra-compact subgraph with
diameters much smaller than the
diameter of this network. In this section we discuss various types
of this condensation.


\subsection{Condensation of edges in equilibrium networks}
\label{ssec:condensation_equilibrium}

{\em Networks with multiple connections.}
We start with rather simple equilibrium uncorrelated networks, where
multiple connections, loops of length one, and other arbitrary
configurations are allowed. There exist a number of more or less
equivalent models of these networks
\cite{Burda:bck01,Bauer:bb02,Dorogovtsev:dms03b,Berg:bl02,Farkas:fdp04}.
In many respects, these networks are equivalent to an equilibrium
non-network system---balls statistically distributed among
boxes---and so that they can be easily treated. On the other hand,
the balls-in-boxes model has a condensation phase transition
\cite{Bialas:bbj97,Burda:bjj02}.

We can arrive at uncorrelated networks with complex degree
distribution in various ways. Here we mention
two equivalent approaches to networks with a fixed number $N$ of
vertices.

(i) Similarly to the balls-in-boxes model , one can define the
statistical weights of the random ensemble members in the factorized
form: $\prod_{i=1}^N p(q_i)$ \cite{Burda:bck01}, where the
``one-vertex'' probability $p(q)$ is the same for all vertices (or
boxes) and depends on the degree of a vertex. If the number of edges
$L$ is fixed, these weights additionally take into account the
following constraint $\sum_i q_i = 2L$. With various $p(q)$ (and the
mean degree $\langle q \rangle = 2L/N$) we can obtain various
complex degree distributions.

(ii) A more ``physical'', equivalent approach is as follows. A
network is treated as an evolving statistical ensemble, where edges
permanently change their positions between vertices
\cite{Dorogovtsev:dms03b}. After
relaxation, this ensemble approaches a final state---an equilibrium
random network. If the rate of relinking factors into the product of
simple, one-vertex-degree preference functions $f(q)$, the resulting
network is uncorrelated. For example, one may choose a random edge
and move it to vertices $i$ and $j$ selected with probability
proportional to the product $f(q_i)f(q_j)$. The form of the
preference function and $\langle q \rangle$ determines the
distribution of connections in this network.

\begin{figure}[t]
\begin{center}
\scalebox{0.30}{\includegraphics[angle=0]{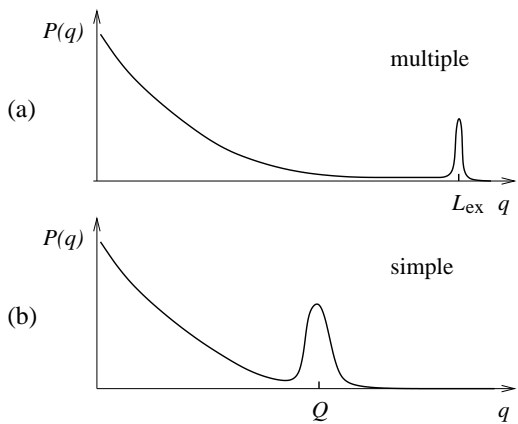}}
\end{center}
\caption{ Schematic plots of the degree distributions of the
equilibrium networks with (a) and without (b) multiple connections
in the condensation phase where the mean degree exceeds the critical
value $q_c$ \cite{Dorogovtsev:dmp05}. The peaks are due to a single
vertex attracting $L_{\text{ex}}=N(\langle q \rangle-q_c)/2$ edges
(a) or due to the highly interconnected core vertices of typical
degree $Q \sim N/N_h(N)$ (b). Note the difference from the rich club
phenomenon, where there are no such peaks in degree distributions.
From \textcite{Dorogovtsev:dmp05}. } \label{f41}
\end{figure}

It turns out that in these equilibrium networks, scale-free degree
distributions can be obtained only if $f(q)$ is a linear function.
Furthermore, the value of the mean degree plays a crucial role. If,
say, $f(q) \cong q + 1-\gamma$ as $q \to \infty$, then three
distinct regimes are possible. (i) When the mean degree is lower
than some critical value $q_c$ (which is determined by the form of
$f(q)$), the degree distribution $P(q)$ is an exponentially
decreasing function. (ii) If $\langle q \rangle=q_c$, then $P(q)
\sim q^{-\gamma}$ is scale-free. (iii) If $\langle q \rangle>q_c$,
then one vertex attracts a finite fraction of all connections, in
sum, $L_{\text{ex}}=N(\langle q \rangle-q_c)/2$ edges, but the other
vertices are described by the same degree distribution as at the
critical point. In other words, at $\langle q \rangle>q_c$, a finite
fraction of edges are condensed on a single vertex, see
Fig.~\ref{f41}(a). One can show that it is exactly one vertex that
attracts these edges and not two or three or several. Notice a huge
number of one-loops and multiple connections attached to this
vertex. We emphasize that a scale-free degree distribution without
condensation occurs only at one point---at the critical mean degree.
This is in contrast to networks growing under the mechanism of the
preferential attachment, where linear preference functions generate
scale-free architectures for wide range of mean degrees.

One can arrive at the condensation of edges in a quite different
way. In the spirit of the work of \textcite{Bianconi:bb01}, who
applied this idea to growing networks, let few vertices, or even a
single vertex, be more attractive than others. Let, for example, the
preference function for this vertex be $gf(q)$, where $f(q)$ is the
preference function for the other vertices, and $g>1$ is a constant
characterising a relative ``strength'' or ``fitness'' of this
vertex. It turned out that as $g$ exceeds some critical value $g_c$,
a condensation of edges on this ``strong'' vertex occurs
\cite{Dorogovtsev:dmbook03}. Interestingly, in general, this
condensation is not accompanied by scale-free organization of the
rest network.

\begin{figure}[t]
\begin{center}
\scalebox{0.23}{\includegraphics[angle=0]{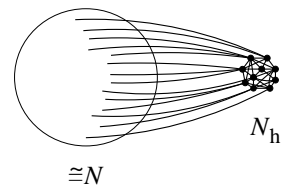}}
\end{center}
\caption{ The structure of a network without multiple connections
when its vertex mean degree exceeds a critical value. The size
$N_h(N)$ of the highly interconnected core varies in the range of
$\sim N^{1/2}$ and $\sim N^{2/3}$ vertices. These vertices are
interconnected by $\sim N$ edges. From \textcite{Dorogovtsev:dmp05}.
} \label{f42}
\end{figure}

{\em Networks without multiple connections.}
If multiple connections and one-loops are forbidden, the structure
of the condensate changes crucially. This difficult problem was
analytically solved in \textcite{Dorogovtsev:dmp05}. The essential
difference from the previous case is only in the structure of the
condensate. It turns out that in these networks, at $\langle q
\rangle>q_c$, a finite fraction of edges, involved in the
condensation, link together a relatively small, highly
interconnected core of $N_h$ vertices, $N_h(N) \ll N$,
Fig.~\ref{f42}. This core, however, is not fully interconnected,
i.e., it is not a clique. (i) If the degree distribution $P(q)$ of
this network decreases slower than any stretched exponential
dependence, e.g., the network is scale-free, then $N_h \sim
N^{1/2}$. (ii) In the case of a stretched exponential $P(q) \sim
\exp(-\text{const}\,q^\alpha)$, $0<\alpha<1$, the core consists of
\begin{equation}
N_h \sim N^{(2-\alpha)/(3-\alpha)} \label{e4.5}
\end{equation}
vertices, that is the exponent of $N_h(N)$ is in the range
$(1/2,2/3)$. 
The connections inside the core are distributed according to the
Poisson law, and the mean degree $\sim N/N_h$ varies in the range
from $\sim N/N^{1/2} \sim N^{1/2}$ to $\sim N/N^{2/3} \sim N^{1/3}$.

In the framework of traditional statistical mechanics, one can also
construct networks with various correlations \cite{Berg:bl02},
directed networks \cite{Angel:ahe06}, and many others.
\textcite{Derenyi:dfp04}, \textcite{Palla:pdf04}, and
\textcite{Farkas:fdp04} constructed a variety of network ensembles,
with statistical weights of members $\propto \prod_i\exp[-E(q_i)]$,
where $E(q)$ is a given one-vertex degree function---``energy'', as
they called it. In particular, in the case $E(q) = -\text{const}\,
q\ln q$, these authors numerically found an additional, first-order
phase transition. They studied a variation of the maximum vertex
degree $q_{\text{max}}$ in a network. As, say, $\langle q \rangle$
reaches $q_c$, a condensation transition takes place, and
$q_{\text{max}}$ approaches the value $\sim N\langle q \rangle$,
i.e., a finite fraction of all edges. Remarkably, at some
essentially higher mean degree, $q_{c2}$, $q_{\text{max}}$ sharply,
with hysteresis, drops to $\sim N^{1/2}$. That is, the network
demonstrates a first order phase transition from the condensation
(``star'') phase to the ``fully connected graph'' regime.






\subsection{Condensation of triangles in equilibrium nets}
\label{ssec:condensation_triangles}

The condensation of triangles in
network models was already observed in the pioneering work of
\textcite{Strauss:s86}. Strauss proposed {\em the exponential
model}, where statistical weights of graphs are
\begin{equation}
W(g) = \exp\Bigl[-\sum_n \beta_n E_n(g)\Bigr] . \label{e4.10}
\end{equation}
Here $E_n(g)$ is a set of some quantities of a graph $g$---a member
of this statistical ensemble, and $\beta_n$ is a set of some
positive constants. The reader may see that many modern studies of
equilibrium networks are essentially based on the exponential model.
Strauss included the quantity $E_3(g)$, that is the number of triangles
in the graph $g$ taken with the minus sign, in the exponential. This
term leads to the presence of a large number of triangles in the
network. On the other, hand, they turn out to be very
inhomogeneously distributed over the network. By simulating this (in
his case, very small) network Strauss discovered that all triangles
merge together forming a clique (fully connected subgraph) in the
network---the condensation of triangles.

\textcite{Burda:bjk04a,Burda:bjk04b}
analytically described and explained this non-trivial phenomenon.
Let us discuss
the idea and results of their theory.
The number of edges, $L$, and the number of triangles, $T$, in a
network are expressed in terms of its adjacency matrix, $\hat{A}$.
Namely, $L=Tr(\hat{A}^2)/2!$ and $T=Tr(\hat{A}^3)/3!$. The partition
function of the Erd\H{o}s-R\'enyi graph is simply
$Z_0=\sum_{\hat{A}} \delta(Tr(\hat{A}^2)-2L)$, where sum is over all
possible adjacency matrices. In the spirit of Strauss, the simplest
generalization of the Erd\H{o}s-R\'enyi ensemble, favoring
triangles,
has the following partition function 
\begin{equation}
Z = \sum_{\hat{A}} \delta(Tr(\hat{A}^2)-2L)e^{G\, Tr(\hat{A}^3)/3!}
= Z_0 \langle e^{G\, Tr(\hat{A}^3)/3!} \rangle_0 , \label{e4.11}
\end{equation}
where the constant $G$ quantifies the tendency to have many
triangles, and $\langle \ldots \rangle_0$ denotes the averaging over
the Erd\H{o}s-R\'enyi ensemble. Equation (\ref{e4.11}) shows the
form of the partition function for the canonical ensemble, i.e.,
with fixed $L$. In the grand canonical formulation, it looks more
invariant: $Z_{gc}= \sum_{\hat{A}} \exp{[-C\,Tr(\hat{A}^2)/2! + G\,
Tr(\hat{A}^3)/3!]}$ (we here do not discuss the constant $C$). In
fact, based on this form, \textcite{Strauss:s86} argued that with
$L/N$ finite and fixed, there exists a configuration, where all
edges belong to a fully connected subgraph and so $Tr(\hat{A}^3)
\sim N^{3/2} \gg N$.
Therefore, as $N \to \infty$, for any positive ``interaction
constant'' $G$, the probability of realization of such a
configuration should go to $1$, which is the stable state of this
theory.

The situation, however, is more delicate. Burda {\em et al.} showed
that apart from this stable condensation state, the network has a
metastable, homogeneous one. These states are separated by a
barrier, whose height approaches infinity as $N\to \infty$. So, in
large networks (with sufficiently small $G$), it is practically impossible to approach the condensation state if we start evolution---relaxation---from a homogeneous configuration. (Recall that Strauss
numerically studied very small networks.)

\begin{figure}[t]
\begin{center}
\scalebox{0.25}{\includegraphics[angle=0]{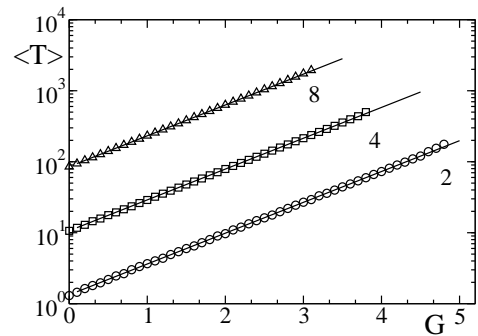}}
\end{center}
\caption{ The mean number of triangles $\langle T \rangle$ as a
function of the parameter $G$ in the metastable state of the network
of $N=2^{14}$ vertices for three values of the mean degree $\langle
q \rangle=2,4,8$. The dots are results of a simulation, and the
lines are theoretical curves $\langle T \rangle = (\langle q
\rangle^3/6)\exp(G) \lesssim N$.
The very right dot in each set corresponds to the threshold value
$G_t(\langle q \rangle)$ above which the network quickly approaches
the condensation state with $\langle T \rangle \sim N^{3/2}$. From
\protect\textcite{Burda:bjk04a}. } \label{f45}
\end{figure}

Assuming small $G$, Burda {\em et al.} used the second equality in
Eq.~(\ref{e4.11}) to make a perturbative analysis of the problem.
They showed that in the ``perturbative phase'', the mean number of
triangles $\langle T \rangle = (\langle q \rangle^3/6)\exp(G)$,
where $\langle q \rangle$ is the mean degree of the network, see
Fig.~\ref{f45}. In this regime, the number of triangles may be
large, $\langle T \rangle \lesssim N$. Above the threshold
$G_t(\langle q \rangle,N) \approx a\ln N + b$, where the
coefficients $a$ and $b$ depend only on $\langle q \rangle$, the
system easily jumps over the barrier and
quickly approaches the condensation state.

\textcite{Burda:bjk04b} generalized this theory to networks with
complex degree distributions using the partition function $Z =
\sum_{\hat{A}} \delta(Tr(\hat{A}^2)-2L)e^{G\,
Tr(\hat{A}^3)}\prod_i^N p(q_i)$, where $q_i$ is the degree of vertex
$i$, and the weight $p(q)$ is given.
In an even more general approach, $Tr(\hat{A}^3)$ in the exponential
should be replaced by a more general perturbation $S(\hat{A})$. Note
that a different perturbation theory for the exponential model was
developed by \textcite{Park:pn04a,Park:pn04b}.







\subsection{Condensation of edges in growing networks}
\label{ssec:condensation_growing}

\textcite{Bianconi:bb01} discovered the condensation phase
transition in
networks, growing under the mechanism of preferential attachment. In
their inhomogeneous network, preference function of vertices had a
random factor (``fitness''): $g_i f(q_i)$ distributed according to a
given function $p(g)$. Bianconi and Barab\'asi indicated a class of
sufficiently long-tailed distributions $p(g)$, for which an
infinitely small fraction of vertices (maximally fitted ones)
attract a finite fraction of edges. In fact, this condensation may be obtained even with a single more fitted vertex ($j$): $g_{i\neq
j}=1$, $g_j=g>1$ \cite{Dorogovtsev:dm01}. In this case, the
condensation on this vertex occurs in large networks of size $t \gg j$, if $g$ exceeds
some critical value $g_c$.

Suppose that the network is a recursive graph, and the preference
function $f(q)$ is linear. Then $g_c=\gamma_0-1$, where $\gamma_0$
is the exponent of the degree distribution of this network with all
equal vertices ($g=1$). Note that if the degree distribution is
exponential ($\gamma_0 \to \infty$), $g_C \to \infty$, and the
condensation is impossible. If $g<g_c$, the degree distribution of
the network is the same as in the ``pure'' network. On the other
hand, the phase with the condensate, $g>g_c$ has the following
characteristics. (i) A finite fraction of edges $d \propto (g-g_c)$
is attached to the ``fittest'' vertex. (ii) The degree distribution
exponent increases: $\gamma=1+g>\gamma_0$. (iii) In the entire
condensation phase, relaxation to the final state (with the fraction
$d$ of edges in the condensate) is very slow, of a power-law kind:
$d_j(t)-d \sim t^{-(g-g_c)/g}$. Here $d_j(t)$ is a condensed
fraction of edges at time $t$.

Bianconi and Barab\'asi called this phenomenon the Bose-Einstein
condensation based on evident parallels (in fact, this term was also
applied sometimes to condensation in equilibrium networks, in the
balls-in-boxes model, and in zero-range processes). We emphasize the
completely classical nature of this condensation.

\textcite{Kim:ktm05} and \textcite{Minnhagen:mrs04} introduced a
wide class of equilibrium and growing networks, where complex
architectures are results of the process of merging and splitting of
vertices. In many of these networks (where vertices differ from each
other only by their degrees) the condensation of edges takes place.
This phenomenon in the networks with aggregation was studied by
\textcite{Alava:ad05}.




\section{CRITICAL EFFECTS IN THE DISEASE SPREADING}
\label{sec:diseases}

The epidemic spreading in various complex networks was quite
extensively studied in recent years, and it is impossible here to
review in detail and even cite numerous works on this issue. In this
section we only explain basic facts on the spread of diseases in
networks, discuss relations to other phenomena in complex networks,
and describe several recent results. The reader may refer to
\textcite{Pastor-Satorras:pvbook04,Pastor-Satorras:pv03}
for a comprehensive introduction to this topic.



\subsection{The SIS, SIR, SI, and SIRS models}

\label{ssec:si_sir_sis}

Four basic models of epidemics are widely used: the SIS, SIR, SI,
and SIRS models, see, e.g., \textcite{Nasell:n02}. S is for
susceptible, I is for infective, and R is for recovered (or
removed). In the network context,
vertices are individuals, which are in one of these three (S,I,R) or
two (S,I) states, and infections spread from vertex to vertex
through edges.
Note that an ill vertex can infect only its nearest neighbors: S$\to$I.

The SIS model describes infections without immunity, where recovered
individuals are susceptible. In the SIR model, recovered individuals
are immune forever, and do not infect. In the SI model, recovery is
absent. In the SIRS model, the immunity is temporary. The SIS, SIR,
and SI models are particular cases of the more general SIRS model.
We will touch upon only first three models.

Here we consider a heuristic approach of
\textcite{Pastor-Satorras:pv01,Pastor-Satorras:pv03}. 
This (a kind of mean-field) theory fairly well describes
the epidemic spreading in complex networks. For a more strict
approach, see, e.g., \textcite{Newman:n02a}, \textcite{Kenah:kr06},
and references therein.

Let a network have only degree-degree correlations, and so it is
defined by the conditional probability $P(q'|q)$, see
Sec.~\ref{ssec:percolation_on_correlated}. Consider the evolution of
the probabilities $i_q(t)$, $s_q(t)$, and $r_q(t)$ that a vertex of
degree $q$ is in the I, S, and R states, respectively. For example,
$i_q(t)=(\text{number of infected vertices of degree $q$})/[NP(q)]$.
As is natural, $i_q(t)+s_q(t)+r_q(t)=1$. Let $\lambda$ be the
infection rate. In other words, a susceptible vertex becomes
infected with the probability $\lambda$ (per unit time) if at least
one of the nearest neighbors is infected. Remarkably, $\lambda$ is
actually the sole parameter in the SIS and SIR models---other
parameters can be easily set to $1$ by rescaling.
Here we list evolution equations for the SIS, SIR, and SI models.
For derivations, see \textcite{Boguna:bpv03}. However, the structure
of these equations is so clear that the reader can easily explain
them himself or herself, exploiting obvious similarities with
percolation.

{\em The SIS model.} In this model, infected vertices become
susceptible with unit rate, $r_q(t)=0$, $s_q(t)=1-i_q(t)$. The
equation is 
\begin{equation}
\frac{d i_q(t)}{dt} = -i_q(t) + \lambda q
[1-i_q(t)]\sum_{q'}P(q'|q)\,i_{q'}(t) . \label{e5.9}
\end{equation}

{\em The SIR model.} In this model, infected vertices become
recovered with unit rate. Two equations describe this system:

\begin{eqnarray}
&& \!\!\!\!\!\!\!\!\!\!\!\!\!\!\!\!\!\frac{d r_q(t)}{dt} = i_q(t) ,
\nonumber
\\[5pt]
&& \!\!\!\!\!\!\!\!\!\!\!\!\!\!\!\!\!\frac{d i_q(t)}{dt} = -i_q(t) +
\lambda q [1{-}i_q(t)]\sum_{q'}\frac{q'{-}1}{q'}P(q'|q)\,i_{q'}(t) .
\label{e5.12}
\end{eqnarray}
Note the factor $(q'-1)/q'$ in the sum. This ratio is due to the
fact that an infected vertex in this model cannot infect back its
infector, and so one of the $q'$ edges is effectively blocked.

{\em The SI model.} Here infected vertices are infected forever,
$s_q(t)=1-i_q(t)$, and the dynamics is described by the following
equation: 
\begin{equation}
\frac{d i_q(t)}{dt} = \lambda q
[1-i_q(t)]\sum_{q'}\frac{q'-1}{q'}P(q'|q)\,i_{q'}(t)
\label{e5.15}
\end{equation}
(compare with Eq.~(\ref{e5.12})). This simplest model has no
epidemic threshold. Moreover, in this model, $\lambda$ may be set to
$1$ without loss of generality.

If a network is uncorrelated, simply substitute
$P(q'|q)=q'P(q')/\langle q \rangle$ into these equations. It is
convenient to introduce $\Theta=\sum_{q'}(q'-1)P(q')i_{q'}\langle q
\rangle$ (for the SIR model) or
$\Theta=\sum_{q'}q'P(q')i_{q'}\langle q \rangle$ (for the SIS model)
and then solve a simple equation for this degree-independent
quantity. We stress that the majority of results on epidemics in
complex networks were obtained by using only Eqs.~(\ref{e5.9}),
(\ref{e5.12}), and (\ref{e5.15}). Note that
one can also analyse these models assuming a degree-dependent
infection rate $\lambda$ \cite{Giuraniuc:ghi06}.




\subsection{Epidemic thresholds and prevalence}
\label{ssec:epidemic_thresholds}

The epidemic threshold $\lambda_c$ is a basic notion in
epidemiology. Let us define the fractions of infected and recovered
(or removed) vertices in the final state as $i(
\infty)=\sum_q P(q)i_q(t {\to} \infty)$ and $r(
\infty)=\sum_q P(q)r_q(t {\to} \infty)$, respectively. Below the
epidemic threshold, $i(
\infty) = r(
\infty) = 0$. 
In epidemiology the fraction $i(t)$ of infected vertices in a network is called the prevalence. 
The 
On the other hand, above the epidemic thresholds, (i)
in the SIS model, $i(
\infty,\lambda{>}\lambda_c^{SIS})$ is finite, and (ii) in the SIR
model, $i(
\infty,\lambda{>}\lambda_c^{SIR}) = 0$ and $r(
\infty,\lambda{>}\lambda_c^{SIR})$ is finite.

The
linearization of Eqs.~(\ref{e5.9}), (\ref{e5.12}), and (\ref{e5.15})
readily provide the epidemic thresholds. The simplest SI model on
any network has no epidemic threshold---all vertices are infected in
the final state, $i_q(t{\to}\infty)=1$.
Here we only discuss results for uncorrelated networks
\cite{Pastor-Satorras:pv01,Pastor-Satorras:pv03}, for correlated
networks, see \textcite{Boguna:bpv03}. The reader can easily check
that the SIS and SIR models have the following epidemic thresholds:
\begin{equation}
\lambda_c^{SIS}= \frac{\langle q \rangle}{\langle q^2 \rangle}, \ \
\
\lambda_c^{SIR} = \frac{\langle q \rangle}{\langle q^2
\rangle-\langle q \rangle}
. \label{e5.18}
\end{equation}
Notice the coincidence of $\lambda_c^{SIR}$ with the percolation
threshold $p_c$ in these networks, Eq.~(\ref{e3.9}). (Recall that
for bond and site percolation problems, $p_c$ is the same.) This
coincidence is not occasional---strictly speaking, the SIR model is
equivalent to dynamic percolation
\cite{Grassberger:g83}. In more simple terms,
the SIR model, in the respect of its final state,
is practically equivalent to the bond percolation problem [see
\textcite{Hastings:h06} for
discussion of some difference, see also discussions in \textcite{Kenah:kr06} and \textcite{Miller:m07}].
Equation~(\ref{e5.18}) shows that general conclusions for
percolation on complex networks are also valid for the SIS and SIR
models.
In particular, (i) the estimates and conclusions for $p_c$ from
Secs. \ref{sssec:percolation_uncorrelated},
\ref{sssec:statistics_finite}, and \ref{sssec:size} are valid for
the SIS and SIR models (simply replace $p_c$ by $\lambda_c^{SIS}$ or
$\lambda_c^{SIR}$), the finite size relations also work; and (ii)
the estimates and conclusions for the size $S$ of the giant
connected component from these sections are also valid for
$i(\infty)$ in the SIS model and for $r(\infty)$ in the SIR model,
i.e., for prevalence.

In particular, \textcite{Pastor-Satorras:pv01} discovered that in
uncorrelated networks with diverging $\langle q^2 \rangle$, the
epidemic thresholds approach zero value, but a finite epidemic
threshold is restored if a network is finite
\cite{May:ml01,Pastor-Satorras:pv02a,Boguna:bpv04}. Similarly to
percolation, the same condition is valid for networks with
degree-degree correlations \cite{Moreno:mv02,Boguna:bpv02}.


The statistics of outbreaks near an epidemic threshold in the SIR model is similar to that for finite connected components near the birth point of a giant component. In particular, at a (SIR) epidemic threshold in a network with a rapidly decreasing degree distribution, the maximum outbreak scales as $N^{2/3}$ and the mean outbreak scales as $N^{1/3}$ \cite{Ben-Naim:bk04}. (In the SIS model, the corresponding quantities behave as $N$ and  $N^{1/2}$.)
These authors also estimated duration of epidemic outbreaks.
At a SIR epidemic threshold in these networks, the maximum duration of an outbreak scales as $N^{1/3}$, the average duration scales as $\ln N$, and the typical duration is of the order of one.


Interestingly, some of results on the disease spreading on complex
networks
were obtained before those for percolation, see the work of
\textcite{Pastor-Satorras:pv01}.
For example, they found that in the SIS and SIR model on the
uncorrelated scale-free network with degree distribution exponent
$\gamma=3$, the final prevalence is proportional to $\exp[-g(\langle
q \rangle)/\lambda)]$. Here $g(\langle q \rangle)$ depends only on
the mean degree. That is, all derivatives of the prevalence over
$\lambda$ equal zero at this specific point (recall the
corresponding result for percolation). Furthermore,
\textcite{Boguna:bp02} fulfilled numerical simulations of the SIS
model on the growing network of \textcite{Callaway:chk01}  and
observed prevalence proportional to
$\exp[-\text{const}/\sqrt{\lambda-\lambda_c}]$, i.e., the
Berezinskii-Kosterlitz-Thouless singularity.

Disease spreading was also studied in many other networks. For
example, for small-world networks, see
\textcite{Moore:mn00a,Newman:n02a,Newman:njz02} and references
therein. For epidemics in networks with high clustering, see
\textcite{Newman:n03b,Petermann:pd04,Serrano:sb06a}. A very popular
topic is various immunization strategies, see
\textcite{Pastor-Satorras:pv02b,Pastor-Satorras:pv03,Dezso:db02,Cohen:chb03b,Gallos:gla07a}, 
and many other works.

Note that the excitation of a system of coupled neurons in response to external stimulus, in principle, may be considered similarly to the disease spreading.
Excitable networks with complex architectures were studied in \textcite{Kinouchi:kc06}, \textcite{Copelli:cc07}, and \textcite{Wu:wxw07}.



\subsection{Evolution of epidemics}
\label{ssec:evolution_of_epidemics}

Equations~(\ref{e5.9}), (\ref{e5.12}), and (\ref{e5.15}) describe
the dynamics of epidemics. Let us discuss this dynamics above an
epidemic threshold, where epidemic outbreaks are giant, that is
involve a finite fraction of vertices in a network
\cite{Moreno:mpv01,Barthelemy:bbp03,Barthelemy:bbp05,Vazquez:v06a}.
The demonstrative SI model is especially easy to analyse.
A characteristic time scale of the epidemic outbreak can be
trivially obtained in the following way
\cite{Barthelemy:bbp03,Barthelemy:bbp05}. Let the initial condition
be uniform, $i_q(t=0)=i_0$. Then in the range of short times the
prevalence $i(t) = \sum_q P(q)i_q(t)$ rises according to the law: 
\begin{equation}
\frac{i(t) - i_0}{i_0} = \frac{\langle q \rangle^2 - \langle q
\rangle}{\langle q^2 \rangle - \langle q \rangle}\,(e^{t/\tau}-1)
\label{e5.21}
\end{equation}
with the time scale 
\begin{equation}
\tau = \frac{\langle q \rangle}{\lambda(\langle q^2 \rangle -
\langle q \rangle)} . \label{e5.24}
\end{equation}
Thus $\tau$ decreases with increasing $\langle q^2 \rangle$.
As is natural, the law~(\ref{e5.21}) is violated at long times, when
$i(t) \sim 1$. Expressions for $\tau$ in the SIS and SIR models are
qualitatively similar to Eq.~(\ref{e5.24}).

\begin{figure}[t]
\begin{center}
\scalebox{0.47}{\includegraphics[angle=0]{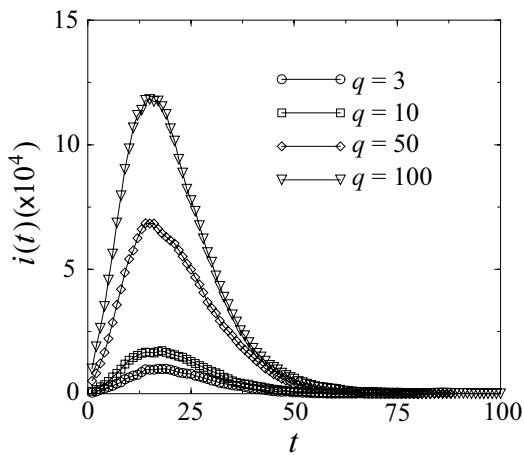}}
\end{center}
\caption{ The evolution of the average fraction of infected vertices in the
SIR model on the Barab\'asi-Albert network of $10^6$ vertices for
various initial conditions.
At $t=0$, randomly chosen vertices of a given degree $q$ are
infected. The spreading rate is $\lambda=0.09$ which is above the
epidemic threshold of this finite network. From
\protect\textcite{Moreno:mpv01}. } \label{f48}
\end{figure}


Notice some difference between the SIR and SIS (or SI) models. In
the SIS and SI models, the fraction of infected vertices $i(t)$
monotonously grows with time until it approaches the final
stationary state. Adversely, in the SIR model, $i(t)$ shows a
peak---outbreak---at $t\sim \tau$ and approaches zero value as $t\to
\infty$. As a result of heterogeneity of a complex network, the
epidemic outbreaks strongly depend on initial conditions, actually
on a first infected individual. Figure~\ref{f48} shows how the
average fraction of infected vertices evolves in the SIR model placed on the
Barab\'asi-Albert network if the first infected individual has
exactly $q$ neighbors \cite{Moreno:mpv01}.
The spreading rate is supposed to be above the epidemic threshold.
If this $q$ is large,
then the outbreak is giant with high probability. On the other hand,
if $q$ is small,
then, as a rule,
the infection disappears after a small outbreak, and the probability of a giant outbreak is low.

When $\langle q^2 \rangle$ diverges ($\gamma \leq 3$),
Eqs.~(\ref{e5.21}) and (\ref{e5.24}) are not applicable.
\textcite{Vazquez:v06a} considered disease spreading in this
situation on a scale-free growing (or causal) tree. Actually he
studied a variation of the SI model, with an ``average generation
time'' $T_G \sim 1/\lambda$.
In this model he analytically found 
\begin{equation}
di(t)/dt \propto t^{\ell_{\text{max}}-1} e^{-t/T_G} , \label{e5.27}
\end{equation}
where $\ell_{\text{max}}(N)$ is the diameter of the network (the
maximum intervertex distance). Vazquez compared this dependence with
his
numerical simulations of the SI model on a generated network and a
real-world one (the Internet at the Autonomous System level). He
concluded that Eq.~(\ref{e5.27}) provides a reasonable fitting to
these results even in rather small networks.




\section{THE ISING MODEL ON NETWORKS}

\label{sec:Ising}

The Ising model, named after the physicist Ernst Ising, is an
extremely simplified mathematical model describing the spontaneous
emergence of order.
Despite its simplicity, this model
is valuable for verification of general theories and assumptions,
such as scaling and universality hypotheses in the theory of
critical phenomena. What is important is that many real systems can
be
approximated by the Ising model.
The Hamiltonian of the model is
\begin{equation}
{\cal H}=-\sum_{i<j}J_{ij}a_{ij}S_{i}S_{j}-\sum_{i}H_{i}S_{i},  \label{Ising}
\end{equation}
where the indices $i$ and $j$ numerate vertices on a network,
$i,j=1,2...N$. $a_{ij}$ is an element of the adjacency matrix:
$a_{ij}=1$ or 0 if vertices $i$ and $j$ are connected or
disconnected, respectively. Network topology is encoded in the
adjacency matrix. In general, couplings $J_{ij}$ and local fields
$H_{i}$ can be random parameters.
What kind of a critical behavior one might expect if we put the
Ising model on the top of a complex network? Is it the standard
mean-field like behavior? A naive answer is yes because a complex
network is an infinite-dimensional system.
Indeed, it is generally accepted that the critical behavior of the
ferromagnetic Ising model on a $d$-dimensional lattice at $d>4$ is
described by the simple mean-field theory which assumes that an
average effective magnetic field $H+Jz_{1}M$ acts on spins, where
$M$ is an average magnetic moment and $z_{1}=\left\langle
q\right\rangle$ is the mean number of the nearest neighbors. An
equation
\begin{equation}
M=\tanh [\beta H+\beta Jz_{1}M]  \label{naive-MF}
\end{equation}
determines $M$. This theory predicts a second order ferromagnetic phase
transition at the critical temperature $T_{\text{MF}}=Jz_{1}$ in zero field
with the standard critical behavior: $M\sim \tau^{\beta }$, $%
\chi =dM/dH\sim \left| \tau\right| ^{-\tilde{\gamma}}$, where
$\tau\equiv T_{\text{MF}}-T$, $\beta =1/2$, and $\tilde{\gamma}=1$.
First investigations of the ferromagnetic Ising model on the
Watts-Strogatz networks revealed the second order phase transition
\cite{Barrat:bw00,Gitterman:g00,Hong:hkc02,Herrero:h02}. This result
qualitatively agreed with the simple mean-field theory.

Numerical simulations of the ferromagnetic Ising model on a growing
Barab\'{a}si-Albert scale-free network \cite{Aleksiejuk:ahs02}
demonstrated that the critical temperature $T_{c}$ increases
logarithmically with increasing $N $: $T_{c}(N)\sim \ln N$.
Therefore, in the thermodynamic limit, the system is ordered at any
finite $T$. The simple mean-field theory fails to explain this
behavior. Analytical investigations
\cite{Dorogovtsev:dgm02b,Leone:lvv02} based on a microscopic theory
revealed that the critical behavior of the ferromagnetic Ising model
on complex networks is richer and extremely far from that expected
from the standard mean-field theory. They showed that the simple
mean-field theory does not take into account the strong
heterogeneity of networks.

In the present section, we look first at exact and approximate
analytical methods (see also Appendices \ref{bethe-thermodynamic},
\ref{bp-algorithm-magnetization}, \ref{replica}) and then consider
critical properties of ferro- and antiferromagnetic, spin-glass and
random-field Ising models on complex networks.

\subsection{Main methods for tree-like networks}

\label{ssec:methods}

\subsubsection{Bethe approach}

\label{sssec:bethe_approach}

The Bethe-Peierls approximation is one of the most powerful methods
for studying cooperative phenomena \cite{Domb:domb60}. It was
proposed by \textcite{Bethe:b35} and then applied by
\textcite{Peierls:p36} to the Ising model. This approximation gives
a basis for developing a remarkably accurate mean-field theory. What
is important, it can be successfully used to study a finite system
with a given quenched disorder.

The list of modern applications of the Bethe-Peierls approximation
ranges from solid state physics, information and computer sciences
\cite{Pearl:p88}, for example, image restoration \cite{Tanaka:t02},
artificial vision \cite{Freeman:fpc00}, decoding of error-correcting
codes \cite{McEliece:mmc98}, combinatorial optimization problems
\cite {Mezard:mz02}, medical diagnosis \cite{Kappen:k02} to social
models.

Let us consider the Ising model Eq.~(\ref{Ising}) on an arbitrary
complex network.
In order to calculate magnetic moment of a spin $S_{i}$, we must
know the total magnetic field $H_{i}^{(t)}$ which acts on this spin.
This gives $M_{i}=\left\langle S_{i}\right\rangle=\tanh [\beta
H_{i}^{(t)}]$, where $\beta =1/T$. $H_{i}^{(t)}$ includes both a
local field $H_{i}$ and fields created by nearest neighboring spins.
The spins interact with their neighbors who in turn interact with
their neighbors, and so on. As a result, in order to calculate
$H_{i}^{(t)}$ we have to account for all spins in the system. It is
a hard work.

Bethe and Peierls proposed to take into account only interactions of
a spin with its nearest neighbors. Interactions of these neighbors
with remaining spins on a network were included in ``mean fields''.
This simple idea reduces the problem of $N$ interacting spins to a
problem of a finite cluster.

\begin{figure}[t]
\begin{center}
\scalebox{0.4}{\includegraphics[angle=270]{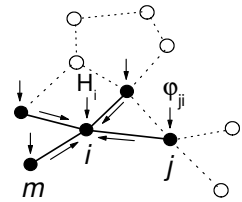}}
\end{center}
\caption{ A cluster on a graph. Within the Bethe-Peierls approach we
choose a cluster consisting of spin $i$ and its nearest neighbors
(closed circles). Cavity fields $\protect\varphi _{j\backslash i}$
(vertical arrows) take into account interactions with remaining
spins (dotted lines and open circles). $H_{i}$ is a local field.
Arrows along edges show fields created by neighboring spins at
vertex $i$. } \label{fig-cluster}
\end{figure}

Consider a cluster consisting of a central spin $S_{i}$ and its
nearest neighbors $S_{j}$, see Fig.~\ref{fig-cluster}. The energy
of the cluster is
\begin{equation}
H_{\text{cl}}=-\sum_{j\in N(i)}J_{ij}S_{i}S_{j}-H_{i}S_{i}-\sum_{j\in
N(i)}\varphi _{j\backslash i}S_{j},  \label{H-cluster}
\end{equation}
where $N(i)$ means all vertices neighboring vertex $i$. Interactions
between the spins $j\in N(i)$ are neglected. They will be
approximately taken into account by the fields $\varphi
_{j\backslash i}$. These fields are called {\em cavity fields}
within the cavity method \cite{Mezard:mp01}. The cavity fields must
be found in a self-consistent way.

\begin{figure}[t]
\begin{center}
\scalebox{0.3}{\includegraphics[angle=270]{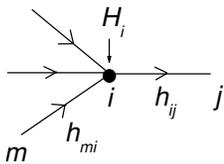}}
\end{center}
\caption{ Diagram representation of Eq.~(\ref{recursion}). An
outgoing message $h_{ij}$ from vertex $i$ to vertex $j$ is
determined by the local field $H_{i}$ and incoming messages to $i$
excluding the message from $j$.} \label{fig-recursion}
\end{figure}

It is easy to calculate the magnetic moments of spins in the cluster.
The magnetic field $H_{i}^{(t)}$ acting on $i$ is
\begin{equation}
H_{i}^{(t)}=H_{i}+\sum_{j\in N(i)}h_{ji},  \label{total field}
\end{equation}
where $h_{ji}$ is an additional field created by a spin $S_{j}$ at
vertex $i$ (see Fig.~\ref{fig-cluster}):
\begin{equation}
\tanh \beta h_{ji}\equiv \tanh \beta J_{ij}\tanh \beta \varphi _{j\backslash
i}.  \label{hij}
\end{equation}
In turn the field $H_{j}^{(t)}$ acting on spin $j$ is
\begin{equation}
H_{j}^{(t)}=\varphi _{j\backslash i}+h_{ij},  \label{total-field-j}
\end{equation}
where the additional field $h_{ij}$\ is created by the central spin
$i$ at vertex $j$. This field is related to the additional fields in
Eq.~(\ref{hij}) as follows:
\begin{equation}
\tanh \beta h_{ij}=\tanh \beta J_{ij}\tanh [\beta (H_{i}+\!\sum_{m\in
N(i)\backslash j}\!h_{mi})],
\label{recursion}
\end{equation}
where $N(i)\backslash j$ means all vertices neighboring vertex $i$,
except $j$. In the framework of the {\em belief-propagation
algorithm} (Sec.~\ref {sssec:belief-propagation}) the additional
fields $h_{ji}$ are called {\em messages}. Using this vivid term, we
can interpret Eq.~(\ref{recursion}) as follows (see Fig.
\ref{fig-recursion}). An outgoing message sent by spin $i$ to
neighbor $j$ is determined by incoming messages which spin $i$
receives from other neighbors $m\in N(i)$, except $j$. Note that if
vertex $i$ is a dead end, then from Eq.~(\ref{recursion}) we obtain
that the message $h_{ij}$ from $i$ to the only neighbor $j$ is
determined by a local field $H_{i}$:
\begin{equation}
\tanh \beta h_{ij}=\tanh \beta J_{ij}\tanh (\beta H_{i}).
\label{dead-h}
\end{equation}


We can choose a cluster in which $S_{j}$ is the central spin. The
field $H_{j}^{(t)}$ is given by the same Eq.~(\ref{total field}).
Comparing Eq.~(\ref{total field}), where $j$ replaces $i$, with
Eq.~(\ref{total-field-j}), we obtain
\begin{equation}
\varphi _{i\backslash j}=H_{i}+\!\sum_{m\in N(i)\backslash j}\!h_{mi}.
\label{fi}
\end{equation}
Equations (\ref{total field})--(\ref{fi}) establish relations between
the fields $\{h_{ij}\}$ and $\{\varphi _{i\backslash j}\}$. All we
need is to solve Eq.~(\ref{recursion}) and find messages
$\{h_{ij}\}$ in a graph. Apart of the local magnetic moments, the
Bethe-Peierls approximation allows one to find a spin correlation
function and the free-energy. These formulas are given in Appendix
\ref{bethe-thermodynamic}.

The Bethe-Peierls approach is exact for a treelike graph and the
fully connected graph. It leads to the same equations as the cavity
method and the exact recursion method (see Sec.~\ref
{ssec:ising_perfect_tree}). The Bethe-Peierls approach is
approximate for graphs with loops due to spin correlations induced
by loops. However, even in this case, it usually leads to remarkably
accurate results. The approach can be improved by using the Kikuchi
``cluster variation method'' \cite {Kikuchi:k51,Domb:domb60,
Yedidia:yfw01}.

How large are loop corrections to the Bethe-Peierls approximation?
There is no clear answer on this important question. Several methods
have recently been proposed for calculating loop corrections
\cite{Yedidia:yfw01,Montanari:mr05,Parisi:ps06,Chertkov:cc06a,Chertkov:cc06b,Rizzo:rwk06},
however this problem is still \vspace{-13pt}unsolved.

\paragraph{Regular Bethe lattice.}

\label{ssssec: regular Bethe}

The Bethe-Peierls approach gives an exact solution of the
ferromagnetic Ising model in an uniform magnetic field on a regular
Bethe lattice with a coordination number $q$ \cite{Baxter:bbook82}.
In this case, all vertices and edges on the lattice are equivalent,
therefore, $M_{i}=M$ and $h_{ij}=h$. From Eqs.~(\ref{total field})
and (\ref{recursion}), we obtain
\begin{eqnarray}
M &=&\tanh [\beta H+\beta qh],  \label{M-regular}\\[5pt]
\tanh \beta h &=&\tanh \beta J\tanh [\beta H+\beta Bh].  \label{h-regular}
\end{eqnarray}
The parameter $B\equiv q-1$ on the right-hand side is the branching
parameter.

At $H=0$, the model undergoes the standard second order phase
transition at a critical point in which $B\tanh \beta J=1$. It gives
the critical temperature
\begin{equation}
T_{\text{BP}}=2J/\ln [(B+1)/(B-1)].  \label{Tc-regular}
\end{equation}
In the limit $q \gg 1$ the critical temperature $T_{\text{MF}}$
tends to $T_{\text{BP}}$, i.e., the simple mean-field approach
Eq.~(\ref{naive-MF}) becomes exact in this limit. At the critical
temperature $T=T_{\text{BP}}$, the magnetic moment $M$ is a
nonanalytic function of $H$: \vspace{-13pt}$M(H)\sim H^{1/3}$.

\paragraph{Fully connected graph.}

\label{ssssec: TAP}

The Bethe-Peierls approximation is exact for the fully connected
graph. For example, consider the spin-glass Ising model with random
interactions $\left| J_{ij}\right| \propto N^{-1/2}$ on the graph
(the Sherrington-Kirkpatrick model). The factor $N^{-1/2}$ gives a
finite critical temperature. In the leading order in $N$, Eqs.~(\ref
{total field}) and (\ref{recursion}) lead to a set of equations for
magnetic moments $M_{i}$:
\begin{equation}
M_{i}=\tanh \Bigl[\beta H+\beta \sum_{j}J_{ij}M_{j}-\beta
^{2}\sum_{j}J_{ij}^{2}M_{i}(1-M_{j}^{2})\Bigr].  \label{TAP}
\end{equation}
These are the TAP equations \cite{Thouless:tap77} which are exact in
the thermodynamic limit.

\subsubsection{Belief-propagation algorithm}

\label{sssec:belief-propagation}

The belief-propagation algorithm is an effective numerical method
for solving inference problems on sparse graphs. It was originally
proposed by \textcite{Pearl:p88} for treelike graphs. Among its
numerous applications are computer vision problems, decoding of high
performance turbo codes and many others, see
\textcite{Frey:98,McEliece:mmc98}. Empirically it was found that it
works surprisingly good even for graphs with loops.
\textcite{Yedidia:yfw01} recently discovered that the
belief-propagation algorithm actually coincides with the
minimization of the Bethe free energy. This discovery renews
interest in the Bethe-Peierls approximation and related methods
\cite{Pretti:pp03,Mooij:mk05,Hartmann:hw05}. The recent progress in
the survey propagation algorithm, which was proposed to solve some
difficult combinatorial optimization problems, is a good example of
interference between computer science and statistical physics
\cite{Mezard:mpz02,Mezard:mz02,Braunstein:bz04}.

In this section we give a physical interpretation of the
belief-propagation algorithm in application to the Ising and other
physical models on a graph.
It enables us to find a general solution of an arbitrary physical
model with discrete or continuous variables on a complex network.

We start with the Ising model on a graph. Consider a spin $i$.
Choose one of its nearest neighbors, say, a spin $j\in N(i)$. We
define a parameter $\mu _{ji}(S_{i})$ as probability to find
spin
$i$ in a state $S_{i}$ under the following conditions: (i) spin $i$
interacts only with spin $j$ while other neighboring spins are
removed; (ii) an local magnetic field $H_{i}$ is zero. We normalize
$\mu _{ji}(S_{i})$ as follows: $\sum_{S_{i}=\pm 1}\mu
_{ji}(S_{i})=1.$ For example, if $\mu _{ji}(+1)=1$ and $\mu
_{ji}(-1)=0$, then the spin $j$ permits the spin state $S_{i}=+1$
and forbids the spin state $S_{i}=-1$. In the same way, we define
probabilities $\mu _{ni}(S_{i})$ for other neighboring spins $n\in
N(i)$. We assume that the probabilities $\mu
_{ji}(S_{i})$ for all $j\in N(i)$ are statistically independent. 
Strictly speaking, this assumption holds true only in a treelike
graph. For a graph with loops this approach is approximate.
In
the belief-propagation algorithm the probabilities $\mu
_{ji}(S_{i})$ are traditionally called {\em messages} (do not mix
with the messages in the Bethe-Peierls approach).

\begin{figure}[t]
\begin{center}
\scalebox{0.25}{\includegraphics[angle=270]{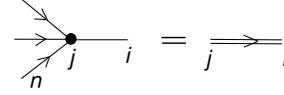}}
\end{center}
\caption{ Diagram representation of the belief-propagation update
rule. Arrows show incoming messages to a vertex $j$. A factor $\exp
(\protect\beta H_{j}S_{j})$ is shown as the closed circle. A solid
line between $j $ and $i$ shows a factor $\exp (\protect\beta
J_{ij}S_{i}S_{j})$. The double line is a new (outgoing) message from
$j$ to $i$.} \label{fig-BP}
\end{figure}

Let us search for an equilibrium state, using an iteration
algorithm. We start from an initial set of non-equilibrium
normalized probabilities $\{\mu _{ji}^{(0)}(S_{i})\}$. Let us choose
two neighboring vertices $i$ and $j$. Using the initial
probabilities, we can calculate a probability to find a spin $j$ in
a state $S_{j}$ under the condition that the state $S_{i}$ is fixed.
This probability is proportional to the product of independent
probabilities which determine the state $S_{j}$. First, we have the
product of all incoming messages $\mu _{nj}^{(0)}(S_{j})$ from
nearest neighboring spins $n$ of $j $, except $i$, because its state
is fixed. This is $\prod_{n\in N(j)\backslash i}\mu
_{nj}^{(0)}(S_{j})$. Second, we have a probabilistic factor $\exp
(\beta H_{j}S_{j})$ due to a local field $H_{j}$. Third, we have a
probabilistic factor $\exp (\beta J_{ij}S_{i}S_{j})$ due to the
interaction between $i$ and $j$. Summing the total product of all
these factors over two possible states $S_{j}=\pm 1$, we obtain a
new probability:
\begin{equation}
A\sum_{S_{j}=\pm 1}e^{\beta H_{j}S_{j}+\beta
J_{ij}S_{i}S_{j}}\prod_{n\in N(j)\backslash i}\mu
_{nj}^{(0)}(S_{j})=\mu _{ji}^{\text{new}}(S_{i}), \label{mu-update}
\end{equation}
where $A$ is a normalization constant. This equation is the standard
update rule of the belief-propagation algorithm. Its diagram
representation is shown in Fig.~\ref{fig-BP}.
We assume that the update procedure converges to a fixed point $\mu
_{ji}^{\text{new}}(S_{i})\rightarrow \mu _{ji}(S_{i})$. Sufficient
conditions for convergence of the belief-propagation algorithm to a
unique fixed point are derived in \textcite{Mooij:mk05,Ihler:ifw05}.
This fixed point determines an equilibrium state of the Ising model
on a given graph. Indeed, we can write $\mu _{ji}(S_{i})$ in a
general form as follows:
\begin{equation}
\mu _{ji}(S_{i})=\exp (\beta h_{ji}S_{i})/[2\cosh \beta h_{ji}],
\label{mu-2}
\end{equation}
where $h_{ji}$ is some parameter. Inserting Eq.~(\ref{mu-2}) into
Eq.~(\ref {mu-update}), we obtain that the fixed point equation is
exactly the recursion equation (\ref{recursion}) in the
Bethe-Peierls approach. This demonstrates a close relationship
between the belief-propagation algorithm and the Bethe-Peierls
approximation.
Local magnetic moments and the Bethe free energy are calculated in
Appendix \ref{bp-algorithm-magnetization}.

One can apply the belief-propagation algorithm to practically
arbitrary physical model with discrete (Potts states) or continues
(many component vectors) local parameters $x_{i}$. Let us introduce
local energies $E_{i}(x_{i})$ and pairwise interaction energies
$E_{ij}(x_{i},x_{j})$. A generalized fixed point equation is
\begin{equation}
A\sum_{x_{j}}e^{-\beta E_{j}(x_{j})-\beta
E_{ij}(x_{i},x_{j})}\prod_{n\in N(j)\backslash i}\mu
_{nj}(x_{j})=\mu _{ji}(x_{i}),  \label{general}
\end{equation}
where $A$ is a normalization constant. If $x_{i}$ is a continuous
variable, then we integrate over $x_{j}$ instead of summing. In
particular, one can show that for the Potts model this equation
leads to the exact recursion equation (\ref{recursion-potts}).


The belief-propagation algorithm was recently applied to study
ferro- and antiferromagnetic, and spin-glass Ising models on the
configuration model \cite {Mooij:mk05}
and the Barab\'{a}si-Albert growing network \cite{Ohkubo:oyt05}.

\subsubsection{Annealed network approach}

\label{sssec: annealed anzats}

In this subsection we describe an improved mean-field theory which
accounts for heterogeneity of a complex network.
Despite its simplicity, usually this approximation gives
surprisingly good results in the critical region.

The main idea of the annealed network approach is to replace a model
on a complex network by a model on a weighted fully connected graph.
Let us consider the Ising model Eq.~(\ref{Ising}) on a graph with
the adjacency matrix $a_{ij}$. We replace $a_{ij}$ by the
probability that vertices $i$ and $j$ with degrees $q_{i}$ and
$q_{j}$ are connected. For the configuration model, this probability
is $q_{i}$ $q_{j}/z_{1}N$, where $z_{1}=\left\langle q\right\rangle
$. We obtain the Ising model on the fully connected graph:
\begin{equation}
{\cal H}_{\text{an}}=-\frac{1}{z_{1}N}\sum_{i<j}J_{ij}q_{i}q_{j}S_{i}S_{j}-%
\sum_{i}H_{i}S_{i}.  \label{annealed}
\end{equation}
where $q_{i}$ plays the role of a ``fitness'' of vertex $i$. The
resulting fully connected graph with these inhomogeneous fitnesses
approximates the original complex network. Assuming that couplings
$J_{ij}$ are finite and using exact equation (\ref{TAP}), we find
magnetic moments:
\begin{equation}
M_{i}=\tanh \Bigl[\beta H_{i}+\frac{\beta q_{i}}{z_{1}N}%
\sum_{j=1}^{N}J_{ij}q_{j}M_{j}\Bigr].  \label{M-ann}
\end{equation}
Note that this set of equations is exact for the model
Eq.~(\ref{annealed}) in the limit $N\rightarrow \infty $. For the
ferromagnetic Ising model with $J_{ij}=J$ in zero field, i.e.,
$H_{i}=0$ for all $i$, the magnetic moment $M_{i}$ is given by
\begin{equation}
M_{i}=\tanh [\beta Jq_{i}M_{w}],  \label{M-annealed}
\end{equation}
where we introduced a weighted magnetic moment $M_{w}\equiv
(z_{1}N)^{-1}\sum_{j}q_{j}M_{j}$ which is a solution of the
equation:
\begin{equation}
M_{w}=\frac{1}{z_{1}}\sum_{q}P(q)q\tanh [\beta JqM_{w}].
\label{w-M-annealed}
\end{equation}
Equations (\ref{M-annealed}) and (\ref{w-M-annealed}) were first
derived by \textcite{Bianconi:b02} for the Barab\'{a}si-Albert
network. They give an approximate mean-field solution of the
ferromagnetic Ising model on an uncorrelated random complex network.

The effective model Eq.~(\ref{annealed}) undergoes a continues phase
transition at a critical temperature $T_{c}/J=z_{2}/z_{1}+1$ which
approaches the exact critical temperature, Eq.~(\ref{Tc-net}), at
$z_{2}/z_{1}\gg 1$. The annealed network approach gives a correct
critical behavior of the ferromagnetic Ising model. It shows that at
$T$ near $T_{c}$ the magnetic moment $M_{i}\propto q_{i}M_{w}$ for
not too large degree $q_{i}$ in agreement with the microscopic
results in Sec.~\ref {sssec:ising_transition}.
However this approach gives wrong results for a cooperative model on
an original network with $z_{2}/z_{1}\rightarrow 1$, i.e., near the
birth point of the giant connected component. It predicts a non-zero
$T_{c}$ contrary to the exact one which tends to 0. At $T=0$ the
annealed network approximation gives the average magnetic moment
$M=1$. The exact calculations in Sec.~\ref{sssec:ising_transition}
give $M<1$ due to the existence of finite clusters with zero
magnetic moment.

The annealed network approximation was used for studying
the ferromagnetic Ising model with degree-degree dependent couplings
\cite{Giuraniuc:ghi05,Giuraniuc:ghi06} and the random-field Ising
model \cite {Lee:ljn06}. In Secs.~\ref{sec:xy model} and
\ref{sec:synchronization} we apply it to the ferromagnetic $XY$ and
Kuramoto models, respectively.


\subsection{The Ising model on a regular tree}

\label{ssec:ising_perfect_tree}

The Ising model Eq.~(\ref{Ising}) on a regular tree can be solved by
using the exact recursion method developed for Bethe lattices and
Cayley trees \cite{Baxter:bbook82}. Recall that by definition, a
Cayley tree is a finite tree while a Bethe lattice is infinite (see
Sec.~\ref{ssec:cayley_versus_bethe}). We will see that, even in the
thermodynamic limit, thermodynamic properties of the ferromagnetic
Ising model on a regular Cayley tree differ from
those for
a regular
Bethe lattice.

\subsubsection{Recursion method}

Let us consider the ferromagnetic Ising model on a regular Cayley
tree (see Fig.~\ref{f1} in Sec.~\ref{ssec:cayley_versus_bethe}). Any
vertex can be considered as a root of the tree. This enables us to
write a magnetic moment $M_{i}$ of spin $i$ and the partition
function $Z$ as follows:
\begin{eqnarray}
M_{i} &=&\frac{1}{Z}\sum\limits_{S_{i}=\pm 1}S_{i}e^{\beta
H_{i}S_{i}}\prod\limits_{j\in N(i)}g_{ji}(S_{i}),  \label{M2} \\[5pt]
Z &=&\sum\limits_{S_{i}=\pm 1}e^{\beta H_{i}S_{i}}\prod\limits_{j\in
N(i)}g_{ji}(S_{i}).  \label{Z}
\end{eqnarray}
Here $g_{ji}(S_{i})$ is a partition function of subtrees growing
from vertex $j$, under the condition that the spin state $S_{i}$ is
fixed: 
\begin{equation}
g_{ji}(S_{i})=\sum_{\{S_{n}=\pm 1\}}\exp [\beta J_{ij}S_{i}S_{j}-\beta
{\cal H}_{j\backslash i}(\{S_{n}\})].  \label{gg1}
\end{equation}
Here ${\cal H}_{j\backslash i}(\{S_{n}\})$ is the interaction energy
of spins, including spin $j$, on the subtrees
except the edge $(ij)$.
The advantage of a regular tree is that we can calculate the
parameters $g_{ji}(S_{i})$ at a given vertex $i$ by using the
following recursion relation:
\begin{equation}
g_{ji}(S_{i})=\sum_{S_{j}=\pm 1}\exp [\beta J_{ij}S_{i}S_{j}+\beta
H_{j}S_{j}]\prod\limits_{m\in N(j)\backslash i}g_{mj}(S_{j}).
\label{g-recursion}
\end{equation}
Note that this equation is equivalent to Eq.~(\ref{mu-update}) at
the fixed point within the belief-propagation algorithm. In order to
shows this we introduce a parameter
\begin{equation}
h_{ji}\equiv \frac{T}{2}\ln [g_{ji}(+1)/g_{ji}(-1)]  \label{xij}
\end{equation}
and obtain $M_{i}=\tanh [(\beta H_{i}+\beta \sum_{j}h_{ji})]$.
According to the Bethe-Peierls approach in
Sec.~\ref{sssec:bethe_approach}, the parameter $h_{ji}$ has the
meaning of the additional field (message) created by vertex $j$ at
nearest neighboring vertex $i$. These fields satisfy the recursion
equation (\ref{recursion}).
In turn, Eq.~(\ref{dead-h}) determines messages which go out
boundary spins of a given tree.
Starting from the boundary spins and using Eq.~(\ref{recursion}), we
can calculate one by one all fields $h_{ij}$ on the Cayley tree and
then find thermodynamic parameters of the Ising model. %

\subsubsection{Spin correlations}

Using the recursion method, one can calculate the spin correlation
function $\left\langle S_{i}S_{j}\right\rangle $ for two spins which
are at a distance $\ell_{ij}$ from each other. We consider the
general case when couplings $J_{ij}$ on a Cayley tree are arbitrary
parameters. In zero field, $\left\langle S_{i}S_{j}\right\rangle $
is equal to a product of parameters $\tanh \beta J_{ij}$ along the
shortest path connecting $i$ to $j$:
\begin{equation}
\left\langle S_{i}S_{j}\right\rangle
=\prod\limits_{m=0}^{\ell_{ij}-1}\tanh \beta J_{m,m+1}.  \label{chi
2}
\end{equation}
Here the index\ $m$ numerates vertices on the shortest path,
$m=0,1,\,\,...\ell_{ij}$, where $m=0$ corresponds to vertex $i$ and
$m= \ell_{ij}$ corresponds to vertex $j$
\cite{Mukamel:m74,Falk:f75,Harris:h75}. This function coincides with
a correlation function of an Ising spin chain \cite{Bedeaux:bso70},
and so spin correlations on a treelike graph have a one-dimensional
character.

An even-spin correlation function $\left\langle
S_{1}S_{2}...S_{2n}\right\rangle $ can also be calculated and
presented as a product of pairwise correlation functions
\cite{Falk:f75,Harris:h75}.
Odd-spin correlation functions are zero in zero field$.$

\subsubsection{Magnetic properties}

The free energy of the ferromagnetic Ising model with $J_{ij}=J>0$
in zero magnetic field, $H=0$, on a regular Cayley tree was
calculated by \textcite{Eggarter:e74}:
\begin{equation}
F=-TL\ln [2\cosh \beta J].  \label{f-tree}
\end{equation}
where $L$ is the number of edges. Moreover this is the exact free
energy of an arbitrary tree with $L$ edges. $F$ is an analytic
function of $T$. Hence there is no phase transition even in the
limit $N\rightarrow \infty $ in contrast to a regular Bethe lattice.
A magnetization is zero at all temperatures except $T=0$.

\begin{figure}[t]
\begin{center}
\scalebox{0.27}{\includegraphics[angle=270]{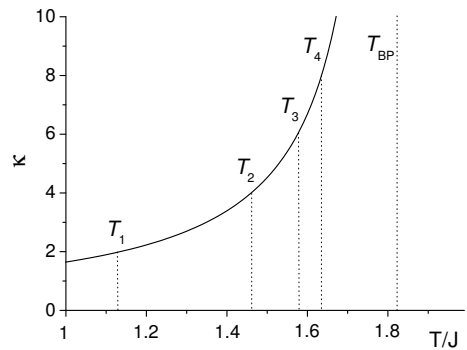}}
\end{center}
\caption{ Exponent $\protect\kappa $ versus $T$ for the
ferromagnetic Ising model on the regular Cayley tree with degree
$q=3$. The critical temperatures $T_{l}$ are shown in dotted lines.
} \label{fig-crit-index}
\end{figure}

\textcite{Muller-Hartmann:mz74} revealed that the ferromagnetic
Ising model on a regular Cayley tree with a branching parameter
$B=q-1\geq 2$ exhibits a new type of a phase transition which is
seen only in the magnetic field dependence of the free energy. The
free energy becomes a nonanalytic function of magnetic field $H>0$
at temperatures below the critical temperature $T_{\text{BP}}$ given
by Eq.~(\ref{Tc-regular}):
\begin{eqnarray}
F(T,H) &=&F(T,H=0)+\sum_{l=1}^{\infty }a_{n}(T)H^{2l}  \nonumber \\
&&+A(T)H^{\kappa }+{\cal O}(H^{\kappa +1}),  \label{F(H)}
\end{eqnarray}
where $a_{n}(T)$ and $A(T)$ are temperature dependent coefficients.
The exponent $\kappa $ depends on $T$: $\kappa =\ln B/\ln [Bt]$,
where $t\equiv \tanh \beta J$. It smoothly increases from 1 to
$\infty $ as temperature varies from 0 to $T_{\text{BP}}$, see
Fig.~\ref{fig-crit-index}. $F(T,H)$ is a continuous function of $H$
at $T=T_{\text{BP}}$. All derivatives of $F$ with respect to $H$ are
finite. Therefore, the phase transition is of the infinite order in
contrast to the second order phase transition on a regular Bethe
lattice (see Sec. \ref{sssec:bethe_approach}). With decreasing $T$
below $T_{\text{BP}}$, the singularity in $F$ is enhanced. The
leading nonanalytic part of $F$ has a form $H^{2l}\left| \ln
H\right| $ at critical temperatures $T_{l}$ given by an equation
$tB^{1-1/2l}=1$ which leads to $T_{1}<T_{2}<\ldots<T_{\infty
}=T_{\text{BP}}$.\ The zero-field susceptibility  $\chi (T)$
diverges as $(1-t^{2}B)^{-1}$ at $T=T_{1}$. Note that this
divergence does not means the appearance of a spontaneous
magnetization. Magnetic properties of the Ising model on a regular
Cayley tree were studied in
\cite{Heimburg:ht74,Matsuda:74,Falk:f75,Melin:mac96,Stosic:ssf98}.


Insight into the origin of the critical points $T_{1}$ and
$T_{\text{BP}}$ may be gained by considering local magnetic
properties of the Cayley tree. Let us apply a small local magnetic
field $\Delta H_{i}$ on a vertex $i$. Due to ferromagnetic coupling
between spins, this field induces a magnetic moment $\Delta {\cal
M}(i)=\beta V(i)\Delta H_{i}$ in a region around $i$, where $V(i)$
is a so-called correlation volume which determines a size of likely
ferromagnetic correlations around $i$ (see Sec.~\ref
{sssec:correlation_volume}).
An exact calculation of $V(i)$ shows that the correlation volume of
the central spin diverges at $T=T_{\text{BP}}$ in the infinite size
limit $N\rightarrow \infty $. The central spin has long-ranged
ferromagnetic correlations with almost all spins except for spins at
a finite distance from the boundary. The correlation volume of a
boundary spin diverges at a lower temperature
$T=T_{1}<T_{\text{BP}}$ simultaneously with the zero-field
susceptibility $\chi (T)=N^{-1}\sum_{i}\beta V(i)$. Therefore
long-ranged spin correlations cover the whole system only at
$T<T_{1}$.

\begin{figure}[t]
\begin{center}
\scalebox{0.22}{\includegraphics[angle=270]{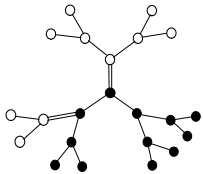}}
\end{center}
\caption{ Domains of flipped spins in the ferromagnetic Ising model
on a regular Cayley tree. Filled and open circles represent spins up
and down. Double lines shows ``frustrated'' edges connecting
antiparallel spins. } \label{fig-metastable-Cayley}
\end{figure}

A specific structure of the Cayley tree leads to the existence of
numerous metastable states \cite{Melin:mac96} which do not exist on
a Bethe lattice. These states have a domain structure (see
Fig.~\ref{fig-metastable-Cayley}) and are stable with respect to
single-spin flips. In order to reverse all spins in a large domain
it is necessary to overcome an energy barrier which is proportional
to the logarithm of the domain size. Therefore, a state with large
domains will relax very slowly to the ground state.
\textcite{Melin:mac96} found that a glassy-like behavior appears at
temperatures below a crossover temperature $T_{g}=2J/\ln [\ln N/\ln
B]$. Notice that $T_{g}\rightarrow 0$ as $N\rightarrow \infty $.
However, $T_{g}$ is finite in a finite Cayley tree. Even if $N$ is
equal to Avogadro's number $6.02\times 10^{23}$, we obtain
$T_{g}\approx
0.46J$ at $q=3$. For comparison, $T_{\text{BP}}\approx 1.8J$ and $%
T_{1}\approx 1.1J$. Large domains of flipped spins may arise at $T<T_{g}$.
This leads to a non-Gaussian form of the magnetization distribution.

\subsection{The ferromagnetic Ising model on uncorrelated networks}

\label{ssec:ferromagnetic_ising_uncorrelated}

Here we show how strong is the influence of network topology on the
critical behavior of the ferromagnetic Ising model. We will see that
when increased network heterogeneity changes the critical behavior
(the ferromagnetic phase transition becomes less sharp) and
simultaneously increases the critical temperature. We also discuss
spin correlations and finite-size effects.

\subsubsection{Derivation of thermodynamic quantities}

\label{sssec:thermodynamic_quantities}

The microscopic theory of the ferromagnetic Ising model on
uncorrelated random networks was developed by using the exact
recursion method \cite{Dorogovtsev:dgm02b}, which is equivalent to
the Bethe-Peierls approximation, and the replica trick
\cite{Leone:lvv02}.

We consider the ferromagnetic Ising model, Eq.~(\ref{Ising}), with
couplings $J_{ij}=J>0$ in uniform magnetic field $H_{i}=H$ within
the Bethe-Peierls approach (see Sec.~\ref{sssec:bethe_approach}). In
this approach, a thermodynamic state of the model is completely
described by additional fields (messages) created by spins. In a
complex network, due to intrinsic heterogeneity, the fields are
random parameters. We introduce a distribution function $\Psi (h)$
of messages $h_{ij}$: $\Psi (h)=(1/Nz_{1})\sum_{i,j}\delta
(h-h_{ij})$, where $Nz_{1}$ is the normalization constant.
If we assume the self-averageness, then, in the limit $N\rightarrow
\infty $, the average over a graph is equivalent to the average over
a statistical network ensemble.

A self-consistent equation for $\Psi (h)$ follows from the recursion
equation (\ref{recursion}) (see also Fig.~\ref{fig-recursion} in
Sec.~\ref{sssec:bethe_approach}):
\begin{eqnarray}
\!\!\!\!\!\!\!\!\!\!\!\!\!\!\!\!\!\Psi (h)
&=&\sum_{q}\frac{P(q)q}{z_{1}}\int \delta \Bigl(h-T\tanh ^{-1}\Bigl[\tanh
\beta J\times  \nonumber \\
&&\tanh \Bigl(\beta H+\beta \sum_{m=1}^{q-1}h_{m}\Bigr)\Bigr]\Bigr)\prod_{m=1}^{q-1}\Psi
(h_{m})dh_{m}.  \label{dh}
\end{eqnarray}
This equation assumes, first, that all incoming messages $\{h_{m}\}$
are statistically independent. This assumption is valid for an
uncorrelated random network. Second, an outgoing message $h$ is sent
along a chosen edge by a vertex of degree $q$ with the probability
$P(q)q/z_{1}$.
If we know $\Psi (h)$, we can find the free energy and other
thermodynamic parameters (see Appendix \ref{bethe-thermodynamic}).
For example, the average magnetic moment is
\begin{equation}
M=\sum_{q}P(q)\int \tanh \Bigl(\beta H+\beta
\sum_{m=1}^{q}h_{m}\Bigr)\prod_{m=1}^{q}\Psi (h_{m})dh_{m}.  \label{Mav}
\end{equation}
%
%
%
The replica trick \cite{Leone:lvv02} leads to the same equations
(see Appendix \ref{replica}).



\begin{table*}
\caption{ Critical behavior of the magnetization $M$, the specific
heat $\delta C$, and the susceptibility $\chi$ in the Ising model on
networks with a degree distribution $P(q)\sim q^{-\gamma }$ for
various values of exponent $\gamma $. $\tau \equiv 1-T/T_c$.
}
  \begin{ruledtabular}
    \begin{tabular}{cccc}
&
$M$ & $
\delta C(T<T_c)$ & $\chi$
\\
\hline
  $\gamma > 5$
& $ \tau^{1/2}$
&
jump at $T_c$ &
\\
  $\gamma =5$
& $
\tau^{1/2}/(\ln \tau^{-1})^{1/2}$ &
$
1/\ln \tau^{-1}$ & $
\tau^{-1}$
\\
$3<\gamma <5$
& $
\tau^{1/(\gamma -3)}$
&
$
\tau^{(5-\gamma)/(\gamma -3)}$ &
\\ \hline
 $\gamma =3$
& $
e^{-2T/\langle q \rangle}$
&
$
T^2e^{-4T/\langle q \rangle}$ &
\\
 $2<\gamma<3$
& $
T^{-1/(3-\gamma )}$ &
$
T^{-(\gamma -1)/(3-\gamma )}$
& $
T^{-1}$
\end{tabular}
  \end{ruledtabular}
\label{t1}
\end{table*}



\subsubsection{Phase transition}

\label{sssec:ising_transition}

In the paramagnetic phase at zero field $H=0$, equation (\ref {dh})
has a trivial solution: $\Psi (h)=\delta (h)$, i.e., all messages
are zero. A non-trivial solution (which describes a
ferromagnetically ordered state) appears below a critical
temperature $T_{c}$:
\begin{equation}
T_{c}=2J/\ln \Bigl(\frac{z_{2}+z_{1}}{z_{2}-z_{1}}\Bigr).  \label{Tc-net}
\end{equation}
This is the exact result for an uncorrelated random network
\cite{Dorogovtsev:dgm02b,Leone:lvv02})

The critical temperature $T_{c}$ can be found from a ``naive''
estimate. As we have noted in Sec.~\ref{sssec:bethe_approach}, the
critical temperature $T_{\text{BP}}$, Eq.~(\ref{Tc-regular}), is
determined by the branching parameter rather than the mean degree.
In a complex network, the branching parameter fluctuates from edge
to edge. The average branching parameter $B$ may remarkably differ
from the mean degree $z_{1}$.
For the configuration model, inserting the average branching
parameter $B=z_{2}/z_{1}$ into Eq.~(\ref{Tc-regular}), we obtain
Eq.~(\ref{Tc-net}). If the parameter $z_{2}$ tends to $z_{1}$, then
$T_{c}\rightarrow 0$. It is not surprising, because at the
percolation threshold we have $z_{2}=z_{1}$, and the giant connected
component disappear.

A general analytical solution of Eq.~(\ref{dh}) for the distribution
function $\Psi (h)$ is unknown.
A correct critical behavior of the Ising model at $T$ near $T_{c}$
can be found by using an ``effective medium'' approximation:
\begin{equation}
\sum_{m=1}^{q-1}h_{m}\approx (q-1)\overline{h}+{\cal O}(q^{1/2}),
\label{anzatz}
\end{equation}
where $\overline{h}\equiv \int h\Psi (h)dh$ is the average field
which can be found self-consistently \cite{Dorogovtsev:dgm02b}.
This approximation takes into account the most ``dangerous'' highly
connected spins in the best way. The ansatz (\ref{anzatz}) is
equivalent to the approximation $\Psi (h)\sim \delta
(h-\overline{h})$ \cite{Leone:lvv02}. At lower temperatures a
finite width of $\Psi (h)$ becomes important.

\begin{figure}[t]
\begin{center}
\scalebox{0.35}{\includegraphics[angle=0]{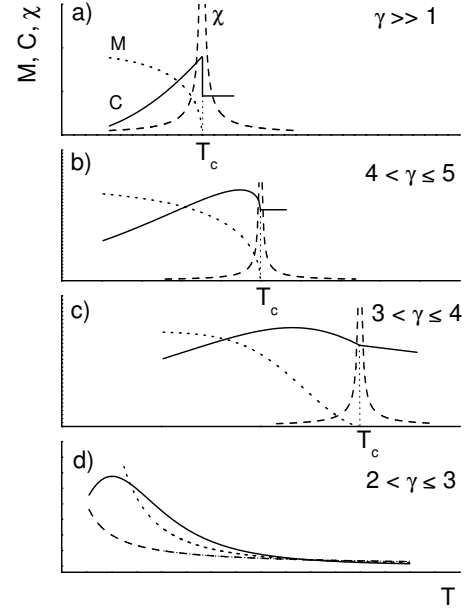}}
\end{center}
\caption{ Schematic representation of the critical behavior of the
magnetization $M$ (dotted lines), the magnetic susceptibility
$\protect\chi $ (dashed lines), and the specific heat $C$ (solid
lines) for the ferromagnetic Ising model on uncorrelated random
networks with a degree distribution $P(q)\sim q^{-\protect\gamma }$.
(a) $\protect\gamma \gg 1$, the standard mean-field critical
behavior. A jump of $C$ disappears when $\gamma \rightarrow 5$. (b)
$4<\protect\gamma \leqslant 5$, the ferromagnetic phase transition
is of second order. (c) $3<\protect\gamma \leqslant 4$, the
transition becomes of higher order. (d) $2<\protect\gamma \leqslant
3$, the transition is of infinite order, and $T_{c}\rightarrow
\infty $ as $N\rightarrow \infty $. } \label{fig-critical-behavior}
\end{figure}

The ferromagnetic Ising model on uncorrelated random networks
demonstrates three classes of universal critical behavior:
\begin{itemize}

\item[(i)] the standard mean-field critical behavior in networks with a
finite fourth moment $\left\langle q^{4}\right\rangle$ (scale-free
networks with the degree distribution exponent $\gamma
>5$);

\item[(ii)] a critical behavior with non-universal critical exponents depending on a
degree distribution in networks with divergent $\left\langle
q^{4}\right\rangle$, but a finite second moment $\left\langle
q^{2}\right\rangle $ (scale-free networks with $3<\gamma \leq 5$);

\item[(iii)] an infinite order phase transition in networks with a
divergent second moment $\left\langle q^{2}\right\rangle$, but a
finite mean degree $\left\langle q\right\rangle$ (scale-free
networks with $2<\gamma \leq 3$).

\end{itemize}
The corresponding critical exponents $(M\sim \tau ^{\beta }$,
$\delta C\sim \tau ^{-\alpha }$, $\chi \sim \tau ^{-\tilde{\gamma
}})$ are reported in the Table \ref{t1}. The evolution of the
critical behavior with increasing heterogeneity is shown
schematically in Fig.~\ref{fig-critical-behavior}.
Notice that the Ising model on a regular random network
demonstrates the standard mean-field critical behavior in the
infinite size limit \cite{Scalettar:s91}. The corresponding exact
solution
is given
in Sec.~\ref{ssssec: regular Bethe}.

The conventional scaling relation between the critical exponents
takes place at $\gamma >3$:
\begin{equation}
\alpha +2\beta +\tilde{\gamma} =2.  \label{scaling}
\end{equation}
Interestingly, the magnetic susceptibility $\chi$ has a universal
critical behavior with the exponent $\tilde{\gamma}=1$ when
$\left\langle q^{2}\right\rangle <\infty$, i.e., at $\gamma
>3$. This agrees with the scaling relation $\tilde{\gamma} / \nu=2-\eta$
if we insert the standard mean-field exponents: $\nu=1/2$ and the
Fisher exponent $\eta=0$ (see Sec.~\ref{ssec:finite_size_scaling}).
When $2<\gamma \leq 3$, the susceptibility $\chi $\ has a
paramagnetic temperature dependence, $\chi \propto 1/T$, at
temperatures $T\gtrsim J$ despite the system is in the ordered
state.

\begin{figure}[t]
\begin{center}
\scalebox{0.3}{\includegraphics[angle=270]{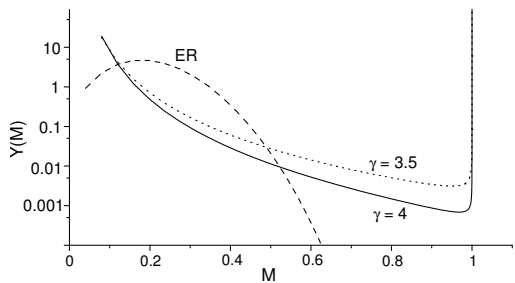}}
\end{center}
\caption{ Distribution function $Y(M)$\ of magnetic moments $M$ in
the ferromagnetic Ising model on the Erd\H{o}s-R\'{e}nyi graph with
mean degree $z_{1}=$ $5$ (dashed line) and scale-free graphs with
$\protect\gamma =4$ and 3.5 (solid and dotted lines) at $T$ close
$T_{c}$, $\protect\beta \overline{h}=0.04$.}
\label{fig-M-distribution}
\end{figure}

At $T<T_{c}$ the ferromagnetic state is strongly heterogeneous
because the magnetic moment $M_{i}$ fluctuates from vertex to
vertex. The ansatz Eq.~(\ref{anzatz}) enables us to find an
approximate distribution function of $M_{i}$:
\begin{equation}
Y(M)\equiv \frac{1}{N}\sum_{i=1}^{N}\delta (M-M_{i})\approx \frac{P(q(M))}{%
\beta \overline{h}(1-M^{2})},  \label{Y(M)-approx}
\end{equation}
where the function $q(M)$ is a solution of an equation $M(q)=\tanh [\beta
\overline{h}q]$.
Near $T_{c}$, low-degree vertices have a small magnetic moment,
$M(q)\sim q\left| T_{c}-T\right| ^{1/2}\ll 1$, while hubs with
degree $q>T/\overline{h}\gg 1$ have $M(q)\sim 1$.
The function $Y(M)$ is shown in Fig.~\ref{fig-M-distribution}. Note
that the distribution of magnetic moments in scale-free networks is
more inhomogeneous than in the Erd\H{o}s-R\'{e}nyi graphs. Moreover
$Y(M)$ diverges at $M\rightarrow 1$.
A local magnetic moment depends on its neighborhood. In particular,
a magnetic moment of a spin, neighboring a hub may differ from a
moment of a spin surrounded by low-degree vertices. Studies of these
correlations are at the very beginning \cite {Giuraniuc:ghi06}.

In the ground state $(T=0$, $H=0)$, an exact distribution function
$\Psi (h) $ converges to a function with two delta peaks:
\begin{equation}
\Psi (h)=x\delta (h)+(1-x)\delta (h-J),  \label{psi-T0}
\end{equation}
where the parameter $x$ is determined by an equation describing
percolation in networks (see
Sec.~\ref{sssec:giant_connected_component}). Equation (\ref{psi-T0})
tells us that in the ground state, spins, which belong to a finite
cluster, have zero magnetic moment while spins in a giant connected
component have magnetic moment 1 because non-zero fields acts on
these spins.
The average magnetic moment is $M=
1-\sum_{q}P(q)x^{q}$. This is exactly the size of the giant connected
component of the network.

\subsubsection{Finite-size effects}

\label{sssection:finite_ size_Ising}

When $2<\gamma \leqslant 3$, a dependence of $T_{c}$ on the size $N$
is determined by the finite-size cutoff $q_{\text{cut}}(N)$ of the
degree distribution in Sec.~\ref{sssec:cutoffs}. We obtain
\begin{eqnarray}
&&T_{c}(N)\approx \frac{z_{1}\ln N}{4},\ \ \ \ \ \ \ \
\ \ \ \ \ \ \ \ \ \ \ \mbox{at}\ \gamma =3,  \label{1} \\
&&T_{c}(N)\approx \frac{(\gamma -2)^{2}z_{1}q_{\text{cut}}^{3-\gamma
}(N)}{(3-\gamma )(\gamma -1)},\ \ \ \ \mbox{at}\ 2<\gamma <3
\label{2}
\end{eqnarray}
\cite{Dorogovtsev:dgm02b,Leone:lvv02,Bianconi:b02}. These estimates
agree with the numerical simulations of
\textcite{Aleksiejuk:ahs02,Herrero:h04}.
Notice that Herrero used the cutoff $q_{\text{cut}}(N)\sim
N^{1/\gamma }$ which leads to $T_{c}\sim N^{z}$ with the exponent
$z=(3-\gamma )/\gamma $.


\subsubsection{Ferromagnetic correlations}

\label{sssec:correlation_volume}

Let us consider spin correlations in the ferromagnetic Ising model
in the paramagnetic state. Recall that the correlation length $\xi $
of spin correlations in the Ising model on a finite-dimensional
lattice diverges at a critical point of a continuous phase
transition. In contrast, in an uncorrelated random complex network,
the correlation length $\xi $ is finite at any temperature.
Indeed, according to Eq.~(\ref{chi 2}), the correlation function
$C(\ell )=\left\langle S_{i}S_{j}\right\rangle $ decays
exponentially with distance $\ell \equiv\ell_{ij}$: $C(\ell
)=e^{-\ell /\xi }$, where the coherence length $\xi \equiv 1/\left|
\ln \tanh \beta J\right|\neq0 $ at non-zero temperature. Moreover,
spin correlations have a one-dimensional character despite a complex
network is an infinite-dimensional system. Strictly speaking, this
is valid
at distances $\ell <$ $\overline{\ell }(N)\sim \ln N$ when a network is
treelike.

In complex networks, the so called {\em correlation volume} rather
than $\xi $ plays a fundamental role (see also
Sec.~\ref{sssec:statistics_finite}). We define a correlation volume
$V(i)$ around a spin $i$ as follows:
\begin{equation}
V(i)\equiv \sum_{j=1}^{N}a_{ij} \left\langle S_{i}S_{j}\right\rangle
. \label{corr-vol}
\end{equation}
It determines the size of likely ferromagnetic fluctuations around
the spin. In the paramagnetic phase, $V(i)$ is expressed through
local network characteristics: $V(i)=\sum_{\ell =0}^{\infty }z_{\ell
}(i)t^{\ell }$, where $t\equiv\tanh \beta J$, and $z_{\ell }(i)$ is
the number of vertices which are at a distance $\ell $ from vertex
$i$, and $z_{0}(i)\equiv 1$.

It is obvious that a correlation volume around a high degree vertex
(hub) is larger than the one around a poorly connected vertex. In a
scale-free network, hubs may form a highly connected cluster (the
rich-club phenomenon \cite{Zhou:zm04,Colizza:cfsv06}). A region of
likely ferromagnetic correlations around the rich-club may be very
large. It grows with decreasing $T$, absorbing small clusters of
correlated spins.
The Erd\H{o}s-R\'{e}nyi network is more homogeneous than a
scale-free network with the same average degree. At high
temperatures there are many small clusters of ferromagnetically
correlated spins. With decreasing $T$ small clusters merge together,
forming larger clusters.

The average correlation volume $\overline{V}$ is related with the
total magnetic susceptibility:
\begin{equation}
\overline{V}\equiv \frac{1}{N}\sum\limits_{i=1}^{N}V(i)=\sum_{\ell
=0}^{\infty }z_{\ell }t^{\ell }=T\chi,  \label{m-corr-v}
\end{equation}
where $z_{\ell }$ is the average number of $\ell $-th nearest
neighbors of a vertex on a given network:
$z_{\ell}=N^{-1}\sum_{i}z_{\ell }(i)$. The average correlation
volume $\overline{V}$ diverges as $\ln N$ in the critical point of a
continuous phase transition. The condition of divergence of the
series in Eq.~(\ref{m-corr-v}) leads to the equation: $B \tanh
\beta_{c}J=1$, where $B\equiv \lim_{\ell \rightarrow \infty
}\lim_{N\rightarrow \infty }[z_{\ell }]^{1/\ell }$ is the average
branching parameter of the network. This criterion for the critical
point is valid for any treelike network \cite{Lyons:l89}, including
networks with degree-degree correlations, growing networks etc.
$B=1$ corresponds to the point of the birth of the giant connected
component. At $B<1$ a network consists of finite clusters, the
correlation volume is finite at all $T$, and there is no phase
transition.

Using Eq.~(\ref{m-corr-v}), we can calculate $\chi $ in the
paramagnetic phase. In the configuration model of uncorrelated
random networks, we have $z_{\ell }=z_{1}(z_{2}/z_{1})^{\ell -1}$.
This gives
\begin{equation}
T\chi =\overline{V}=1+\frac{z_{1}t}{1-z_{2}t/z_{1}}.  \label{chi
t-av}
\end{equation}
So $\chi $ diverges as $\left| T-T_{c}\right| ^{-1}$.

Equation (\ref{chi 2}) for the correlation function $\left\langle
S_{i}S_{j}\right\rangle $ is not valid for scale-free networks with
$2<\gamma <3$ due to numerous loops. How do spin correlations decay
in this case?
\textcite{Dorogovtsev:dgm05} found that in these networks the pair
correlation function $\left\langle S_{i}S_{j}\right\rangle $ between
the second and more distant neighbors vanishes in the limit
$N\rightarrow \infty $. Only pair correlations between nearest
neighbors are observable in this limit.


\subsubsection{Degree-dependent interactions}

\label{sssec:degree_dependent_interactions}

\textcite{Giuraniuc:ghi05,Giuraniuc:ghi06} studied analytically and
numerically a ferromagnetic Ising model on a scale-free complex
network with a topology dependent coupling: $J_{ij}=Jz_{1}^{2\mu
}(q_{i}q_{j})^{-\mu }$, where a constant $J>0$, $\mu $ is a tunable
parameter, $q_{i}$ and $q_{j}$ are degrees of neighboring vertices
$i$ and $j$. The authors demonstrated that the critical behavior of
the model on a scale-free network with degree distribution exponent
$\gamma$ is equivalent to the critical behavior of the ferromagnetic
Ising model with a constant coupling $J$ on a scale-free network
with renormalized degree distribution exponent $\gamma '=(\gamma
-\mu )/(1-\mu ).$ Therefore the critical exponents can be
obtained, replacing $\gamma $ by $\gamma '$ in Table \ref{t1}.
Varying $\mu $ in range $[2-\gamma ,1]$ allows us to explore the
whole range of the universality classes represented in Table
\ref{t1}. For example, the ferromagnetic Ising model with $J_{ij}=J$
on a scale-free network with $\gamma =3$ undergoes an infinite order
phase transition while the model with the degree dependent coupling
for $\mu =1/2$ undergoes a second order phase transition with the
critical behavior corresponding to $\gamma '=5.$

\subsection{The Ising model on small-world networks}

\label{ssec:Ising_small-world}

The phase transition in the ferromagnetic Ising model on
small-world networks strongly resembles that in the percolation
problem for these nets, Sec.~\ref{ssec:percolation_small-world}.
This system was extensively studied by \textcite{Barrat:bw00},
\textcite{Gitterman:g00}, \textcite{Pekalski:p01}, and many other
researchers. Here we mostly discuss small-world networks based on
one-dimensional lattices, with a fraction $p$ of shortcuts. Let us
estimate the critical temperature $T_c(p)$ assuming for the sake
of simplicity only nearest-neighbor interactions in the
one-dimensional lattice. The reader may easily see that if $p$ is
small, this network has a locally tree-like structure. At small
$p$, the mean branching parameter in this graph is
$B=1+cp+O(p^2)$, where $c$ is some model dependent constant.
Substituting $B$ into Eq.~(\ref{Tc-regular}), we arrive at
\begin{equation}
T_{c}(p)\sim \frac{J}{\left| \ln p\right| },  \label{Tc-p}
\end{equation}
where $J$ is the ferromagnetic coupling. \textcite{Barrat:bw00}
arrived at this result using the replica trick. Exact calculations
of \textcite{Lopes:lpd04} confirmed this formula.

Far from the critical temperature, the thermodynamic quantities of
this system are close to those of the $d$-dimensional substrate
lattice. However, in the vicinity of the critical temperature the
ordinary mean-field picture is valid. Two circumstances naturally
explain these traditional mean-field features. (i) In the
interesting range of small $p$, the small-world networks effectively
have a locally tree-like structure (short loops due the lattice are
not essential). (ii) The small-world networks have rapidly
decreasing degree distributions. As we have explained, this
architecture leads to the traditional mean-field picture of critical
phenomena. The
region of temperatures around $T_c(p)$, where this mean-field
picture is realized, is narrowed as $p$ decreases.
\textcite{Lopes:lpd04} obtained the specific heat as a function of
temperature and $p$ and showed that its jump at the critical point
approaches zero as $p\to 0$.
\textcite{Roy:rb06} demonstrated numerically that the Ising model
on the Watts-Strogatz network is self-averaging in the limit
$N\rightarrow \infty$, i.e., the average over this ensemble is
equivalent to the average over a single Watts-Strogatz network.
With increasing
network size $N$, the distributions of the
magnetization, the specific heat and the critical temperature of
the Ising model in the ensemble of different realizations of the
Watts-Strogatz network approach the $\delta$-function. The size
dependence of these parameters agrees with the finite scaling
theory in Sec.~\ref{ssec:finite_size_scaling}.

\textcite{Hastings:h03} investigated the Ising model on the
$d$-dimensional small-world. He found that for any $d$, the shift of
the critical temperature is $T_c(p)-T_c(p=0) \sim
p^{1/\tilde{\gamma}}$, where $\tilde{\gamma}$ is the susceptibility
exponent at $p=0$, $\chi(T,p=0)\sim |T-T_c(0)|^{-\tilde{\gamma}}$.
Compare this shift with the similar shift of the percolation
threshold in the same network,
Sec.~\ref{ssec:percolation_small-world}. Simulations of
\textcite{Herrero:h02,Zhang:zn06} confirmed this prediction.

In their simulations, \textcite{Jeong:jhk03} studied the Ising model
with specific interactions placed on the ordinary one-dimensional
small-world network. In their system the ferromagnetic interaction
between two neighboring spins, say, spins $i$ and $j$, is
$|i-j|^{-\alpha}$. $|i-j|$ is a distance measured along the chain.
Surprisingly, a phase transition was revealed only at $\alpha=0$, no
long-range order for $\alpha>0$ was observed at any non-zero
temperature. \textcite{Chatterjee:cs06} performed numerical
simulations of the ferromagnetic Ising model placed on a
one-dimensional small-world network, where vertices, say, $i$ and
$j$, are connected by a shortcut with probability $\sim
|i-j|^{-\alpha}$ (Kleinberg's network, see
Sec.~\ref{ssec:small-world_networks}). They observed a phase
transition at least at $\alpha<1$. In both these studies, the small
sizes of simulated networks made difficult to arrive at reliable
conclusions. On the other hand, these two systems were not studied
analytically.

\subsection{Spin glass transition on networks}

\label{ssec:spin_glass}

Despite years of efforts, the understanding of spin glasses is still
incomplete.
The nature of the spin-glass state is well understood for the
infinite-range Sherrington-Kirkpatrick model
\cite{Mezard:mpv87,Binder:by86}.
The basic property of the spin-glass model is that a huge number of
pure thermodynamic states with non-zero local magnetic moments
$M_{i}$ spontaneously emerge below a critical temperature. This
corresponds to the replica symmetry breaking.

Investigations of a spin-glass Ising model on treelike networks
began very soon after the discovery of spin glasses.
\textcite{Viana:vb85} proposed the so called dilute Ising
spin-glass model which is equivalent to the Ising model on the Erd\H{o}%
s-R\'{e}nyi graph (the reader will find a review of early
investigations in \textcite{Mezard:mp01}). Most of
studies
considered a spin-glass on random regular and
Erd\H{o}s-R\'{e}nyi networks. A spin-glass on the Watts-Strogatz
and scale-free networks
only recently drew attention.

Here we first review recent studies of the spin-glass Ising model on
complex networks. Then we consider a pure antiferromagnetic Ising
model, which becomes a spin-glass when placed on a complex network,
and discuss relationships of this model with famous NP-complete
problems (MAX-CUT and vertex cover).

\subsubsection{The Ising spin glass}

\label{sssec:ising_spin_glass}

The spin-glass state arises due to frustrations. The nature of
frustrations in the Sherrington-Kirkpatrick model and a spin-glass
model on a finite dimensional lattice is clear.
On the other hand,
for
an uncorrelated random network, the nature of frustrations is not
so clear because such a network has a treelike structure in the
thermodynamic limit.
How do frustrations appear in this case? In order to answer this
question we recall that locally tree-like networks usually have
numerous long loops of typical length $O(\ln N)$, see Sec.~\ref{ssec:loops}. It turns out that
frustrations in a network are
due to
these long loops.

Two main methods are used to study the spin-glass Ising model on a
random network. These are the replica trick and the cavity method
\cite{Mezard:mp01}.
Early investigations of a spin glass on a Bethe lattice assumed that
there is only one pure thermodynamic state, and the replica symmetry
is unbroken. This assumption led to unphysical results such as, for
example, a negative specific heat.
The order parameter of the Sherrington-Kirkpatrick model is an
overlap $\left\langle S_{\alpha }S_{\beta }\right\rangle $ between
spins of two replicas $\alpha $ and $\beta $. The spin-glass\ Ising
model on a random network requires multi-spin overlaps $\left\langle
S_{\alpha }S_{\beta }S_{\gamma }\right\rangle ,\left\langle
S_{\alpha }S_{\beta }S_{\gamma }S_{\delta }\right\rangle $ and
higher \cite{Viana:vb85,Mottishaw:m87,Goldschmidt:gd90,Kim:krkk05}.
This makes this model more complex.

Many evidences have been accumulated indicating that a spin-glass
state may exist in the spin-glass Ising model on a Bethe lattice
\cite{Mezard:mp01}.
It means that this model has many pure thermodynamic states at low
temperatures. In order to obtain a complete description of a
spin-glass state it is necessary to solve the recursion equations
(\ref{recursion}) and find the distribution function $\Psi _{\alpha
}(h)$
of the messages for every pure state $\alpha$. It is a difficult
mathematical problem which is equivalent to search for a solution
with replica symmetry breaking.
In order to find an approximate solution, a one step
replica-symmetry breaking approximation was developed
\cite{Mezard:mp01,Pagnani:ppr03,Castellani:ckr05}. This
approximation assumes that a space of pure states has a simple
cluster structure (a set of clusters).
Numerical simulations of the spin-glass Ising model on a random
regular network demonstrated that this approximation gives better
results than the replica symmetric solution. A similar result was
obtained for the Watts-Strogatz network
\cite{Nikoletopoulos:nccshw04}. Unfortunately the space of pure
states is probably more complex, and a solution with a complete
replica symmetry breaking is necessary.

The phase diagram of the Ising spin glass on the Erd\H{o}s-R\'{e}nyi
graphs was studied by
\textcite{Kanter:ks87,Kwon:kt88,Castellani:ckr05}, and
\textcite{Hase:has06}. The diagram looks like the phase diagram of
the Sherrington-Kirkpatrick model. The exact critical temperature of
the spin-glass transition, $T_{\text{SG}}$, on a treelike complex
network can be found without the replica trick. The criterion of
this transition is the divergence of the spin-glass susceptibility:
\begin{equation}
\chi _{\text{SG}}=\frac{1}{N}\sum\limits_{i=1}^{N}\sum\limits_{j=1}^{N}\left%
\langle S_{i}S_{j}\right\rangle ^{2}.  \label{chi-SG-1}
\end{equation}
Using Eq.~(\ref{chi 2}) for the correlation function $\left\langle
S_{i}S_{j}\right\rangle $, we find that $\chi _{\text{SG}}$ diverges
at a critical temperature $T_{\text{SG}}$ determined by the
following equation:
\begin{equation}
B\int \tanh ^{2}(\beta _{\text{SG}}J_{ij})P(J_{ij})dJ_{ij}=1,  \label{Tc-SG}
\end{equation}
where $B$ is the average branching parameter.

If the distribution function $P(J_{ij})$ is asymmetric, and the mean
coupling $\overline{J}=\int J_{ij}P(J_{ij})dJ_{ij}$ is larger than a
critical value, then a ferromagnetic phase transition occurs at a
higher critical temperature $T_{c}$ than $T_{\text{SG}}$. The
criterion of the ferromagnetic phase transition is the divergence of
the magnetic susceptibility $\chi $:
\begin{equation}
B\int \tanh (\beta _{c}J_{ij})P(J_{ij})dJ_{ij}=1.  \label{P-F-Tc}
\end{equation}
In a multicritical point, we have $T_{c}=T_{\text{SG}}$. Equations
(\ref{Tc-SG})--(\ref{P-F-Tc}) generalize the results obtained by
the replica trick and others methods for regular random graphs, the Erd\H{o}%
s-R\'{e}nyi networks, and the static and configuration models of
uncorrelated complex networks
\cite{Viana:vb85,Thouless:t86,Baillie:bjj95,Kim:krkk05,Mooij:mk05,Ostilli:o06a,Ostilli:o06b}.

It is well-known that if $\overline{J}$ exceeds a critical value,
the Sherrington-Kirkpatrick model at low temperatures undergoes a
phase transition from a ferromagnetic state into a so called {\em
mixed} state in which ferromagnetism and spin-glass order coexist.
The coexistence of ferromagnetism and spin-glass order in the
spin-glass Ising model on a random regular graph with degree $q$ was
considered by \textcite{Liers:lphj03,Castellani:ckr05}.
\textcite{Castellani:ckr05} studied a zero-temperature phase diagram
of the spin-glass Ising model with a random coupling $J_{ij}$ which
takes values $\pm J$ with probabilities $(1\pm \rho )/2$. They found
that at $\rho $ exceeding a critical value $\rho _{c}(q)$ the
spin-glass Ising model is in a replica symmetric ferromagnetic
state. For $\rho <\rho _{c}(q)$, the replica symmetric state becomes
unstable. The system goes into a mixed state with a broken replica
symmetry. In particular, for degree $q=3$, $ \rho _{c}(q=3)=5/6$. At
$q\gg 1$, $ \rho _{c}(q)\sim \ln q/\sqrt{q}$. The one-step symmetry
breaking solution showed that the mixed state exists in a range
$\rho _{\text{F}}<\rho <\rho _{c}(q)$. At $\rho <\rho _{\text{F}}$
the ground state is a nonmagnetic spin-glass state.
\textcite{Liers:lphj03} studied numerically a spin-glass model with
a Gaussian coupling $J_{ij}$. They did not observe a mixed state in
contrast to \textcite{Castellani:ckr05}.

A strong effect of the network topology on the spin-glass phase
transition was recently revealed by \textcite{Kim:krkk05} in the
Ising spin-glass model with $J_{ij}=\pm J$ on an uncorrelated scale
free network. These authors used a replica-symmetric
perturbation approach of \textcite{Viana:vb85}. It turned out that
in a scale-free network with $3<\gamma <4$, the critical behavior of
the spin-glass order parameter at $T$ near $T_{\text{SG}}$
differs from the critical behavior of the Sherrington-Kirkpatrick
model and depends on $\gamma$. For the paramagnetic-ferromagnetic
phase transition, a deviation from the standard critical behavior
takes place at $\gamma <5$ similarly to the ferromagnetic Ising
model in Sec.~\ref{sssec:ising_transition}. Critical temperatures of
the ferromagnetic and spin-glass phase transitions approach infinity
in the thermodynamic limit at $2<\gamma <3$. These transitions
become of infinite order.

\subsubsection{The antiferromagnetic Ising model and MAX-CUT problem}

\label{sssec:antiferromagnetic_ising}

The antiferromagnetic (AF) Ising model becomes non-trivial on a
complex network. As we will see, the model is a spin glass. We here
also discuss a mapping of the ground state problem onto the MAX-CUT
problem.

Consider the pure AF model on a graph:
\begin{equation}
E=\frac{J}{2}\sum_{i,j}a_{ij}S_{i}S_{j},  \label{AF-Ising}
\end{equation}
where $J>0$. The search for the ground state is equivalent to
coloring a graph in two colors (colors correspond to spin states
$S=\pm 1$) in such a way that no two adjacent vertices have the same
color. Let us first consider a bipartite network, that is a network
without odd loops. It is obvious that this network is 2-colorable.
The ground state energy of the AF model on a bipartite graph is
$E_{0}=-JL$, where $L$ is the total number of edges in the graph. An
uncorrelated complex network with a giant connected component cannot
be colored with 2 colors due to numerous odd loops. So the ground
state energy, $E_{0}$, of the AF model on a random graph is higher
than $-JL$ due to frustrations produced by odd loops.

\begin{figure}[t]
\begin{center}
\scalebox{0.27}{\includegraphics[angle=270]{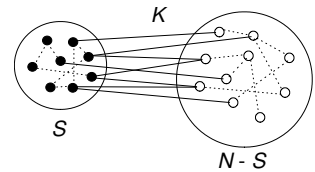}}
\end{center}
\caption{
Partition of vertices in a graph into two sets consisting of $S$
and $N-S$ vertices, and $K$ edges (solid lines) in the cut. Dotted lines
show edges inside the sets.}
\label{fig-max-cut}
\end{figure}

The ground state problem can be mapped to the MAX-CUT problem which
belongs to the class of NP-complete optimization problems. Let us
divide vertices of a graph (of $N$ vertices and $L$ edges) into two
sets in such a way that the number $K$ of edges which connect these
sets is maximum, see Fig.~\ref{fig-max-cut}. If we define spins at
vertices in one set as spins up and spins in the other set as spins
down, then the maximum cut gives a minimum energy $E_{0}$ of the AF
model. Indeed, $K$ edges between two sets connect antiparallel spins
and give a negative contribution $-JK$ into $E_{0}$. The remaining
$L-K$ edges connect parallel spins and give a positive contribution
$J(L-K)$. The ground state energy, $E_{0}=J(L-2K)$, is minimum when
$K$ is maximum.

The maximum cut of the Erd\H{o}s-R\'{e}nyi graph with high
probability is
\begin{equation}
K_{c}\equiv \max K=L/2+AN\sqrt{z_{1}}+o(N)  \label{max-cut}
\end{equation}
for mean degree $z_{1}>>1$ \cite{Kalapala:km02,Coppersmith:cghs04}.
Here $A$ is a constant with lower and upper bounds $0.26<A<\sqrt{\ln
2}/2\approx 0.42$. Recall that $L=z_{1}N/2$. The estimation of
$K_{c}$ is given in Appendix \ref{max-cut-ER}. Thus the ground state
energy is
\begin{equation}
E_{0}/N=-2JA\sqrt{z_{1}}.  \label{E0}
\end{equation}
The fraction of ``frustrated'' edges, i.e., edges which connect
``unsatisfied'' parallel spins, is $(L-K_{c})/L=1/2-2A/\sqrt{z_{1}}$.
Thus almost half of edges are frustrated. We found that this result
is valid not only for classical random graphs but also for arbitrary
uncorrelated random network.

Interestingly, the lower bound of the ground state energy
Eq.~(\ref{E0}) is quite similar to the lower bound for the ground
state energy of the random energy model introduced by
\textcite{Derrida:d81}. This model approximates to spin-glass in any
dimensions. Replacing the mean degree $z_{1}$ in Eq.~(\ref{E0}) by
degree of a $D$-dimensional cubic lattice, $2D$, we obtain the
ground state energy of Derrida's model: $E_{0}/N=-J\sqrt{2D\ln 2}$.
(We are grateful to M. Ostilli for attracting our attention to this
fact.)

Despite the seeming simplicity, the pure AF model on complex
networks is not well studied yet. We assume that this model is the
usual spin glass. On the other hand, the analysis of
\textcite{Mooij:mk05} revealed an antiferromagnetic phase transition
in the model on an uncorrelated random network at a critical point
$(z_{2}/z_{1})\tanh \beta J=1$, i.e., at the critical temperature
$T_{\text{BP}}$ in Eq.~(\ref{Tc-net}). If this result is correct,
then, as temperature  decreases, the AF model may undergo a phase
transition from an antiferromagnetic state into a spin-glass state.
The structure of pairwise spin correlations in this system is
non-trivial. The correlations between two spins separated by
distance $\ell$ are characterized by their average value
$\overline{C}(\ell )=z_{\ell }^{-1}\sum_{ij}\left\langle
S_{i}S_{j}\right\rangle \delta_{\ell ,\ell_{ij}}$. We expect that at
least for locally tree-like networks, the spin correlations are
antiferromagnetic at all distances smaller than the mean intervertex
separation $\overline{\ell}(N)$. These correlations should be
present in the spin-glass phase and even in some range of
temperatures above the spin-glass transition. Antiferromagnetic
correlations of this kind were observed in numerical simulations by
\textcite{Bartolozzi:bslw05}.

\textcite{Holme:hlek03} used the antiferromagnetic Ising model to
study the bipartivity of real-world networks (professional
collaborations, on-line interactions and so on). In order to measure
the bipartivity, they proposed to put the AF model on the top of the
network and calculate a fraction of edges between spins with
opposite signs in the ground state. We have explained that this
procedure is equivalent to finding the maximum cut of the graph. The
larger is this fraction the closer is the network to bipartite.
Measuring bipartivity allows one to reveal the bipartite nature of
seemingly one-partite networks. Note that only their one-mode
projections are usually studied, while most of real-world networks
are actually multipartite.

\subsubsection{Antiferromagnet in a magnetic field, the hard-core gas model, and vertex covers}
\label{ssec:vertex_cover}

Here  we discuss relations between an antiferromagnetic Ising model,
the hard-core gas model and the vertex cover problem on classical
random graphs. On complex networks these problems are
poorly studied.

\paragraph{The vertex cover problem.}

\begin{figure}[t]
\begin{center}
\scalebox{0.22}{\includegraphics[angle=270]{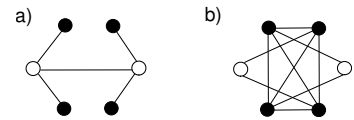}}
\end{center}
\caption{ Vertex cover of a graph. a) Open circles form a minimum
vertex cover of the graph. Every edge has at least one endpoint
which belongs to the vertex cover. The closed circles form the
maximum independent set of the graph. b) The complement of the same
graph (we add the missing edges and remove the already existing
edges). Closed circles form the maximum clique.}
\label{fig-vertex-cover}
\end{figure}

This problem is one of the basic NP-complete optimization problems.
A {\em vertex cover} of a graph is a set of vertices with the
property that every edge of the graph has at least one endpoint
which belongs to this set. In general, there are many different
vertex covers of a graph. We look for a vertex cover of a minimum
size, see Fig.~\ref{fig-vertex-cover}.
\textcite{Weigt:wh00} proposed a vivid picture for this problem:
``Imagine you are director of an open-air museum situated in a large
park with numerous paths. You want to put guards on crossroads to
observe every path, but in order to economize cost you have to use
as few guards as possible.''


Let us find size of a minimum vertex cover of the
Erd\H{o}s-R\'{e}nyi graph of $N$ vertices, $L=z_{1}N/2$ edges and
mean degree $z_{1}$.
We denote the number of vertices in a vertex cover as $N_{vc}=xN$.
The parameter $x$ can be interpreted as the probability that a
randomly chosen vertex is covered, i.e., it belongs to the vertex
cover.
An edge can be between every pair of vertices with the same
probability. So the probability that a randomly chosen edge connects
two vertices which do not belong to the vertex cover is $(1-x)^{2}$.
With the conjugate probability $1-(1-x)^{2}=2x(1-x)$, an edge has at
least one covered
endpoint. There are $%
{N \choose N_{vc}}%
$ ways to choose $N_{vc}$ vertices from $N$ vertices.
Only a small fraction of the partitions, $[2x(1-x)]^{L}$, are vertex
covers. Thus the number of possible vertex covers is
\begin{equation}
{\frak N}_{vc}(x)=%
{N \choose N_{vc}}%
[2x(1-x)]^{L}\equiv e^{N\Xi (x)}.  \label{vc1}
\end{equation}
Using the estimate Eq.~(\ref{binomial}), we obtain
\begin{equation}
\Xi (x)=-(1-x)\ln (1-x)-x\ln x+\frac{z_{1}}{2}\ln [2x(1-x)].  \label{vc2}
\end{equation}
The threshold fraction $x_{c}$ is determined by the condition: $\Xi
(x_{c})=0 $. It gives $x_{c}(z_{1})\approx 1-2\ln z_{1}/z_{1}+O(\ln
\ln z_{1})$ at $z_{1}\gg 1$. The exact asymptotics was found by
\textcite{Frieze:f90}:
\begin{equation}
x_{c}(z_{1})=1-\frac{2}{z_{1}}(\ln z_{1}-\ln \ln z_{1}-\ln 2+1)+o(\frac{1}{%
z_{1}}).  \label{vc-threshold}
\end{equation}
At $x<x_{c}$,\ with high probability there is no vertex cover of
size $xN<x_{c}N$, while at $x>x_{c}$ there are exponentially many
different covers of size $xN>x_{c}N$. The appearance of many vertex
covers looks like a phase transition which occurs at the threshold
parameter $x=x_{c}$.

The exact threshold $x_{c}(z_{1})$ and the number of minimum vertex
covers were calculated for the Erd\H{o}s-R\'{e}nyi graph by using a
statistical mechanics analysis of ground state properties of a
hard-core model (see below) and the replica method. The replica
symmetric solution gives an exact result in the interval
$1<z_{1}\leqslant e$:
\begin{equation}
x_{c}(z_{1})=1-\frac{2W(z_{1})+W(z_{1})^{2}}{2z_{1}},  \label{vc-xc}
\end{equation}
where $W(x)$ is the Lambert-function defined by an equation $W\exp
W=x$ \cite{Weigt:wh00}, see also
\textcite{Weigt:wz06}. The
same result was derived by \textcite{Bauer:bg01a,Bauer:bg01b},
using the leaf algorithm. Note that the giant connected component
of the Erd\H{o}s-R\'{e}nyi graph disappear at $z_{1}<1$. The
presence of the replica symmetry indicates that in the interval
$1<z_{1}\leqslant e$ the degeneracy of the minimum vertex covers
is trivial in the following sense. One can interchange a finite
number of covered and uncovered vertices in order to receive
another minimum vertex cover.
Many non-trivial minimum vertex covers appear at mean degrees
$z_{1}>e$. The replica symmetry is broken and Eq.~(\ref{vc-xc}) is
not valid.
For this case the threshold $x_{c}(z_{1})$ and the degeneracy of the
minimum vertex cover were calculated by using the one-step replica
symmetry breaking in \textcite{Weigt:wh00,Weigt:wh01,Zhou:z03}.
Minimum vertex covers form a single cluster at $z_{1}\leqslant e$,
while they are arranged in many clusters at $z_{1}>e$.
As a result, the typical running time of an algorithm for finding a
vertex cover at $z_{1}\leqslant e$ is polynomial while the time
grows exponentially with the graph size at $z_{1}>e$
\cite{Barthel:bh04}.

The vertex cover problem on correlated scale-free networks was
studied by \textcite{Vazquez:vw03}. It turned out
that increase of likewise degree-degree correlations (assortative mixing) increases
the computational complexity of this problem in comparison with an
uncorrelated scale-free network having the same degree distribution.
If the assortative correlations exceed a critical threshold, then
many nontrivial vertex covers appear.

Interestingly, the minimum vertex cover problem is essentially
equivalent to another NP-hard optimization problem---{\em the
maximum clique problem}. Recall that a clique is a subset of
vertices in a given graph such that each pair of vertices in the
subset are linked. In order to establish the equivalence of these
two optimization problems, it is necessary to introduce the notion
of the {\em complement} or {\em inverse} of a graph. The complement
of a graph $G$ is a graph $\overline{G}$ with the same vertices such
that two vertices in $\overline{G}$ are connected if and only if
they are not linked in $G$. In order to find the complement of a
graph, we must add the missing edges, and remove the already
existing edges.
One can prove that vertices, which do not belong to the maximum
clique in $\overline{G}$, form the minimum vertex cover in $G$ (see
Fig.~\ref{fig-vertex-cover}).

A generalization of the vertex cover problem to hypergraphs can be
found in \vspace{-13pt}\textcite{Mezard:mt07}.

\paragraph{The hard-core gas model.}

Let us treat uncovered vertices as particles, so that we assign a
variable $\nu =1 $ for uncovered and $\nu =0$ for covered vertices.
Hence there are $\sum_{i}\nu _{i}=N-N_{vc}$ particles on the graph.
We also introduce a repulsion between particles such that only one
particle can occupy a vertex (the exclusion principle). A repulsion
energy between two nearest neighboring particles is $J>0$. Then we
arrive at the so called hard-core gas model with the energy
\begin{equation}
E=\frac{J}{2}\sum_{i,j}a_{ij}\nu _{i}\nu _{j},  \label{vc3}
\end{equation}
where $a_{ij}$ are the adjacency matrix elements. If the number of
particles is not fixed, and there is a mass exchange with a
thermodynamic bath, then we add a chemical potential $\mu
>0$. This results in the Hamiltonian of the hard-core gas model:
${\cal H} =E-\mu \sum_{i=1}^{N} \nu _{i}.$

In the ground state of this model, particles occupy vertices which
do not belong to a minimum vertex cover. Their number is equal to
$(1-x_{c})N$, where $x_{c}$ is the fraction of vertices in the
minimum vertex cover. The ground state energy is $E_{0}=0$ because
configurations in which two particles occupy two nearest neighboring
vertices, are energetically unfavorable. In other words, particles
occupy the maximum subset of vertices in a given graph such that no
two vertices are adjacent. In graph theory, this subset is called
the maximum {\em independent set} (see Fig.~\ref{fig-vertex-cover}).
Unoccupied vertices form the minimum vertex cover of the graph. Thus
finding the minimum vertex cover (or equivalently, the maximum
independent set) of a graph is equivalent to finding the maximum
clique of the complement of this graph.

The ground state of the hard-core model is degenerate if there are
many minimum vertex covers (or equivalently, many maximum
independent sets). The reader may see that searching for the
ground state is exactly equivalent to the minimum vertex cover
\vspace{-13pt}problem.

\paragraph{Antiferromagnet in a random field.}

Let us consider the following antiferromagnetic Ising model
\cite{Zhou:z03,Zhou:z05}:
\begin{equation}
E=\frac{J}{2}\sum_{i,j}a_{ij}S_{i}S_{j}-\sum_{i=1}^{N}S_{i}H_{i}+JL.
\label{vc5}
\end{equation}
Here $J>0$, $H_{i}=-Jq_{i}$ is a degree dependent local field, where
$q_{i}$ is degree of vertex $i$. $L$ is the number of edges in a
graph.
The negative local fields force spins to be in the state $-1$,
however the antiferromagnetic interactions compete with these
fields.

Consider a spin $S_{i}$ surrounded by $q_{i}$ nearest neighbors $j$
in the state $S_{j}=-1$. The energy of this spin is
\begin{equation}
\Bigl(J\sum_{j\in N(i)}S_{j}-H_{i}\Bigr)S_{i}=0\times S_{i}=0  \label{vc7}
\end{equation}
in any state $S_{i}=\pm 1$. Therefore this spin is
effectively free. 
Positions of ``free'' spins on a graph are not quenched. If one of the
neighboring spins flips up, then the state $S_{i}=-1$ becomes
energetically favorable.

Let us apply a small uniform magnetic field $\mu $, $0<\mu \ll J$.
At $T=0$, all ``free'' spins are aligned along $\mu $, i.e., they are
in the state $+1$. One can prove that the spins $S = +1$ occupy
vertices which belong to the maximum independent set, while the
spins $S =-1$ occupy the minimum vertex cover of a given graph. For
this let us make the transformation $S_{i}=2\nu _{i}-1$, where $\nu
_{i}=0,1$ for spin states $\mp 1$, respectively. Then the
antiferromagnetic model Eq.~(\ref{vc5}) is reduced to the hard-core
gas model where the external field $\mu $ corresponds to the
chemical potential of the particles.
The energy of the degenerate ground state is $E_{0}=0$. All these
pure states have the same energy $E_{0}=0$, the same average
magnetic moment $M=1-2x_{c}$ but correspond to different non-trivial
minimum vertex covers.

\begin{figure}[t]
\begin{center}
\scalebox{0.29}{\includegraphics[angle=270]{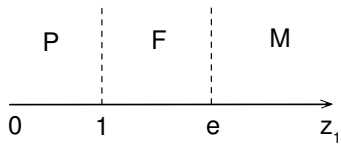}}
\end{center}
\caption{ Phase diagram of the antiferromagnetic Ising model
Eq.~(\ref{vc5}) at $T=0$. P, F and M denote paramagnetic,
ferromagnetic and mixed (spin-glass) phases, respectively. At mean
degree $z_{1}>e$, in the mixed phase, ferromagnetism and spin-glass
order coexist.}
\label{fig-AF-phase}
\end{figure}

The exact mapping of the AF model Eq.~(\ref{vc5}) onto the vertex
cover problem
leads to the zero-temperature phase diagram shown in
Fig.~\ref{fig-AF-phase}. The model is in a paramagnetic state at
small degree $0<z_{1}<1$ because in this case the network is below
the percolation threshold and consists of finite clusters. Above the
percolation threshold, at $1<z_{1}<e$, the ground state is
ferromagnetic with an average magnetic moment $M=1-2x_{c}(z_{1})$,
where $x_{c}(z_{1})$ is given by Eq.~(\ref{vc-xc}). The replica
symmetry is unbroken at $z_{1}<e$. Many pure states appear
spontaneously and the replica symmetry is broken at $z_{1}>e$. In
this case the AF model is in a mixed phase in which ferromagnetism
and spin-glass order coexist. At $z_{1}\gg 1$ the magnetic moment
$M$ is determined by $x_{c}(z_{1})$\ from Eq.~(\ref{vc-threshold}).

\subsection{The random-field Ising model}

\label{ssec:random-field}

The random-field Ising model is probably one of the simplest models
showing a dramatic influence of a quenched disorder (random fields)
on a collective behavior of a system with an exchange interaction
\cite{Lacour:lt74,Imry:im75}. Despite its simplicity, the
random-field model was an object of intensive and controversial
investigations during the last three decades. The energy of this
model is
\begin{equation}
E=-\frac{J}{2}\sum_{i,j}a_{ij}S_{i}S_{j}-H\sum_{i}S_{i}-\sum_{i}H_{i}S_{i},
\label{RFIM}
\end{equation}
where $J>0$, $H$ is a uniform field, and $H_{i}$ is a random field.
In most cases, the distribution function of the random field is
either Gaussian,
\begin{equation}
P_{\text{RF}}(H_{i})=\frac{1}{\sqrt{2\pi }\sigma }\exp \Bigl[-\frac{H_{i}^{2}}{%
2\sigma ^{2}}\Bigr],  \label{Gaussian-RF}
\end{equation}
or bimodal,
\begin{equation}
P_{\text{RF}}(H_{i})=\frac{1}{2}\delta (H_{i}-H_{0})+\frac{1}{2}\delta
(H_{i}+H_{0}).  \label{bimodal-RF}
\end{equation}
The parameters $\sigma $ and $H_{0}$ characterize a strength of
random fields.



The search for the ground state of the random-field model on a graph
is related with a famous optimization problem of a maximum flow
through the graph (\textcite{Picard:pr75}; see also
\textcite{Hartmann:hw05}). This problem belongs to the class P, that
is it may be solved in time bound by a polynomial in the graph size.

\subsubsection{Phase diagram}

\label{sssec:RFIM-phases}

\begin{figure}[t]
\begin{center}
\scalebox{0.27}{\includegraphics[angle=270]{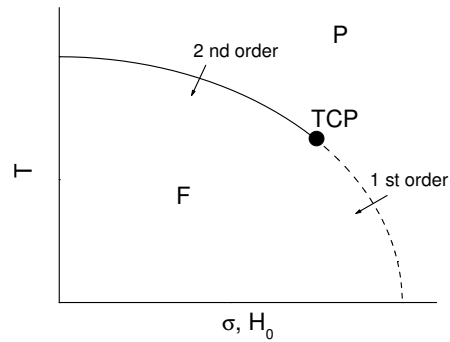}}
\end{center}
\caption{ Phase diagram of the random-field model on the fully
connected graph. For the Gaussian distribution, the phase transition
from the para- (P) to ferromagnetic (F) phase is of second order in
the $T-\protect\sigma $ plane. For the bimodal distribution, there
is a tricritical point (TCP) in the $T-H_{0}$ plane. The phase
transition is of second order (solid line) above TCP and first order
(dashed line) below TCP. } \label{fig-RFIM}
\end{figure}

The random-field model is exactly solved on the fully connected
graph (all-to-all interaction) \cite{Schneider:sp77,Aharony:a78}. In
this case we replace the coupling $J$ by $J/N$ in Eq.~(\ref{RFIM}).
The average magnetic moment is
\begin{equation}
M=\int \tanh [\beta (JM+H+H_{i})]P_{\text{RF}}(H_{i})dH_{i}.  \label{M-RF}
\end{equation}
For the Gaussian distribution, the phase transition from the para-
to ferromagnetic state is a mean-field second order phase
transition. Sufficiently strong random fields suppress the phase
transition at $\sigma >\sigma _{c}=J[2/\pi ]^{1/2}$, and the system
is in a disordered state at all $T$. The phase diagram of the
random-field model with the bimodal distribution of random field is
shown in Fig.~\ref{fig-RFIM}.
\textcite {Bruinsma:84} found that the random-field model with the
bimodal distribution on a regular Bethe lattice has a rich ground
state structure.


\subsubsection{Hysteresis on a fully connected graph}

\label{sssec:RFIM-hysteresys-FCG}

The random-field model demonstrates a peculiar hysteresis phenomena
at $T=0$
which may be relevant for understanding out-of-equilibrium phenomena
in many complex systems \cite{Sethna:sdm01}.

We start with a physical picture of the hysteresis which is valid
for any network. Let all spins be in the state $-1$. This initial
state corresponds to an applied field $H=-\infty $. An adiabatic
increase of the magnetic field results in a series of the so called
discrete Barkhausen jumps (avalanches) of a finite size
\cite{Sethna:sdkkrs93,Percovic:pds95}. A spin avalanche can be
initiated by a single spin flip. Indeed, if the total magnetic field
$H+H_{i}$ at vertex $i$ becomes larger than the energy of the
ferromagnetic interaction of the spin with neighboring spins, then
the spin turns up. This spin flip can stimulate flips of neighboring
spins, if they are energetically favorable. In turn the neighboring
spins may stimulate flips of their neighbors and so on. As a result
we observe an avalanche.
If $H$ is smaller than a critical field $H_{c}(\sigma )$, then the
average avalanche size is finite. At $H=H_{c}(\sigma )$ a
macroscopic avalanche takes place, and the magnetization has a jump
$\Delta M$.

\begin{figure}[t]
\begin{center}
\scalebox{0.33}{\includegraphics[angle=270]{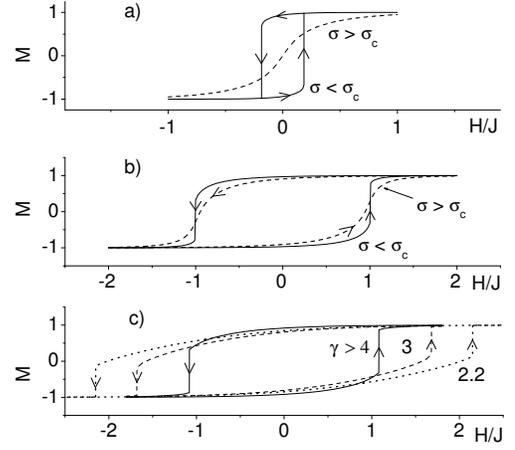}}
\end{center}
\caption{Hysteresis in the ferromagnetic Ising model with Gaussian
random fields (magnetization $M$ versus $H$). (a) Fully
connected graph: solid line, $\protect%
\sigma =0.5 <\protect\sigma _{c};$ dashed line, $\protect\sigma \geqslant \protect%
\sigma _{c}$. (b) Random regular network with degree $q=4$: solid line, $\protect%
\sigma =1.7<\protect\sigma _{c}$, dashed line, $\protect%
\sigma =2\geqslant \protect\sigma _{c}$. (c) Uncorrelated scale-free
networks for $\protect \sigma = 1.7$: solid line, $\protect \gamma
\geq 4$, $\langle q \rangle \approx 4$; dashed line, $\protect
\gamma=3$, $\langle q \rangle \approx 4$; dotted line, $\protect
\gamma=2.2$, $\langle q \rangle = 5.3$.} \label{fig-hysteresis-1}
\end{figure}

The exact properties of the hysteresis on the fully connected graph
at $T=0$ were found by \textcite{Sethna:sdkkrs93}. The dependence of
the magnetization $M$ on $H$ along a hysteresis loop follows from
Eq.~(\ref{M-RF}):
\begin{equation}
M=2\int_{-MJ-H}^{\infty }P_{\text{RF}}(H_{i})dH_{i}-1.  \label{M-RF-T0}
\end{equation}
The analysis of this equation for the Gaussian distribution of
random fields shows that the critical field $H_{c}(\sigma )$ is
non-zero at small strengths $\sigma <\sigma _{c}$. There is no
hysteresis at a sufficiently large strength of the random field,
$\sigma >\sigma _{c}=J[2/\pi ]^{1/2}$, see
Fig.~\ref{fig-hysteresis-1}.  The magnetization has a universal
scaling behavior near the critical point ($\sigma _{c}$,
$H_{c}(\sigma_{c})$):
\begin{equation}
M(r,h)=\left| \sigma -\sigma _{c}\right| ^{\beta }G(h/\left| \sigma
-\sigma _{c}\right| ^{\beta \delta }), \label{M-scaling}
\end{equation}
where $h=H-H_{c}(\sigma _{c})$. $\beta =1/2$ and $\delta =3$ are the
mean-field critical exponents. $G(x)$ is a scaling function.


\subsubsection{Hysteresis on a complex network}

\label{sssec:RFIM-hysteresis-RRG}

Another approach applied to zero-temperature hysteresis on the
random regular graph was developed by \textcite{Dhar:dss97}.
Here we generalize this approach to the configuration model of an
uncorrelated random network with a given degree distribution $P(q)$.

As above, we suppose that all spins are in the initial state $-1$ at
$H=-\infty $. Then the applied field is adiabatically increased. Let
$P^{\ast }$ be the conventional probability that if a spin at an end
of a randomly chosen edge is down, then for the other end spin, it
will be energetically favorable to flip up.
$P^{\ast }$ satisfies the equation:
\begin{equation}
P^{\ast }=\sum_{q}\frac{P(q)q}{z_1}\sum_{n=0}^{q-1}%
{q-1 \choose n}%
[P^{\ast }]^{n}[1-P^{\ast }]^{q-1-n}p_{n}(H),  \label{P-conv}
\end{equation}
where $q$ is vertex degree. The $n$-th term in the sum is the
probability that $n$ neighbors of a spin turn up simultaneously with
the spin while the other $q-n-1$ neighboring spins remain in the
state $-1$. The parameter
\begin{equation}
p_{n}(H)\equiv \int\limits_{-H+(q-2n)J}^{\infty }P_{\text{RF}}(H_{i})dH_{i}
\label{pnH}
\end{equation}
is the probability to find a vertex with a random field $H_{i}>-H+(q-2n)J$.
Knowing $P^{\ast }$, we can calculate the fraction of spins which
turn up at an applied field $H$:
\begin{equation}
N_{\uparrow }(H)=\sum_{q}P(q)\sum_{n=0}^{q}%
{q \choose n}%
[P^{\ast }]^{n}[1-P^{\ast }]^{q-n}p_{n}(H). \label{N_hysteresis}
\end{equation}
It gives the magnetization: $M(H)=2N_{\uparrow }(H)-1$. Note that
Eqs.~(\ref{P-conv}) and (\ref{N_hysteresis}) resemble
Eqs.~(\ref{e3.113}) and (\ref{e3.112}) describing the $k$-core
architecture of networks.

Hysteresis was only studied in detail for a random regular network
(all vertices have the same degree, i.e., $P(q)=\delta _{q,k}$),
see Fig.~\ref{fig-hysteresis-1}. In this case there is hysteresis
without a jump of the magnetization if the strength $\sigma $\ of
Gaussian random fields is larger than a critical strength $\sigma
_{c}$, in contrast to the fully connected graph where hysteresis
disappears at $\sigma > \sigma _{c}$. The critical field $H_{c}$
of the magnetization jump does not depend on $q>3$. Monte Carlo
simulations of the random-field model on a random regular network
made by \textcite{Dhar:dss97} confirmed this analytical approach.
A numerical solution of Eqs.~(\ref{pnH}) and (\ref{N_hysteresis})
shows that the random-field model on uncorrelated scale-free
networks has a similar hysteresis behavior (see
Fig.~\ref{fig-hysteresis-1}). Note that the critical field $H_{c}$
depends on the degree distribution exponent $\gamma$ only when $2<
\gamma < 4$.



A similar hysteresis phenomenon was found numerically in the
antiferromagnetic Ising model on growing scale-free and
Erd\H{o}s-R\'{e}nyi networks within zero-temperature field-driven
dynamics of spins \cite{Tadic:tmk05,Malarz:mak07,Hovorka:hf07}. It
was shown that the network topology influences strongly properties
of hysteresis loops. In this model, it is the network
inhomogeneity that plays the role of disorder similar to the
random fields.

\subsubsection{The random-field model at $T=0$}

\label{sssec:RFIM-networks}

Critical properties of the random-field model at $T=0$ on scale-free
networks were studied numerically and analytically by
\textcite{Lee:ljn06} by using a mean-field approximation which is
equivalent to the annealed network approximation in Sec. \ref{sssec:
annealed anzats}. These authors found that a critical behavior near
a phase transition from a disordered state into the ferromagnetic
state depends on
degree distribution exponent $\gamma $. If the
distribution function of random fields is concave at $H_{i}=0$
(i.e., $P''_{\text{RF}}(H_{i}=0)<0$, similar to the Gaussian
distribution) then the spontaneous magnetization $M$ emerges below a
critical strength $\sigma _{c}$: $M\propto \left| \sigma _{c}-\sigma
\right| ^{\beta }$, where $\beta (\gamma
>5)=1/2$, $\beta (3<\gamma \leqslant 5)=1/(\gamma -3)$. In the case
of the convex distribution function, i.e.,
$P''_{\text{RF}}(H_{i}=0)>0$, 
the phase transition is of the first order at all $\gamma >3$. When
$2<\gamma \leqslant 3$, the random-field model is in the
ferromagnetic state for an arbitrary strength and 
an arbitrary distribution function of random fields. This effect is
quite similar to the effect found in the ferromagnetic Ising and
Potts models in Secs.~\ref{sssec:ising_transition}, \ref{sec:Potts},
and \ref {sec:phenomenology}).

\textcite{Son:sjn06} proposed to use the random-field model as a
tool for extracting a community structure in complex networks. In
sociophysics, the random-field Ising model
is used for describing the emergence of a collective opinion.



\subsection{The Ising model on growing networks}
\label{ssec:ising_growing}

In this section we assume that a spin system on a growing network approaches equilibrium much faster than the network changes, and the adiabatic approximation works.
So we discuss the following circle of problems: a network is grown up to an infinite size and then the Ising model is placed on it.





\subsubsection{Deterministic graphs with BKT-like transitions}
\label{sssec:deterministic_bkt}

As is natural, the use of deterministic graphs
dramatically facilitates the analysis of any problem. Surprisingly,
very often results obtained in this way appear to be qualitatively
similar to conclusions for models on random networks.
Various graphs similar to those shown in Fig.~\ref{f3} allow one to
effectively apply the real space renormalization group technique.
For example, \textcite{Andrade:ah04} studied the Ising model on the
graph shown in Fig.~\ref{f3}(c)---``the Apollonian network''---and
observed features typical for the Ising model on random scale-free
network with exponent $\gamma<3$.


\begin{figure}[t]
\begin{center}
\scalebox{0.2}{\includegraphics[angle=0]{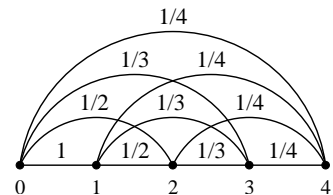}}
\end{center}
\caption{The deterministic fully connected graph
\protect\cite{Costin:ccd90}, which is equivalent to the asymmetric
annealed network. The values of the Ising coupling are shown on the
edges. } \label{f75}
\end{figure}

More interestingly, the Ising model on some deterministic graphs
shows the BKT-like singularities which was already discovered in the
1990s by \textcite{Costin:ccd90} and \textcite{Costin:cc91}. In
network context, their model was studied in \textcite{Bauer:bkd05}.
This network substrate is an asymmetric annealed network, which is
actually an annealed version of the random recursive graph. Vertices
are labelled $i=0,1,2,\ldots,t$, as in a growing network. Each
vertex, say vertex $i$ have a single connection of unit strength to
``older'' vertices. One end of this edge is solidly fixed at vertex
$i$, while the second end frequently hops at random among vertices
$0,1,\ldots,i-1$, which just means the specific asymmetric
annealing. The resulting network is equivalent to the fully
connected graph with a specific large scale inhomogeneity of the
coupling (see Fig.~\ref{f75}).

\begin{figure}
\begin{center}
\scalebox{0.25}{\includegraphics[angle=0]{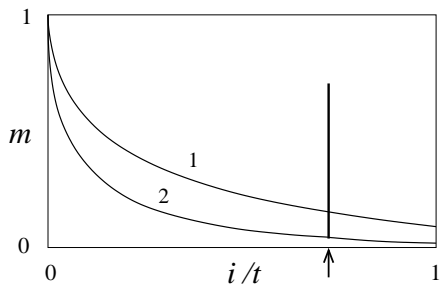}}
\end{center}
\caption{The magnetization profile for the ferromagnetic Ising model
on the graph shown in Fig.~\protect\ref{f75}. $i$ labels vertices
starting from the ``oldest'' one, and $t$ is the network size. Curve
$1$ is valid both for $T<T_c$ with an arbitrary homogeneous applied
field $H$ and for $T>T_c$, $H\neq 0$.
Curve $2$ describes the profile when an external field is applied to
a single spin, while $T>T_c$. The arrow indicates the point of
application of the local magnetic field. The mean magnetic moment of
this vertex is very distinct from others. } \label{f76}
\end{figure}

The ferromagnetic Ising model on this network is described by the
Hamiltonian 
\begin{equation}
{\cal H} = -\sum_{0\leq i<j\leq t} \frac{s_i s_j}{j} - \sum_{i=0}^t
H_i s_i , \label{e6.100}
\end{equation}
where $H_i$ are local magnetic fields. The mean-field theory, exact
for this Hamiltonian, indicates the presence of a phase transition
in this system. Figure~\ref{f76} shows an inhomogeneous distribution
of the magnetization $m(i)$ over the network. Only in the normal
phase, without field, $m(i)=0$. Otherwise, the oldest spin turns out
to be strictly directed, $m(i=0)=1$, and the profile is
non-analytic: $m(i) \cong 1 - \text{const} (i/t)^{2/T}$.
Earlier, \textcite{Coulomb:cb03} observed a resembling effect
studying a giant connected component in random growing networks. The
full magnetization $M(T)$ demonstrates the BKT-kind behavior near
the phase transition:
\begin{equation}
M(T) \propto \exp \left( -\frac{\pi }{2}\,\sqrt{\frac{T_c}{T_c-T}}
\right) . \label{e6.101}
\end{equation}
Note that the BKT singularity, Eq.~(\ref{e6.101}), and the specific
non-analyticity of $m(i)$ at $i=0$ are closely related.

The distribution of the linear response, $\sum_i\partial
m(i)/\partial H_j |_{H=0}$, to a local magnetic field, which may be
also called the distribution of correlation volumes, in this model
is very similar to the size distribution of connected components in
growing networks with the BKT-like transition. It has a power-law
decay in the whole normal phase. Exactly the same decay has the
distribution of correlations $\partial m(i)/\partial H_j|_{H=0}$ in
this phase \cite{Khajeh:kdm07}.

We may generalize the inhomogeneity of the interaction in the
Hamiltonian to a power law, $\propto j^{-\alpha}$, with an arbitrary
exponent. (For brevity, we omit the normalization---the sum of the
coupling strengths must grow proportionally to the size of the
network.) One may show that in this model the BKT-singularity exists
only when $\alpha=1$. For $\alpha>1$, phase ordering is absent at
any nonzero temperature as in the one-dimensional Ising model, and
for $0<\alpha<1$, there is a quite ordinary second order transition.

Let us compare this picture with the well-studied ferromagnetic
Ising model for a spin chain with regular long-range interactions
$\propto |i-j|^{-\alpha}$ (see, e.g., \textcite{Luijten:lb97}). In
this model,
(i) for $\alpha>2$, $T_c=0$, similarly to the one-dimensional Ising
model; (ii) at $\alpha=2$, there is a transition resembling the BKT
one;
(iii) for $1<\alpha<2$, there is a transition at finite $T_c$. The
reader may see that in both models, there exist boundary values of
exponent $\alpha$, where BKT-kind phenomena take place. In very
simple terms, these special values of $\alpha$ play the role of
lower critical dimensions. (Recall that the BKT transitions in solid
state physics occur only at a lower critical dimension.) These
associations show that the BKT singularities in these networks are
less strange and unexpected than one may think at first sight.


\textcite{Khajeh:kdm07} solved the $q$-state Potts model on this
network and, for all $q \geq 1$ arrived at results quite similar to
the Ising model, i.e., $q=2$. Recall that $q=1$ corresponds to the
bond percolation model, and that the traditional mean-field theory
on lattices gives a first order phase transition if $q>2$. Thus,
both the first and the second order phase transitions transformed
into the BKT-like one on this network.

\textcite{Hinczewski:hb05} found another deterministic graph, on
which the Ising model shows the BKT-like transition, so that this
singularity is widespread in evolving networks with large-scale
inhomogeneity.



\subsubsection{The Ising model on growing random networks}
\label{sssec:ising_growing_scale-free}

There is still no analytical solution of the Ising model on growing
random networks. \textcite{Aleksiejuk:ahs02} and their numerous
followers simulated the Ising model on the very specific
Barab\'asi-Albert network, where degree-degree correlations are
virtually absent. So, the resulting picture is quite similar to the
Ising model on an uncorrelated scale-free network with degree
distribution exponent $\gamma=3$. In general, the growth results in
a wide spectrum of structural correlations, which may dramatically
change the phase transition.

Based on
known results for the percolation (the one-state Potts model), see
Sec.~\ref{ssec:giant_in_growing}, we expect the following picture
for the Ising model on recursive growing graphs. If each new vertex
has a single connection, the recursive graph is a tree, and so the
ferromagnetic ordering takes place only at zero temperature.
Now let a number of connections of
new vertices be greater than $1$, so that these networks are not
trees. (i) If new vertices are attached to randomly chosen ones,
there will be the Berezinskii-Kosterlitz-Thouless critical
singularity. (ii) If the mechanism of the growth is the preferential
attachment, then the critical feature is less exotic, more similar
to that for uncorrelated networks.


\section{THE POTTS MODEL ON NETWORKS}

\label{sec:Potts}

The Potts model is related to a number of outstanding problems in
statistical and mathematical physics \cite{Wu:w82,Baxter:bbook82}.
The bond percolation and the Ising model are only particular cases
of the $p$-state Potts model. The bond percolation is equivalent to
one-state Potts model (\textcite{Kasteleyn:kf69,Fortuin:fk72}, see
also \textcite{Lee:lgk04c}). The Ising model is exactly the
two-state Potts model. Here we first look at critical properties of
the Potts model and then consider its applications for coloring a
random graph and for extracting communities.

\subsection{Solution for uncorrelated networks}

\label{ssec:solution_for_uncorrelated}

The energy of the Potts model with $p$ states is
\begin{equation}
E=-\frac{1}{2}\sum_{i,j}J_{ij}a_{ij}\delta _{\alpha _{i},\alpha
_{j}}-H\sum_{i}\delta _{\alpha _{i},1}\,,  \label{Hamilt}
\end{equation}
where $\delta _{\alpha ,\beta }=0,1$ if $\alpha \neq \beta $ and $\alpha
=\beta $, respectively. Each vertex $i$ can be in any of $p$ states: $\alpha
_{i}=1,2,\ldots ,p$.
The ``magnetic field'' $H>0$ distinguishes the state $\alpha =1$. The
$\alpha $-component of the magnetic moment of vertex $i$ is defined
as follows:
\begin{equation}
M_{i}^{(\alpha )}=\frac{p\left\langle \delta _{\alpha _{i},\alpha
}\right\rangle -1}{p-1}.  \label{M-potts}
\end{equation}
In the paramagnetic phase at zero magnetic field, $M_{i}^{(\alpha )}=0$ for
all $\alpha $. In an ordered state $M_{i}^{(\alpha )}\neq 0$.

Exact equations for magnetic moments of the Potts model on a
treelike complex network (see Appendix~\ref{state_Potts}) were
derived by \textcite{Dorogovtsev:dgm04} by using the recursion
method which, as we have demonstrated, is equivalent to the
Bethe-Peierls approximation and the belief-propagation algorithm .
It was shown that the ferromagnetic $p$-state Potts model with
couplings $(J_{ij}=J>0)$ on the configuration model has the critical
temperature
\begin{equation}
T_{\text{P}}=J/\ln \Big[\frac{B+p-1}{B-1}\Big]\,.  \label{Tc-Potts}
\end{equation}
where $B=z_{2}/z_{1}$ is the average branching parameter.
Interestingly, $T_{\text{P}}$ has different meanings for
$p=1,2$ and $p\geqslant 3$. In the case $p=1$, the critical
temperature $T_{\text{P}}$ determines the percolation threshold (see
Appendix~\ref{state_Potts}). When $p=2$, $T_{\text{P}}$ is equal to
the exact critical temperature Eq.~(\ref{Tc-net}) of the
ferromagnetic phase transition in the Ising model (it is only
necessary to rescale $J\rightarrow 2J$). For $p\geqslant 3$,
$T_{\text{P}}$ gives the lower temperature boundary of the
hysteresis phenomenon at the first order phase transition.

\subsection{A first order transition}

\label{ssec:first_order}

In the standard mean-field theory, the ferromagnetic Potts model
with $J_{ij}=J>0$ undergoes a first order phase transition for all
$p\geqslant 3$ \cite{Wu:w82}.
In order to study critical properties of the Potts model on a
complex network, we need to solve Eq.~(\ref{recursion-potts}) which
is very difficult to do analytically. An approximate solution based
on the ansatz Eq.~(\ref{anzatz}) was obtained by
\textcite{Dorogovtsev:dgm04}.
It turned out that in uncorrelated random networks with a finite
second moment $\left\langle q^{2}\right\rangle $ (which corresponds
to scale-free networks with $\gamma
>3$) a first order phase transition occurs at a critical temperature $T_c$ 
if the number of Potts states $p\geqslant 3$.
In the region $T_{\text{P}}<T<T_{c}$, two metastable thermodynamic
states with magnetic moments $M=0$ and $M\neq 0$ coexist.
This leads to hysteresis phenomena which are typical for a first
order phase transition. At $T<T_{\text{P}}$, only the ordered state
with $M\neq 0$ is stable.


\begin{figure}[t]
\begin{center}
\scalebox{0.27}{\includegraphics[angle=0]{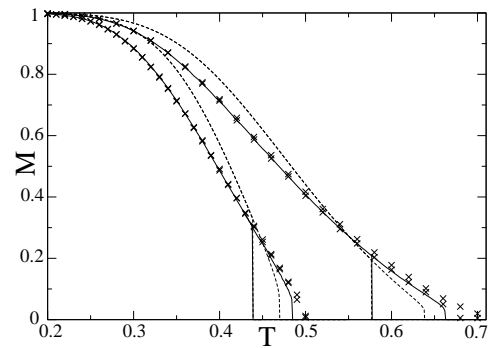}}
\end{center}
\caption{ Magnetic moment $M$\ versus $T$ for the ferromagnetic
Potts model on an uncorrelated scale-free network with  degree
$z_{1}=10$. Leftmost curves:  $\protect\gamma =4$; rightmost curves,
$\protect\gamma =3.5$. Numerical simulations and exact numerical
solution \protect\cite {Ehrhardt:em05} are shown in crosses and
solid lines. Dotted lines, an approximate solution
\protect\cite{Dorogovtsev:dgm04}. Vertical lines, the lower
temperature boundary $T_{\text{P}}$ of the hysteresis region. From
\protect\textcite{Ehrhardt:em05}.} \label{fig-potts}
\end{figure}


When $\gamma $ tends to 3 from above,
$T_{c}$ increases while the jump of the magnetic moment at the first
order phase transition tends to zero.
The influence of the network heterogeneity becomes dramatic when
$2<\gamma \leqslant 3$ and the second moment $\left\langle
q^{2}\right\rangle $ diverges: instead of a first order phase
transition, the $p$-state Potts model with $p\geqslant 3$ undergoes
an infinite order phase transition at the critical temperature
$T_{c}(N)/J\approx z_{2}/(z_{1}p)\gg 1$, similarly to the Ising
model in Sec.~\ref {sssec:ising_transition}. In the limit
$N\rightarrow \infty $, the Potts model is ordered at any finite
$T$.

\textcite{Ehrhardt:em05} used a population dynamics algorithm to
solve numerically Eq.~(\ref{recursion-potts}) for uncorrelated
scale-free networks. The exact numerical calculations and numerical
simulations of the Potts model confirmed that a first order phase
transition occurs at $p\geqslant 3$ when 
$\gamma >3$.
Some results obtained by \textcite{Ehrhardt:em05} are represented in
Fig.~\ref{fig-potts}, where they are compared with the approximate
solution. As one could expected, the approximate solution gives poor
results for vertices with small degree. For graphs with a large
minimum degree (say $q_{0}=10$) the approximate solution agrees well
with the exact calculations and numerical simulations.

A simple mean-field approach to the Potts model on uncorrelated
scale-free networks was used by \textcite{Igloi:it02}. Its
conclusions essentially deviate from the exact results.
\textcite{Karsai:kdi07} studied the ferromagnetic large-$p$ state Potts model on evolving networks and described finite-size scaling in these systems.

\subsection{Coloring a graph}
\label{ssec:antiferromagnetic_potts}

Coloring random graphs is a remarkable problem in combinatorics
\cite {Garey:gj79} and statistical physics \cite{Wu:w82}. Given a
graph, we want to know if this graph can be colored with $p$ colors
in such a way that no two neighboring vertices have the same color.
A famous theorem states that four colors is sufficient to color a
planar graph, such as a political map
\cite{Appel:ak77a,Appel:ak77b}. Coloring a graph is not only
beautiful mathematics but it also has important applications. Good
examples are scheduling of registers in the central processing unit
of computers, frequency assignment in mobile radios, and pattern
matching. Coloring a graph is a NP complete problem. The time needed
to properly color a graph grows exponentially with the graph size.

How many colors do we need to color a graph? Intuitively it is
clear that any graph can be colored if we have a large enough
number of colors, $p$. The minimum needed number of colors is
called the ``chromatic number'' of the graph.
The chromatic number is determined by the graph structure.
It is also interesting to find the number of ways one can color a graph.

The coloring problem
was extensively
investigated for
classical random graphs.
There exists a critical degree $c_{p}$
above which
the graph becomes uncolorable by
$p$ colors with high probability. This transition is the so called $p$-COL/UNCOL
transition. Only graphs with average degree $z_{1}\equiv
\left\langle q\right\rangle <c_{p}$ may be colored with $p$ colors.
For larger $z_{1}$ we need more colors.

In order to estimate the threshold degree $c_{p}$ for the
Erd\H{o}s-R\'{e}nyi graph, one can use the
so-called
first-moment method (annealed computation, in other words).
Suppose that $p$ colors are assigned randomly to vertices. It
means that a vertex may have any color with equal probability
$1/p$. The probability that two ends of a randomly chosen edge
have different colors is $1-1/p$. We can color $N$ vertices of the
graph in $p^{N}$ different ways. However only a small fraction
$(1-1/p)^{L}$ of these configurations have the property that all
$L=z_{1}N/2$ edges connect vertices of different colors. Hence the
number of $p$-colorable configurations is
\begin{equation}
{\frak N}(z_{1})=p^{N}(1-1/p)^{L}\equiv \exp [N\Xi (p)].
\label{q-colors}
\end{equation}
If $\Xi (p)\geqslant 0$, then with high probability there is at
least one $p$-colorable configuration. At $p\gg 1$, this condition
leads to the threshold average degree $c_{p}\sim 2p\ln p-\ln p$. The
exact threshold $c_{p}\sim 2p\ln p-\ln p+o(1)$ was found by
\textcite{Luczak:l91}, see also \textcite{Achlioptas:anp05}.

The coloring problem was reconsidered by methods of statistical mechanics of disordered systems, and a complex structure of
the colorable phase was revealed \cite{Mulet:mpw02,Braunstein:bmpwz03,Krzakala:kpw04,Mezard:mmz05}.
It was found that the colorable phase itself contains several different phases.
These studies used the equivalence of this problem to
the problem of finding the ground state of the Potts model, Eq.~(\ref{Hamilt}), with $p$ states (colors) and antiferromagnetic interactions $J_{ij}=-J<0$ in zero field. Within this approach, the graph is $p$-colorable if in the ground state the endpoints of all edges are in different Potts states. The corresponding ground state energy is $E=0$. The degeneracy of this ground state means that there are several ways for coloring a graph. In the case of a $p$-uncolorable graph, the ground state energy of the antiferromagnetic Potts model is positive due to a positive contribution from pairs of neighboring vertices having the same color.

It was shown that
if the mean degree
$z_1$ is sufficiently small, then it is easy to find a solution of
the problem by using usual computational algorithms. In these
algorithms, colors of one or several randomly chosen vertices are
changed one by one. For example, the Metropolis algorithm gives
an exponentially fast relaxation from an arbitrary initial
set of vertex colors
to a correct solution \cite{Svenson:sn99}.
On the other hand, for higher mean degrees (of course, still below $c_p$), these algorithms can approach a solution only in non-polynomial times---``computational hardness''.
The computational hardness is related to the presence of a hierarchy of numerous ``metastable'' states with a positive energy,
which can dramatically slow down or even trap any simple numerical algorithm.

The mentioned works focused on the structure of the space of solutions for coloring a graph. (A solution here is a proper coloring of a graph.) It was found that this structure qualitatively varies with the mean degree. In general, the space of solutions is organized as a set of disjoint clusters---``pure states''. Each of these clusters consists of solutions which can be approached from each other by changing colors of only $o(N)$ vertices.
On the other hand,
to transform a solution belonging to one cluster into a solution in another cluster, we have to change colors of $O(N)$ vertices, i.e., of a
finite fraction of vertices.
Clearly, if a network consists of only bare vertices ($z_1=0$), the space of solutions consists of a single cluster. However, above some threshold value of a mean degree, this space becomes highly clustered. The structure and statistics of these clusters at a given $z_1$ determine whether the coloring problem is computationally hard or not.

\begin{figure}[t]
\begin{center}
\scalebox{0.4}{\includegraphics[angle=270]{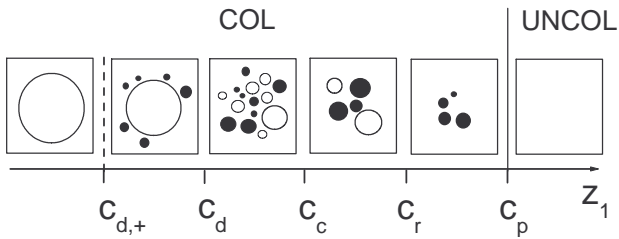}}
\end{center}
\caption{
Schematic phase diagram and structure of solutions for
coloring the Erd\H{o}s-R\'{e}nyi graphs in $p$-colors versus the
average degree $z_{1}$.
(i) $z_{1}<c_{d,+}$, solutions form one
large connected cluster without frozen variables (open circle).
(ii) $c_{d,+}<z_{1}<c_{d}$, in addition to a large cluster, small
disjoint clusters with frozen variables (black circles) appear.
They include however an exponentially small fraction of solutions.
(iii) $c_{d}<z_{1}< c_{c}$, solutions are arranged in
exponentially many clusters of different sizes with and without
frozen variables.
Exponentially many clusters without frozen variables dominate.
(iv) $c_{c}< z_{1}< c_{r}$, there are a
finite number
statistically dominating large clusters.
These clusters does not contain frozen variables.
(v) $c_{r}<z_{1}< c_{p}$,
dominating clusters contain frozen variables.
Above
$c_{p}$, a graph is $p$-uncolorable.
$c_{d,+}$ coincides with the $2$-core birth point.
$c_d$, $c_c$, and $c_r$ correspond to so-called clustering, condensation, and rigidity (freezing) transitions, respectively.
Adapted from \textcite{Zdeborova:zk07}.
}
\label{fig-color-space}
\end{figure}

The statistics
of clusters in a full range of mean degrees was obtained in \textcite{Krzakala:kmr07} and \textcite{Zdeborova:zk07}. Their results indicate a chain of topologically different phases inside the colorable phase, see Fig.~\ref{fig-color-space}.
An important notion in this kind of problems is {\em a frozen variable}. (A variable here is a vertex.) By definition, a frozen variable (vertex) has the same color in all solutions of a given cluster. Figure~\ref{fig-color-space} demonstrates that the clusters with frozen variables are statistically dominating in the range $c_r<z_1<c_p$.
Remarkably, the computational hardness was observed only in this region, although the replica symmetry breaking was found in the essentially wider range $c_d<z_1<c_p$.


Coloring the Watts-Strogatz small-world networks was studied
numerically by \textcite{Walsh:99}. He found that it is easy to
color these networks at small and large densities of shortcuts,
$p$. However it is hard to color them in the intermediate region
of $p$.

\subsection{Extracting communities}

\label{ssec:extracting_communities}

It is a matter of common experience that a complex system or a data
set may consist of clusters, communities or groups. A common
property of a network having a community structure is that edges are
arranged denser within a community and sparser between communities.
If a system is small, we can reveal a community structure by eye.
For a large network we need a special method \cite{Newman:n03a}.
Statistical physics can provide useful tools for this purpose. In
particular, the Potts model has interesting applications which are
ranged from extracting species of flowers, collective listening
habits, communities in a football league to a search of groups of
configurations in a protein folding network.

\textcite{Blatt:bwd96} showed that a search for clusters in a data
set can be mapped to extracting superparamagnetic clusters in the
ferromagnetic Potts model formed in the following way: each point
in the data set is represented as a point in a $d$-dimensional
space. The dimensionality $d$ is determined by the number of
parameters we use to describe a data point, e.g., color, shape,
size etc. A Potts spin is assigned to each of these points. The
strength of short-range ferromagnetic interaction between nearest
neighboring spins is calculated following a certain rule: the
larger the distance between two neighboring points in the space
the smaller is the strength. The energy of the model is given by
Eq.~(\ref {Hamilt}). \textcite{Blatt:bwd96} used Monte Carlo
simulations of the Potts model at several temperatures in order to
reveal clusters of spins with strong ferromagnetic correlations
between neighboring spins. Clusters of aligned spins showed a
superparamagnetic behavior at low temperatures. They were
identified as clusters in the data set. An analysis of real data,
such as Iris data and data taken from a satellite image of the
Earth, demonstrated a good performance of the method.

\textcite{Lambiotte:la05} studied a complex bipartite network formed
by musical groups and listeners. Their aim was to uncover collective
listening habits. These authors represented individual musical
signatures of people as Potts vectors. A scalar product of the
vectors characterized a correlation between music tastes. This
investigation found that collective listening habits do not fit the
usual genres defined by the music industry.

\textcite{Reichardt:rb04,Reichardt:rb06} proposed to map the
communities of a network onto the magnetic domains forming the
ground state of the $p$-state Potts model. In this approach, each
vertex in the network is assigned a Potts state $\alpha =1,2,...p$.
Vertices, which are in the same Potts state $\alpha$, belong to the
same community $\alpha$. The authors used the following Hamiltonian:
\begin{equation}
{\cal H}=-\frac{1}{2}\sum_{i,j}a_{ij} \delta_{\alpha _{i},\alpha
_{j}}+\frac{\lambda}{2}\sum_{\alpha=1}^{p}n_{s} (n_{s} -1),
\label{H-RB-Potts}
\end{equation}
where $a_{ij}$ are the elements of the adjacency matrix of the
network, $n_{s}$ is the number of vertices in the community
$\alpha$, i.e., $n_{s}=\sum_{i} \delta_{\alpha _{i},\alpha }$. The
number of possible states, $p$, is chosen large enough to take into
account all possible communities. $\lambda $ is a tunable parameter.
The first sum in Eq.~(\ref{H-RB-Potts}) is the energy of the
ferromagnetic Potts model. It favors merging vertices into one
community. The second repulsive term is minimal when the network is
partitioned into as many communities as possible. In this approach
the communities arise as domains of aligned Potts spins in the
ground state which can be found by Monte Carlo optimization. 

At $\lambda =1$ the energy Eq.~(\ref{H-RB-Potts})
is proportional to the modularity measure $Q$, 
namely ${\cal H}=-QL$ where $L$ is the total number of edges in the network.
Thus the ground
state of the model Eq.~(\ref{H-RB-Potts}) corresponds to the maximum
modularity $Q$. The modularity measure was introduced in
\textcite{Newman:ng04,Clauset:cnm04}. 
For a given partition of a network into communities, the modularity is the difference between
the fraction of edges within communities and the expected fraction
of such edges under 
an appropriate {\em null model} of the network (a random network model assuming the absence of a modular structure):  
\begin{equation}
Q = \sum_{\alpha}\Bigl(\frac{L_{\alpha}}{L}- \frac{L_{\alpha}^{\text{exp}}}{L}\Bigr)
=\frac{1}{2L}\sum_{\alpha}\sum_{i,j}(a_{ij}-p_{ij})\delta_{\alpha _{i},\alpha
_{j}}
.
\label{modular}
\end{equation}
Here $L_{\alpha}$ and $L_{\alpha}^{\text{exp}}$ are the numbers of edges within community $\alpha$ in the 
network and in its null model, respectively; $p_{ij}$ is the probability that vertices $i$ and $j$ are connected in the null model. \textcite{Reichardt:rb04,Reichardt:rb06} used
the configuration model as the null model, i.e., $p_{ij}=q_{i}q_{j}/2L$ where $q_{i}$ and $q_{j}$ are degrees of vertices $i$ and $j$ respectively.
Tuning $\lambda $ and $p$, one can find a partition
of a given network into communities such that a density of edges
inside communities is maximal when compared to one in a completely
random network. If however the size distribution of communities is
sufficiently broad, then it is not easy to find an optimal value of
the parameter $\lambda$.
Searching for small communities and the resolution limit of this method
are discussed in \textcite{Kumpula:ksk07}. Interestingly, finding the partition of a complex network into communities, such that it
maximizes the modularity measure, is an NP-complete problem
\cite{Brandes:bdg06}.

\textcite{Reichardt:rb04,Reichardt:rb06} applied the Potts model
Eq.~(\ref{H-RB-Potts}) to study a community structure of real world
networks, such as a US college football network and a large
protein folding network.

\textcite{Guimera:gsa04} proposed another approach to the problem of
extracting communities based on a specific relation between the
modularity measure $Q$ and the ground state energy of a Potts model
with multiple interactions.

\section{THE XY MODEL ON NETWORKS}
\label{sec:xy model}

The $XY$ model describes interacting planar rotators. The energy of
the $XY$ model on a graph is
\begin{equation}
{\cal H}=-\frac{J}{2}\sum_{i,j}a_{ij}\cos (\theta _{i}-\theta _{j})
\label{XY-model}
\end{equation}
where $a_{ij}$ are the elements of the adjacency matrix of the
graph, $\theta _{i}$ is the phase of a rotator at vertex $i$, $J$ is
the coupling strength. Unlike the Ising and Potts models with
discrete spins, the $XY$ model is described by continuous local
parameters and belongs to the class of models with continuous
symmetry.

A one-dimensional $XY$ model has no phase transition.
On a two-dimensional regular lattice, this model $(J>0)$ undergoes
the unusual Berezinskii-Kosterlitz-Thouless phase transition. On
a $d$-dimensional lattice at $d>4$ and the fully connected graph
\cite{Antoni:as95} the ferromagnetic phase transition in the $XY$
model is of second order with the standard mean-field critical
exponents.

The study of the $XY$ model on complex networks is motivated by
several reasons. In principal, the continuous symmetry may lead to a
new type of critical behavior in complex networks.
Moreover the ferromagnetic $XY$ model is close to the Kuramoto model
which is used for describing the synchronization phenomenon in
Sec.~\ref{ssec:kuramoto-fcg}.

\subsection{The XY model on small-world networks}

\label{ssec:xy_small-world}

\begin{figure}[t]
\begin{center}
\scalebox{0.6}{\includegraphics[angle=0]{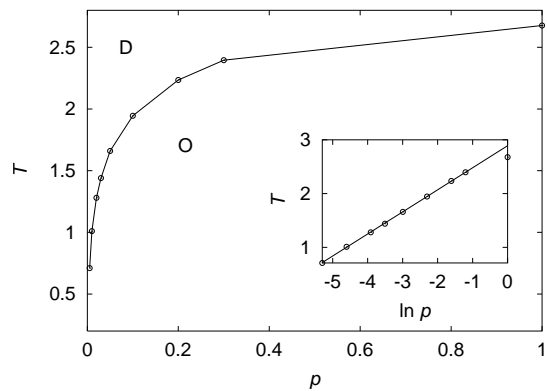}}
\end{center}
\caption{ $p-T$ phase diagram of the ferromagnetic $XY$ model on
the Watts-Strogatz network. $p$ is the fraction of shortcuts. $D$
and $O$ denote the disordered and ordered phases, respectively. The
inset shows that the critical temperature is well approximated by a
function $0.41\ln p+2.89$. From \protect\textcite{Kim:khh01}. }
\label{fig-XY-diagram}
\end{figure}

There
were a few
studies
of the $XY$ model on complex
networks. \textcite{Kim:khh01} carried out Monte-Carlo simulations
of the ferromagnetic $XY$ model on the Watts-Strogatz small-world
network generated from a ring of $N$ vertices. They measured the
order parameter $r=\left| N^{-1}\sum_{j}\exp (i\theta _{j})\right|$.
By using the standard finite-size scaling analysis they showed that
the phase transition appears even at a tiny fraction of shortcuts,
$p$. The transition is of second order with the standard mean-field
critical exponent $\beta =1/2$ (similar to the phase transition in
the Ising model in Sec.~\ref {ssec:Ising_small-world}).
The phase diagram of the $XY$ model is shown in
Fig.~\ref{fig-XY-diagram}. There is no phase transition at $p=0$
because the system is one-dimensional. Surprisingly, the dependence
of the critical temperature $T_{c}$ on $p$ was well fitted by a
function $T_{c}(p)/J=0.41\ln p+2.89$ in contrast to
$T_{c}(p)/J\propto 1/\left| \ln p\right| $ for the Ising model. The
origin of this difference is unclear. Dynamical Monte-Carlo
simulations of \textcite {Medvedyeva:mhm03} confirmed the mean-field
nature of the phase transition.
These authors found that, at $T$ near $T_{c}$, the characteristic
time $\tau $ scales as $\tau \sim \sqrt{N}$ as it should be for
networks with rapidly decreasing degree distribution (see the theory
of finite-size scaling in Sec.~\ref{ssec:finite_size_scaling}).


\subsection{The XY model on uncorrelated networks}

\label{ssec:xy_uncorrelated}


An exact solution of
the $XY$ model
on tree-like networks, in principle,
can obtained in the framework of the belief propagation algorithm,
see Eq.~(\ref{general}). Another analytical approach based on the
replica theory and the cavity method was developed
by
\textcite{Coolen:csc05,Skantzos:sch05}, see also \textcite{Skantzos:sh07} for the dynamics of a related model. We here consider the
influence of network topology on the critical behavior of the
$XY$ model, using the annealed network approximation from
Sec.~\ref{sssec: annealed anzats}.
Recall that in this approach, the element of adjacency matrix
$a_{ij}$ is replaced by the probability that two vertices $i$ and
$j$ with degrees $q_{i}$ and $q_{j}$, are linked. In this way, for
the configuration model, we obtain the $XY$ model with a degree
dependent coupling on the fully connected graph:
\begin{equation}
{\cal H}_{\text{MF}}=-\frac{J}{Nz_{1}}\sum_{i<j}q_{i}q_{j}\cos
(\theta _{i}-\theta _{j}).  \label{annealed-XY}
\end{equation}
This model is solved exactly by using a weighted complex order
parameter
\begin{equation}
\widetilde{r}e^{i\psi }=\frac{1}{Nz_{1}}\sum_{j=1}^{N}q_{j}e^{i\theta _{j}}.
\label{XY-op}
\end{equation}
The phase $\psi $ determines the direction along which rotators are
spontaneously aligned.
The order parameter $\widetilde{r}$ is a solution of an equation:
\begin{equation}
\widetilde{r}=\frac{1}{z_{1}}\sum_{q}P(q)q\frac{I_{1}(\widetilde{r}q\beta J)%
}{I_{0}(\widetilde{r}q\beta J)},  \label{XY-op2}
\end{equation}
where $I_{0}(x)$ and $I_{1}(x)$ are the modified Bessel functions.
An analysis of Eq.~(\ref{XY-op2}) shows that the ferromagnetic $XY$
model has the same critical behavior as the ferromagnetic Ising
model.
Notice that the usually used order parameter $r=\left|
N^{-1}\sum_{j}e^{i\theta _{j}}\right| $ has the same critical
behavior as $\widetilde{r}$. The critical temperature of the
continuous phase transition is $T_{c}=J \langle q^{2} \rangle
/2z_{1}$. It is finite in a complex network with a finite second
moment $\langle q^{2} \rangle$, and diverges if $\langle q^{2}
\rangle \rightarrow \infty $. In the latter case, the $XY$ model is
in the ordered state at any finite $T$. The annealed network
approximation predicts that the $XY$ model on the Watts-Strogatz
small-world network has the standard mean-field critical behavior.
This agrees with the numerical simulations in \textcite{Kim:khh01}
and \textcite{Medvedyeva:mhm03}.

\section{PHENOMENOLOGY OF CRITICAL PHENOMENA IN NETWORKS}

\label{sec:phenomenology}

Why do critical phenomena in networks differ so strongly from those
in usual substrates and what is their common origin? Why do all
investigated models demonstrate universal behavior when
$\left\langle q^{2}\right\rangle $ diverges? In order to answer
these questions and analyze results of simulations and experiments
from a general point of view, we need a general theory which is not
restricted by specific properties of any model.

In the phenomenological approach, the origin of interactions and
nature of interacting objects are irrelevant.
In this section, we consider a phenomenological theory of
cooperative phenomena in networks proposed by
\textcite{Goltsev:gdm03}. This theory is based on concepts of the
Landau theory of continuous phase transitions and leads to the
conclusion that the universal critical behavior in networks is
determined by two factors: (i) the structure of a network and (ii)
the symmetry of a given model.

\subsection{Generalized Landau theory}
\label{ssec:landau_theory}

Let us consider a system of interacting objects. Interactions or
links between these objects form a net. We assume that some kind of
order can emerge. This ordered phase may be characterized by some
quantitative characteristic $x$ while it will vanish in a disordered
phase above a critical point.
We assume that the thermodynamic potential $\Phi $ of the system is
not only a function of the
order parameter $x$ but also depends on the degree distribution: 
\begin{equation}
\Phi (x,H)=-Hx+\sum_{q}^{\infty }P(q)\phi (x,qx)\,.  \label{assump0}
\end{equation}
Here $H$ is a field conjugated with $x$. Equation~(\ref{assump0}) is
not obvious {\it a priori}. The function $\phi (x,qx)$ can be
considered as a contribution of vertices with $q$ connections. There
are arguments in favor of this assumption. Let us consider the
interaction of an arbitrary vertex with $q$ neighboring vertices. In
the framework of a mean-field approach, $q$ neighbors with a
spontaneous ``moment'' $x$ produce an effective field $qx$ acting on
this vertex.

We impose only general restrictions on $\phi (x,y)$:
\begin{itemize}
\item[(i)] $\phi (x,y)$ is
a smooth function of $x$ and $y$ and can be represented as a series
in powers of both $x$ and $y$. Coefficients of this series are
functions of ``temperature'' $T$ and ``field'' $H$.
\item[(ii)] $\Phi (x,H)$ is finite
for any finite average degree $\left\langle q\right\rangle $.
\end{itemize}

A network topology affects analytical properties of $\Phi $. If the
distribution function $P(q)$ has a divergent moment $\left\langle
q^{p}\right\rangle $, then we have
\begin{equation}
\Phi (x,H)=-Hx+\sum_{n=2}^{p-1}f_{n}x^{n}+x^{p}s(x)\,,  \label{sFL}
\end{equation}
where $s(x)$ is a non-analytic function. The specific form of $s(x)$
is determined by the asymptotic behavior of $P(q)$ at $q\gg 1$. It
is the nonanalytic term that can lead to a deviation from the
standard mean-field behavior.



Following Landau, we assume that near the critical temperature the
coefficient $f_{2}$ can be written as $a(T-T_{c})$ where $a$ is
chosen to be positive for the stability of the disordered phase. The
stability of the ordered phase demands that either $f_{3}>0$ or if
$f_{3}=0,$ then $f_{4}>0$. The order parameter $x$ is determined by
the condition that $\Phi (x,H)$ is minimum: $ d\Phi (x,H)/dx=0$.

If degree distribution exponent $\gamma $ is non-integer, then
the leading nonanalytic term in $\Phi (x)$ is $x^{\gamma -1}$. If
$\gamma $ is integer, then the leading nonanalytic term is
$x^{\gamma -1}\left| \ln x\right| $. Interestingly, this
nonanaliticity looks like that of the free energy for the
ferromagnetic Ising model in magnetic field on a Cayley tree (see
Sec.~\ref{ssec:ising_perfect_tree}).

We must be take into account a symmetry of the system. When $\Phi
(x,H)=\Phi (-x,-H)$ and the coefficient $f_{4}$ is positive, we
arrive at the critical behavior which describes the ferromagnetic
Ising model on equilibrium uncorrelated random networks
in Sec.~\ref{sssec:ising_transition}. In a network with
$\left\langle q^{4}\right\rangle <\infty$ a singular term in $\Phi $
is irrelevant, and we have the usual $x^{4}$- Landau theory which
leads to the standard mean-field phase transition. The singular term
$x^{\gamma -1}$ becomes relevant for $2<\gamma \leq 5$ (this term is
$x^{4}\left| \ln x\right|$ at $ \gamma =5$). Critical exponents are
given in Table \ref{t1}. At the critical point $T=T_{c} $ the order
parameter $x$ is a non-analytic function of $H$: $x\propto
H^{1/\delta }$, where $\delta (\gamma >5)=3$ and $\delta (3<\gamma
<5)=\gamma -2$.

If the symmetry of the system permits odd powers of $x$ in $\Phi $
and $f_{3} $ is positive, then the phenomenological approach gives a
critical behavior which was found for percolation on uncorrelated
random networks in Sec. \ref{sssec:percolation_uncorrelated}. Note
that when $\gamma
>4$, a singular term $x^{\gamma -1}$ is irrelevant. It becomes
relevant for $2<\gamma \leq 4$ (this term is $x^{3}\left| \ln
x\right|$ at $\gamma =4$).

At $2<\gamma \leqslant 3$, the thermodynamic potential has a
universal form, independent on the symmetry:
\begin{equation}
\Phi (x,H)=-Hx+Cx^{2}-Ds(x),  \label{Free-3}
\end{equation}
where $s(x)=x^{2}\left| \ln x\right| $ for $\gamma =3$, and
$s(x)=x^{\gamma -1}$ for $2<\gamma <3$.
We can choose $C\propto T^{2}$ and $D\propto T$, then the
phenomenological theory gives a correct temperature behavior of the
ferromagnetic Ising model.

When $f_{3}<0$ (or $f_{4}<0$\ if $f_{3}=0$), the phenomenological
theory predicts a first-order phase transition for a finite
$\left\langle q^{2}\right\rangle $. This corresponds, e.g., to the
ferromagnetic Potts model with $p\geq 3$ states (see Sec.
\ref{sec:Potts}).



The phenomenological approach agrees with the microscopic theory and
numerical simulations of the ferromagnetic Ising, Potts, $XY$, spin
glass, Kuramoto and the random-field Ising models, percolation and
epidemic spreading on various uncorrelated random networks. These
models have also been studied on complex networks with different
clustering coefficients, degree correlations, etc. It seems that
these characteristics are not relevant, or at least not essentially
relevant, to critical behavior. When the tree ansatz for complex
networks gives exact results, the phenomenology leads to the same
conclusions. In these situations the critical fluctuations are
Gaussian.
We strongly suggest that the critical fluctuations are Gaussian in
all networks with the small-world effect, as is natural for
infinite-dimensional objects.


\subsection{Finite-size scaling}
\label{ssec:finite_size_scaling}

Based on the phenomenological theory one can get scaling exponents
for finite-size scaling phenomena in complex networks. Let
$\Phi(m,\tau,H,N)$ be a
thermodynamic potential per vertex, where $\tau$ is the deviation
from a critical point. According to
the standard scaling hypothesis (in its finite-size scaling form),
in the critical region, 
\begin{equation}
N\Phi(m,\tau,H,N) = f_\phi(m N^x, \tau N^y, H N^z) , \label{e8.003}
\end{equation}
where
$f_\phi(a,b,c)$ is a scaling function. Note that there is exactly
$N$ on the left-hand side of this relation and not an arbitrary
power of $N$.
Formally substituting $\Phi(m,\tau,H)=A\tau m^2 + B
m^{\Delta(\gamma)}-Hm$,
one can find exponents $x$, $y$, and $z$. As was explained, $\Delta$
may be (i) $\text{min}(4,\gamma-1)$, as, e.g., in the Ising model, or
(ii) $\text{min}(3,\gamma-1)$, as, e.g., in percolation.
This
naive
substitution, however, does not allow one to obtain a proper scaling
function, which must be analytical, as is natural. The derivation of
the scaling function demands more rigorous calculations.


As a result, for the two classes of theories listed above, we arrive
at the following scaling forms of the order parameter: 
\begin{eqnarray}
&& \!\!\!\!\!\!\!\!\!\!\!\!\!\!\!\!\!\!\text{(i) for} \ \gamma\geq
5, \ m(\tau,H,N){=}N^{-1/4}\!f(N^{1/2}\tau,N^{3/4}\!H), \label{e8.1}
\\[5pt]
&& \!\!\!\!\!\!\!\!\!\!\!\!\!\!\!\!\!\!\!\!\text{(ii) for}\
\gamma\geq 4, \  m(\tau,H,N){=}N^{-1/3}\!f(N^{1/3}\tau,N^{2/3}\!H),
\label{e8.2}
\end{eqnarray}
and for more small $3<\gamma<5$ (i) or $3<\gamma<4$ (ii), $m$ scales
as 
\begin{equation}
\!m(\tau,\!H,\!N){=}N^{-1/(\gamma-1)}\!f(N^{(\gamma-3)/(\gamma-1)}\tau,N^{(\gamma-2)/(\gamma-1)}H)
. \label{e8.3}
\end{equation}
\textcite{Hong:hhp07} obtained these scaling relations (without
field) by using other
arguments and confirmed them simulating
the Ising model on the static and the configuration models of
uncorrelated networks.
Their idea may be formally reduced to the following steps.
Recall a relevant standard scaling relation from the physics of
critical phenomena in lattices. The standard form is usually written
for dimension $d$ lower than the upper critical dimension $d_u$.
So, rewrite this scaling relation for $d>d_u$: substitute $d_u$ for
$d$ and use the mean-field values of the critical exponents which should be obtained as follows.  
For networks, in this relation, formally substitute $\nu=1/2$ for
the correlation length exponent and $\eta=0$ for the Fisher
exponent, use the susceptibility exponent
$\tilde{\gamma}=\nu(2-\eta)=1$, exponent $\beta=\beta(\gamma)$ (see
Sec.~\ref{ssec:landau_theory}), and 
\begin{equation}
d_u(\gamma) = \frac{2\Delta(\gamma)}{\Delta(\gamma)-2}
\label{e8.400000}
\end{equation}
\cite{Hong:hhp07}. This procedure allows one to easily derive
various scaling relations. We have used it in
Sec.~\ref{sssec:statistics_finite}. 

Finite-size scaling of this kind works in a wide class of models and
processes on networks. \textcite{Hong:hhp07} also applied these ideas to
the contact process on networks. Earlier, \textcite{Kim:khh01} and
\textcite{Medvedyeva:mhm03} studied the finite size scaling by
simulating the $XY$ model on the Watts-Strogatz network. In their
work, in particular, they investigated {\em the dynamic finite-size
scaling}. 
In the framework of our phenomenology, we can easily reproduce their
results and generalize them to scale-free networks.
Let us assume the relaxational dynamics of the order parameter:
$\partial m/\partial t = -\partial\Phi(m)/\partial m$.
In dynamical models, the
scaling hypothesis also implies the scaling time variable,
$t_{\text{scal}}=tN^{-s}$, which means that the relaxation time
diverges as $N^s$ at the critical point. For brevity, we only find
the form of this scaling variable, which actually resolves the
problem. In terms of scaling variables, the dynamic equation for the
order parameter must not contain $N$. With this condition, passing
to the scaling variables $mN^x$ and $tN^{-s}$ in the dynamic
equation, we immediately get
$s=y$, which means that time scales with $N$ exactly in the same way
as $1/\tau$. So, for the indicated two classes of theories, (i) and
(ii), the time scaling variable is of the following form: 
\begin{eqnarray}
&& \!\!\!\!\!\!\!\!\!\!\!\!\!\!\!\!\!\!\!\!\!\!\!\!\!\!\text{in
theory (i),}  \ \,\text{for} \ \gamma\geq 5, \ \ \
t_{\text{scal}}=tN^{-1/2}, \nonumber
\\
&& \!\!\!\!\!\!\!\!\!\!\!\!\!\!\!\!\!\!\!\!\!\!\!\!\!\!\!\text{in
theory (ii),}  \ \text{for} \ \gamma\geq 4, \ \  \
t_{\text{scal}}=tN^{-1/3},
\label{e8.4}
\end{eqnarray}
and
\begin{eqnarray}
&& \!\!\!\!\!\!\!\!\text{in theory\,(i),} \ \text{for} \ 3{<}\gamma
{<} 5, \ \text{and in theory\,(ii),} \  \text{for} \ 3 {<} \gamma
{<} 4, \nonumber
\\
&& \ \ \ \ \ \ \ \ \ \ \ \ \ \ \ \ \ \ \
t_{\text{scal}}=tN^{-(\gamma-3)/(\gamma-1)}
. \label{e8.5}
\end{eqnarray}

Finally, we recommend that
the reader
refer to \textcite{Gallos:gsh07b} for the finite-size scaling in
scale-free networks with fractal properties.
For description of these networks, see \textcite{Song:shm05,Song:shm06,Song:sghm07} and also \textcite{Goh:gsk06}.



\section{SYNCHRONIZATION ON NETWORKS}
\label{sec:synchronization}

Emergence of synchronization in a system of coupled individual
oscillators is an intriguing phenomenon. Nature gives many
well-known examples: synchronously flashing fireflies, crickets that
chirp in unison, two pendulum clocks mounted on the same wall
synchronize their oscillations, synchronous neural activity, and
many others. Different dynamical models were proposed to describe
collective synchronization, see, for example, monographs and reviews
of \textcite{Pikovsky:prk01}, \textcite{Strogatz:sbook03},
\textcite{Strogatz:s00}, \textcite{Acebron:abvrs05}, and
\textcite{Boccaletti:blm06}.

Extensive investigations aimed at searching for network
architectures which optimize synchronization. First (mostly
numerical) studies of various dynamical models have already revealed
that the ability to synchronize can be improved in small-world
networks
\cite{Lago-Fernandez:lhcs00,Gade:gh00,Jost:jj01,Hong:hck02,Barahona:bp02,Wang:wc02}.
On the other hand, an opposite effect was observed in
synchronization dynamics of pulse-coupled oscillators
\cite{Guardiola:gd-glp00}, where homogeneous systems synchronize
better.

We here consider the effect of the network topology on the
synchronization in the Kuramoto model and a network of coupled
dynamical systems. These two models represent two different types
of synchronization phenomena. The interested reader will find the
discussion of this effect for coupled map lattices in
\textcite{Gade:gh00,Lind:lgh04,Jost:jj01,Atay:ajw04,Grinstein:gl05,Huang:hplyy06},
for networks of Hodgkin-Huxley neurons---in
\textcite{Lago-Fernandez:lhcs00,Kwon:km02}, for pulse-coupled
oscillators---in \textcite{Denker:dtdwg04,Timme:twg04}, and for
the Edwards-Wilkinson model---in \textcite{Kozma:khk04}.

\subsection{The Kuramoto model}

\label{ssec:kuramoto-fcg}

\begin{figure}[t]
\begin{center}
\scalebox{0.36}{\includegraphics[angle=270]{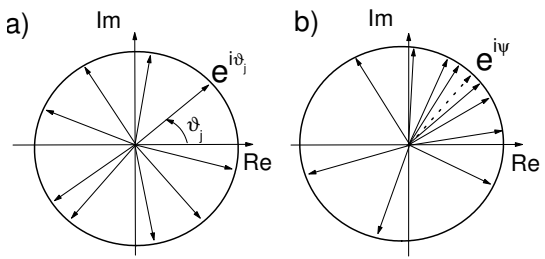}}
\end{center}
\caption{ Schematic view of phases in the Kuramoto model. (a)
Incoherent phase. Unit length vectors representing individual states
are randomly directed in the complex plane. (b) Coherent phase. The
individual states condense around a direction $\protect\psi $. }
\label{fig-synchronization}
\end{figure}

The Kuramoto model is a classical paradigm for a spontaneous
emergence of collective synchronization
\cite{Kuramoto:kbook84,Strogatz:s00,Acebron:abvrs05}. It describes
collective dynamics of $N$ coupled phase oscillators with phases
$\theta _{i}(t)$, $i=1,2,...N$, running at natural frequencies
$\omega _{i}$:
\begin{equation}
\stackrel{.}{\theta }_{i}=\omega _{i}+J\sum\limits_{j=1}^{N}a_{ij}\sin
(\theta _{j}-\theta _{i}),  \label{Kuramoto}
\end{equation}
where $a_{ij}$ is the adjacency matrix of a network. $J$ is the
coupling strength. The frequencies $\omega _{i}$ are distributed
according to a distribution function $g(\omega )$. It is usually
assumed that $g(\omega )$ is unimodular and symmetric about its mean
frequency $\Omega $. It is convenient to use a rotating frame and
redefine $\theta _{i}\rightarrow \theta _{i}-\Omega t$ for all $i$.
In this frame we can set the mean of $g(\omega )$ to be zero. The
state of oscillator $j$ can be characterized by a complex exponent
$\exp (i\theta _{j})$ which is represented by a vector of unit
length in the complex plane (see Fig.~\ref{fig-synchronization}).


The Kuramoto model is solved exactly for the fully connected graph
(all-to-all interaction), i.e., $a_{ij}=1$ for all $i\neq j$, with
rescaling $J\rightarrow J/N$. When $J<J_{c}$, there is no collective synchronization between the rotations of individual oscillators. Nonetheless some finite clusters of synchronized oscillators may exist.
Collective synchronization between oscillators emerges
spontaneously above a critical coupling $J_{c}$ if $N\rightarrow
\infty $. The global state of the model is characterized by the
following average:
\begin{equation}
r(t)e^{i\psi (t)}=\frac{1}{N}\sum_{j=1}^{N}e^{i\theta _{j}},
\label{r-kuramoto}
\end{equation}
where $r(t)$ is the order parameter which measures the phase
coherence, and $\psi (t)$ is the average phase. Simulations show
that if we start from any initial state, then at $J<J_{c}$ in the
incoherent phase, $r(t)$ decays to a tiny jitter of the order of
$O(N^{-1/2})$. On the other hand, in the coherent phase ($J>J_{c}$),
the parameter $r(t)$ decays to a finite value $r(t\rightarrow \infty
)=r<1$. At $J$ near $J_{c}$, the order parameter $r\propto $ $\left|
J-J_{c}\right| ^{\beta }$ with $\beta =1/2$. In the original frame,
$\psi (t)$ rotates uniformly at the mean frequency $\Omega $.
Substituting Eq.~(\ref{r-kuramoto}) into Eq.~(\ref{Kuramoto}) gives
\begin{equation}
\stackrel{.}{\theta }_{i}=\omega _{i}+Jr\sin (\psi -\theta _{i}).
\label{p-kuramoto}
\end{equation}
The steady solution of this equation shows that at $J>J_{c}$, a
finite fraction of synchronized oscillators emerges.\ These
oscillators rotate coherently at frequency $\Omega $ in the original
frame. In the rotating frame, they have individual frequencies
$\left| \omega _{i}\right| \leqslant Jr$ and their phases are locked
according to the equation: $\sin \theta _{i}=\omega _{i}/Jr$, where
we set $\psi =0$. Others oscillators, having individual frequencies
$\left| \omega _{i}\right| >Jr$, are ``drifting''. Their phases are
changed non-uniformly in time. The order parameter $r$ satisfies the
self-consistent equation:
\begin{equation}
r=\int\limits_{-Jr}^{Jr}\sqrt{1-\frac{\omega
^{2}}{J^{2}r^{2}}}\,\,g(\omega )d\omega ,  \label{r-self}
\end{equation}
which gives the critical coupling $J_{c}=2/[\pi g(0)]$. Note that
the order of the synchronization phase transition in the Kuramoto
model depends on the distribution $g(\omega)$. In particular, it can
be of first order if the natural frequencies are uniformly
distributed \cite{Tanaka:tlo97}.

The Kuramoto model on finite networks and lattices shows synchronization if the coupling is sufficiently strong. 
Is it possible to observe collective synchronization in the Kuramoto model on an infinite regular lattice? 
For sure, there is no synchronization in a
one-dimensional system with a short-ranged coupling. 
According to \textcite{Hong:hpc04,Hong:hpc05}, phase and frequency ordering is absent also in two-dimensional ($d=2$) lattices;  
frequency ordering is possible only in three-, four-, and higher-dimensional lattices, while phase ordering is possible only when $d>4$.
The value of the upper critical dimension of the Kuramoto model is 
still under discussion \cite{Acebron:abvrs05}. Simulations in
\textcite{Hong:hcpt07} 
indicate the mean-field behavior of the Kuramoto model at $d>4$.

\subsection{Mean-field approach}

\label{ssec:kuramoto-mean_field}

The Kuramoto model was recently investigated numerically and
analytically on complex networks of different architectures.
We here first look at analytical studies and then discuss
simulations though the model was first studied numerically in
\textcite{Hong:hck02}.

Unfortunately no exact results for the Kuramoto model on complex
networks are obtained yet. A finite mean degree and a strong
heterogeneity of a complex network make difficult to find an
analytical solution of the model.
\textcite{Ichinomiya:i04,Ichinomiya:i05} and \textcite{Lee:l05}
developed a simple mean-field theory
which is actually equivalent to the annealed network approximation
in Sec.~\ref{sssec: annealed anzats}.
Using this approximation, we arrive at the Kuramoto model with a
degree dependent coupling on the fully connected graph:
\begin{equation}
\stackrel{.}{\theta }_{i}=\omega _{i}+\frac{Jq_{i}}{Nz_{1}}%
\sum\limits_{j=1}^{N}q_{j}\sin (\theta _{j}-\theta _{i}).
\label{anneled-Kuramoto}
\end{equation}
This effective model can be easily solved exactly. Introducing a
weighted order parameter
\begin{equation}
\widetilde{r}(t)e^{i\widetilde{\psi }(t)}=\frac{1}{Nz_{1}}%
\sum_{j=1}^{N}q_{j}e^{i\theta _{j}},  \label{r-annealed}
\end{equation}
one can write Eq.~(\ref{anneled-Kuramoto}) as follows:
\begin{equation}
\stackrel{.}{\theta }_{i}=\omega _{i}+J\widetilde{r}q_{i}\sin (\widetilde{%
\psi }-\theta _{i}).  \label{p-a-kuramoto}
\end{equation}
The steady solution of this equation shows that in the coherent
state, oscillators with individual frequencies $\left| \omega
_{i}\right| \leqslant J\widetilde{r}q_{i}$ are synchronized. Their
phases are locked and depend on vertex degree: $\sin \theta
_{i}=\omega _{i}/(J\widetilde{r}q_{i})$, where we set
$\widetilde{\psi }=0$. This result shows that hubs with degree
$q_{i}\gg 1$ synchronize more easy than oscillators with low
degrees. The larger the degree $q_{i}$ the larger the probability
that an individual
frequency $\omega _{i}\ $of an oscillator $i$ falls into the range $[-J%
\widetilde{r}q_{i},J\widetilde{r}q_{i}]$. Other oscillators are
drifting. $\widetilde{r}$ is a solution of the equation:
\begin{equation}
\widetilde{r}=\sum_{q}\frac{P(q)q}{z_{1}}\int\limits_{-J\widetilde{r}q}^{J%
\widetilde{r}q}\sqrt{1-\frac{\omega
^{2}}{(J\widetilde{r}q)^{2}}}\,\,\,g(\omega )d\omega .
\label{r-a-self}
\end{equation}
Spontaneous synchronization with $\widetilde{r}>0$ emerges above the
critical coupling
\begin{equation}
J_{c}=\frac{2z_{1}}{\pi g(0)\left\langle q^{2}\right\rangle }  \label{Jc}
\end{equation}
which strongly depends on the degree distribution. $J_{c}$ is finite
if the second moment $\left\langle q^{2}\right\rangle $ is finite.
Note that at a fixed mean degree $z_{1}$, $J_{c}$ decreases (i.e.,
the network synchronizes easily) with increasing $\left\langle
q^{2}\right\rangle $---increasing heterogeneity.
Similarly to percolation, if the moment $\left\langle
q^{2}\right\rangle $ diverges (i.e., $2<\gamma\leq 3$),
the synchronization threshold $J_{c}$ is absent, and the
synchronization is robust against random failures. In finite
networks, the critical coupling is finite, $J_{c}(N)\propto
1/q_{\text{cut}}^{3-\gamma }(N)$, and is determined by the
size-dependent cutoff $q_{\text{cut}}(N)$ in
Sec.~\ref{sssec:cutoffs}.

Another important result, which follows from Eq.~(\ref{r-a-self}),
is that the network topology strongly influences the critical
behavior of the order parameter $\widetilde{r}$. \textcite{Lee:l05}
found that the critical singularity of this parameter is described
by the standard mean-field critical exponent $\beta =1/2$ if an
uncorrelated network has a finite fourth moment $\left\langle
q^{4}\right\rangle $, i.e., $\gamma >5$. If $3<\gamma <5$, then
$\beta =1/(\gamma -3)$. Note that the order parameters $r$,
Eq.~(\ref{r-kuramoto}), and $\widetilde{r}$, Eq.~(\ref{r-annealed}),
have the same critical behavior. Thus, with fixed $z_{1}$, the
higher heterogeneity of a network, the better its sinchronizability
and the smoother the phase transition. The critical behavior of the
Kuramoto model is similar to one found for the ferromagnetic Ising
model in Sec.~\ref{ssec:ferromagnetic_ising_uncorrelated} and
confirms the phenomenological theory described in
Sec.~\ref{sec:phenomenology}. A finite-size scaling analysis of the
Kuramoto model in complex networks was  carried out by
\textcite{Hong:hpt07}. Within the mean-field theory, they found that
the order parameter $\widetilde{r}$
has the finite-size scaling behavior,
\begin{equation}
\widetilde{r}=N^{-\beta
\overline{\nu}}f((J-J_{c})N^{1/\overline{\nu}}), \label{r-scaling}
\end{equation}
with the critical exponent $\beta$ found above. 
Remarkably, the
critical exponent $\overline{\nu}$ is different from that of the
Ising model in Sec.~\ref{ssec:finite_size_scaling}, namely,
$\overline{\nu}=5/2$ at $\gamma>5$, $\overline{\nu}=(2 \gamma -
5)/(\gamma - 3)$ at $4< \gamma <5$, $\overline{\nu}=(\gamma -
1)/(\gamma - 3)$ at $3< \gamma <4$. 
Simulations of the Kuramoto model carried out by
\textcite{Hong:hpt07} agree with these analytical results.

The mean-field theory of synchronization is based on the assumption
that every oscillator ``feels'' a ``mean field''
created by nearest neighbors. This assumption is valid if the
average degree $z_{1}$ is sufficiently large.
In order to improve the mean-field theory,
\textcite{Restrepo:roh05} introduced a local order parameter at
vertex $n$,
\begin{equation}
r_{n}e^{i\psi _{n}}=\sum_{m}a_{nm}e^{i\theta _{m}},  \label{l-r}
\end{equation}
and found it by using intuitive arguments. In their approach the
critical coupling $J_c$ is inversely proportional to the maximum
eigenvalue $\lambda _{\max }$ of the adjacency matrix $a_{ij}$.
However, in an uncorrelated random complex network, the maximum
eigenvalue $\lambda _{\max }$ is determined by the cutoff
$q_{cut}(N)$ of the degree distribution, $\lambda _{\max }\approx
q_{cut}^{1/2}(N)$
\cite{Krivelevich:ks03,Dorogovtsev:dgms03,Chung:clv03}. In
scale-free networks ($\gamma <\infty$), the cutoff diverges in the
limit $N\rightarrow \infty $. Therefore this approach
predicts $J_{c}=0$ in the thermodynamic limit even for a scale-free
network with $\gamma >3$ in sharp contrast to the approach of
\textcite{Ichinomiya:i04} and \textcite{Lee:l05}.

\textcite{Oh:olkk07} studied the Kuramoto model with asymmetric
degree dependent coupling $Jq_{i}^{-\eta }a_{ij}$ instead of
$Ja_{ij}$ in Eq.~(\ref {Kuramoto}) by using the mean-field theory.
They found that tuning the exponent $\eta $ changes the critical
behavior of collective synchronization. On scale-free networks, this
model has a rich phase diagram in the plane $(\eta ,\gamma )$. In
the case $\eta =1 $, the critical coupling $J_{c}$ is finite even in
a scale-free network with $2<\gamma <3$ contrary to $J_{c}=0$ for
the symmetric coupling which corresponds to $\eta =0$. Note that the
influence of the degree dependent coupling is similar to the effect
of degree dependent interactions on the phase transition in the
ferromagnetic Ising model
(see Sec.~\ref{sssec:degree_dependent_interactions}).

\subsection{Numerical study of the Kuramoto model}
\label{ssec:kuramoto-numerics}

The Kuramoto model was investigated numerically on various networks.
\textcite{Hong:hck02} studied numerically the model on the
Watts-Strogatz network generated from a one-dimensional regular
lattice. They observed that collective synchronization emerges even
for a tiny fraction of shortcuts, $p$, which make the
one-dimensional lattice to be a small world. The critical coupling
$J_{c}$ is well approximated as follows: $J_{c}(p)\approx 2/[\pi
g(0)]+ap^{-1}$, where $a$ is a constant.
As one might expect, the synchronization phase transition is of
second order with the standard critical exponent $\beta =0.5$.

\begin{figure}[t]
\begin{center}
\scalebox{0.16}{\includegraphics[angle=0]{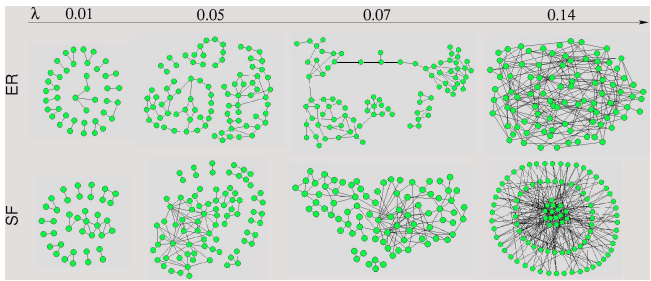}}
\end{center}
\caption{ Synchronization patterns of Erd\H{o}s-R\'{e}nyi (ER) and
scale-free (SF) networks for several values of coupling
$\protect\lambda$ (or $J$ in our notations). From
\protect\textcite{Gomez-Gardenes:g-gma07a}.}
\label{fig-synch-clusters}
\end{figure}

The evolution of synchronization in the Kuramoto model on the
Erd\H{o}s-R\'{e}nyi and scale-free networks was recently studied
by \textcite{Gomez-Gardenes:g-gma07a,Gomez-Gardenes:g-gma07b}.
These authors solved numerically Eq.~(\ref {Kuramoto}) for
$N=1000$ coupled phase oscillators and demonstrated that (i) the
synchronization on a scale-free network ($\gamma =3$) appears at a
smaller critical coupling $J_{c}$ than the one on the
Erd\H{o}s-R\'{e}nyi network (with the same average degree as the
scale-free network), (ii) the synchronization phase transition on
the Erd\H{o}s-R\'{e}nyi network is sharper than the transition on
the scale-free network. This critical behavior agrees
qualitatively with the mean-field theory.
\textcite{Gomez-Gardenes:g-gma07a,Gomez-Gardenes:g-gma07b}
calculated a fraction of synchronized pairs of neighboring
oscillators for several values of the coupling $J$ and revealed an
interesting difference in the synchronization patterns between the
Erd\H{o}s-R\'{e}nyi and scale-free networks (see
Fig.~\ref{fig-synch-clusters}). In a scale-free network,
a central core of synchronized oscillators formed
by hubs
grows
with $J$
by
absorbing small
synchronized clusters.
In contrast, in the Erd\H{o}s-R\'{e}nyi network numerous small
synchronized clusters homogeneously spread over the graph. As $J$
approaches $J_{c}$, they progressively merge together and form
larger clusters.


\textcite{Moreno:mp04} carried out numerical study of the Kuramoto
model on the Barab\'{a}si-Albert network of size $N=5\times 10^{4}
$. They found that the critical coupling is finite, though small.
Surprisingly, the measured critical exponent was close to the
standard mean-field value, $\beta \sim 0.5$, contrary to
an
infinite order phase transition and zero $J_{c}$ predicted by the
mean-field theory in the limit $N\rightarrow \infty $. A reason of
this discrepancy is unclear.

A community (modular) structure of complex networks makes a strong
effect on 
synchronization.
In such networks, oscillators inside a community are synchronized
first because edges within a community are arranged denser than
edges between communities. On the other hand, intercommunity edges
stimulate the global synchronization. The role of network motifs
for the synchronization in the Kuramoto model was first studied
numerically by \textcite{Moreno:mvp04}. \textcite{Oh:orh05} solved
numerically the dynamical equations Eq.~(\ref{Kuramoto}) with the
asymmetric degree dependent coupling $Jq_{i}^{-1}a_{ij}$ for two
real networks---the yeast protein interaction network and the
Internet at the Autonomous system level. These networks have
different community structures. In the yeast protein network,
communities are connected diversely while in the Internet
communities are connected mainly to the North America continent.
It turned out that for a given coupling $J$, the global synchronization for the yeast
network is stronger than that for the Internet.
These numerical calculations showed that the
distributions of phases of oscillators inside communities in the
yeast network overlap each other. This corresponds to the mutual
synchronization of the communities. In contrast, in the Internet,
the phase distributions inside communities do not overlap, the
communities are coupled weaker and synchronize independently.
A modular structure produces a similar effect on synchronization
of coupled-map networks \cite{Huang:hplyy06}.

\textcite{Arenas:adp06a,Arenas:adp06b} showed that the evolution of a
synchronization pattern reveals different topological scales at
different time scales in a complex network with nested communities.
Starting from random initial conditions, highly interconnected
clusters of oscillators synchronize first. Then larger and larger
communities do the same up to the global coherence.
Clustering produces a similar effect.
\textcite{McGraw:mm06} studied numerically the synchronization on
the Barabasi-Albert networks of size $N=1000$ with low and high
clustering coefficients (networks with a high clustering coefficient
were generated by using the method proposed by \textcite{Kim:k04}).
These authors found that in a clustered network the synchronization
emerges at a lower coupling $J$ than a network with the same degree
distribution but with a lower clustering coefficient. However, in
the latter network the global synchronization is stronger.

\textcite{Timme:t06} simulated the Kuramoto model on directed
networks and observed a topologically induced transition from
synchrony to disordered dynamics. This transition may be a general
phenomenon for different types of dynamical models of
synchronization on directed networks.

Synchronization of coupled oscillators in the Kuramoto model to an
external periodic input, called {\em pacemaker}, was studied for
lattices, Cayley trees and complex networks by
\textcite{Yamada:y02,Kori:km04,Kori:km06,Radicchi:rm06}. This
phenomenon is called {\em entrainment}. The pacemaker is assumed
to be coupled with a finite number of vertices in a given network.
Entrainment appears above a critical coupling strength $J_{cr}$.
\textcite{Kori:km04} showed that $J_{cr}$ increases exponentially
with increasing the mean shortest path distance $\cal L$ from the
pacemaker to all vertices in the network, i.e., $J_{cr} \sim
e^{a{\cal L}}$. In a complex network, $\cal L$ is proportional to
the mean intervertex distance $\overline{\ell}(N)$ which, in turn,
is typically proportional to $\ln N$, see
Sec.~\ref{ssec:characteristics}. This leads to $J_{cr} \sim
N^{b}$, where $b$ is a positive exponent. It was shown that
frequency locking to the pacemaker strongly depends on its
frequency and the network architecture.

\subsection{Coupled dynamical systems}

\label{ssec:coupled systems}

Consider $N$ identical dynamical systems. An individual system is
described by a vector dynamical variable ${\bf x}_{i}(t)$,
$i=1,...N$. The individual dynamics is governed by the equation:
$\dot{{\bf x}}_{i}={\bf F}({\bf x}_{i})$, where ${\bf F}$ is a
vector function. These dynamical systems are coupled by edges and
their dynamics is described by the equation:
\begin{equation}
\dot{{\bf x}}_{i}={\bf F}({\bf x}_{i})-J\sum_{j}L_{ij}{\bf H}({\bf x}%
_{j}),  \label{dynamics1}
\end{equation}
where $J$ is the coupling strength, ${\bf H}({\bf x}_{j})$ is an
output function which determines the effect of vertex $j$ on
dynamics of vertex $i$. The network topology is encoded in the
Laplacian matrix $L_{ij}=q_{i}\delta _{ij}-a_{ij}$, where $a_{ij}$
is the adjacency matrix, and $q_{i}$ is degree of vertex $i$. The
Laplacian matrix is a zero-row-sum matrix, i.e., $\sum_{j}L_{ij}=0$
for all $i$. This property has the following consequence. Any
solution of the equation $\dot{{\bf s}}={\bf F}({\bf s})$ is also a
solution of Eq.~(\ref{dynamics1}), ${\bf x}_{i}={\bf s}(t)$, i.e.,
dynamical systems evolve \vspace{-13pt}coherently.

\subsubsection{Stability criterion.}

\label{sssec:stability_criterion}

We use the spectral properties of $L$ in order to determine the
stability of the fully synchronized state against small
perturbations, ${\bf x}_{i}={\bf s}(t)+{\bf \eta }_{i}$. The
Laplacian has nonnegative eigenvalues which can be ordered as
follows, $0=\lambda _{1}<\lambda _{2}\leqslant ...\leqslant \lambda
_{N}$. The zero eigenvalue corresponds to the uniform eigenfunction,
$f_{i}^{(0)}=1$ for all $i$ (the synchronized state). The remaining
eigenfunctions $f_{i}^{(\lambda )}$ with $\lambda \geqslant \lambda
_{2}$ are transverse to $f_{i}^{(0)}$. Representing a perturbation
as a sum of the transversal modes, ${\bf \eta }_{i}=\sum_{\lambda
\geqslant \lambda _{2}}{\bf \eta }_{\lambda }f_{i}^{(\lambda )}$, we
find the master stability equation from Eq.~(\ref{dynamics1}):
\begin{equation}
\dot{{\bf \eta }}_{\lambda }=[D{\bf F}({\bf s})-\alpha D{\bf H}({\bf
s})]{\bf \eta }_{\lambda },  \label{master}
\end{equation}
where $\alpha =J\lambda $. $D{\bf F}$ and $D{\bf H}$ are the
Jacobian matrices. If the largest Lyapunov exponent $\Lambda (\alpha
)$ of this equation is negative, then the fully synchronized state
is stable \cite {Pecora:pc98}. $\Lambda (\alpha )$ is called the
master stability function. This function is known for various
oscillators such as R\"{o}ssler, Lorenz, or double-scroll chaotic
oscillators. Equation~(\ref{master}) is valid if the coupling matrix
$L_{ij}$ is diagonalizable. A generalization of the master stability
equation for non-diagonalizable networks (e.i., for the case of a non-symmetric coupling matrix) is given in
\textcite{Nishikawa:nm06a,Nishikawa:nm06b}.

Thus we have the following criterion of the stability: the
synchronized state is stable if and only if $\Lambda (J\lambda
_{n})<0$ for all $n=2,...N$. In this case, a small perturbation
${\bf \eta }_{\lambda }$ converges to zero exponentially as
$t\rightarrow \infty $. The condition $\Lambda (J\lambda
_{1})=\Lambda (0)<0$ determines the dynamical stability of the
solution ${\bf s}(t)$ to the individual dynamics.

Usually, the function $\Lambda (\alpha )$ is negative in a bound
region $\alpha _{1}<\alpha <\alpha _{2}$. Therefore, a network is
synchronizable if simultaneously $J\lambda _{2}>\alpha _{1}$ and
$J\lambda _{N}<\alpha _{2}$. This is equivalent to the following
condition:
\begin{equation}
\frac{\lambda _{N}}{\lambda _{2}} < \frac{\alpha _{2}}{\alpha _{1}}
\label{criterion_synch}
\end{equation}
\cite{Barahona:bp02}. Note that $\lambda_{2}$ and $\lambda _{N}$ are
completely determined by the network topology, while $\alpha _{1}$
and $\alpha _{2}$ depend on the specific dynamical functions ${\bf
F}$ and ${\bf H}$. The value of $\alpha _{2}/\alpha $
typically
ranges from 5
to 100 for various chaotic oscillators. The criterion
Eq.~(\ref{criterion_synch}) implies the existence of the interval
$(\alpha _{1}/\lambda _{2},\alpha _{2}/\lambda _{N})$ of the
coupling strength $J$\ where the synchronization is stable. The
smaller the eigenratio $\lambda _{N}/\lambda _{2}$, the larger this
interval and the better synchronizability. If $J<\alpha _{1}/\lambda
_{2}$, then modes with the small eigenvalues $\lambda <\alpha _{1}/J
$ break down synchronization. If $J>\alpha _{2}/\lambda _{N}$, then
modes with the large eigenvalues $\lambda
>\alpha _{2}/J$ lead away from the synchronized state.

The spectrum of the Laplacian on the fully connected graph is
simple: $\lambda _{1}=0$ and $\lambda _{2}=...=\lambda _{N}=N$. The
eigenratio $\lambda _{N}/\lambda _{2}$ is equal to 1. It corresponds
to the highest possible synchronizabilty. In the $d$-dimensional
cubic lattice of
side length $l=N^{1/d}$, the minimum eigenvalue
$\lambda _{2}$ of the Laplacian is small: $\lambda _{2}\propto
l^{-2}$. On the other hand, the largest eigenvalue $\lambda _{N}$ is
finite: $\lambda _{N}\sim d$. Therefore, the eigenratio $\lambda
_{N}/\lambda _{2}$ diverges as $N\rightarrow \infty $. It means that
the complete synchronization is impossible in an infinite
$d$-dimensional lattice \cite{Wang:wc02,Hong:hkcp04}. Only a finite
lattice can be \vspace{-13pt}synchronized.

\subsubsection{Numerical study.}

\label{sssec:coupled_systems_numerics}

\begin{figure}[t]
\begin{center}
\scalebox{0.55}{\includegraphics[angle=0]{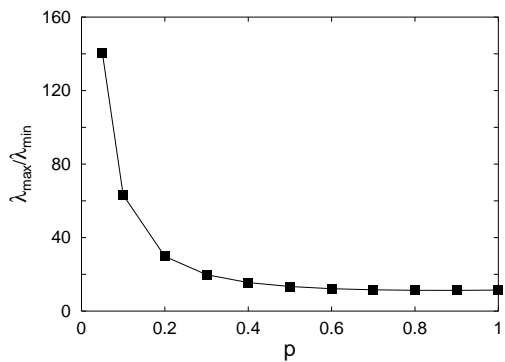}}
\end{center}
\caption{ Ratio $\protect\lambda _{N}/\protect\lambda _{2}$ versus
the fraction of shortcuts, $p$, for the Watts-Strogatz network
generated from a ring. Adapted from \protect\textcite{Hong:hkcp04}.
} \label{fig-eigenratio}
\end{figure}

Synchronization of coupled dynamical systems on various complex
networks was extensively studied numerically. It turned out that the
random addition of a small fraction of shortcuts, $p$, to a regular
cubic lattice leads to a synchronizable network
\cite{Barahona:bp02,Wang:wc02,Hong:hkcp04}. For example, a ring of
$N$ vertices with shortcuts is always synchronizable if $N$ is
sufficiently large. The shortcuts decrease sharply the ratio
$\lambda _{N}/\lambda _{2}$ (see Fig.~\ref{fig-eigenratio}) until
the network becomes synchronizable. The heuristic reason for this
effect lies in the fact that adding shortcuts leads to the
Watts-Strogatz network with the small-world effect. The average
shortest path between two vertices chosen at random becomes very
small
compared to the original regular lattice. In other words,
the small world effect improves synchronizability of the
Watts-Strogatz network
compared with a regular lattice.


Synchronization is also enhanced in other complex networks. One can
show that the minimum eigenratio $\lambda _{N}/\lambda _{2}$ is
achieved for the Erd\H{o}s-R\'{e}nyi graph. In scale-free networks,
the eigenratio $\lambda _{N}/\lambda _{2}$ increases with decreasing
degree distribution exponent $\gamma $,
and so
synchronizability
becomes worse. This effect was
explained by the
increase of heterogeneity
\cite{Nishikawa:nmlh03,Motter:mzk05a,Motter:mzk05b}.
It was found that a suppression of synchronization is related to the
increase of the load on vertices.
Importantly, the eigenratio $\lambda _{N}/\lambda _{2}$ increases
strongly with $N$. \textcite{Kim:km07}, see also
\textcite{Motter:m07}, found that the largest eigenvalue $\lambda
_{N}$ in a uncorrelated scale-free network is determined by the
cutoff of the degree distribution: $\lambda
_{N}=q_{\text{cut}}+1$. The eigenvalue $\lambda _{2}$ is
nearly size-independent and is ensemble averageable.
(The last statement means that
as $N
\rightarrow \infty$, the ensemble distribution of $\lambda _{2}$
converges to a peaked distribution.)
This leads to
\begin{equation}
\frac{\lambda _{N}}{\lambda _{2}} \sim \text{min}
[N^{1/(\gamma -1)},N^{1/2}]
,
\label{e8c4-100}
\end{equation}
see Sec.~\ref{sssec:cutoffs}.
Therefore, it is difficult or even impossible to synchronize a large
scale-free network with sufficiently small $\gamma $. These
analytical results agree with numerical calculations of the
Laplacian spectra of uncorrelated scale-free networks.

Another way to enhance synchronization is to use a network with
asymmetric or weighted couplings.
\textcite{Motter:mzk05a,Motter:mzk05b,Motter:mzk05c} considered an
asymmetric degree dependent coupling matrix $q_{i}^{-\eta }L_{ij}$
instead of $L_{ij}$ in Eq.~(\ref {dynamics1}), where $\eta $ is a
tunable parameter. Their numerical and analytical calculations
demonstrated that if $\eta =1$, then in a given network topology the
synchronizability is maximum and does not depend on the network
size. In this case, the eigenratio $\lambda _{N}/\lambda _{2}$ is
quite insensitive to the form of the degree distribution.
Interestingly, in a random network the eigenvalues $\lambda _{2}$
and $\lambda _{N}$ of the normalized Laplacian matrix
$q_{i}^{-1}L_{ij}$ achieve 1 as $\lambda _{2}=1-O(1/\sqrt{\langle q
\rangle})$ and $\lambda _{N}=1+O(1/\sqrt{\langle q \rangle})$ in the
limit of a large mean degree $\langle q \rangle \gg 1$
\cite{Chung:cbook97}. Therefore in this limit the eigenratio
$\lambda _{N}/\lambda _{2}$ is close to 1, and the system is close
to the highest possible synchronizability.

Note that apart the synchronization, network spectra have
numerous applications
to structural properties of networks and processes in them.
For results on Laplacian
spectra
of
complex networks and their applications, see, e.g.,
\textcite{Chung:cbook97,Dorogovtsev:dgms03,Kim:km07,Motter:m07} and
references therein.


\textcite{Chavez:chahb05} found that a further enhancement of
synchronization in scale-free networks can be achieved by scaling
the coupling strength to the load of each edge. Recall that the load
$l_{ij}$ of an edge $ij$ is the number of shortest paths which go
through this edge. The authors replaced the Laplacian $L_{ij}$ to a
zero row-sum matrix with off-diagonal elements $-l_{ij}^{\alpha
}/\sum_{j\in N_{i}}l_{ij}^{\alpha }$, where $\alpha $ is a tunable
parameter. This weighting procedure used a global information of
network pathways.
\textcite{Chavez:chahb05} demonstrated that varying the parameter
$\alpha $, one may efficiently get better synchronization.
A similar improvement was obtained by using a different, local
weighting procedure based on the degrees of the nearest neighbors
\cite{Motter:mzk05c}.
In networks with inhomogeneous couplings between oscillators,
the intensity of a vertex
is defined
as the total strength of input couplings. \textcite{Zhou:zmk06} showed that the synchronizability in weighted random networks is enhanced
as vertex intensities become more homogeneous.

The effect of degree correlations in a network on synchronization of
coupled dynamical systems was revealed by \textcite{Bernardo:bgs05}.
These authors studied assortatively mixed scale-free networks. Their
degree correlated networks were generated by using the method
proposed by \textcite{Newman:n03d}. They showed that disassortative
mixing
(connections between high-degree and low-degree vertices are more
probable) enhances synchronization in both weighted and unweighted
scale-free networks
compared to uncorrelated networks. However the synchronization in a
correlated network depends on the weighting procedure
\cite{Chavez:chm06}.

\begin{figure}[t]
\begin{center}
\scalebox{0.41}{\includegraphics[angle=270]{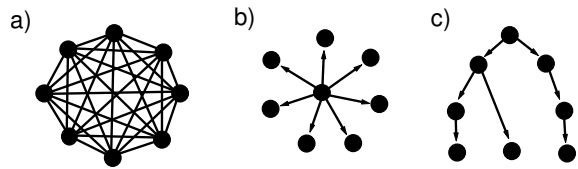}}
\end{center}
\caption{Examples of graphs with
optimal synchronizability: (a)
a fully connected graph; (b) a directed star; (c) a hierarchical
directed random graph. }
\label{f-new4}%
\end{figure}

Above we showed that the fully connected graph gives the optimal
synchronization. However this graph is cost-is-no-object and
uncommon in nature. Which other
architectures
maximize the
synchronizability of coupled dynamical systems?
\textcite{Nishikawa:nm06a,Nishikawa:nm06b} came to the conclusion
that the most optimal networks are directed and non-diagonalizable.
Among the optimal networks they found a subclass of hierarchical
networks with the following properties: (i) these networks embed an
oriented spanning tree (i.e., there is a node from which all other
vertices of the network can be reached by following directed links);
(ii) there are no directed loops, and (iii) the total sum of input
couplings at each vertex is the same for all vertices. Examples of
optimal network topologies are shown in Fig.~\ref{f-new4}.

\section{SELF-ORGANIZED CRITICALITY PROBLEMS ON NETWORKS}
\label{sec:self-organized}

In this section we discuss avalanche processes in models defined on
complex networks and other related phenomena.



\subsection{Sandpiles and avalanches}
\label{ssec:sandpiles}

The sandpile dynamics on the Erd\H{o}s-R\'enyi random graphs was
studied since \cite{Bonabeau:b95} but no essential difference from
high-dimensional lattices was found.
\textcite{Goh:glk03,Lee:lgk04a,Lee:lgk04b} investigated a variation
of the famous Bak-Tang-Wiesenfeld (BTW) model on scale-free
uncorrelated networks and observed an effect of the network
architecture on the self-organized criticality (SOC) phenomenon. Let
us discuss these results.

The model is defined as follows. For each vertex $i$, a threshold
$a_i=q_i^{1-\eta}$ is defined, where $0\leq \eta \leq 1$, so that
$a_i \leq q_i$. A number of grains at vertex $i$ is denoted by
$h_i$.

\begin{itemize}

\item[(i)]
A grain is added to a randomly chosen vertex $i$, and $h_i$
increases by $1$.

\item[(ii)]
If the resulting $h_i<a_i$,
go to (i). On the other hand, if $h_i\geq a_i$, then $h_i$ is
decreased by $\lceil a_i \rceil$, the smallest integer greater or
equal to $a_i$. That is, $h_i \to h_i-\lceil a_i \rceil$. These
$\lceil a_i \rceil$ toppled grains jump to $\lceil a_i \rceil$
randomly chosen nearest neighbors of vertex $i$: $h_j \to h_j+1$.

\item[(iii)]
If for all these $\lceil a_i \rceil$ vertices, the resulting
$h_j<a_j$, then the ``avalanche'' process finishes. Otherwise, the
vertices with $h_j\geq a_j$ are  updated in parallel (!), $h_j\to
\lceil a_j \rceil$, their randomly chosen neighbors receive grains,
and so on until the avalanche stops. Then repeat (i).

\end{itemize}
Note that the particular, ``deterministic''
case of $\eta=0$, where all nearest neighbors of an activated vertex
receive grains (as in the BTW model) essentially differs from the
case of $\eta>0$, where $\lceil a_i \rceil<q_i$.

As is usual in SOC problems, the statistics of avalanches was
studied: the size distribution ${\cal P}_s(s)\sim s^{-\tau}$ for the
avalanches (the ``size'' is here the total number of toppling events
in an avalanche) and the distribution ${\cal P}_t(t) \sim
t^{-\delta}$ of their durations. (The distribution of the avalanche
area---the number of vertices involved---is quite similar to ${\cal
P}_s(s)$.) Taking into account the tree-like structure of
uncorrelated networks, one can see that (i) an avalanche in this
model is a branching process, avalanches are trees, (ii) the
duration of an avalanche $t$ is the distance from its root to its
most remote vertex, and (iii) the standard technique for branching
processes is applicable to this problem.

The basic characteristic of the avalanche tree is the distribution
of branching, $p(q)$. According to
\textcite{Lee:lgk04a,Lee:lgk04b,Goh:glk05}, $p(q)=p_1(q)p_2(q)$. The
first factor is the probability that 
$q-1 < a \leq q$, 
that is
$q$ grains will
fall from 
a vertex
in the act of toppling. $p_2(q)$ is the probability that before the
toppling, the vertex has exactly $q-1$ grains. The assumption that
the distribution of $h$ is homogeneous gives the estimate $p_2(q)
\sim 1/q$. As for $p_1(q)$, one must take into account that (i) the
degree distribution of an end of an edge is $q P(q)/\langle
q\rangle$, (ii) $P(q) \sim q^{-\gamma}$, and (iii) $a=q^{1-\eta}$.
As a result, $p_1(q) \sim q^{-(\gamma-1-\eta)/(1-\eta)}$. Thus, the
distribution of branching is $p(q) \sim q^{-(\gamma-2\eta)/(1-\eta)}
\equiv q^{-\gamma'}$. One can see that if $p_2(q)=1/q$, then $\sum_q
qp(q)=1$.

\textcite{Goh:glk03,Lee:lgk04a,Lee:lgk04b} applied the standard
technique to the branching process with this $p(q)$ distribution and
arrived at power-law size and duration distributions, which
indicates the presence of a SOC phenomenon for the assumed threshold
$a=q^{1-\eta}$. They obtained exponents $\tau$ and $\delta$. With
these exponents, one can easily find the dynamic exponent
$z=(\delta-1)/(\tau-1)$ (the standard SOC scaling relation), which
in this case coincides with the fractal dimension of an avalanche.
The results are as follows. There is a threshold value,
$\gamma_c=3-\eta$, which separates two regimes:
\begin{eqnarray}
&& \!\!\!\!\!\!\!\!\!\!\!\!\!\!\!\!\!\!\text{if} \ \gamma>3{-}\eta, \
\ \ \text{then} \  \tau=3/2, \ \ \delta = z = 2,
\\[5pt]
&& \!\!\!\!\!\!\!\!\!\!\!\!\!\!\!\!\!\!\!\!\text{if} \
2{<}\gamma{<}3{-}\eta, \ \ \text{then} \ \tau =
\frac{\gamma{-}2\eta}{\gamma{-}1{-}\eta}, \   \delta = z =
\frac{\gamma{-}1{-}\eta}{\gamma{-}2}
. \label{e8.10}
\end{eqnarray}

It is easy to understand these results for the fractal dimension of
an avalanche, $z$. The reader may check that this $z$ exactly
coincides with the fractal dimension of equilibrium connected trees
with the degree (or branching) distribution equal to
$p(q) \sim q^{-\gamma'}$, see Sec.~\ref{ssec:equilibrium_trees}.



For the numerical study of the BTW model on small-world networks,
see \textcite{de_Arcangelis:dh02}. The BTW model is one of numerous
SOC models. There were a few studies of other SOC models on complex
networks. For example, for the Olami-Feder-Christensen model on
various networks, see \textcite{Caruso:clp06,Caruso:cpl07}, and for a Manna type
sandpile model on small-world networks, see
\textcite{Lahtinen:lkk05}. The Bak-Sneppen model on networks was
studied in \textcite{Kulkarni:kas99}, \textcite{Moreno:mv01},
\textcite{Masuda:mgk05}, and \textcite{Lee:lhl05}.


\subsection{Cascading failures}
\label{ssec:cascading}

Devastating power blackouts are in the list of most impressive
large-scale accidents in artificial networks. In fact, a blackout is
a result of an avalanche of overload failures in power grids. A very
simple though representative model of a cascade of overload failures
was proposed by \textcite{Motter:ml02}. The load of a vertex in this
model is betweenness centrality---the
number of the shortest
paths between other vertices, passing through the vertex, Sec.~\ref{ssec:characteristics}.
Note that frequently the betweenness centrality is simply called
load \cite{Goh:gkk01}.

For every vertex $i$ in this model, a limiting load---{\em
capacity}---is introduced: 
\begin{equation}
c_i=(1+\alpha)b_{0i} , \label{e10.46}
\end{equation}
where $b_{0i}$ is the load (betweenness centrality) of this vertex
in the undamaged network.
The constant $\alpha\geq 0$ is a ``tolerance parameter'' showing how
much an initial load can be exceeded. A cascading failure in this
models looks as follows.

\begin{itemize}

\item[(i)]
Delete a vertex. This leads to the redistribution of loads of the
other vertices: $b_{0i} \to b'_{0i}$.

\item[(ii)]
Delete all overloaded vertices, that is the vertices with
$b'_{0i}>c_i$.

\item[(iii)]
Repeat this procedure until no overloaded vertices remain.

\end{itemize}

In their simulations of various networks, Motter and Lai measured
the ratio $G=N_{\text{after}}/N$, where $N$ and $N_{\text{after}}$
are, respectively, the original number of vertices in a network and
the size of its largest connected component after the cascading
failure. (Assume that the original network coincides with its giant
connected component.) Resulting $G(\alpha)$ depend on (i) the
architecture of a network, (ii) the parameter $\alpha$, and (iii)
characteristics of the first failing vertex, e.g., on its degree.

In a random regular graph,
for any $\alpha>0$, $G$
is $1$, and only if $\alpha=0$, the network will be completely
destroyed, $G=0$. On the other hand, in networks with heavy-tailed
degree distributions, $G$ strongly depends on the degree (or the
load) of the first removed vertex. Motter and Lai used a scale-free
network with $\gamma=3$ in their simulation. Let us briefly discuss
their results.
$\alpha=0$
gives $G=0$ for any starting vertex in any network, while $\alpha\to
\infty$ results in $G=1$. The question is actually about the form of
the monotonously growing curve $G(\alpha)$. When the first removed
vertex is chosen at random, the cascade is large ($G$ strongly
differs from $1$) only at small $\alpha$, and $G(\alpha)$ rapidly
grows from $0$ to $1$. If the first vertex is chosen from ones of
the highest degrees, then $G$ gently rises with $\alpha$, and
cascades may be giant even at rather large $\alpha$.

\textcite{Lee:lgk05} numerically studied the statistics of the
cascades in this model defined on a scale-free network with $2<
\gamma \leq 3$ and found that in this case, there is a critical
point
$\alpha_c \approx 0.15$.
At $a<a_c$, there are giant avalanches, and at $\alpha>\alpha_c$,
the avalanches are finite. These authors observed that at the
critical point, the size distribution of avalanches has a power-law
form, ${\cal P}(s) \sim s^{-\tau}$, where exponent $\tau \approx
2.1(1)$ in the whole range $2< \gamma \leq 3$.

This model can be easily generalized: $\alpha$ may be defined as a
random variable, instead of betweenness centrality other
characteristics may be used, etc.---see, e.g., \textcite{Motter:m04}
or, for a model with overloaded links, \textcite{Moreno:mpv03} and
\textcite{Bakke:bhk06}. Note
that there are other approaches to cascading failures. For example,
\textcite{Watts:w02} proposed a model where, in simple terms,
cascading failures were treated as a kind of epidemic outbreaks.




\subsection{Congestion}
\label{ssec:congestion}

Here we only touch upon basic models of jamming and congestion
proposed by physicists. \textcite{Ohira:os98} put forward a quite
simple model of congestion. Originally it was defined on a lattice
but it can be easily generalized to arbitrary network geometries.

The vertices in this model are of two types---hosts and routers.
Hosts send packets at some rate $\lambda$ to other (randomly chosen)
hosts, so that every packet has its own target. Each packet passes
through a chain of routers storing and forwarding packets. There is
a restriction: the routers can forward not more than one packet per
time step. The routers are supposed to have infinite buffer space,
where a queue of packets is stored. The packet at the head of the
queue is sent first. A router sends a packet to that its neighboring
router which is the closest to the target. If there occur more than
one such routers, then one of them is selected by some special
rules. For example, one may choose the router with the smallest flow
of packets through it.

In their simulations Ohira and Sawatari studied the average time a
packet needs to rich its target
versus the packet injection rate $\lambda$. It turned out that this
time strongly rises above some critical value $\lambda_c$, which
indicates the transition to the congestion phase.
The observations of these authors suggest that it is a continuous
transition, without a jump or hysteresis. The obvious reason for
this jamming transition is the limited forwarding capabilities of
routers---one packet per time step.

\textcite{Sole:sv01} investigated this transition in the same model.
They numerically studied the time-series dynamics of the number of
packets at individual routers, and found a set of power laws at the
critical point. In particular, they observed a $1/f$-type power
spectrum of these series and a power-law distribution of queue
lengths.
[Similar critical effects were found in an analytically treatable model of traffic in networks with hierarchical branching, see \textcite{Arenas:adg01}.]
They proposed the following idea.
Since the traffic is most efficient at $\lambda_c$, the Internet
self-organizes to operate at criticality.
This results in various self-similar scaling phenomena in the
Internet traffic.

These attractive ideas became the subject of strict criticism from
computer scientists \cite{Willinger:wgj02}. Let us dwell on this
criticism, all the more so that it was from the discoverers of the
scaling properties of the Internet traffic \cite{Leland:ltw94}.
They wrote: ``self-similar scaling has been observed in networks
with low, medium, or high loads, and any notion of a ``magical''
load scenario where the network has to run at critical rate
$\lambda_c$ to show self-similar traffic characteristics is
inconsistent with the measurements''. They
listed very simple alternative reasons for these self-similar
phenomena. This criticism was, in fact, aimed at a wide circle of
self-organized criticality models of various aspects of the real
Internet, proposed by physicists. Willinger {\em et al.} stressed
that these models ``are only evocative; they are not explanatory''.
In their definition, an evocative model only ``can reproduce the
phenomenon of interest but does not necessarily capture and
incorporate the true underlying cause''. On the other hand, an
explanatory model ``also captures the causal mechanisms (why and
how, in addition to what).''
Ask yourself: how many explanatory models of real networks were
proposed?

\begin{figure}[t]
\begin{center}

\scalebox{0.31}{\includegraphics[angle=270]{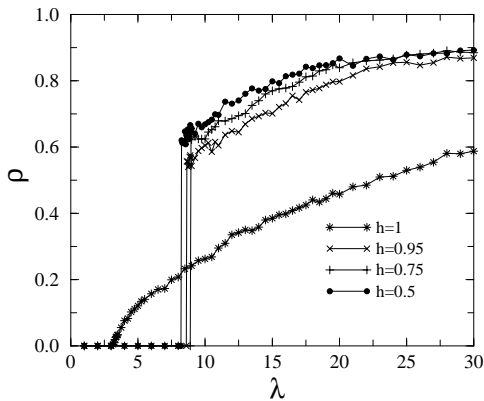}}
\end{center}
\caption{Order parameter $\rho$ versus the packet injection rate
$\lambda$ for varios $h$ in the model of
\protect\textcite{Echenique:egm05}. From
\protect\textcite{Echenique:egm05}. } \label{f105}
\end{figure}

\textcite{Guimera:gdv02} developed an analytical approach where search and congestion problems were interrelated. In their simple theory the mean queue length at vertices of a network was related to a search cost in this network. The latter is the mean number of steps needed to find a target vertex.
In this approach, minimizing the mean queue length is reduced to minimizing the search cost.
This approach was used to find optimal network architectures with minimum congestion.

\textcite{Echenique:egm05} introduced a model of network traffic
with a protocol allowing one to prevent and relieve congestion.
In their model, routers
forward packets, taking into account the queue lengths at their
neighbors. Namely, a packet is sent to that neghboring router $j$,
which has the minimum value 
\begin{equation}
\delta_j \equiv h\ell_{jt}+(1-h)c_j . \label{e10.49}
\end{equation}
Here $\ell_{jt}$ is the length of the shortest path from router $j$
to the target of the packet, $c_j$ is the queue length at the
router, and the parameter $h$ is in the range $0\leq h \leq 1$.
Echenique {\em et al.} performed the numerical simulations by using
the map of a real Internet network but their results should be also
valid for other architectures. As an order parameter for congestion
they used the ratio: $\rho = ($the number of packets that have not
reached their targets during the observation$)/($the total number of
packets generated during this time period).
It turned out that if the parameter $h$ is smaller than $1$, then
the transition to the congestion phase occurs at an essentially higher rate 
$\lambda_c$. Furthermore, when $h<1$, the order parameter emerges
with a jump as in a first order phase transition, while at $h=1$ the
transition resembles a usual second order phase transition, see
Fig.~\ref{f105}. Remarkably, the locations of these transitions, as
well as the whole curves $\rho(\lambda)$, practically coincide at
the studied $h=0.95,\ 0.75,\ 0.5$. On the other hand, the congestion
$\rho$ at $h<1$ is much higher than at $h=1$ at the same
$\lambda>\lambda_c$.
The routing protocol of Echenique {\em et al.} was explored and generalized in a number of studies.
For one of possible generalizations see \textcite{Liu:lmz06} and \textcite{Zhang:hlt07}.

Another approach to network traffic, treating this process
in terms of specific diffusion of packets, was developed by
\textcite{Tadic:tt04,Tadic:ttr04,Tadic:trt06}, see also \textcite{Wang:wwy06}. 
The theory of this kind of traffic was elaborated by \textcite{Fronczak:ff07}.  
\textcite{Danila:dye06} studied routing based on local information. They considered ``routing rules with different degrees of congestion awareness, ranging from random diffusion to rigid congestion-gradient driven flow''.
They found that the strictly congestion-gradient driven routing easily leads to jamming.
\textcite{Carmi:ccd06} presented a physical solution of the problem of effective routing with minimal memory resources.
\textcite{Toroczkai:tb04} investigated the influence of network
architectures on the congestion. \textcite{Helbing:hsl07} described
the generation of oscillations in network flows.
\textcite{Rosvall:rms04} discussed how to use limited information to
find the optimal routes in a network. For the problem of
optimization of network flows, see also \textcite{Gourley:gj06} and
references therein.










\section{OTHER PROBLEMS AND APPLICATIONS}
\label{sec:other}

In this section we briefly review a number of critical effects and processes in networks, which have been missed in the previous sections.






\subsection{Contact and reaction-diffusion processes}
\label{ssec:contact}

\subsubsection{Contact process}
\label{contact_process}

The contact process \cite{Harris:h74} 
is in a wide class of
models exhibiting non-equilibrium phase transitions, for example,
the SIS model of epidemics,   
which belong to the directed percolation universality class 
\cite{Grassberger:gt79}, 
see the review of \textcite{Hinrichsen:h00}. The contact process on a
network is defined as follows. An initial population of particles
occupies vertices in a network. Each vertex can be occupied by only
one particle (or be empty). At each time step $t$, a particle on an
arbitrary chosen vertex either (i) disappears with a probability
$p$\, or (ii) creates with the probability $1-p$\, a new particle at
an arbitrary chosen unoccupied neighboring vertex.

Let us introduce an average density $\rho_{q}(t)$ of particles at
vertices with degree $q$.
The time evolution of $\rho_{q}(t)$ is given by the mean-field rate
equation:
\begin{equation}
\frac{d\rho_{q}(t)}{dt}=-p \rho_{q}(t)+(1-p)q [1-\rho_{q}(t)]
\sum_{q'}\rho_{q'}(t)\frac{P(q'|q)}{q'}, \label{cp-1}
\end{equation}
where $P(q'|q)$ is the conditional probability that a vertex of
degree $q$ is connected to a vertex of degree $q'$
\cite{Castellano:cp06}. The first and second terms in
Eq.~(\ref{cp-1}) describe disappearance and creation of particles,
respectively, at vertices with degree $q$. The factor $1/q'$ shows
that a new particle is created with the same probability at any
(unoccupied) nearest neighboring vertex of a vertex with degree
$q'$. Recall that in uncorrelated networks, $P(q'|q)=q'P(q')/\langle
q \rangle$.

Equation (\ref{cp-1}) shows that if the probability $p$ is larger
than a critical probability $p_c$, then any initial population of
particles disappears at $t \rightarrow \infty$, because particles
disappear faster then they are created. This is the so called {\em
absorbing phase}. When $p<p_c$, an initial population of particles
achieves a state with a non-zero average density:
\begin{equation}
\rho=\sum_{q} P(q)\rho_{q}(t\rightarrow\infty)\propto
\epsilon^\beta, \label{cp-3}
\end{equation}
where $\epsilon = p_c -p$. This is {\em the active phase}. In the
configuration model of uncorrelated random networks, the critical
probability $p_c=1/2$ does not depend on the degree distribution
while the critical exponent $\beta$ does. In networks with a finite
second moment $\left\langle q^{2}\right\rangle$ we have $\beta =1$.
If $\left\langle q^{2}\right\rangle \rightarrow \infty$, then
$\beta$ depends on the asymptotic behavior of the degree
distribution at $q \gg 1$. If the network is scale-free with $2<
\gamma \leq 3$, the exponent $\beta$ is $1/ (\gamma-2)$. This
critical behavior occurs in the infinite size limit, $N \rightarrow
\infty$. In a finite network, $\rho$ is very small but finite at all
$p>0$ and it is necessary to use the finite-size scaling theory.


\textcite{Ha:hhp06} and \textcite{Hong:hhp07} applied the mean-field
finite-size scaling theory to the contact process on finite networks
(see Sec.~\ref{ssec:finite_size_scaling}). They showed that near the
critical point $p_c$ the average density $\rho$ behaves as
$\rho(\epsilon,N)=N^{-\beta/\overline{\nu}}f(\epsilon
N^{1/\overline{\nu}})$, where $f(x)$ is a scaling function, the
critical exponent $\beta$ is the same as above. The critical
exponent $\overline{\nu}$ depends on degree distribution:
$\overline{\nu} (\gamma >3)=2$, and $\overline{\nu} (2<\gamma \leq
3) =(\gamma -1)/(\gamma -2)$. The authors carried out Monte Carlo
simulations of the contact process on the configuration model of
uncorrelated scale-free networks of size to $N=10^7$. These
simulations agreed well with the predictions of the mean-field
scaling theory in contrast to earlier calculations of
\textcite{Castellano:cp06,Castellano:cp07}.

Based on the phenomenological theory of equilibrium critical
phenomena in complex networks (Sec.~\ref{ssec:landau_theory}),
\textcite{Hong:hhp07} proposed a phenomenological mean-field
Langevin equation which describes the average density of particles
in the contact process on uncorrelated scale-free networks near the
critical point: 
\begin{equation}
\frac{d\rho(t)}{dt}= \epsilon \rho - b\rho ^2 -d \rho^{\gamma -1}
+\sqrt{\rho} \eta(t), 
\label{cp-2}
\end{equation}
where $\eta (t)$ is the Gaussian noise, $b$ and $d$ are constants.
Note that the contact process contains the so-called multiplicative
noise $\sqrt{\rho} \eta(t)$, in contrast to an equilibrium process
with a thermal Gaussian noise (see, e.g.,
\textcite{Hinrichsen:h00}). Neglecting the noise in
Eq.~(\ref{cp-2}), in the steady state one can obtain the critical
behavior of $\rho$ and a finite-size scaling behavior of the
relaxation rate (Sec.~\ref{ssec:finite_size_scaling}). As is natural, when a degree distribution is rapidly decreasing, this finite-size scaling coincides with that for the contact process on high-dimensional lattices \cite{Lubeck:lj05}. 

The time evolution of the average density $\rho (t)$ was studied by
\textcite{Castellano:cp06,Castellano:cp07} and
\textcite{Hong:hhp07}. 
When $p\neq p_c$, in an infinite network, an initial population of particles exponentially relaxes to a steady distribution. The relaxation time $t_c$ is finite.  
At the critical point, $p = p_c$, the
characteristic time $t_{c}$ diverges, and an initial distribution
decays as $\rho (t) \sim t^{-\theta}$. Exponent $\theta=1$ for an
uncorrelated complex network with a finite second moment
$\left\langle q^{2}\right\rangle$, and $\theta =1/(\gamma -2)$ for a
scale-free network with $2< \gamma <3$. In a finite network,
$t_{c}(N)$ is finite even at the critical point. 
In
uncorrelated networks with $\left\langle q^{2}\right\rangle <\infty$,
the characteristic relaxation time is 
\begin{equation}
t_{c} \sim 
\sqrt{\frac{N \langle q\rangle ^2}{\langle q^2\rangle}}
\label{ecp-2p}
\end{equation}
\cite{Castellano:cp07}. 
Note that when exponent $\gamma>3$, the phenomenological approach based on Eq.~(\ref{cp-2}) also leads to $t_c \propto N^{1/2}$.  
The size dependence of $t_{c}$ in the range $2< \gamma
<3$, where $\langle q^{2}\rangle$ depends on $N$, is 
still under discussion, see \textcite{Castellano:cp07,Hong:hhp07}. 

\textcite{Giuraniuc:ghi06} considered the contact process with a
degree dependent rate of creation of particles 
assumed 
to be
proportional to $(q_{i}q_{j})^{-\mu}$, where $\mu$ is a tunable
parameter, $q_{i}$ and $q_{j}$ are degrees of neighboring vertices.
Using a mean-field approximation which is equivalent to the annealed
network approximation, they showed that this degree dependent rate
changes the critical behavior of the contact process in scale-free
networks. The result is the shift of degree distribution exponent
$\gamma$ to $\gamma '=(\gamma-\mu)/(\gamma -1)$. This effect is
similar to the Ising model with degree dependent interactions in
Sec.~\ref{sssec:degree_dependent_interactions}.
For finite-size scaling in contact processes with this degree-dependent rate of creation, see \textcite{Karsai:kji06}.

\begin{figure}[t]
\begin{center}
\scalebox{0.31}{\includegraphics[angle=-90]{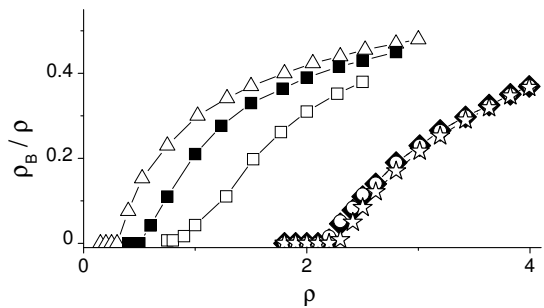}}
\end{center}
\caption{ Relative density of particles $B$ versus the total density
of $A$ and $B$ particles in the reaction-diffusion model in
scale-free networks at $\mu/\lambda =2$. Rightmost curves ($A$
particles are non-diffusing): stars, $N=10^{4}$, $\gamma=3$; closed
diamonds, $N=10^{4}$, $\gamma=2.5$; open circles, $N=10^{5}$,
$\gamma=2.5$. Leftmost curves (both $A$ and $B$ particles are
diffusing, $\gamma=2.5$): open squares, $N=10^{3}$; closed squares,
$N=10^{4}$; open triangles, $N=10^{5}$. Adapted from
\textcite{Colizza:gpv07}.} \label{rd}
\end{figure}
\subsubsection{Reaction-diffusion processes}
\label{rection-diffusion}

Reaction-diffusion processes on uncorrelated random complex networks
were studied by \textcite{Colizza:gpv07}. Consider the following
process for particles of two types, $A$ and $B$. In an initial
state, particles are distributed randomly over vertices of a
network. There may be an arbitrary number of these particles at any
vertex. Suppose that only particles at the same vertex may react and
transform to other particles. The rules of this transformation are
the following:
\begin{itemize}

\item[(i)] Each particle $B$ can spontaneously turn into a $A$ particle at the same vertex at a
rate $\mu$: $B \rightarrow A$.

\item[(ii)] Particles $A$ and $B$ can transform into two $B$
particles at the same vertex at a rate $\lambda$: $A+B \rightarrow
2B$.

\item[(iii)] Particles $B$ can hop to neighboring vertices at the unit
rate.

\end{itemize}
These reactions preserve the total number of particles in the
system. The steady state of this process strongly depends on a
supposed behavior of particles $A$.

If $A$ particles are non-diffusing and the total density of $A$ and
$B$ particles, $\rho$, is smaller than the critical density $\rho_c
= \mu / \lambda$, then $B$ particles disappear in the limit
$t\rightarrow\infty$ (this is the absorbing phase). At $\rho
>\rho_c$ there is a non-zero density of $B$ particles, $\rho_{B}$,
in the steady state (this is the active phase), see Fig.~\ref{rd}.
\textcite{Colizza:gpv07} showed that $\rho_c$ and the critical
behavior do not depend on the degree distribution.

If $A$ particles can also hop, then the phase transition into the
active phase occurs at a degree dependent critical density
$\rho_c=\langle q \rangle^2 \mu /(\langle q^{2} \rangle \lambda)$.
In networks with divergent $\langle q^{2} \rangle$, $\rho_c$ is zero
in the limit $N\rightarrow\infty$, see Fig.~\ref{rd}. A similar
disappearance of the critical threshold was observed in percolation
and the spread of diseases.

Network topology strongly affects dynamics of the
diffusion-annihilation process. This process is defined in the
following way. Identical particles diffuse in a network. If two
particles are at the same vertex, they annihilate ($A+A \rightarrow
\varnothing$).
\textcite{Catanzaro:cbp05} within the mean-field theory showed that
in infinite uncorrelated random networks the average density of
particles, $\rho(t)$, decreases as $t^{-\alpha}$ at large times,
where the exponent $\alpha = 1$ for a network with a finite second
moment $\langle q^{2} \rangle$, and $\alpha = 1/(\gamma -2)$ for an
uncorrelated scale-free network with degree distribution exponent
$2<\gamma<3$ (i.e., with divergent $\langle q^{2} \rangle$).
However, in a finite scale-free network, there is a crossover to the
traditional mean field behavior $1/t$ at times $t>t_{c}(N)$, where
the crossover time $t_{c}(N)$ increases with increasing $N$. Thus
the non-mean-field behavior with $\alpha= 1/(\gamma -2)$ may be
observed only in a sufficiently large network (see
\textcite{Avraham:ag07} for a discussion of kinetics of coalescence,
$A+A\rightarrow A$, and annihilation,$A+A\rightarrow \varnothing$,
beyond the mean-field approximation in the Bethe lattice). This
agrees with numerical simulations of
\textcite{Gallos:ga04,Catanzaro:cbp05}.




\subsection{Zero-range processes}
\label{ssec:zero-range}

Zero-range process describes non-equilibrium dynamics of
condensation of interacting particles in lattices and networks.
This process is closely related to the balls-in-boxes model
\cite{Bialas:bbj97} and equilibrium network ensembles
\cite{Burda:bck01,Dorogovtsev:dms03b,Angel:ael05,Angel:ahe06}
discussed in Sec.~\ref{ssec:condensation_equilibrium}. The
interested reader will find a review of several applications
of this model in \textcite{Evans:eh05}.

In the zero-range process, identical particles hop between vertices
on a graph with a rate $u(n)$ which depends on the number of
particles, $n$, at the vertex of departure. The total number of
particles is conserved. In fact, an interaction between particles on
the same vertex is encoded in the function $u(n)$. The case
$u(n)\propto n$ corresponds to noninteracting particles. If $u(n)$
increases faster than $n$, then we deal with a local repulsion. If
$u(n)$ decreases with $n$, then it assumes a local attraction.
Emergence of the condensation depends on the hop rate $u(n)$ and the
network structure.

The system evolves from an initial distribution of particles to a
steady state. At a certain condition, the condensation of a finite
fraction of particles occurs onto a single vertex. Note that this
non-equilibrium phase transition occurs even in a one-dimensional
lattice. In the steady state the distribution of particles over
vertices can be found exactly. The probability that vertices
$i=1,2,...N$ are occupied by $n_{1},n_{2}, ...\, n_{N}$ particles is
\begin{equation}
{\cal P}(n_{1},n_{2},...\,n_{N})= A \prod_{i=1}^{N} f_{i}(n_{i}),
\label{zr-1}
\end{equation}
where $A$ is a normalization constant, the function
$f_{i}(n)\equiv \prod_{m=1}^{n} [\omega _{i}/u(m)]$ for $n\geq 1$,
and $f_{i}(0)\equiv 1$ \cite{Evans:eh05}. The parameters $\omega
_{i}$ are the steady state weights of a single random walker which
moves on a given network. In simple terms, the frequency of visits
of the walker to a vertex is proportional to its weight. The
weights satisfy the equation $\omega _{i}= \sum_{j}\omega
_{j}T_{ji}$, where $T_{ji}$ is a rate of particle hops from vertex
$j$ to neighboring vertex $i$. Using the function
Eq.~(\ref{zr-1}), one can find exact mean occupation numbers of
vertices.

Let us first consider a homogeneous system where all vertices have
the same degree. The condensation is absent if
$u(n\rightarrow\infty) \rightarrow \infty$. In the steady state all
vertices have the same average occupation number (this is the so
called {\em fluid phase}). The condensation occurs if $u(n)$ decays
asymptotically as $u(\infty)(1+b/n)$ with $b>2$. In this case, the
steady state with the condensate emerges when the concentration of
particles $\rho$ is larger than a critical concentration $\rho_c$
determined by the function $u(n)$. In the condensed phase, a finite
fraction of particles, $\rho- \rho_c$, occupies a single vertex
chosen at random. All other vertices are occupied uniformly with the
mean occupation number $\rho_c$. If $u(n\rightarrow\infty)=0$, then
$\rho_c=0$.

The zero-range process in uncorrelated scale-free networks with
degree distribution exponent $\gamma>2$ was studied by
\textcite{Noh:nsl05,Noh:n05}. The authors considered the case when
the function $u(n) = n^\delta$. A particle can hop with the same
probability to any nearest neighboring vertex, i.e., the transition
probability $T_{ij}=1/q_{i}$, where $q_{i}$ is degree of departure
vertex $i$. In this case $\omega _{i} = q_{i}$. It was shown that if
$\delta > \delta_{c}= 1/(\gamma - 2)$, then the steady state is the
fluid phase at any density of particles. If $\delta \leq
\delta_{c}$, then the critical concentration $\rho_c=0$.
At $\rho >0$, in the steady state, almost all particles are
condensed not at a single vertex but a set of vertices with degrees
exceeding $q_{c} \equiv [q_{\text{cut}}(N)]^{1-\delta /
\delta_{c}}$. These vertices form a vanishingly small fraction of
vertices in the network in the limit $N\rightarrow \infty$. Note
that these results were obtained for the cutoff
$q_{\text{cut}}(N)=N^{1/(\gamma -1)}$ (see the discussion of
$q_{\text{cut}}(N)$ in Sec.~\ref{sssec:cutoffs}). When $\delta =0$,
then $q_{c}=q_{\text{cut}}(N)$ and all particles condense at a
vertex with the highest degree $q_{\text{cut}}(N)$.
See \textcite{Tang:tlz06} for specifics of condensation in a zero-range process in weighted scale-free networks.

The steady state in the zero-range process on a scale-free network
is completely determined by the degree distribution. The
topological structure plays no role (i.e., it does not matter
whether vertices are arranged in a finite dimensional system or
form a small world).
It is
assumed that the network structure may influence relaxation dynamics
of the model, unfortunately, no exact results are known.
\textcite{Noh:n05} studied the evolution of an initial distribution
of particles to the steady state and estimated the relaxation time
$\tau$. In an uncorrelated random scale-free network, the relaxation
time is $\tau \sim N^z$, where $z= \gamma/(\gamma-1) - \delta$,
while in clustered scale-free networks, this exponent is $z= 1 - \delta$.
This estimate agrees with numerical simulations in
\textcite{Noh:nsl05,Noh:n05}. Note that the scaling relation $\tau
\sim N^z$ is also valid for a $d$-dimensional lattice. In this case
the exponent $z$ depends on both the dimension $d$ and the
probability distribution of hopping rates. Particularly, $z=2$ for a
$d>2$-regular lattice \cite{Evans:eh05}.

In a finite network, the condensate at a given vertex exists a
finite
time $\tau_m(N)$.
After ``melting'' at this vertex, the condensate
appears at another vertex, then---at another one, and so on.
For a homogeneous network, e.g., for a random regular graph,
$\tau_{m} \sim N^{z'} \gg \tau \sim N^{z}$, where $z'>z$.
\textcite{Bogacz:bbj07a,Bogacz:bbj07b,Waclaw:wbb07}
argued that in heterogeneous systems the typical melting time of
the condensate, $\tau_{m}(N)$, increases exponentially with $N$,
i.e., $\tau_{m}(N) \sim e^{cN}$, in contrast to a homogeneous
system.
The zero-range process relaxes 
slowly to the condensed phase in comparison to a relaxation time
to the equilibrium state in the ferromagnetic Ising model. (The
relaxation time of the Ising model
is finite at all
temperatures except the critical one,
at which it scales with $N$ as $N^{s}$, see Sec.~\ref{ssec:finite_size_scaling}.)
As soon as the condensate
is formed, it
exists for an exponentially long time
at a vertex of the network:
$\tau_{m} \sim e^{cN} \gg \tau \sim N^{z}$.

\subsection{The voter model}
\label{ssec:voter}


According to \textcite{Sood:sr05}, ``the voter model is perhaps the simplest and most completely solved example of cooperative behavior''. In this model, each vertex is in one of two states---spin up or spin down. In {\em the vertex update version} of the model, the evolution is defined as follows. At each time step,

\begin{itemize}

\item[(i)]
choose a vertex at random and

\item[(ii)]
ascribe to this vertex the state of its randomly chosen neighbor.

\end{itemize}
The evolution in the voter model starts with some random configuration of up and down spins, say, with a fraction of $n_0$ spins up.
One can see that this evolution is determined by random annihilation of chaotic interfaces between ``domains'' with up and down spins. In a finite system, there is always a chance that the system will reach an {\em absorbing state}, where all spins up (or down). However, on the infinite regular lattices of dimensionality greater than $2$, the voter model never reaches the absorbing states staying in {\em the active state} for ever.
For the voter model on the finite regular lattices of dimensionality greater than $2$, the mean time to reach consensus is $\tau_N \sim N$ \cite{ben-Avraham:bcm90,Krapivsky:k92}. Here $N$ is the total number of vertices in a lattice.

On the other hand, the infinite one-dimensional voter model evolves to consensus.
\textcite{Castellano:cvv03} and \textcite{Vilone:vc04} studied the voter model on the Watts-Strogatz small-world networks and found that even a small concentration of shortcuts makes consensus unreachable in the infinite networks. This is quite natural, since these networks are infinite-dimensional objects.

It is important that the average fraction of spins up, $n$, conserves in the voter model on regular lattices, i.e., $n(t)=n_0=\text{const}$. Here the averaging is
over all initial spin configurations and over all evolution histories. \textcite{Suchecki:sem05b,Suchecki:sem05a} found that on random networks, $n(t)$ is not conserved.
Instead, the following weighted quantity conserves:
\begin{equation}
\tilde{n} = \sum_q \frac{qP(q)}{\langle q \rangle} n(q)
,
\label{voter-010}
\end{equation}
where $n(q)$ is the average fraction of spins up among vertices of degree $q$.
Thus, $\tilde{n}(t)=\tilde{n}_0=\text{const}$, where $\tilde{n}_0\equiv \tilde{n}(t=0)$. Note that $\tilde{n}$ is actually the probability that an end vertex of a randomly chosen edge is in state up.

Based on this conservation, \textcite{Sood:sr05} arrived at the following physical picture for the voter model on uncorrelated complex networks. Consensus is unreachable if these networks are infinite. In the finite networks, the mean time to reach consensus is finite. The evolution consists of two stages. The first is a short initial transient to an active state where at any particular evolution history, the fraction of vertices of a given degree with spin up is approximately $\tilde{n}_0$. In the slow second stage, coarsening develops, and the system has an increasing chance to approach consensus. The mean time to reach consensus is
\begin{equation}
\tau_N = N \frac{\langle q \rangle^2}{\langle q^2 \rangle}
\left[(1-\tilde{n}_0)\ln(1-\tilde{n}_0)^{-1} + \tilde{n}_0\ln\tilde{n}_0^{-1}\right]
.
\label{voter-020}
\end{equation}
So, the theory of Sood and Redner gives (a) $\tau_N \sim N$ for
uncorrelated networks with a converging second moment of a degree
distribution, (b) $\tau_N \sim N/\ln N$ in the case of the degree
distribution $P(q) \sim q^{-3}$, and (c) $\tau_N$ growing slower
than $N$ if $\langle q^2 \rangle$ diverges, i.e., if the degree
distribution exponent is less than $3$. In the last case,
this size dependence (a power of $N$ with exponent less than $1$)
is determined by a specific model-dependent cutoff of the degree
distribution,
$q_{\text{cut}}(N)$.

Interestingly, in the second version of the voter model---edge
update---the average fraction of up vertices is conserved as well
as the mean magnetization. In the edge update voter model, at each
time step, an end vertex of a randomly chosen edge adopts the
state of the second end. In this model, the evolution of the
system on a complex network is qualitatively the same as on
high-dimensional regular lattices, and $\tau_N \sim N$
\cite{Suchecki:sem05b,Suchecki:sem05a}.

Other basic types of spin dynamics are also widely discussed.
\textcite{Castellano:clb05} studied a difference between the voter
dynamics and the Glauber-Metropolis zero-temperature dynamics on
networks \cite{Zhou:zl05,Castellano:cp06b}. In the Glauber-Metropolis dynamics in application to the
Ising model at zero temperature, at each time step, a randomly
chosen spin gets an energetically favorable value, $+1$ or $-1$.
In contrast to the evolution due to the interface annihilation in
the voter model, in the Glauber-Metropolis dynamics, domain walls
shorten diminishing surface tension. \textcite{Svenson:s01} showed
numerically that in infinite random networks, the
Glauber-Metropolis dynamics of the Ising model at zero temperature
does not reach the ground state. \textcite{Haggstrom:h02}
rigorously proved that this is true at least in the case of the
Gilbert model of classical random graphs. Thus, this kind of
dynamics can result in consensus only in finite networks, as in
the voter model. Nonetheless, Castellano, {\em et al.} found that
the voter and Glauber-Metropolis dynamics provide markedly
different relaxation of spin systems on random networks. For the
Glauber-Metropolis dynamics, the time dependence of the
probability that a system does not yet reach consensus essentially
deviates from exponential relaxation, typical for the voter
dynamics.

For detailed discussion of the voter model on complex networks in context of
opinion formation, see \textcite{Wu:wh04}.
For other nonequilibrium 
phenomena
in
complex networks modeling social interactions, see, e.g.,
\textcite{Klemm:ket03}, \textcite{Antal:akr05}, and \textcite{Baronchelli:bdb06}.

A few numerical studies were devoted to non-equilibrium
phase transitions in the ferromagnetic Ising model on directed complex networks
with possible application to
processes in social,
economic, and biological systems.
In the directed Ising model, the interactions between spins
are asymmetric and directed, so a Hamiltonian formulation is impossible.
Each spin is affected only by those of its nearest
neighboring spins which are connected to this spin by, say, outgoing edges.
Using a directed Watts-Strogatz network generated from a square
lattice, \textcite{Sanchez:slr02} found that a ferromagnetic phase
transition in this system
is continuous at a sufficiently small density of the
shortcuts.
This transition, however, becomes of the first order
above a critical concentration of the shortcuts.
\textcite{Lima:ls06} carried out simulations of
the ferromagnetic Ising model on a directed Barab\'asi-Albert
network at $T=0$ and found that different
dynamics algorithms lead to different final states of the spin system.
These first investigations
demonstrate
a strong
influence of
a directed
network structure on the non-equilibrium dynamics.
However, these systems are not understood as yet.


\subsection{Co-evolution models}
\label{ssec:co-evolution}

We mostly discuss systems where a cooperative model does not
influence its network substrate. \textcite{Holme:hn06} described a
very interesting contrasting situation, where an evolving network
and interacting agents on it strongly influence each other. The
model of Holme and Newman, in essence, is an adaptive voter model
and may be formulated as follows. There is a sparse network of $N$
vertices with a mean degree $\langle q \rangle$. Each vertex may
be in one of $G$ states---``opinions'', where $G$ is a large
number (which is needed for a sharp phase transition). Vertices
and connections evolve: at each time step, choose a random vertex
$i$ in state $g_i$. If the vertex is isolated, do nothing.
Otherwise,

\begin{itemize}

\item[(i)]
with probability $\phi$, reattach the other end of a randomly chosen
edge of vertex $i$ to a randomly chosen vertex with the same opinion
$g_i$; or

\item[(ii)]
with probability $1-\phi$, ascribe the opinion $g_i$ to a randomly
chosen nearest neighbor $j$ of vertex $i$.

\end{itemize}

\noindent Due to process (i), vertices with similar opinions become
connected---agents influence the structure of the network.
Due to process (ii), opinions of neighbors change---the network
influences agents.

Suppose that the initial state is the classical random graph with
vertices in random states. Let the mean degree be greater than $1$,
so that the giant connected component is present. This system
evolves to a final state consisting of
a set of connected components, with all vertices in each of the
components being in coinciding states---internal consensus. Of
course, vertices in different connected components may be in
different states.
In their simulation, Holme and Newman studied the structure of this
final state at various values of the parameter $\phi$.
In more precise terms, they investigated the
resulting size distribution $P(s)$ of the connected components.

If $\phi=0$, the connections do not move, and the structure of the
final network coincides with the original one, with the giant
connected component. This giant component is destroyed by process
(i) if the probability $\phi$ is sufficiently high, and at $\phi
\sim 1$, the network is segregated into a set of finite connected
components, each one of about $N/G$ vertices. It turns out that at
some critical value $\phi_c=\phi_c(\langle q \rangle, N/G)$ there is
a sharp transition, where the giant connected component disappears.
At the critical point, $P(s)$ seems to have a power-law form with a
nonstandard exponent. There is a principle difference
from the usual birth of the giant connected component in random
networks---in this evolving system, the phase transition is
nonequilibrium. In particular, this transition depends on the
initial state of the system. We expect that models of this kind will
attract much interest in the future, see works of
\textcite{Caldarelli:ccg06,Zimmermann:zes04,Ehrhardt:emv06,Gil:gz06,Zanette:z07,Kozma:kb07},
and the review of \textcite{Gross:gb07}.
\textcite{Allahverdyan:ap06} and \textcite{Biely:bht07} considered somewhat related problems where spins at vertices and edges interacted with each other.




\subsection{Localization transitions}
\label{ssec:localization}

In this subsection we
briefly discuss two quite different localization problems---quantum
and classical.

\subsubsection{Quantum localization}
\label{sssec:quantum localization}

Here we touch upon the transition from localized states of an
electron on the Erd{\H o}s-R\'enyi graph, that is the quantum
percolation problem. The set of corresponding eigenfunctions
$\psi(i,E)$, where $E$ is the energy described by the hopping Hamiltonian,
obeys the equations: 
\begin{equation}
E\psi(i,E) = \sum_j a_{ij} \psi(j,E) , \label{e9.40}
\end{equation}
where $a_{ij}$ are elements of the adjacency matrix.
So, the quantum percolation problem is in fact the problem of the
structure of eigenvectors of the adjacency matrix and its spectrum.

In the phase with delocalized states, the spectrum ($|E|<C\langle q
\rangle$, where $C$ is some positive constant) is organized as
follows \cite{Harris:h82}. All states with $E_c(\langle q
\rangle)<|E|<C\langle q \rangle$ are localized, where $E_c$ is the
mobility edge energy. On the other hand, in the range
$|E|<E_c(\langle q \rangle)$, both localized and extended are
present. At the localization threshold, $E_c$ becomes zero, and, as
is natural, all the states are localized in the localization phase.
This picture allows one to find the localization threshold by
investigating only the zero energy states, since extended states
first emerge at zero energy.

\textcite{Harris:h82} (see also references therein) explained how to
distinguish localized and extended states in the spectrum and how to
relate the quantum percolation problem to classical percolation. It
is important that he showed that the delocalization point,
$q_{\text{deloc}}$, does not coincide with the classical percolation
threshold (i.e., the point of the birth of the giant connected
component, which is $\langle q \rangle=1$ in the Erd{\H o}s-R\'enyi
model). \textcite{Bauer:bg01b} showed that the localization phase is
at $\langle q \rangle<q_{\text{deloc}}=1.421529\ldots$, and above
$q_{\text{deloc}}$ the conducting phase is situated.
They also revealed another, relocalization transition at a higher
mean degree, $q_{\text{reloc}}=3.154985\ldots$. This intriguing
relocalization was observed only in this work.

For numerical study of quantum localization in scale-free networks,
which is a pretty difficult task, see \textcite{Sade:skh05}. This
problem was not studied analytically.


\subsubsection{Biased random walks}
\label{sssec:biased}

Let a classical particle randomly walk on a graph. It is well known
that on $d$-dimensional lattices, (i) if $d \leq 2$, a walk is
recurrent, that is
a drunkard almost surely will get back to his
home---``localization''; and (ii) if $d > 2$, a walk is transient,
that is with finite probability, it goes to infinity without
returning to a starting point. Thus the dimension $d=2$ may be
interpreted as a ``localization transition''.

\textcite{Lyons:l90} found and analytically described a very similar
transition in random networks, see also \textcite{Lyons:lpp96}.
Actually he considered random growing trees with a given
distribution of branching, but \textcite{Sood:sg07} showed that in
networks with locally tree-like structure, nearly the same
conclusions hold. For brevity, let the network be uncorrelated.
Consider a random walk started from a randomly chosen vertex $0$,
assuming that there is an ``exponential'' bias in the direction of
vertex $0$. One may easily arrange this bias by labelling all
vertices in the network by their shortest path distance to the
starting vertex. Suppose that the probabilities of a jump of the
walker from vertex $i$ ($\ell$ steps from vertex $0$) to its nearest
neighbors at distances $\ell-1$, $\ell$, or $\ell+1$ are related in
the following way: 
\begin{equation}
\frac{p(i;\ell\to\ell{-}1)}{p(i;\ell\to\ell)} =
\frac{p(i;\ell\to\ell)}{p(i;\ell\to\ell{+}1)} = \sqrt{\lambda} .
\label{e9.400}
\end{equation}
Then the localization transition is at $\lambda_c$, coinciding with
the mean branching coefficient $B$, which is, as we know,
$B=z_2/z_1=(\langle q^2 \rangle-\langle q \rangle)/\langle q \rangle$
for the configuration model. It is exactly the same critical point
as was observed in cooperative models on this network, see
Secs.~\ref{sssec:statistics_finite} and
\ref{sssec:correlation_volume}, which indicates a close relation
between these two classes of problems.

For $1 \leq \lambda < \lambda_c$, the average return time grows
proportionally to $N^\epsilon$ with exponent 
$\epsilon = \ln(B/\lambda)/\ln B$, which is the analytical result of \textcite{Benichou:bv07}.  
At $\lambda = \lambda_c$, this time is
$\propto \ln N$ as in the unbiased random walks on the chain of
length $\ln N$ with the reflecting boundaries. Finally, for $\lambda
> \lambda_c$, above the critical bias, the mean return time
approaches a finite value at large $N$. 
Remarkably, the average return time coincides with the mean correlation volume $\overline{V}$, Sec.~\ref{sssec:statistics_finite}, if the parameter $b$ characterizing the decay of correlations is taken to be $b=1/\lambda$.

\textcite{Sood:sg07} measured the distribution of return times and
found that due to the absence of small loops in the network, returns
with sufficiently short odd times are virtually absent for any bias.
In other words, in this range of times, a walker may get back to the
starting vertex only by the same way he walked away.






\subsection{Decentralized search}
\label{ssec:Decentralized}

Recall that in the Watts-Strogatz small-world networks 
with variation of the number
of shortcuts, there is a smooth crossover from a lattice (large
world) to a small world. 
In marked contrast to this 
are Kleinberg's networks described in Sec.~\ref{ssec:small-world_networks} as well as the long-range percolation problem. 
In these systems there is 
a sharp transition between the lattice and small-world geometries at some special
value of the control parameter---exponent $\alpha$,
which depends on the dimensionality $d$ the lattice substrate
\cite{Benjamini:bb01,Biskup:b03,Martel:mn04}.
In these works actually a closely related long-range percolation problem was analysed.
Assuming that the
number of shortcuts is $O(N)$,
i.e., the network is sparse, gives
the following mean intervertex distances:

\begin{itemize}

\item[(i)]
for $\alpha<d$, $\overline{\ell}(N) \sim \ln N$;

\item[(ii)] for $d<\alpha<2d$, $\overline{\ell}(N) \sim (\ln N)^{\delta(\alpha)}$, where $\delta(\alpha)\cong\ln2/\ln(2d/\alpha)>1$;

\item[(iii)]
for $\alpha>2d$, $\overline{\ell}(N) \sim c(\alpha)N$, where
$c(\alpha)$ depends only on $\alpha$.

\end{itemize}

\noindent Thus, there is a sharp transition from a ``large world''
to a ``small world'' at $\alpha=2d$. (Note, however, that
\textcite{Moukarzel:mm02} presented heuristic and numerical
arguments that this transition is at $\alpha=d$, and
$\overline{\ell}\sim N^{\mu(\alpha)}$ for $d<\alpha<2d$, where
$0<\mu(\alpha)<1/d$. The reason for this difference between two
groups of results is not clear.) For
other networks with a similar transition, see
\textcite{Hinczewski:hb05} and \textcite{Holme:h07}.

\begin{figure}[t]
\begin{center}
\scalebox{0.31}{\includegraphics[angle=0]{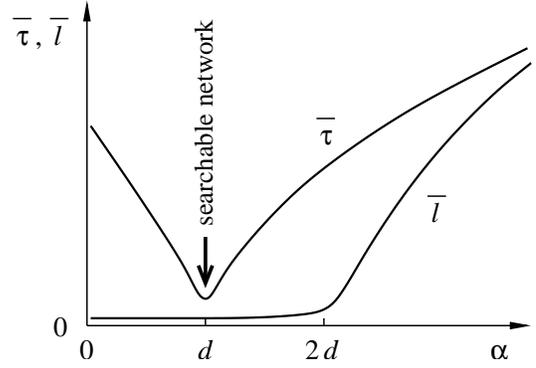}}
\end{center}
\caption{Schematic plot of the mean intervertex distance
$\overline{\ell}(N,\alpha)$ and the mean search time
$\overline{\tau}(N,\alpha)$ vs. exponent $\alpha$ for sparse
Kleinberg's network of fixed large size $N$, based on a $d$
dimensional lattice. The network with $\alpha=d$ is searchable.
} \label{f109}
\end{figure}

The sparse network with exponent $\alpha$ equal to $d$ is unique in
the following respect described by
\textcite{Kleinberg:k06,Kleinberg:k99,Kleinberg:k00}. Kleinberg
asked: how many steps in average,
$\overline{\tau}(N)>\overline{\ell}(N)$, it will take to
approach/find a target from an arbitrary vertex by using the fast
``decentralized search greedy algorithm''? This algorithm exploits
some information about geographic positions of vertices:
at each step, move to the nearest neighbor (including the neighbors
through shortcuts) geographically closest to the target [for other search algorithms based on local information, see \textcite{Adamic:alh03}]. In
particular, for $d=2$:

\begin{itemize}

\item[(i)] for $0 \leq \alpha < 2$, $\overline{\tau}(N) \sim N^{(2-\alpha)/6}$;

\item[(ii)] for $\alpha=2$, $\overline{\tau}(N) \sim \ln^2 N$ (which is also valid for general $\alpha=d$);

\item[(iii)] for $\alpha>2$, $\overline{\tau}(N) \sim N^{(\alpha-2)/(2\alpha-2)}$.

\end{itemize}

\noindent That is, $\alpha=d$ gives the
best search performance (with this algorithm). In this respect, the
network with $\alpha=d$ may be called ``searchable'' (see
Fig.~\ref{f109}). Remarkably, $\overline{\tau}(N) \sim \overline{\ell}(N)$ in the searchable networks.

A similar phenomenon was observed also on trees with added shortcuts
\cite{Watts:wdn02}. In this situation, the probability that a
shortcut connects a pair of vertices separated by $r$ steps on the
tree should be taken not power-law but exponential, proportional to
$\exp(-r/\xi)$. With the same greedy algorithm, using ``geographic
positions'' of the vertices on the underlining tree, this network
appears to be searchable at special values of the parameter $\xi$.
Interestingly, the ferromagnetic Ising model placed on this network
has long-range order only at zero temperature at any positive $\xi$
\cite{Woloszyn:wsk06}.

\textcite{Dorogovtsev:dkm07} exactly described a transition from a small world to a large one in growing trees with a power-law aging. Remarkably, they found that $\overline{\ell}(N) \sim \ln^2 N$ at the point of this transition similarly to a searchability point in Kleinberg's networks. This suggests that the tree ansatz works at a searchability point of Kleinberg's networks.

\subsection{Graph partitioning}
\label{ssec:partitioning}

The size of this article does not allow us to touch upon each of
studied transitions in various networks. In the end of this section,
we only mention a phase transition found by
\textcite{Paul:pcs07}. They studied the following problem: partition
a graph by removing a fraction $1-p$ of edges in a way minimizing
the size of the largest partition, $S$. 
This problem is related to the optimal immunization strategy for a complex network. 
In the random regular graph
with the coordination number $q$, all partitions are small if
$p<p_c=2/q$, where $p_c$ does not coincide with the usual
percolation threshold, $1/(q-1)$.
On the other side of the threshold, the largest partition turns out
to be
giant, $S \sim N$. Moreover, in contrast to percolation, as the
fraction $p$ of retained edges decreases, a sequence of jumps in
$S$---a sequence of ``transitions''---takes place.



\section{SUMMARY AND OUTLOOK}
\label{sec:summary}

\vspace{-37pt}$\phantom{.}$


\subsection{Open problems}
\label{ssec:open}

We would like to indicate a few directions of particular interest
among those discussed in this article. The first one is the
synchronization in the Kuramoto model on complex networks, for which
there is no solid theory.
The second direction is the co-evolving networks and interacting
systems defined on them \cite{Holme:hn06,Pacheco:ptm07}. We did not
discussed a number of interesting NP optimization problems which
were studied by tools of statistical physics but were considered
only for classical random graphs. Among them, there were sparse
graph error correcting codes (see, e.g., \textcite{Montanari:m05}
and references therein), phase transitions in random satisfiability
problems
\cite{Mezard:mpz02,Mertens:mmz03,Achlioptas:anp05,Krzakala:kmr07},
and combinatorial auctions \cite{Galla:glm06}. Note that the
coloring graph problem and minimum vertex covers were also not
analysed for complex networks. Finally, we add to our list the tough
but, we believe, doable problem of finding a replica-symmetry
breaking solution for a spin glass on a complex network.

Real-life networks are finite, loopy (clustered) and correlated.
Most of them are out of equilibrium.
A solid theory of correlation phenomena in complex networks must
take into account finite-size effects, loops, degree correlations
and other structural peculiarities. We described two successful
analytical approaches to cooperative phenomena in infinite networks.
The first was based on the tree ansatz, and the second was the
generalization of the Landau theory of phase transitions. What is
beyond these approaches?

Several first methodical studies aiming at strict accounting for
loops were performed recently, see \textcite{Montanari:m05},
\textcite{Montanari:mr05}, and
\textcite{Chertkov:cc06a,Chertkov:cc06b}. The approximations and
loop expansions proposed in these works were not applied to complex
networks yet. Rather, it is a tool for future work.
It is still unknown when and how loops change cooperative phenomena in complex networks.

The tree ansatz usually fails in finite networks. In this respect,
the problem of a finite size network is closely related to the
problem of loops. It is technically difficult to go beyond intuitive
estimates of finite-size effects demonstrated in
Sec.~\ref{sssec:size}, and the finite-size scaling conjecture. The
strict statistical mechanics theory of finite networks is still not
developed.

Despite some number of interesting results, cooperative models on
growing networks are poorly understood. As a rule, it is still
impossible to predict the type of a critical phenomenon in an
interacting system of this kind. The effect of structural
correlations in a complex network on collective phenomena is also a
little studied problem.

\subsection{Conclusions}
\label{ssec:conclusions}

We have reviewed recent progress in critical phenomena in complex
networks.
In more precise terms, we have considered critical effects in a wide
range of cooperative models placed on various networks and network
models. We have demonstrated a number of diverse critical effects
and phenomena, which greatly differ from those in lattices. It turns
out, however, that each of these  phenomena in networks, in
principle, can be explained in the framework of a
unified approach. This unified view has been presented in this
article.

We have shown that in simple terms, the brand new appearance of
critical phenomena is determined by the combination of
two factors---the small-world effect and a strong heterogeneity and
complex architecture of networks. The compactness of networks leads
to Gaussian critical fluctuations, and in this respect, the theory
of phase transitions in networks is even more simple than in
low-dimensional lattices. On the other hand, the complex
organization of connections makes
these critical phenomena far more rich and
strayed from those predicted by the traditional mean-field theories.

It was claimed only four years ago that ``the study of complex
networks is still in its infancy'' \cite{Newman:n03a}. Now
the baby has come of age.
Nonetheless, we have indicated a wide circle of open problems and
challenging issues. We stress that in contrast to the impressive
progress in understanding the basic principles and nature of the
critical phenomena in networks, progress in the application of these
ideas to real-world networks is rather modest (though see, e.g., the
work of \textcite{Colizza:cbb07}). There is much to be done in this
direction.

Complex networks are
ultimately
compact, maximally disordered, and heterogeneous substrates for
interacting systems. Importantly, these network systems are among
the fundamental structures of nature. The phenomena and processes in
these highly nontraditional systems remarkably
differ from those in ordered and disordered lattices and fractals.
This is why the
study of these
intriguing effects
will lead to a new
understanding
of
a wide circle of
natural, artificial, and social systems.



%
%






\section*{Acknowledgments}

We thank A.~N. Samukhin, M. Alava, A.-L. Barab\'asi,  M. Bauer, O. B\'enichou, G. Bianconi, M. Bogu\~n\'a, B. Bollob\'as, S. Bornholdt, Z. Burda, S.
Coulomb, D. Dhar, M.~E. Fisher, M. Hase, S. Havlin, B. Kahng,  T. Kaski, E.
Khajeh, P.~L. Krapivsky, F. Krzakala, A. Krzewicki, D. Krioukov, S.~N. Majumdar,
S. Maslov, D. Mukamel, M.~E.~J. Newman, J.~D. Noh, J.~G. Oliveira,
M. Ostilli, J. Pacheco, H. Park, R. Pastor-Satorras, M. Peltomaki,
A.~M. Povolotsky, J.~J. Ramasco, S. Redner, O. Riordan, G.~J.
Rodgers, M. Rosvall, B.~N. Shalaev, K. Sneppen, B. S\"oderberg, B.
Tadi\'c, A. Vespignani, T. Vicsek, B. Waclaw, M. Weigt, L. Zdeborov\'a, and A. Zyuzin for
numerous helpful discussions and conversations
on the topic of this work. We particularly thank J.~G. Oliveira for
numerous comments and remarks on the manuscript of this article.
This work was supported by the POCI program, projects FAT/46241/2002,
MAT/46176/2003, FIS/61665/2004,
and BIA-BCM/62662/2004, and by the DYSONET program.




\appendix

\section{BETHE-PEIERLS APPROACH: THERMODYNAMIC PARAMETERS}
\label{bethe-thermodynamic}

The Bethe-Peierls approximation in Sec.~\ref{sssec:bethe_approach}
allows us to calculate a number of important thermodynamic
parameters. The correlation function $C_{ij}\equiv \left\langle
S_{i}S_{j}\right\rangle $ between two neighboring spins is
\begin{equation}
C_{ij}=\tanh \Bigl\{\beta J_{ij}+\tanh ^{-1}\Bigl[\frac{\tanh \beta
h_{ji}\tanh \beta h_{ij}}{\tanh ^{2}\beta J_{ij}}\Bigr]\Bigr\}.  \label{Corr}
\end{equation}
Notice that $C_{ij}$ is completely determined by the messages
$h_{ij}$ and $h_{ji}$ which two neighboring spins, $i$ and $j$, send
to each other. Knowing $C_{ij}$ and $M_{i}$, we find the internal
energy
\begin{equation}
E=-\sum_{(ij)}J_{ij}a_{ij}C_{ij}-\sum_{i}H_{i}M_{i}  \label{E}
\end{equation}
and the free energy
\cite{Mezard:mp01}
\begin{equation}
F=\sum\limits_{(ij)}F_{(ij)}^{(2)}-\sum\limits_{i}(q_{i}-1)F_{i}^{(1)},
\label{F}
\end{equation}
where
\begin{eqnarray}
\!\!\!\!\!\!\!\!\!\!
F_{i}^{(1)} &=&-T\ln \Bigl\{\sum_{S_{i}=\pm 1}\exp \Bigl[\beta
\Bigl(H_{i}+\!\sum_{j\in
N(i)}\!h_{ji}\Bigr)S_{i}\Bigr]\Bigr\},
\label{F1}
\\[5pt]
\!\!\!\!\!\!\!\!\!\!
F_{(ij)}^{(2)} &=&-T\ln \Bigl\{\sum_{S_{i},S_{j}=\pm 1}\exp \Bigl[\beta
J_{ij}S_{i}S_{j}+\beta \varphi _{i\backslash j}S_{i}
\nonumber
\\[5pt]
\!\!\!\!\!\!\!\!\!\!
&&
+\beta \varphi _{j\backslash i}S_{j}\Bigr]\Bigr\}.  \label{F2}
\end{eqnarray}
The free energy $F$ satisfies the thermodynamic relations: $\partial
(\beta F)/\partial \beta =E$, $\partial F/\partial H_{i}=-M_{i}$,
and the extremum condition $\partial F/\partial h_{ji}=0$.


\section{BELIEF-PROPAGATION ALGORITHM: MAGNETIC MOMENT AND THE BETHE FREE ENERGY}
\label{bp-algorithm-magnetization}

\begin{figure}[t]
\begin{center}
\scalebox{0.25}{\includegraphics[angle=270]{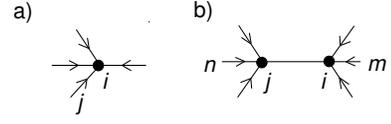}}
\end{center}
\caption{ Diagram representation of the beliefs (a) $b_{i}$ and (b)
$b_{ji}$. Notations are explained in Fig.~\ref{fig-BP}.}
\label{fig-BP-2}
\end{figure}

Using the belief-propagation algorithm discussed in
Sec.~\ref{sssec:belief-propagation}, we can easily calculate a local
magnetic moment:
\begin{equation}
M_{i}=\sum_{S_{i}=\pm 1}S_{i}b_{i}(S_{i}),  \label{m-b}
\end{equation}
where $b_{i}(S_{i})$ is the probability of finding a spin $i$ in
state $S_{i}$. This probability is normalized, $\sum_{S_{i}=\pm
1}b_{i}(S_{i})=1$, and related to the fixed point probabilities
$\{\mu _{ji}(S_{i})\}$ and the probabilistic factor $\exp (\beta
H_{i}S_{i})$, see Fig.~\ref{fig-BP-2}:
\begin{equation}
b_{i}(S_{i})=Ae^{\beta H_{i}S_{i}}\prod_{j\in N(i)}\mu _{ji}(S_{i}),
\label{1-belief}
\end{equation}
where $A$ is a normalization constant.


The correlation function $C_{ij}=\left\langle
S_{i}S_{j}\right\rangle $ is determined by the probability
$b_{ij}(S_{i},S_{j})$ to simultaneously find neighboring spins $i$
and $j$ in spin states $S_{i}$ and $S_{j}$:
\begin{equation}
C_{ij}=\sum_{S_{i},S_{j}=\pm 1}S_{i}S_{j}b_{ij}(S_{i},S_{j}),
\label{corr-b}
\end{equation}
where
\begin{equation}
\sum_{S_{j}=\pm 1}b_{ij}(S_{i},S_{j})=b_{i}(S_{i}).
\label{1-2-belief}
\end{equation}
In the belief-propagation algorithm, the probabilities $b_{i}$ and
$b_{ij}$ are called ``beliefs''. Using Fig.~\ref{fig-BP-2}, we obtain
\begin{eqnarray}
\!\!\!\!\!\!\!\!\!\!\!\!\!\!b_{ij}(S_{i},S_{j}) &=&Ae^{\beta
H_{i}S_{i}+\beta J_{ij}S_{i}S_{j}+\beta
H_{j}S_{j}}   \nonumber \\[5pt]
&&\times \prod_{n\in N(j)\backslash i}\mu _{nj}(S_{j})\prod_{n\in
N(i)\backslash j}\mu _{mi}(S_{i}).  \label{2-belief}
\end{eqnarray}
At the fixed point, we find that $C_{ij}$ is given by
Eq.~(\ref{Corr}).

\textcite{Yedidia:yfw01} proved that at the fixed point
the beliefs $\{b_{i},b_{ij}\} $ give a local minimum of the Bethe
free energy $F_{\text{B}}$:
\begin{eqnarray}
\!\!\!\!\!\!\!\!\!\!\!\!\!F_{\text{B}}(\{b_{i},b_{ij}\})
&=&\sum_{(ij)}\sum_{S_{i},S_{j}=\pm 1}b_{ij}(S_{i},S_{j})\ln
\frac{b_{ij}(S_{i},S_{j})}{\phi _{ij}(S_{i},S_{j})}
\nonumber \\[5pt]
\!\!\!\!\!\!\!\!\!\!\!\!\!\!\!&&-\sum_{i}(q_{i}-1)\sum_{S_{i}=\pm
1}b_{i}(S_{i})\ln \frac{b_{i}(S_{i})}{\psi _{i}(S_{i})},
\label{F-Bethe}
\end{eqnarray}
where $\psi _{i}(S_{i})=\exp (\beta H_{i}S_{i})$, $\phi
_{ij}(S_{i},S_{j})=\exp (\beta H_{i}S_{i}+\beta
J_{ij}S_{i}S_{j}+\beta H_{j}S_{j})$.


\section{REPLICA TRICK}
\label{replica}

The replica trick is a powerful mathematical method which allows one
to average over a quenched disorder. We first introduce a
statistical network ensemble, describe an average over a network
ensemble, and then develop a replica approach for the ferromagnetic
Ising model on an uncorrelated random network.

Let us consider the Erd\H{o}s-R\'{e}nyi graph of $N$ vertices. The
probability $P_{a}(a_{ij})$ that an edge between vertices $i$ and
$j$ is present ($a_{ij}=1$) or absent ($a_{ij}=0$) is
\begin{equation}
P_{a}(a_{ij})=\frac{z_{1}}{N}\delta
(a_{ij}-1)+\Bigl(1-\frac{z_{1}}{N}\Bigr)\delta (a_{ij})  \label{ad-matrix}
\end{equation}
where $a_{ij}$ are the adjacency matrix elements, $z_{1}\equiv
\left\langle q\right\rangle $ is average degree.
Given that the matrix elements are independent and uncorrelated
random parameters, the probability of realization of a graph with a
given adjacency matrix $a_{ij}$, is the product of probabilities
$P_{a}(a_{ij})$ over all pairs of vertices:
\begin{equation}
{\cal
P}(\{a_{ij}\})=\prod_{i=1}^{N-1}\prod_{j=i+1}^{N}P_{a}(a_{ij}).
\label{ER-graph}
\end{equation}
The average of a physical quantity $A(\{a_{ij}\})$ over the network
ensemble is
\begin{equation}
\left\langle A\right\rangle _{\text{en}}=\int A(\{a_{ij}\}){\cal P}%
(\{a_{ij}\})\prod_{i=1}^{N-1}\prod_{j=i+1}^{N}da_{ij}\text{.}
\label{en-av}
\end{equation}

In the configuration model with a given degree distribution $P(q)$
the probability of the realization of a given graph is
\begin{equation}
{\cal P}(\{a_{ij}\})=\frac{1}{{\cal N}}\prod_{i=1}^{N-1}%
\prod_{j=i+1}^{N}P_{a}(a_{ij})\prod_{i=1}^{N}\delta
\Bigl(\sum_{j}a_{ij}-q_{i}\Bigr). \label{config-model}
\end{equation}
The delta-functions fix degrees of the vertices. ${\cal N}$ is a
normalization factor:
\begin{equation}
{\cal N}=\exp \Bigl[N\sum_{q}P(q)\ln (z_{1}^{q}/q!)-Nz_{1}\Bigr].
\label{norma}
\end{equation}
$P_{a}(a_{ij})$ is given by the same Eq.~(\ref{ad-matrix}).


In the static model of a complex network (see
Sec.~\ref{sssec:static_model}), a 
desired degree $d_i$ 
is assigned to
each vertex $i$. The probability that vertices $i$ and $j$ are
linked is equal to $p_{ij}$. With the
probability $(1-p_{ij})$ the edge $(ij)$ is absent. We have
\begin{equation}
P_{a}(a_{ij})=p_{ij}\delta (a_{ij}-1)+(1-p_{ij})\delta (a_{ij}),
\label{static-model}
\end{equation}
where 
$p_{ij}=1-\exp (-N d_i d_j/N\langle d \rangle)$. 
The probability ${\cal P}(\{a_{ij}\})$ is given by
Eq.~(\ref{ER-graph}).

The replica trick is usually used for calculating an average free
energy $\left\langle F\right\rangle _{\text{av}}=-\left\langle T\ln
Z\right\rangle _{\text{av}}$, where $\left\langle ...\right\rangle
_{\text{av}}$ is an average over a quenched disorder. The replica
trick is based on the identity:
\begin{equation}
\left\langle \ln Z\right\rangle _{\text{av}}=\lim_{n\rightarrow 0}\frac{%
\left\langle Z^{n}\right\rangle _{\text{av}}-1}{n}=\lim_{n\rightarrow 0}%
\frac{\ln \left\langle Z^{n}\right\rangle _{\text{av}}}{n}.
\label{RT}
\end{equation}

Let us demonstrate averaging over the statistical ensemble Eq.~(\ref
{config-model}) for the configuration model:
\begin{equation}
\left\langle Z^{n}\right\rangle _{\text{en}}=\int
Z^{n}\prod_{i=1}^{N}\delta
\Bigl(\sum_{j}a_{ij}-q_{i}\Bigr)\prod_{i=1}^{N-1}\prod_{j=i+1}^{N}P_{a}(a_{ij})da_{ij}.
\end{equation}
We consider the ferromagnetic Ising model with $J_{ij}=J$ in a
uniform field $H$, placed on the configuration model
\cite{Leone:lvv02}. Using an integral representation of the
constrains
\begin{equation}
\delta \Bigl(\sum_{j}a_{ij}-q_{i}\Bigr)=\int\limits_{-\infty }^{\infty
}\frac{d\psi _{i}}{2\pi }e^{i(\sum_{j}a_{ij}-q_{i})\psi _{i}},
\end{equation}
we integrate over $a_{ij}$ with the probability function
$P_{a}(a_{ij})$ given by Eq.~(\ref{ad-matrix}):
\begin{eqnarray}
&& \int \exp [\beta Ja_{ij}{\bf S}_{i}{\bf S}_{j}+ia_{ij}(\psi
_{i}+\psi _{j})]P_{a}(a_{ij})da_{ij} \nonumber \\[5pt]
\!\!\!\!\!\!\!\!\!\!\!\!\!\!\!\!\!\!\!\!\!\!\!\!\!\!&&=1+
\frac{z_{1}}{N}(e^{\beta
J{\bf S}_{i}{\bf S}_{j}+i(\psi _{i}+\psi _{j})}-1\Bigr)\nonumber \\[5pt]
&&\approx \exp \Bigl[\frac{z_{1}}{N}\Bigl(e^{\beta J{\bf S}_{i}{\bf S}%
_{j}+i(\psi _{i}+\psi _{j})}-1\Bigl)\Bigl],
\end{eqnarray}
where ${\bf S}_{i}\equiv (S_{i}^{1},S_{i}^{2},...,S_{i}^{n})$, ${\bf
S}_{i}{\bf S}_{j}\equiv \sum_{\alpha }S_{i}^{\alpha }S_{j}^{\alpha
}$. $\alpha =1,2,...,n$ is the replica index. Note that one can
simultaneously integrate over random couplings $J_{ij}$ and random
fields $H_{i}$.

In the limit $N \gg 1$ we obtain
\begin{eqnarray}
\!\!\!\!\!\!\!\!\!\!\!\!\!\!\!\!\!\!\!\!\!\!\!\!\!\!\!\!\left\langle Z^{n}\right\rangle _{\text{en}}=\frac{1}{{\cal N}}%
\sum_{\{S_{i}^{\alpha }=\pm 1\}}\int \Bigl(\prod_{i}\frac{d\psi
_{i}}{2\pi }e^{-iq_{i}\psi _{i}}\Bigr)\times \nonumber \\[5pt]
\!\!\!\!\!\!\!\!\!\!\exp \Bigl[\frac{z_{1}}{2N}\!\sum_{ij}e^{\beta J{\bf S}_{i}{\bf S}%
_{j}+i(\psi _{i}+\psi _{j})}+\beta \sum_{i}{\bf
HS}_{i}-\frac{1}{2}Nz_{1}\Bigr], \label{Zn}
\end{eqnarray}
where ${\bf HS}=\sum_{\alpha }HS^{\alpha }$.

Let us introduce a functional order parameter:
\begin{equation}
\rho ({\bf \sigma })=\frac{1}{N}\sum_{i}\delta ({\bf \sigma }-{\bf S}%
_{i})e^{i\psi _{i}},
\end{equation}
where ${\bf \sigma }=(\sigma ^{1},\sigma ^{2},...,\sigma ^{n})$.
There is an identity:
\begin{equation}
\frac{1}{N}\sum_{ij}e^{\beta J{\bf S}_{i}{\bf S}_{j}+i(\psi
_{i}+\psi
_{j})}=N\sum_{{\bf \sigma }_{1},{\bf \sigma }_{2}}\rho ({\bf \sigma }%
_{1})\rho ({\bf \sigma }_{2})e^{\beta J{\bf \sigma }_{1}{\bf \sigma
}_{2}}.
\end{equation}
We use the functional Hubbard-Stratonovich transformation:
\begin{eqnarray}
&&\!\!\!\!\!\!\exp \Bigl[\frac{Nz_{1}}{2}\!\sum_{\{{\bf \sigma }_{1},{\bf \sigma
}_{2}=\pm
1\}}\rho ({\bf \sigma }_{1})\rho ({\bf \sigma }_{2})e^{\beta J{\bf \sigma }%
_{1}{\bf \sigma }_{2}}\Bigr]
\nonumber
\\[5pt]
&=&\int D\widehat{\rho }({\bf \sigma })\exp \Bigl[-\frac{Nz_{1}}{2}\sum_{{\bf %
\sigma }_{1},{\bf \sigma }_{2}}\widehat{\rho }({\bf \sigma }_{1})C({\bf %
\sigma }_{1},{\bf \sigma }_{2})\widehat{\rho }({\bf \sigma }_{2})
\nonumber
\\[5pt]
&&+Nz_{1}\sum_{{\bf \sigma }}\widehat{\rho }({\bf \sigma })\rho
({\bf \sigma })\Bigr].  \label{H-S-transf}
\end{eqnarray}
Here $C({\bf \sigma }_{1},{\bf \sigma }_{2})$ is an inverse function to $%
e^{\beta J{\bf \sigma }_{1}{\bf \sigma }_{2}}$:
\begin{equation}
\sum_{{\bf \sigma }_{1}}C({\bf \sigma },{\bf \sigma }_{1})e^{\beta J{\bf %
\sigma }_{1}{\bf \sigma }_{2}}=\delta ({\bf \sigma }-{\bf \sigma
}_{2}).
\end{equation}
The transformation Eq.~(\ref{H-S-transf}) enables us to integrate
over variables $\psi _{i}$ in Eq.~(\ref{Zn}):
\begin{eqnarray}
\left\langle Z^{n}\right\rangle _{\text{en}} \!\!\!\!&=&\!\!\!\!\int \frac{D\widehat{\rho }(%
{\bf \sigma })}{{\cal N}}\exp \Bigl\{-\frac{Nz_{1}}{2}\sum_{{\bf \sigma }_{1},%
{\bf \sigma }_{2}}\widehat{\rho }({\bf \sigma }_{1})C({\bf \sigma }_{1},{\bf %
\sigma }_{2})\widehat{\rho }({\bf \sigma }_{2})  \nonumber \\[5pt]
&&\!\!\!\!\!\!\!\!\!\!\!\!\!\!\!\!\!\!\!\!+N\sum_{q}P(q)\ln \Bigl[\sum_{{\bf S}}\frac{1}{q!}z_{1}^{q}\widehat{\rho }^{q}(%
{\bf S})e^{\beta {\bf HS}}\Bigr]-\frac{1}{2}Nz_{1}\Bigr\}. \label{Zn2}
\end{eqnarray}
In the thermodynamic limit $N\rightarrow \infty $, the functional
integral over $\widehat{\rho }({\bf \sigma })$ is calculated by
using the saddle point method. The saddle point equations are
\begin{eqnarray}
&&\widehat{\rho }({\bf S})=\sum_{{\bf \sigma }}\rho ({\bf \sigma })e^{\beta J%
{\bf \sigma S}},  \label{s-p-2} \\[5pt]
&&\rho ({\bf S})=\sum_{q}\frac{P(q)q}{z_{1}}\frac{\widehat{\rho }^{q-1}({\bf S}%
)e^{\beta {\bf HS}}}{\sum_{{\bf S}}\widehat{\rho }^{q}({\bf S})e^{\beta {\bf %
HS}}}.  \label{s-p-1}
\end{eqnarray}
Equation (\ref{Zn2}) gives the replica free energy per vertex:
\begin{eqnarray}
-n\beta FN^{-1} &=&N^{-1}\ln \left\langle Z^{n}\right\rangle _{\text{av}}=-z_{1}\sum_{%
{\bf \sigma }}\widehat{\rho }({\bf \sigma })\rho ({\bf \sigma })+\frac{1}{2}%
z_{1}  \nonumber \\[5pt]
&&+\frac{z_{1}}{2}\sum_{{\bf \sigma }_{1},{\bf \sigma }_{2}}\rho ({\bf %
\sigma }_{1})\rho ({\bf \sigma }_{2})e^{\beta J{\bf \sigma }_{1}{\bf \sigma }%
_{2}}  \nonumber \\[5pt]
&&+\sum_{q}P(q)\ln \Bigl[\sum_{{\bf S}}\widehat{\rho }^{q}({\bf S})e^{\beta {\bf %
HS}}\Bigr].  \label{F-replica}
\end{eqnarray}
A replica symmetric solution of the saddle-point equations
(\ref{s-p-2}) and (\ref{s-p-1}) can be written in a general form:
\begin{eqnarray}
\rho ({\bf S})=\int dh\Phi (h)\frac{e^{\beta {\bf hS}}}{(2\cosh
\beta h)^{n}}, \label{e-c100a} \\[5pt]
\widehat{\rho }({\bf S})=\int dh\Psi (h)\frac{e^{\beta {\bf
hS}}}{(2\cosh \beta h)^{n}}, \label{e-c100b}
\end{eqnarray}
where ${\bf hS}=h\sum_{\alpha =1}^{n}S^{\alpha }$. Substituting the
replica symmetric solution into the saddle point equations
(\ref{s-p-2}) and (\ref {s-p-1}), we obtain
\begin{eqnarray}
&&\!\!\!\!\!\!\!\!\!\Phi (h)=\sum_{q}\frac{P(q)q}{z_{1}}\!\int \!\delta
\Bigl(h-\!\sum_{m=1}^{q-1}h_{m}{-}H\Bigr)\!\prod_{m=1}^{q-1}\!\Psi
(h_{m})dh_{m}, \nonumber \\[5pt]
&&\!\!\!\!\!\!\!\!\!\Psi (h)=\!\int \!\!\delta (h-\!T\tanh ^{-1}[\tanh
\beta J\tanh \beta y])\Phi (y)dy. \label{psi-1}
\end{eqnarray}
Substituting $\Phi (h)$ into the equation for $\Psi (h)$, we obtain
the self-consistent equation (\ref{dh}) derived by using the
Bethe-Peierls approximation. $\Psi (h)$ is actually the distribution
function of additional fields (messages) in a network.


\section{MAX-CUT ON THE ERD\H{O}S-R\'{E}NYI GRAPH}
\label{max-cut-ER}

One can prove the validity of the upper bound Eq.~(\ref{max-cut})
for the maximum cut $K_{c}$ on the Erd\H{o}s-R\'{e}nyi graph, using
the so called first-moment method.

We divide the Erd\H{o}s-R\'{e}nyi graph into two sets of $S$ and
$N-S$ vertices.
The probability that a randomly chosen edge has endpoints from
different sets is $Q=2(S/N)(1-S/N)$.
The number of ways to divide a graph by a cut of $K$ edges is
\begin{equation}
{\frak N}(K)=%
{L \choose K}%
\sum_{S=0}^{N}%
{N \choose S}%
Q^{K}(1-Q)^{L-K}.  \label{max-cut-3}
\end{equation}
Here $%
{L \choose K}%
$ is the number of ways to choose $K$ edges from $L$ edges, $%
{N \choose S}%
$ is the number of ways to choose $S$ vertices from $N$ vertices,
$Q^{K}(1-Q)^{L-K}$ is the probability that there are $K$ edges in
the cut, and the $L-K$ remaining edges do not belong to the cut. The
main contribution to ${\frak N}(K)$ is given by terms with $S\approx
N/2$, i.e., $Q\approx 1/2$. So,
\begin{equation}
{\frak N}(K)\approx
{L \choose K}%
2^{N-L}=e^{L\Xi (\alpha )},  \label{max-cut-4}
\end{equation}
where $\alpha =K/L$. Using the entropy bound on the binomial,
\begin{equation}
{L \choose K}%
\leqslant \exp [-L(1-\alpha )\ln (1-\alpha )-L\alpha \ln \alpha ],
\label{binomial}
\end{equation}
we find
\begin{equation}
\Xi (\alpha )=-(1-\alpha )\ln (1-\alpha )-\alpha \ln \alpha
+(z_{1}/2-1)\ln 2.  \label{max-cut-5}
\end{equation}
The maximum cut $K_{c}$ is given by the condition $\Xi (\alpha
_{c})=0$. In the limit $N\rightarrow \infty $, there is no cut with
a size $K>\alpha _{c}L=K_{c}$ while there are exponentially many
cuts at $K<K_{c}$. This condition at $z_{1}\gg 1$ leads to
Eq.~(\ref{max-cut}) with the upper bound $A=\sqrt{\ln 2}/2$.

\section{EQUATIONS OF STATE OF THE POTTS MODEL ON A NETWORK}
\label{state_Potts}

\textcite{Dorogovtsev:dgm04} showed that for the ferromagnetic
$p$-state Potts model Eq.~(\ref{Hamilt}) on an uncorrelated random
graph, the magnetic moments $M_{i}\equiv M_{i}^{(1)}$ along the
magnetic field $H$ and the additional fields (messages) $h_{ij}$ are
determined by the following equations:
\begin{eqnarray}
M_{i} &=&\frac{1-\exp [-\beta (H+\sum_{j\in
N(i)}h_{ji})]}{1+(p-1)\exp
[-\beta (H+\sum_{j\in N(i)}h_{ji})]}\,,  \label{Mi-potts} \\[5pt]
\text{ }h_{ij} &=&T\ln \left\{ \frac{e^{\beta J_{ij}}+(p-1)e^{-\beta
\varphi _{i\backslash j}}}{1+(e^{\beta J_{ij}}+p-2)e^{-\beta \varphi
_{i\backslash j}}}\right\} \,,  \label{recursion-potts}
\end{eqnarray}
where $\varphi _{i\backslash j}=H+\sum_{m\in N(i)\backslash
j}h_{mi}$ is the cavity field.
These equations unify the percolation, the ferromagnetic Ising model
and a first order phase transition on uncorrelated complex networks.
They are exact in the limit $N\rightarrow \infty $.

For the one-state ferromagnetic Potts model in zero field, Eq.~(\ref
{recursion-potts}) takes a simple form:
\begin{equation}
x_{ij}=1-r+r\!\prod_{m\in N(i)\backslash j}x_{mi}, \label{perc}
\end{equation}
where $x_{ij}\equiv \exp (-h_{ij})$, the coupling $J_{ij}=J>0$, and
$r\equiv (1-e^{-\beta J})$.
In the configuration model,
the parameters $x_{mi}$ are statistically independent. Averaging
over the network ensemble and introducing the parameter $x\equiv
\left\langle x_{ij}\right\rangle _{\text{en}}$, we arrive at
Eq.~(\ref{e3.6}) describing bond percolation on uncorrelated
networks. The critical temperature $T_{\text{P}}$ in
Eq.~(\ref{Tc-Potts}) determines the percolation threshold
$r(T=T_{\text{P}})=z_{1}/z_{2}$ in agreement with Eq.~(\ref{e3.9}).

When $p=2$, equation (\ref{recursion-potts}) is reduced to
Eq.~(\ref{recursion}) for the Ising model. It is only necessary to
rescale $J\rightarrow 2J$, $H\rightarrow 2H$, and $h\rightarrow 2h$.

\bibliographystyle{apsrmp}
\bibliography{review-critical16imp}


\begin{thebibliography}{42}
\expandafter\ifx\csname
natexlab\endcsname\relax\def\natexlab#1{#1}\fi
\expandafter\ifx\csname bibnamefont\endcsname\relax
  \def\bibnamefont#1{#1}\fi
\expandafter\ifx\csname bibfnamefont\endcsname\relax
  \def\bibfnamefont#1{#1}\fi
\expandafter\ifx\csname citenamefont\endcsname\relax
  \def\citenamefont#1{#1}\fi
\expandafter\ifx\csname url\endcsname\relax
  \def\url#1{\texttt{#1}}\fi
\expandafter\ifx\csname urlprefix\endcsname\relax\def\urlprefix{URL
}\fi \providecommand{\bibinfo}[2]{#2}
\providecommand{\eprint}[2][]{\url{#2}}


\bibitem[{\citenamefont{Acebr\'{o}n} \emph{et~al.}(2005)\citenamefont{Acebr\'{o}n, Bonilla, P\'{e}rez
Vicente, Ritort, and Spigler}}]
{Acebron:abvrs05}
\bibinfo{author}{\bibnamefont{Acebr\'{o}n},~\bibfnamefont{J.~A.}},
\bibinfo{author}{\bibfnamefont{L.~L.}~\bibnamefont{Bonilla}},
\bibinfo{author}{\bibfnamefont{C.~J.}~\bibnamefont{P\'{e}rez Vicente}},
\bibinfo{author}{\bibfnamefont{F.}~\bibnamefont{Ritort}},
and
\bibinfo{author}{\bibfnamefont{R.}~\bibnamefont{Spigler}},
\bibinfo{year}{2005},
\bibinfo{title}{``The Kuramoto model: A simple paradigm for synchronization phenomena},''
\bibinfo{journal}{Rev. Mod. Phys.} \textbf{\bibinfo{volume}{77}},
\bibinfo{pages}{137}.


\bibitem[{\citenamefont{Achlioptas} \emph{et~al.}(2005)\citenamefont{Achlioptas, Naor, and Peres}}]
{Achlioptas:anp05}
\bibinfo{author}{\bibnamefont{Achlioptas},~\bibfnamefont{D.}},
\bibinfo{author}{\bibfnamefont{A.}~\bibnamefont{Naor}},
and
\bibinfo{author}{\bibfnamefont{Y.}~\bibnamefont{Peres}},
\bibinfo{year}{2005},
\bibinfo{title}{``Rigorous location of phase transitions in hard optimization problems},''
\bibinfo{journal}{Nature} \textbf{\bibinfo{volume}{435}},
\bibinfo{pages}{759}.


\bibitem[{\citenamefont{Adamic} \emph{et~al.}(2003)\citenamefont{Adamic, Lukose, and Huberman}}]
{Adamic:alh03}
\bibinfo{author}{\bibnamefont{Adamic},~\bibfnamefont{L.~A.}},
\bibinfo{author}{\bibfnamefont{R.~M.}~\bibnamefont{Lukose}},
and
\bibinfo{author}{\bibfnamefont{B. Ã.}~\bibnamefont{Huberman}},
\bibinfo{year}{2003},
\bibinfo{title}{``Local search in unstructured networks},''
in \emph{\bibinfo{booktitle}{Handbook of Graphs and Networks}},
edited by
\bibinfo{editor}{\bibfnamefont{S.}~\bibnamefont{Bornholdt}}
and
\bibinfo{editor}{\bibfnamefont{H.~G.}~\bibnamefont{Schuster}}
(\bibinfo{publisher}{Wiley-VCH GmbH \& Co., Weinheim}), p.
\bibinfo{pages}{295}.


\bibitem[{\citenamefont{Aharony}(1978)}]
{Aharony:a78}
\bibinfo{author}{\bibnamefont{Aharony},~\bibfnamefont{A.}}
\bibinfo{year}{1978},
\bibinfo{title}{``Tricritical points in systems with random fieldsl},''
\bibinfo{journal}{Phys. Rev. B} \textbf{\bibinfo{volume}{18}},
\bibinfo{pages}{3318}.


\bibitem[{\citenamefont{Aizenman and Lebowitz}(1988)}]
{Aizenman:al88}
\bibinfo{author}{\bibnamefont{Aizenman},~\bibfnamefont{M.}}
and
\bibinfo{author}{\bibfnamefont{J. L.}~\bibnamefont{Lebowitz}},
\bibinfo{year}{1988},
\bibinfo{title}{``Metastability effects in bootstrap percolation},''
\bibinfo{journal}{J. Phys. A} \textbf{\bibinfo{volume}{21}},
\bibinfo{pages}{3801}.





\bibitem[{\citenamefont{Alava and Dorogovtsev}(2005)}]
{Alava:ad05}
\bibinfo{author}{\bibnamefont{Alava},~\bibfnamefont{M.~J.}}
and
\bibinfo{author}{\bibfnamefont{S.~N.}~\bibnamefont{Dorogovtsev}},
\bibinfo{year}{2005},
\bibinfo{title}{``Complex networks created by aggregation},''
\bibinfo{journal}{Phys. Rev. E} \textbf{\bibinfo{volume}{71}},
\bibinfo{pages}{036107}.


\bibitem[{\citenamefont{Albert and Barab{\'a}si}(2002)}]
{Albert:ab02}
\bibinfo{author}{\bibnamefont{Albert},~\bibfnamefont{R.}}
and
\bibinfo{author}{\bibfnamefont{A.-L.}~\bibnamefont{Barab{\'a}si}},
\bibinfo{year}{2002},
\bibinfo{title}{``Statistical mechanics of complex networks},''
\bibinfo{journal}{Rev. Mod. Phys.} \textbf{\bibinfo{volume}{74}},
\bibinfo{pages}{47}.


\bibitem[{\citenamefont{Albert} \emph{et~al.}(2000)\citenamefont{Albert, Jeong, and Barab\'{a}si}}]
{Albert:ajb00}
\bibinfo{author}{\bibnamefont{Albert},~\bibfnamefont{R.}},
\bibinfo{author}{\bibfnamefont{H.}~\bibnamefont{Jeong}},
and
\bibinfo{author}{\bibfnamefont{A.-L.}~\bibnamefont{Barab\'{a}si}},
\bibinfo{year}{2000},
\bibinfo{title}{``Error and attack tolerance in complex networks},''
\bibinfo{journal}{Nature} \textbf{\bibinfo{volume}{406}},
\bibinfo{pages}{378}.


\bibitem[{\citenamefont{Albert} \emph{et~al.}(1999)\citenamefont{Albert, Jeong, and Barab\'{a}si}}]
{Albert:ajb99}
\bibinfo{author}{\bibnamefont{Albert},~\bibfnamefont{R.}},
\bibinfo{author}{\bibfnamefont{H.}~\bibnamefont{Jeong}},
and
\bibinfo{author}{\bibfnamefont{A.-L.}~\bibnamefont{Barab\'{a}si}},
\bibinfo{year}{1999},
\bibinfo{title}{``Diameter of the world-wide web},''
\bibinfo{journal}{Nature} \textbf{\bibinfo{volume}{401}},
\bibinfo{pages}{130}.



\bibitem[{\citenamefont{Aleksiejuk} \emph{et~al.}(2002)\citenamefont{Aleksiejuk, Holyst, and Stauffer}}]
{Aleksiejuk:ahs02}
\bibinfo{author}{\bibnamefont{Aleksiejuk},~\bibfnamefont{A.}},
\bibinfo{author}{\bibfnamefont{J.~A.}~\bibnamefont{Holyst}},
and
\bibinfo{author}{\bibfnamefont{D.}~\bibnamefont{Stauffer}},
\bibinfo{year}{2002},
\bibinfo{title}{``Ferromagnetic phase transition in Barab\'asi-Albert networks},''
\bibinfo{journal}{Physica A} \textbf{\bibinfo{volume}{310}},
\bibinfo{pages}{260}.


\bibitem[{\citenamefont{Allahverdyan and Petrosyan}(2006)}]
{Allahverdyan:ap06}
\bibinfo{author}{\bibnamefont{Allahverdyan},~\bibfnamefont{A.~E.}}
and
\bibinfo{author}{\bibfnamefont{K.~G.}~\bibnamefont{Petrosyan}},
\bibinfo{year}{2006},
\bibinfo{title}{``Statistical networks emerging from link-node interactions},''
\bibinfo{journal}{Europhys. Lett.} \textbf{\bibinfo{volume}{75}},
\bibinfo{pages}{908}.




\bibitem[{\citenamefont{Alvarez-Hamelin} \emph{et~al.}(2006)\citenamefont{Alvarez-Hamelin, Dall\'{}Asta, Barrat, and Vespignani}}]
{Alvarez-Hamelin:adb05a}
\bibinfo{author}{\bibnamefont{Alvarez-Hamelin},~\bibfnamefont{J.~I.}},
\bibinfo{author}{\bibfnamefont{L.}~\bibnamefont{Dall\'{}Asta}},
\bibinfo{author}{\bibfnamefont{A.}~\bibnamefont{Barrat}},
and
\bibinfo{author}{\bibfnamefont{A.}~\bibnamefont{Vespignani}},
\bibinfo{year}{2006},
\bibinfo{title}{``$k$-core decomposition: a tool for the visualization of large scale networks},''
\bibinfo{journal}{Advances in Neural Information Processing Systems (Canada)} \textbf{\bibinfo{volume}{18}},
\bibinfo{pages}{41}. 


\bibitem[{\citenamefont{Alvarez-Hamelin} \emph{et~al.}(2005b)\citenamefont{Alvarez-Hamelin, Dall\'{}Asta, Barrat, and Vespignani}}]
{Alvarez-Hamelin:adb05b}
\bibinfo{author}{\bibnamefont{Alvarez-Hamelin},~\bibfnamefont{J.~I.}},
\bibinfo{author}{\bibfnamefont{L.}~\bibnamefont{Dall\'{}Asta}},
\bibinfo{author}{\bibfnamefont{A.}~\bibnamefont{Barrat}},
and
\bibinfo{author}{\bibfnamefont{A.}~\bibnamefont{Vespignani}},
\bibinfo{year}{2005b},
\bibinfo{title}{``$k$-core decomposition: a tool for the analysis of large scale Internet graphs},''
\eprint{cs.NI/0511007}.


\bibitem[{\citenamefont{Andrade and Herrmann}(2005)}]
{Andrade:ah04}
\bibinfo{author}{\bibnamefont{Andrade},~\bibfnamefont{R.~F.~S.}}
and
\bibinfo{author}{\bibfnamefont{H.~J.}~\bibnamefont{Herrmann}},
\bibinfo{year}{2005},
\bibinfo{title}{``Magnetic models on Apollonian networks},''
\bibinfo{journal}{Phys. Rev. E} \textbf{\bibinfo{volume}{71}},
\bibinfo{pages}{056131}.


\bibitem[{\citenamefont{Andrade} \emph{et~al.}(2005)\citenamefont{Andrade, Herrmann, Andrade, and da Silva}}]
{Andrade:aha05}
\bibinfo{author}{\bibnamefont{Andrade Jr.},~\bibfnamefont{J.~S.}},
\bibinfo{author}{\bibfnamefont{H. J.}~\bibnamefont Herrmann{}},
\bibinfo{author}{\bibfnamefont{R.~F.~S.}~\bibnamefont{Andrade}},
and
\bibinfo{author}{\bibfnamefont{L.~R.}~\bibnamefont{da Silva}},
\bibinfo{year}{2005},
\bibinfo{title}{``Apollonian networks},''
\bibinfo{journal}{Phys. Rev. Lett.} \textbf{\bibinfo{volume}{94}},
\bibinfo{pages}{018702}.


\bibitem[{\citenamefont{Angel} \emph{et~al.}(2005)\citenamefont{Angel, Evans, Levine and Mukamel}}]
{Angel:ael05}
\bibinfo{author}{\bibnamefont{Angel},~\bibfnamefont{A. G.}},
\bibinfo{author}{\bibfnamefont{M. R.}~\bibnamefont{Evans}},
\bibinfo{author}{\bibfnamefont{E.}~\bibnamefont{Levine}},
and
\bibinfo{author}{\bibfnamefont{D.}~\bibnamefont{Mukamel}},
\bibinfo{year}{2005},
\bibinfo{title}{``Critical phase in nonconserving zero-range process and rewiring networks},''
\bibinfo{journal}{Phys. Rev. E} \textbf{\bibinfo{volume}{72}},
\bibinfo{pages}{046132}.



\bibitem[{\citenamefont{Angel} \emph{et~al.}(2006)\citenamefont{Angel, Hanney, and Evans}}]
{Angel:ahe06}
\bibinfo{author}{\bibnamefont{Angel},~\bibfnamefont{A. G.}},
\bibinfo{author}{\bibfnamefont{T.}~\bibnamefont{Hanney}},
and
\bibinfo{author}{\bibfnamefont{M. R.}~\bibnamefont{Evans}},
\bibinfo{year}{2006},
\bibinfo{title}{``Condensation transitions in a model for a directed network with weighted links},''
\bibinfo{journal}{Phys. Rev. E} \textbf{\bibinfo{volume}{73}},
\bibinfo{pages}{016105}.


\bibitem[{\citenamefont{Antal} \emph{et~al.}(2005)\citenamefont{Antal, Krapivsky, and Redner}}]
{Antal:akr05}
\bibinfo{author}{\bibnamefont{Antal},~\bibfnamefont{T.}},
\bibinfo{author}{\bibfnamefont{P.~L.}~\bibnamefont{Krapivsky}},
and
\bibinfo{author}{\bibfnamefont{S.}~\bibnamefont{Redner}},
\bibinfo{year}{2005},
\bibinfo{title}{``Dynamics of social balance on networks},''
\bibinfo{journal}{Phys. Rev. E} \textbf{\bibinfo{volume}{72}},
\bibinfo{pages}{036121}.


\bibitem[{\citenamefont{Antoni and Ruffo}(1995)}]
{Antoni:as95}
\bibinfo{author}{\bibnamefont{Antoni},~\bibfnamefont{M.}}
and
\bibinfo{author}{\bibfnamefont{S.}~\bibnamefont{Ruffo}},
\bibinfo{year}{1995},
\bibinfo{title}{``Clustering and relaxation in Hamiltonian long-range dynamics},''
\bibinfo{journal}{Phys. Rev. E} \textbf{\bibinfo{volume}{52}},
\bibinfo{pages}{2361}.


\bibitem[{\citenamefont{Appel and Haken}(1977a)}]
{Appel:ak77a}
\bibinfo{author}{\bibnamefont{Appel},~\bibfnamefont{K.}}
and
\bibinfo{author}{\bibfnamefont{W.}~\bibnamefont{Haken}},
\bibinfo{year}{1977a},
\bibinfo{title}{``Every planar map is four colorable. Part I. Discharging},''
\bibinfo{journal}{Illinois J. Math.} \textbf{\bibinfo{volume}{21}},
\bibinfo{pages}{429}.


\bibitem[{\citenamefont{Appel and Haken}(1977b)}]
{Appel:ak77b}
\bibinfo{author}{\bibnamefont{Appel},~\bibfnamefont{K.}}
and
\bibinfo{author}{\bibfnamefont{W.}~\bibnamefont{Haken}},
\bibinfo{year}{1977b},
\bibinfo{title}{``Every planar map is four colorable. Part II.
Reducibility},''
\bibinfo{journal}{Illinois J. Math.} \textbf{\bibinfo{volume}{21}},
\bibinfo{pages}{491}.


\bibitem[{\citenamefont{de Arcangelis and Herrmann}(2002)}]
{de_Arcangelis:dh02}
\bibinfo{author}{\bibnamefont{de Arcangelis},~\bibfnamefont{L.}}
and
\bibinfo{author}{\bibfnamefont{H.}~\bibnamefont{Herrmann}},
\bibinfo{year}{2002},
\bibinfo{title}{``Self-organized criticality on small world networks},''
\bibinfo{journal}{Physica A} \textbf{\bibinfo{volume}{308}},
\bibinfo{pages}{545}.


\bibitem[{\citenamefont{Arenas} \emph{et~al.}(2001)\citenamefont{Arenas, D\'iaz-Guilera, and Guimera}}]
{Arenas:adg01}
\bibinfo{author}{\bibnamefont{Arenas},~\bibfnamefont{A.}},
\bibinfo{author}{\bibfnamefont{A.}~\bibnamefont{D\'iaz-Guilera}},
and
\bibinfo{author}{\bibfnamefont{R.}~\bibnamefont{Guimera}},
\bibinfo{year}{2001},
\bibinfo{title}{``Communication in networks with hierarchical branching},''
\bibinfo{journal}{Phys. Rev. Lett.} \textbf{\bibinfo{volume}{86}},
\bibinfo{pages}{3196}.


\bibitem[{\citenamefont{Arenas} \emph{et~al.}(2006a)\citenamefont{Arenas, D\'iaz-Guilera, and P\'erez-Vicente}}]
{Arenas:adp06a}
\bibinfo{author}{\bibfnamefont{Arenas}~\bibnamefont{A.}},
\bibinfo{author}{\bibfnamefont{A.}~\bibnamefont{D\'iaz-Guilera}},
and
\bibinfo{author}{\bibfnamefont{C.~J.}~\bibnamefont{P\'erez-Vicente}},
\bibinfo{year}{2006a},
\bibinfo{title}{``Synchronization reveals topological scales in complex networks},''
\bibinfo{journal}{Phys. Rev. Lett.} \textbf{\bibinfo{volume}{96}},
\bibinfo{pages}{114102}.


\bibitem[{\citenamefont{Arenas} \emph{et~al.}(2006b)\citenamefont{Arenas, D\'iaz-Guilera, and Perez-Vicente}}]
{Arenas:adp06b}
\bibinfo{author}{\bibnamefont{Arenas},~\bibfnamefont{A.}},
\bibinfo{author}{\bibfnamefont{A.}~\bibnamefont{D\'iaz-Guilera}},
and
\bibinfo{author}{\bibfnamefont{C.~J.}~\bibnamefont{Perez-Vicente}},
\bibinfo{year}{2006b},
\bibinfo{title}{``Synchronization processes in complex networks},''
\bibinfo{journal}{Physica D} \textbf{\bibinfo{volume}{224}},
\bibinfo{pages}{27}.


\bibitem[{\citenamefont{Aronson} \emph{et~al.}(1998)\citenamefont{Aronson, Frieze, and Pittel}}]
{Aronson:afp98}
\bibinfo{author}{\bibnamefont{Aronson},~\bibfnamefont{J.}},
\bibinfo{author}{\bibnamefont{A.},~\bibfnamefont{Frieze}},
and
\bibinfo{author}{\bibnamefont{B.~G.},~\bibfnamefont{Pittel}},
\bibinfo{year}{1998},
\bibinfo{title}{``Maximum matchings is sparse random graphs: Karp-Sipser revisited},''
\bibinfo{journal}{Rand. Struct. Algor.} \textbf{\bibinfo{volume}{12}},
\bibinfo{pages}{111}.


\bibitem[{\citenamefont{Atay} \emph{et~al.}(2004)\citenamefont{Atay, Jost, and Wende}}]
{Atay:ajw04}
\bibinfo{author}{\bibnamefont{Atay},~\bibfnamefont{F.~M.}},
\bibinfo{author}{\bibfnamefont{J.}~\bibnamefont{Jost}},
and
\bibinfo{author}{\bibfnamefont{A.}~\bibnamefont{Wende}},
\bibinfo{year}{2004},
\bibinfo{title}{``Delays, Connection Topology, and Synchronization
of Coupled Chaotic Maps},''
\bibinfo{journal}{Phys. Rev. Lett.} \textbf{\bibinfo{volume}{92}},
\bibinfo{pages}{144101}.

\bibitem[{\citenamefont{ben-Avraham and Glasser}(2007)}]
{Avraham:ag07}
\bibinfo{author}{\bibnamefont{ben-Avraham},~\bibfnamefont{D.}}
and
\bibinfo{author}{\bibfnamefont{M.~L.}~\bibnamefont{Glasser}},
\bibinfo{year}{2007},
\bibinfo{title}{``Diffusion-limited one-species reactions in the Bethe lattice},''
\bibinfo{journal}{J. Phys. C} \textbf{\bibinfo{volume}{19}},
\bibinfo{pages}{065107}.


\bibitem[{\citenamefont{Baillie} \emph{et~al.}(1995)\citenamefont{Baillie, Janke, Johnston, and Plech\'a\v{c}}}]
{Baillie:bjj95}
\bibinfo{author}{\bibnamefont{Baillie},~\bibfnamefont{C.}},
\bibinfo{author}{\bibfnamefont{W.}~\bibnamefont{Janke}},
\bibinfo{author}{\bibfnamefont{D.}~\bibnamefont{Johnston}},
and
\bibinfo{author}{\bibfnamefont{P.}~\bibnamefont{Plech\'a\v{c}}},
\bibinfo{year}{1995},
\bibinfo{title}{``Spin glasses on thin graphs},''
\bibinfo{journal}{Nucl. Phys. B} \textbf{\bibinfo{volume}{450}},
\bibinfo{pages}{730}.


\bibitem[{\citenamefont{Bakke} \emph{et~al.}(2006)\citenamefont{Bakke, Hansen, and Kert\'esz}}]
{Bakke:bhk06}
\bibinfo{author}{\bibnamefont{Bakke},~\bibfnamefont{J.~O.~H.}},
\bibinfo{author}{\bibfnamefont{A.}~\bibnamefont{Hansen}},
and
\bibinfo{author}{\bibfnamefont{J.}~\bibnamefont{Kert\'esz}},
\bibinfo{year}{2006},
\bibinfo{title}{``Failure and avalanches in complex networks},''
\bibinfo{journal}{Europhys. Lett.} \textbf{\bibinfo{volume}{76}},
\bibinfo{pages}{717}.


\bibitem[{\citenamefont{Barab{\'a}si and Albert}(1999)}]
{Barabasi:ba99}
\bibinfo{author}{\bibnamefont{Barab{\'a}si},~\bibfnamefont{A.-L.}}
and
\bibinfo{author}{\bibfnamefont{R.}~\bibnamefont{Albert}},
\bibinfo{year}{1999},
\bibinfo{title}{``Emergence of scaling in complex networks},''
\bibinfo{journal}{Science} \textbf{\bibinfo{volume}{286}},
\bibinfo{pages}{509}.


\bibitem[{\citenamefont{Barab\'asi} \emph{et~al.}(2001)\citenamefont{Barab\'asi, Ravasz, and Vicsek}}]
{Barabasi:brv01}
\bibinfo{author}{\bibnamefont{Barab\'asi},~\bibfnamefont{A.-L.}},
\bibinfo{author}{\bibfnamefont{E.}~\bibnamefont{Ravasz}},
and
\bibinfo{author}{\bibfnamefont{T.}~\bibnamefont{Vicsek}},
\bibinfo{year}{2001},
\bibinfo{title}{``Deterministic scale-free networks},''
\bibinfo{journal}{Physica A} \textbf{\bibinfo{volume}{299}},
\bibinfo{pages}{559}.


\bibitem[{\citenamefont{Barahona and Pecora}(2002)}]
{Barahona:bp02}
\bibinfo{author}{\bibnamefont{Barahona},~\bibfnamefont{M.}}
and
\bibinfo{author}{\bibfnamefont{L.}~\bibnamefont{Pecora}},
\bibinfo{year}{2002},
\bibinfo{title}{``Synchronization in small-world systems},''
\bibinfo{journal}{Phys. Rev. Lett.} \textbf{\bibinfo{volume}{89}},
\bibinfo{pages}{054101}.


\bibitem[{\citenamefont{Baronchelli} \emph{et~al.}(2007)\citenamefont{Baronchelli, Dall'Asta, Barrat, and Loreto}}]
{Baronchelli:bdb06}
\bibinfo{author}{\bibnamefont{Baronchelli},~\bibfnamefont{A.}},
\bibinfo{author}{\bibfnamefont{L.}~\bibnamefont{Dall'Asta}},
\bibinfo{author}{\bibfnamefont{A.}~\bibnamefont{Barrat}},
and
\bibinfo{author}{\bibfnamefont{V.}~\bibnamefont{Loreto}},
\bibinfo{year}{2007},
\bibinfo{title}{``Non-equilibrium phase transition in negotiation dynamics},'' 
\bibinfo{journal}{Phys. Rev. E} \textbf{\bibinfo{volume}{76}},
\bibinfo{pages}{051102}. 


\bibitem[{\citenamefont{Barrat and Weigt}(2000)}]
{Barrat:bw00}
\bibinfo{author}{\bibnamefont{Barrat},~\bibfnamefont{A.}}
and
\bibinfo{author}{\bibfnamefont{M.}~\bibnamefont{Weigt}},
\bibinfo{year}{2000},
\bibinfo{title}{``On the properties of small-world models},''
\bibinfo{journal}{Eur. Phys. J. B} \textbf{\bibinfo{volume}{13}},
\bibinfo{pages}{547}.


\bibitem[{\citenamefont{Barthel and Hartmann}(2004)}]
{Barthel:bh04}
\bibinfo{author}{\bibnamefont{Barthel},~\bibfnamefont{W.}}
and
\bibinfo{author}{\bibfnamefont{A.~K.}~\bibnamefont{Hartmann}},
\bibinfo{year}{2004},
\bibinfo{title}{``Clustering analysis of the ground-state structure of
the vertex-cover problem},''
\bibinfo{journal}{Phys. Rev. E} \textbf{\bibinfo{volume}{70}},
\bibinfo{pages}{066120}.




\bibitem[{\citenamefont{Barthelemy} \emph{et~al.}(2004)\citenamefont{Barthelemy, Barrat, Pastor-Satorras, and Vespignani}}]
{Barthelemy:bbp03}
\bibinfo{author}{\bibnamefont{Barthelemy},~\bibfnamefont{M.}},
\bibinfo{author}{\bibfnamefont{A.}~\bibnamefont{Barrat}},
\bibinfo{author}{\bibfnamefont{R.}~\bibnamefont{Pastor-Satorras}},
and
\bibinfo{author}{\bibfnamefont{A.}~\bibnamefont{Vespignani}},
\bibinfo{year}{2004},
\bibinfo{title}{``Velocity and hierarchical spread of epidemic outbreaks in scale-free networks},''
\bibinfo{journal}{Phys. Rev. Lett.} \textbf{\bibinfo{volume}{92}},
\bibinfo{pages}{178701 }.


\bibitem[{\citenamefont{Barthelemy} \emph{et~al.}(2005)\citenamefont{Barthelemy, Barrat, Pastor-Satorras, and Vespignani}}]
{Barthelemy:bbp05}
\bibinfo{author}{\bibnamefont{Barthelemy},~\bibfnamefont{M.}},
\bibinfo{author}{\bibfnamefont{A.}~\bibnamefont{Barrat}},
\bibinfo{author}{\bibfnamefont{R.}~\bibnamefont{Pastor-Satorras}},
and
\bibinfo{author}{\bibfnamefont{A.}~\bibnamefont{Vespignani}},
\bibinfo{year}{2005},
\bibinfo{title}{``Dynamical patterns of epidemic outbreaks in complex heterogeneous networks},''
\bibinfo{journal}{J. Theor. Biology} \textbf{\bibinfo{volume}{235}},
\bibinfo{pages}{275}.


\bibitem[{\citenamefont{Bartolozzi} \emph{et~al.}(2006)\citenamefont{Bartolozzi, Surungan, Leinweber,
and Williams}}]
{Bartolozzi:bslw05}
\bibinfo{author}{\bibnamefont{Bartolozzi},~\bibfnamefont{M.}},
\bibinfo{author}{\bibfnamefont{T.}~\bibnamefont{Surungan}},
\bibinfo{author}{\bibfnamefont{D.~B.}~\bibnamefont{Leinweber}},
and
\bibinfo{author}{\bibfnamefont{A.~G.}~\bibnamefont{Williams}},
\bibinfo{year}{2006},
\bibinfo{title}{``Spin glass behavior of the antiferromagnetic Ising model on a scale-free network},''
\bibinfo{journal}{Phys. Rev. B} \textbf{\bibinfo{volume}{73}},
\bibinfo{pages}{224419}.


\bibitem[{\citenamefont{Bauer and Bernard}(2002)}]
{Bauer:bb02}
\bibinfo{author}{\bibnamefont{Bauer},~\bibfnamefont{M.}}
and
\bibinfo{author}{\bibfnamefont{D.}~\bibnamefont{Bernard}},
\bibinfo{year}{2002},
\bibinfo{title}{``Maximal entropy random networks with given degree distribution},''
\eprint{cond-mat/0206150}.


\bibitem[{\citenamefont{Bauer} \emph{et~al.}(2005)\citenamefont{Bauer, Coulomb, and Dorogovtsev}}]
{Bauer:bkd05}
\bibinfo{author}{\bibnamefont{Bauer},~\bibfnamefont{M.}},
\bibinfo{author}{\bibfnamefont{S.}~\bibnamefont{Coulomb}},
and
\bibinfo{author}{\bibfnamefont{S.~N.}~\bibnamefont{Dorogovtsev}},
\bibinfo{year}{2005},
\bibinfo{title}{``Phase transition with the Berezinskii--Kosterlitz--Thouless singularity in the Ising model on a growing network},''
\bibinfo{journal}{Phys. Rev. Lett.} \textbf{\bibinfo{volume}{94}},
\bibinfo{pages}{200602}.


\bibitem[{\citenamefont{Bauer and Golinelli}(2001a)}]
{Bauer:bg01a}
\bibinfo{author}{\bibnamefont{Bauer},~\bibfnamefont{M.}}
and
\bibinfo{author}{\bibfnamefont{O.}~\bibnamefont{Golinelli}},
\bibinfo{year}{2001a},
\bibinfo{title}{``Core percolation in random graphs: a critical phenomena analysis},''
\bibinfo{journal}{Eur Phys. J. B} \textbf{\bibinfo{volume}{24}},
\bibinfo{pages}{339}.


\bibitem[{\citenamefont{Bauer and Golinelli}(2001b)}]
{Bauer:bg01b}
\bibinfo{author}{\bibnamefont{Bauer},~\bibfnamefont{M.}}
and
\bibinfo{author}{\bibfnamefont{O.}~\bibnamefont{Golinelli}},
\bibinfo{year}{2001b},
\bibinfo{title}{``Exactly solvable model with two conductor-insulator transitions driven by impurities},''
\bibinfo{journal}{Phys. Rev. Lett.} \textbf{\bibinfo{volume}{86}},
\bibinfo{pages}{2621}.


\bibitem[{\citenamefont{Baxter}(1982)}]
{Baxter:bbook82}
\bibinfo{author}{\bibnamefont{Baxter}, \bibfnamefont{R.~J.}},
\bibinfo{year}{1982}, \emph{\bibinfo{title}{Exactly Solved Models in
Statistical Mechanics}}
(\bibinfo{publisher}{Academic Press, London}).


\bibitem[{\citenamefont{Bedeaux} \emph{et~al.}(1970)\citenamefont{Bedeaux, Shuler, and Oppenheim}}]
{Bedeaux:bso70}
\bibinfo{author}{\bibnamefont{Bedeaux},~\bibfnamefont{D.}},
\bibinfo{author}{\bibfnamefont{K.~E.}~\bibnamefont{Shuler}},
and
\bibinfo{author}{\bibfnamefont{I.}~\bibnamefont{Oppenheim}},
\bibinfo{year}{1970},
\bibinfo{title}{``Decay of correlations. III. Relaxation of
spin correlations and distribution functions in the one-dimensional
Ising lattice},''
\bibinfo{journal}{J. Stat. Phys.} \textbf{\bibinfo{volume}{2}},
\bibinfo{pages}{1}.






\bibitem[{\citenamefont{ben-Avraham} \emph{et~al.}(1990)\citenamefont{ben-Avraham, Considine, Meakin, Redner, and Takayasu}}]
{ben-Avraham:bcm90}
\bibinfo{author}{\bibnamefont{ben-Avraham},~\bibfnamefont{D.}},
\bibinfo{author}{\bibfnamefont{D.~B.}~\bibnamefont{Considine}},
\bibinfo{author}{\bibfnamefont{P.}~\bibnamefont{Meakin}},
\bibinfo{author}{\bibfnamefont{S.}~\bibnamefont{Redner}},
and
\bibinfo{author}{\bibfnamefont{H.}~\bibnamefont{Takayasu}},
\bibinfo{year}{1990},
\bibinfo{title}{``Saturation transition in a monomer-monomer model of heterogeneous catalysis},''
\bibinfo{journal}{J. Phys. A} \textbf{\bibinfo{volume}{23}},
\bibinfo{pages}{4297}.


\bibitem[{\citenamefont{Ben-Naim and Krapivsky}(2004)}]
{Ben-Naim:bk04}
\bibinfo{author}{\bibnamefont{Ben-Naim},~\bibfnamefont{E.}}, 
and
\bibinfo{author}{\bibfnamefont{P.~L.}~\bibnamefont{Krapivsky}},
\bibinfo{year}{2004},
\bibinfo{title}{``Size of outbreaks near the epidemic threshold},''
\bibinfo{journal}{Phys. Rev. E} \textbf{\bibinfo{volume}{69}},
\bibinfo{pages}{050901 (R)}. 


\bibitem[{\citenamefont{Bender and Canfield}(1978)}]
{Bender:bc78}
\bibinfo{author}{\bibnamefont{Bender},~\bibfnamefont{E.~A.}}
and
\bibinfo{author}{\bibfnamefont{E.~R.}~\bibnamefont{Canfield}},
\bibinfo{year}{1978},
\bibinfo{title}{``The asymptotic number of labeled graphs with given degree sequences},''
\bibinfo{journal}{J. Combin. Theor. A} \textbf{\bibinfo{volume}{24}},
\bibinfo{pages}{296}. 


\bibitem[{\citenamefont{B\'enichou and Voituriez}(2007)}]
{Benichou:bv07}
\bibinfo{author}{\bibnamefont{B\'enichou},~\bibfnamefont{O.}} 
and
\bibinfo{author}{\bibfnamefont{R.}~\bibnamefont{Voituriez}},
\bibinfo{year}{2007},
\bibinfo{title}{``Comment on `Localization Transition of Biased Random Walks on Random Networks'},''
\bibinfo{journal}{Phys. Rev. Lett.} \textbf{\bibinfo{volume}{99}},
\bibinfo{pages}{209801}. 


\bibitem[{\citenamefont{Benjamini and Berger}(2001)}]
{Benjamini:bb01}
\bibinfo{author}{\bibnamefont{Benjamini},~\bibfnamefont{I.}}
and
\bibinfo{author}{\bibfnamefont{N.}~\bibnamefont{Berger}},
\bibinfo{year}{2001},
\bibinfo{title}{``The diameter of long-range percolation clusters on finite cycles},''
\bibinfo{journal}{Rand. Struct. Algor.} \textbf{\bibinfo{volume}{19}},
\bibinfo{pages}{102}.


\bibitem[{\citenamefont{Berezinskii}(1970)}]
{Berezinskii:b70}
\bibinfo{author}{\bibnamefont{Berezinskii},~\bibfnamefont{V.~L.}},
\bibinfo{year}{1970},
\bibinfo{title}{``Destruction of long range order in one-dimensional and twodimensional systems having a continuous symmetry group: I. Classical systems},''
\bibinfo{journal}{ZhETF} \textbf{\bibinfo{volume}{59}},
\bibinfo{pages}{907}
[\bibinfo{journal}{Sov. Phys. JETP} \textbf{\bibinfo{volume}{32}},
\bibinfo{pages}{493}].


\bibitem[{\citenamefont{Berg and L\"assig}(2002)}]
{Berg:bl02}
\bibinfo{author}{\bibnamefont{Berg},~\bibfnamefont{J.}}
and
\bibinfo{author}{\bibfnamefont{M.}~\bibnamefont{L\"assig}},
\bibinfo{year}{2002},
\bibinfo{title}{``Correlated random networks},''
\bibinfo{journal}{Phys. Rev. Lett.} \textbf{\bibinfo{volume}{89}},
\bibinfo{pages}{228701}.




\bibitem[{\citenamefont{Bernardo} \emph{et~al.}(2007)\citenamefont{Bernardo, Garofalo, and Sorrentino}}]
{Bernardo:bgs05}
\bibinfo{author}{\bibnamefont{Bernardo},~\bibfnamefont{M.}},
\bibinfo{author}{\bibfnamefont{F.}~\bibnamefont{Garofalo}},
and
\bibinfo{author}{\bibfnamefont{F.}~\bibnamefont{Sorrentino}},
\bibinfo{year}{2007},
\bibinfo{title}{``Synchronizability and synchronization dynamics of weighed and unweighed scale free networks with degree mixing},''
\bibinfo{journal}{Int. J. Bifurcation and Chaos} \textbf{\bibinfo{volume}{17}},
\bibinfo{pages}{7}.


\bibitem[{\citenamefont{Bethe}(1935)}]
{Bethe:b35}
\bibinfo{author}{\bibnamefont{Bethe},~\bibfnamefont{H.~A.}},
\bibinfo{year}{1935},
\bibinfo{title}{``Statistical theory of superlattices},''
\bibinfo{journal}{Proc. Roy. Soc. London A} \textbf{\bibinfo{volume}{150}},
\bibinfo{pages}{552}.


\bibitem[{\citenamefont{Bialas} \emph{et~al.}(2000)\citenamefont{Bialas, Bogacz, Burda, and Johnston}}]
{Bialas:bbb00}
\bibinfo{author}{\bibnamefont{Bialas},~\bibfnamefont{P.}},
\bibinfo{author}{\bibfnamefont{L.}~\bibnamefont{Bogacz}},
\bibinfo{author}{\bibfnamefont{B.}~\bibnamefont{Burda}},
and
\bibinfo{author}{\bibfnamefont{D.}~\bibnamefont{Johnston}},
\bibinfo{year}{2000},
\bibinfo{title}{``Finite size scaling of the balls in boxes model},''
\bibinfo{journal}{Nucl. Phys. B} \textbf{\bibinfo{volume}{575}},
\bibinfo{pages}{599}.


\bibitem[{\citenamefont{Bialas} \emph{et~al.}(1997)\citenamefont{Bialas, Burda, and Johnston}}]
{Bialas:bbj97}
\bibinfo{author}{\bibnamefont{Bialas},~\bibfnamefont{P.}},
\bibinfo{author}{\bibfnamefont{Z.}~\bibnamefont{Burda}},
and
\bibinfo{author}{\bibfnamefont{D.}~\bibnamefont{Johnston}},
\bibinfo{year}{1997},
\bibinfo{title}{``Condensation in the backgammon model},''
\bibinfo{journal}{Nucl. Phys. B} \textbf{\bibinfo{volume}{493}},
\bibinfo{pages}{505}.


\bibitem[{\citenamefont{Bialas} \emph{et~al.}(2003)\citenamefont{Bialas, Burda, Jurkiewicz, and Krzywicki}}]
{Bialas:bbj02}
\bibinfo{author}{\bibnamefont{Bialas},~\bibfnamefont{P.}},
\bibinfo{author}{\bibfnamefont{Z.}~\bibnamefont{Burda}},
\bibinfo{author}{\bibfnamefont{J.}~\bibnamefont{Jurkiewicz}},
and
\bibinfo{author}{\bibfnamefont{A.}~\bibnamefont{Krzywicki}},
\bibinfo{year}{2003},
\bibinfo{title}{``Tree networks with causal structure},''
\bibinfo{journal}{Phys. Rev. E} \textbf{\bibinfo{volume}{67}},
\bibinfo{pages}{066106}.




\bibitem[{\citenamefont{Bianconi}(2002)}]
{Bianconi:b02}
\bibinfo{author}{\bibnamefont{Bianconi},~\bibfnamefont{G.}},
\bibinfo{year}{2002},
\bibinfo{title}{``Mean field solution of the Ising model on a Barab\'asi-Albert network},''
\bibinfo{journal}{Phys. Lett. A} \textbf{\bibinfo{volume}{303}},
\bibinfo{pages}{166}.




\bibitem[{\citenamefont{Bianconi and Barab\'asi}(2001)}]
{Bianconi:bb01}
\bibinfo{author}{\bibnamefont{Bianconi},~\bibfnamefont{G.}}
and
\bibinfo{author}{\bibfnamefont{A.-L.}~\bibnamefont{Barab\'asi}},
\bibinfo{year}{2001},
\bibinfo{title}{``Bose-Einstein condensation in complex networks},''
\bibinfo{journal}{Phys. Rev. Lett.} \textbf{\bibinfo{volume}{86}},
\bibinfo{pages}{5632}.




\bibitem[{\citenamefont{Bianconi and Capocci}(2003)}]
{Bianconi:bc03}
\bibinfo{author}{\bibnamefont{Bianconi},~\bibfnamefont{G.}}
and
\bibinfo{author}{\bibfnamefont{A.}~\bibnamefont{Capocci}},
\bibinfo{year}{2003},
\bibinfo{title}{``Number of loops of size $h$ in growing scale-free networks},''
\bibinfo{journal}{Phys. Rev. Lett.} \textbf{\bibinfo{volume}{90}},
\bibinfo{pages}{078701}.


\bibitem[{\citenamefont{Bianconi} \emph{et~al.}(2007)\citenamefont{Bianconi, Gulbahce, and Motter}}]
{Bianconi:bgm07}
\bibinfo{author}{\bibnamefont{Bianconi},~\bibfnamefont{G.}},
\bibinfo{author}{\bibfnamefont{N.}~\bibnamefont{Gulbahce}},
and
\bibinfo{author}{\bibfnamefont{A.~E.}~\bibnamefont{Motter}},
\bibinfo{year}{2007},
\bibinfo{title}{``Local structure of directed networks},''
\eprint{arXiv:0707.4084 [physics]}.






\bibitem[{\citenamefont{Bianconi and Marsili}(2005a)}]
{Bianconi:bm05a}
\bibinfo{author}{\bibnamefont{Bianconi},~\bibfnamefont{G.}}
and
\bibinfo{author}{\bibfnamefont{M.}~\bibnamefont{Marsili}},
\bibinfo{year}{2005a},
\bibinfo{title}{``Loops of any size and Hamilton cycles in random scale-free networks},''
\bibinfo{journal}{J. Stat. Mech.} 
\bibinfo{pages}{P06005}.



 
\bibitem[{\citenamefont{Biely} \emph{et~al.}(2007)\citenamefont{Biely, Hanel, and Thurner}}]
{Biely:bht07}
\bibinfo{author}{\bibnamefont{Biely},~\bibfnamefont{C.}},
\bibinfo{author}{\bibfnamefont{R.}~\bibnamefont{Hanel}},
and
\bibinfo{author}{\bibfnamefont{S.}~\bibnamefont{Thurner}},
\bibinfo{year}{2007},
\bibinfo{title}{``Solvable spin model on dynamical networks},''
\eprint{arXiv:0707.3085 [cond-mat]}.


\bibitem[{\citenamefont{Binder and Young}(1986)}]
{Binder:by86}
\bibinfo{author}{\bibnamefont{Binder},~\bibfnamefont{K.}}
and
\bibinfo{author}{\bibfnamefont{A.~P.}~\bibnamefont{Young}},
\bibinfo{year}{1986},
\bibinfo{title}{``Spin glasses: Experimental facts, theoretical concepts, and
open questions},''
\bibinfo{journal}{Rev. Mod. Phys.} \textbf{\bibinfo{volume}{58}},
\bibinfo{pages}{801}.


\bibitem[{\citenamefont{Biskup}(2004)}]
{Biskup:b03}
\bibinfo{author}{\bibnamefont{Biskup},~\bibfnamefont{M.}},
\bibinfo{year}{2004},
\bibinfo{title}{``On the scaling of the chemical distance in long-range percolation models},''
\bibinfo{journal}{Ann. Probab.} \textbf{\bibinfo{volume}{32}},
\bibinfo{pages}{2938}.


\bibitem[{\citenamefont{Blatt} \emph{et~al.}(1996)\citenamefont{Blatt, Wiseman, and Domany}}]
{Blatt:bwd96}
\bibinfo{author}{\bibnamefont{Blatt},~\bibfnamefont{M.}},
\bibinfo{author}{\bibfnamefont{S.}~\bibnamefont{Wiseman}},
and
\bibinfo{author}{\bibfnamefont{E.}~\bibnamefont{Domany}},
\bibinfo{year}{1996},
\bibinfo{title}{``Superparamagnetic clustering of data},''
\bibinfo{journal}{Phys. Rev. Lett.} \textbf{\bibinfo{volume}{76}},
\bibinfo{pages}{3251}.




\bibitem[{\citenamefont{Boccaletti} \emph{et~al.}(2006)\citenamefont{Boccaletti, Latora, Moreno, Chavez, and Hwang}}]
{Boccaletti:blm06}
\bibinfo{author}{\bibnamefont{Boccaletti},~\bibfnamefont{S.}},
\bibinfo{author}{\bibfnamefont{V.}~\bibnamefont{Latora}},
\bibinfo{author}{\bibfnamefont{Y.}~\bibnamefont{Moreno}},
\bibinfo{author}{\bibfnamefont{M.}~\bibnamefont{Chavez}},
and
\bibinfo{author}{\bibfnamefont{D.-U.}~\bibnamefont{Hwang}},
\bibinfo{year}{2006},
\bibinfo{title}{``Complex networks: Structure and dynamics},''
\bibinfo{journal}{Phys. Rep.} \textbf{\bibinfo{volume}{424}},
\bibinfo{pages}{175}.


\bibitem[{\citenamefont{Bogacz} \emph{et~al.}(2007a)\citenamefont{Bogacz, Burda, Janke, and Waclaw}}]
{Bogacz:bbj07a}
\bibinfo{author}{\bibnamefont{Bogacz},~\bibfnamefont{L.}},
\bibinfo{author}{\bibfnamefont{Z.}~\bibnamefont{Burda}},
\bibinfo{author}{\bibfnamefont{W.}~\bibnamefont{Janke}},
and
\bibinfo{author}{\bibfnamefont{B.}~\bibnamefont{Waclaw}},
\bibinfo{year}{2007a},
\bibinfo{title}{``Balls-in-boxes condensation on networks},''
\bibinfo{journal}{Chaos} \textbf{\bibinfo{volume}{17}},
\bibinfo{pages}{026112}. 


\bibitem[{\citenamefont{Bogacz} \emph{et~al.}(2007b)\citenamefont{Bogacz, Burda, Janke, and Waclaw}}]
{Bogacz:bbj07b}
\bibinfo{author}{\bibnamefont{Bogacz},~\bibfnamefont{L.}},
\bibinfo{author}{\bibfnamefont{Z.}~\bibnamefont{Burda}},
\bibinfo{author}{\bibfnamefont{W.}~\bibnamefont{Janke}},
and
\bibinfo{author}{\bibfnamefont{B.}~\bibnamefont{Waclaw}},
\bibinfo{year}{2007b},
\bibinfo{title}{``Free zero-range processes on networks},''
\eprint{arXiv:0705.0549 [cond-mat]}.



\bibitem[{\citenamefont{Bogu\~n\'a and Pastor-Satorras}(2002)}]
{Boguna:bp02}
\bibinfo{author}{\bibnamefont{Bogu\~n\'a},~\bibfnamefont{M.}}
and
\bibinfo{author}{\bibfnamefont{R.}~\bibnamefont{Pastor-Satorras}},
\bibinfo{year}{2002},
\bibinfo{title}{``Epidemic spreading in correlated complex networks},''
\bibinfo{journal}{Phys. Rev. E} \textbf{\bibinfo{volume}{66}},
\bibinfo{pages}{047104}.


\bibitem[{\citenamefont{Bogu\~n\'a and Pastor-Satorras}(2003)}]
{Boguna:bp03}
\bibinfo{author}{\bibnamefont{Bogu\~n\'a},~\bibfnamefont{M.}}
and
\bibinfo{author}{\bibfnamefont{R.}~\bibnamefont{ Pastor-Satorras}},
\bibinfo{year}{2003},
\bibinfo{title}{``Class of correlated random networks with hidden variables},''
\bibinfo{journal}{Phys. Rev. E} \textbf{\bibinfo{volume}{68}},
\bibinfo{pages}{036112}.


\bibitem[{\citenamefont{Bogu\~n\'a} \emph{et~al.}(2003a)\citenamefont{Bogu\~n\'a, Pastor-Satorras, and Vespignani}}]
{Boguna:bpv02}
\bibinfo{author}{\bibnamefont{Bogu\~n\'a},~\bibfnamefont{M.}},
\bibinfo{author}{\bibfnamefont{R.}~\bibnamefont{Pastor-Satorras}},
and
\bibinfo{author}{\bibfnamefont{A.}~\bibnamefont{Vespignani}},
\bibinfo{year}{2003a},
\bibinfo{title}{``Absence of epidemic threshold in scale-free networks with degree correlations },''
\bibinfo{journal}{Phys. Rev. Lett.} \textbf{\bibinfo{volume}{90}},
\bibinfo{pages}{028701}.


\bibitem[{\citenamefont{Bogu\~n\'a} \emph{et~al.}(2003b)\citenamefont{Bogu\~n\'a, Pastor-Satorras, and Vespignani}}]
{Boguna:bpv03}
\bibinfo{author}{\bibnamefont{Bogu\~n\'a},~\bibfnamefont{M.}},
\bibinfo{author}{\bibfnamefont{R.}~\bibnamefont{Pastor-Satorras}},
and
\bibinfo{author}{\bibfnamefont{A.}~\bibnamefont{Vespignani}},
\bibinfo{year}{2003b},
\bibinfo{title}{``Epidemic spreading in complex networks with degree correlations},''
\bibinfo{journal}{Lect. Notes Phys.} \textbf{\bibinfo{volume}{625}},
\bibinfo{pages}{127}.


\bibitem[{\citenamefont{Bogu\~n\'a} \emph{et~al.}(2004)\citenamefont{Bogu\~n\'a, Pastor-Satorras, and Vespignani}}]
{Boguna:bpv04}
\bibinfo{author}{\bibnamefont{Bogu\~n\'a},~\bibfnamefont{M.}},
\bibinfo{author}{\bibfnamefont{R.}~\bibnamefont{Pastor-Satorras}},
and
\bibinfo{author}{\bibfnamefont{A.}~\bibnamefont{Vespignani}},
\bibinfo{year}{2004},
\bibinfo{title}{``Cut-offs and finite size effects in scale-free networks},''
\bibinfo{journal}{Eur. Phys. J. B} \textbf{\bibinfo{volume}{38}},
\bibinfo{pages}{205}.


\bibitem[{\citenamefont{Bogu\~n\'a and Serrano}(2005)}]
{Boguna:bs05}
\bibinfo{author}{\bibnamefont{Bogu\~n\'a},~\bibfnamefont{M.}},
and
\bibinfo{author}{\bibfnamefont{M.~A.}~\bibnamefont{Serrano}},
\bibinfo{year}{2005},
\bibinfo{title}{``Generalized percolation in random directed networks},''
\bibinfo{journal}{Phys. Rev. E} \textbf{\bibinfo{volume}{72}},
\bibinfo{pages}{016106}.


\bibitem[{\citenamefont{Bollob\'as}(1980)}]
{Bollobas:b80}
\bibinfo{author}{\bibnamefont{Bollob\'as},~\bibfnamefont{B.}},
\bibinfo{year}{1980},
\bibinfo{title}{``A probabilistic proof of an asymptotic formula for the number of labelled regular graphs},''
\bibinfo{journal}{Eur. J. Comb.} \textbf{\bibinfo{volume}{1}},
\bibinfo{pages}{311}.


\bibitem[{\citenamefont{Bollob\'{a}s}(1984)}]
{Bollobas:b84}
\bibinfo{author}{\bibnamefont{Bollob\'{a}s},~\bibfnamefont{B.}},
\bibinfo{year}{1984},
\bibinfo{title}{``The evolution of sparse graphs},''
in \emph{\bibinfo{booktitle}{Graph Theory and Combinatorics: Proc.
Cambridge Combinatorial Conf. in honour of Paul Erd\H os}}, edited
by
\bibinfo{editor}{\bibfnamefont{B.}~\bibnamefont{Bollob\'{a}s}}
(\bibinfo{publisher}{Academic Press, New York}), p.
\bibinfo{pages}{35}.







\bibitem[{\citenamefont{Bollob\'as and Riordan}(2003)}]
{Bollobas:br03}
\bibinfo{author}{\bibnamefont{Bollob\'as},~\bibfnamefont{B.}}
and
\bibinfo{author}{\bibfnamefont{O.}~\bibnamefont{Riordan}},
\bibinfo{year}{2003},
\bibinfo{title}{``Mathematical results on scale-free graphs},''
in \emph{\bibinfo{booktitle}{Handbook of Graphs and Networks}},
edited by
\bibinfo{editor}{\bibfnamefont{S.}~\bibnamefont{Bornholdt}}
and
\bibinfo{editor}{\bibfnamefont{H.~G.}~\bibnamefont{Schuster}}
(\bibinfo{publisher}{Wiley-VCH GmbH \& Co., Weinheim}), p.
\bibinfo{pages}{1}.


\bibitem[{\citenamefont{Bollob\'as and Riordan}(2005)}]
{Bollobas:br05}
\bibinfo{author}{\bibnamefont{Bollob\'as},~\bibfnamefont{B.}}
and
\bibinfo{author}{\bibfnamefont{O.}~\bibnamefont{Riordan}},
\bibinfo{year}{2005},
\bibinfo{title}{``Slow emergence of the giant component in the growing $m$-out graph},''
\bibinfo{journal}{Random Struct. Algor.} \textbf{\bibinfo{volume}{27}},
\bibinfo{pages}{1}.


\bibitem[{\citenamefont{Bonabeau}(1995)}]
{Bonabeau:b95}
\bibinfo{author}{\bibfnamefont{Bonabeau}~\bibnamefont{E.}},
\bibinfo{year}{1995},
\bibinfo{title}{``Sandpile dynamics on random graphs},''
\bibinfo{journal}{J. Phys. Soc. Japan} \textbf{\bibinfo{volume}{64}},
\bibinfo{pages}{327}.



\bibitem[{\citenamefont{Borgs} \emph{et~al.}(2001)\citenamefont{Borgs, Chayes, Kesten, and Spencer}}]
{Borgs:bck01}
\bibinfo{author}{\bibnamefont{Borgs},~\bibfnamefont{C.}},
\bibinfo{author}{\bibfnamefont{J. T.}~\bibnamefont{Chayes}},
\bibinfo{author}{\bibfnamefont{H.}~\bibnamefont{Kesten}},
and
\bibinfo{author}{\bibfnamefont{J.}~\bibnamefont{Spencer}},
\bibinfo{year}{2001},
\bibinfo{title}{``The birth of the infinite cluster: Finite-size scaling in percolation},''
\bibinfo{journal}{Comm. Math. Phys.} \textbf{\bibinfo{volume}{224}},
\bibinfo{pages}{153}.




\bibitem[{\citenamefont{Brandes} \emph{et~al.}(2006)\citenamefont{Brandes, Delling, Gaertler, G\"{o}rke, Hoefer, Nikoloski, and Wagner}}]
{Brandes:bdg06}
\bibinfo{author}{\bibnamefont{Brandes},~\bibfnamefont{U.}},
\bibinfo{author}{\bibfnamefont{D.}~\bibnamefont{Delling}},
\bibinfo{author}{\bibfnamefont{M.}~\bibnamefont{Gaertler}},
\bibinfo{author}{\bibfnamefont{R.}~\bibnamefont{G\"{o}rke}},
\bibinfo{author}{\bibfnamefont{M.}~\bibnamefont{Hoefer}},
\bibinfo{author}{\bibfnamefont{Z.}~\bibnamefont{Nikoloski}},
and
\bibinfo{author}{\bibfnamefont{D.}~\bibnamefont{Wagner}},
\bibinfo{year}{2006},
\bibinfo{title}{``Maximizing Modularity is hard},''
\eprint{arXive:physics/0608255}.


\bibitem[{\citenamefont{Braunstein} \emph{et~al.}(2003a)\citenamefont{Braunstein, Buldyrev, Cohen, Havlin, and Stanley}}]
{Braunstein:bbc03}
\bibinfo{author}{\bibnamefont{Braunstein},~\bibfnamefont{L.~A.}},
\bibinfo{author}{\bibfnamefont{S.~V.}~\bibnamefont{Buldyrev}},
\bibinfo{author}{\bibfnamefont{R.}~\bibnamefont{Cohen}},
\bibinfo{author}{\bibfnamefont{S.}~\bibnamefont{Havlin}},
and
\bibinfo{author}{\bibfnamefont{H.~E.}~\bibnamefont{Stanley}},
\bibinfo{year}{2003a},
\bibinfo{title}{``Optimal paths in disordered complex networks},''
\bibinfo{journal}{Phys. Rev. Lett.} \textbf{\bibinfo{volume}{91}},
\bibinfo{pages}{168701}.


\bibitem[{\citenamefont{Braunstein}\emph{et~al.}(2003b)\citenamefont{Braunstein, Mulet, Pagnani, Weigt, and Zecchina}}]
{Braunstein:bmpwz03}
\bibinfo{author}{\bibnamefont{Braunstein},~\bibfnamefont{A.}},
\bibinfo{author}{\bibfnamefont{R.}~\bibnamefont{Mulet}},
\bibinfo{author}{\bibfnamefont{A.}~\bibnamefont{Pagnani}},
\bibinfo{author}{\bibfnamefont{M.}~\bibnamefont{Weigt}},
and
\bibinfo{author}{\bibfnamefont{R.}~\bibnamefont{Zecchina}},
\bibinfo{year}{2003b},
\bibinfo{title}{``Polynomial
iterative algorithms for coloring and analyzing random graphs},''
\bibinfo{journal}{Phys. Rev. E} \textbf{\bibinfo{volume}{68}},
\bibinfo{pages}{036702}.


\bibitem[{\citenamefont{Braunstein} \emph{et~al.}(2004)\citenamefont{Braunstein, Buldyrev, Sreenivasan, Cohen, Havlin, and Stanley}}]
{Braunstein:bbs04}
\bibinfo{author}{\bibfnamefont{Braunstein}~\bibnamefont{L.~A.}},
\bibinfo{author}{\bibfnamefont{S.~V.}~\bibnamefont{Buldyrev}},
\bibinfo{author}{\bibfnamefont{S.}~\bibnamefont{Sreenivasan}},
\bibinfo{author}{\bibfnamefont{R.}~\bibnamefont{Cohen}},
\bibinfo{author}{\bibfnamefont{S.}~\bibnamefont{Havlin}},
and
\bibinfo{author}{\bibfnamefont{H.~E.}~\bibnamefont{Stanley}},
\bibinfo{year}{2004},
\bibinfo{title}{``The optimal path in a random network},''
\bibinfo{journal}{Lecture Notes in Physics} \textbf{\bibinfo{volume}{650}},
\bibinfo{pages}{127}.


\bibitem[{\citenamefont{Braunstein and Zecchina}(2004)}]
{Braunstein:bz04}
\bibinfo{author}{\bibnamefont{Braunstein},~\bibfnamefont{A.}}
and
\bibinfo{author}{\bibfnamefont{R.}~\bibnamefont{Zecchina}},
\bibinfo{year}{2004},
\bibinfo{title}{``Survey propagation as local equilibrium equations},''
\bibinfo{journal}{J. Stat. Mech.}
\bibinfo{pages}{P06007}.



\bibitem[{\citenamefont{Bruinsma}(1984)}]
{Bruinsma:84}
\bibinfo{author}{\bibnamefont{Bruinsma},~\bibfnamefont{R.}}
\bibinfo{year}{1984},
\bibinfo{title}{``Random-field Ising model on a Bethe lattice},''
\bibinfo{journal}{Phys. Rev. B} \textbf{\bibinfo{volume}{30}},
\bibinfo{pages}{289}.




\bibitem[{\citenamefont{Burda} \emph{et~al.}(2001)\citenamefont{Burda, Correia, and Krzywicki}}]
{Burda:bck01}
\bibinfo{author}{\bibnamefont{Burda},~\bibfnamefont{Z.}},
\bibinfo{author}{\bibfnamefont{J.~D.}~\bibnamefont{Correia}},
and
\bibinfo{author}{\bibfnamefont{A.}~\bibnamefont{Krzywicki}},
\bibinfo{year}{2001},
\bibinfo{title}{``Statistical ensemble of scale-free random graphs},''
\bibinfo{journal}{Phys. Rev. E} \textbf{\bibinfo{volume}{64}},
\bibinfo{pages}{046118}.


\bibitem[{\citenamefont{Burda} \emph{et~al.}(2002)\citenamefont{Burda, Johnston, Jurkiewicz, Kaminski, Nowak, Papp,
and I. Zahed}}] {Burda:bjj02}
\bibinfo{author}{\bibnamefont{Burda},~\bibfnamefont{Z.}},
\bibinfo{author}{\bibfnamefont{D.}~\bibnamefont{Johnston}},
\bibinfo{author}{\bibfnamefont{J.}~\bibnamefont{Jurkiewicz}},
\bibinfo{author}{\bibfnamefont{M.}~\bibnamefont{Kaminski}},
\bibinfo{author}{\bibfnamefont{M.~A.}~\bibnamefont{ Nowak}},
and
\bibinfo{author}{\bibfnamefont{G.}~\bibnamefont{Papp}},
\bibinfo{year}{2002},
\bibinfo{title}{``Wealth condensation in Pareto macro-economies},''
\bibinfo{journal}{Phys. Rev. E} \textbf{\bibinfo{volume}{65}},
\bibinfo{pages}{026102}.


\bibitem[{\citenamefont{Burda} \emph{et~al.}(2004a)\citenamefont{Burda, Jurkiewicz, and Krzywicki}}]
{Burda:bjk04a}
\bibinfo{author}{\bibnamefont{Burda},~\bibfnamefont{Z.}},
\bibinfo{author}{\bibfnamefont{J.}~\bibnamefont{Jurkiewicz}},
and
\bibinfo{author}{\bibfnamefont{A.}~\bibnamefont{Krzywicki}},
\bibinfo{year}{2004a},
\bibinfo{title}{``Network transitivity and matrix models},''
\bibinfo{journal}{Phys. Rev. E} \textbf{\bibinfo{volume}{69}},
\bibinfo{pages}{026106}.


\bibitem[{\citenamefont{Burda} \emph{et~al.}(2004b)\citenamefont{Burda, Jurkiewicz, and Krzywicki}}]
{Burda:bjk04b}
\bibinfo{author}{\bibnamefont{Burda},~\bibfnamefont{B.}},
\bibinfo{author}{\bibfnamefont{J.}~\bibnamefont{Jurkiewicz}},
and
\bibinfo{author}{\bibfnamefont{A.}~\bibnamefont{Krzywicki}},
\bibinfo{year}{2004b},
\bibinfo{title}{``Perturbing general uncorrelated networks},''
\bibinfo{journal}{Phys. Rev. E} \textbf{\bibinfo{volume}{70}},
\bibinfo{pages}{026106}.


\bibitem[{\citenamefont{Burda and Krzywicki}(2003)}]
{Burda:bk03}
\bibinfo{author}{\bibnamefont{Burda},~\bibfnamefont{Z.}}
and
\bibinfo{author}{\bibfnamefont{A.}~\bibnamefont{Krzywicki}},
\bibinfo{year}{2003},
\bibinfo{title}{``Uncorrelated random networks},''
\bibinfo{journal}{Phys. Rev. E} \textbf{\bibinfo{volume}{6}},
\bibinfo{pages}{046118}.




\bibitem[{\citenamefont{Caldarelli}(2007)}]
{Caldarelli:cbook07}
\bibinfo{author}{\bibnamefont{Caldarelli},~\bibfnamefont{G.}},
\bibinfo{year}{2007},
\emph{\bibinfo{title}{Scale-Free Networks: Complex Webs in Nature and Technology}} (\bibinfo{publisher}{Oxford Finance Series---Oxford University Press, Oxford}).


\bibitem[{\citenamefont{Caldarelli} \emph{et~al.}(2006)\citenamefont{Caldarelli, Capocci, and Garlaschelli}}]
{Caldarelli:ccg06}
\bibinfo{author}{\bibnamefont{Caldarelli},~\bibfnamefont{G.}},
\bibinfo{author}{\bibfnamefont{A.}~\bibnamefont{Capocci}},
and
\bibinfo{author}{\bibfnamefont{D.}~\bibnamefont{Garlaschelli}},
\bibinfo{year}{2006},
\bibinfo{title}{``Self-organized network evolution coupled to extremal dynamics},''
\eprint{cond-mat/0611201}.


\bibitem[{\citenamefont{Caldarelli} \emph{et~al.}(2002)\citenamefont{Caldarelli, Capocci, De Los Rios, and Mu\~noz}}]
{Caldarelli:ccd02}
\bibinfo{author}{\bibnamefont{Caldarelli},~\bibfnamefont{G.}},
\bibinfo{author}{\bibfnamefont{A.}~\bibnamefont{Capocci}},
\bibinfo{author}{\bibfnamefont{P.}~\bibnamefont{De Los Rios}},
and
\bibinfo{author}{\bibfnamefont{M.~A.}~\bibnamefont{Mu\~noz}},
\bibinfo{year}{2002},
\bibinfo{title}{``Scale-free networks from varying vertex intrinsic fitness},''
\bibinfo{journal}{Phys. Rev. Lett.} \textbf{\bibinfo{volume}{89}},
\bibinfo{pages}{258702}.


\bibitem[{\citenamefont{Callaway} \emph{et~al.}(2001)\citenamefont{Callaway, Hopcroft, Kleinberg, Newman, and Strogatz}}]
{Callaway:chk01}
\bibinfo{author}{\bibnamefont{Callaway},~\bibfnamefont{D.~S.}},
\bibinfo{author}{\bibfnamefont{J.~E.}~\bibnamefont{Hopcroft}},
\bibinfo{author}{\bibfnamefont{J.~M.}~\bibnamefont{Kleinberg}},
\bibinfo{author}{\bibfnamefont{M.~E.~J.}~\bibnamefont{Newman}},
and
\bibinfo{author}{\bibfnamefont{S.~H.}~\bibnamefont{Strogatz}},
\bibinfo{year}{2001},
\bibinfo{title}{``Are randomly grown graphs really random?},''
\bibinfo{journal}{Phys. Rev. E} \textbf{\bibinfo{volume}{64}},
\bibinfo{pages}{041902}.


\bibitem[{\citenamefont{Callaway} \emph{et~al.}(2000)\citenamefont{Callaway, Newman, Strogatz and Watts}}]
{Callaway:cns00}
\bibinfo{author}{\bibnamefont{Callaway},~\bibfnamefont{D.~S.}},
\bibinfo{author}{\bibfnamefont{M.~E.~J.}~\bibnamefont{Newman}},
\bibinfo{author}{\bibfnamefont{S.~H.}~\bibnamefont{Strogatz}},
and
\bibinfo{author}{\bibfnamefont{D.~J.}~\bibnamefont{Watts}},
\bibinfo{year}{2000},
\bibinfo{title}{``Network robustness and fragility: Percolation on random graphs},''
\bibinfo{journal}{Phys. Rev. Lett.} \textbf{\bibinfo{volume}{85}},
\bibinfo{pages}{5468}.


\bibitem[{\citenamefont{Carmi} \emph{et~al.}(2006a)\citenamefont{Carmi, Cohen, and Dolev}}]
{Carmi:ccd06}
\bibinfo{author}{\bibnamefont{Carmi},~\bibfnamefont{S.}},
\bibinfo{author}{\bibfnamefont{R.}~\bibnamefont{Cohen}},
and
\bibinfo{author}{\bibfnamefont{D.}~\bibnamefont{Dolev}},
\bibinfo{year}{2006a},
\bibinfo{title}{``Searching complex networks efficiently with minimal information},''
\bibinfo{journal}{Europhys. Lett.} \textbf{\bibinfo{volume}{74}},
\bibinfo{pages}{1102}.


\bibitem[{\citenamefont{Carmi} \emph{et~al.}(2006b)\citenamefont{Carmi, Havlin, Kirkpatrick, Shavitt, and Shir}}]
{Carmi:chk06}
\bibinfo{author}{\bibnamefont{Carmi},~\bibfnamefont{S.}},
\bibinfo{author}{\bibfnamefont{S.}~\bibnamefont{Havlin}},
\bibinfo{author}{\bibfnamefont{S.}~\bibnamefont{Kirkpatrick}},
\bibinfo{author}{\bibfnamefont{Y.}~\bibnamefont{Shavitt}},
and
\bibinfo{author}{\bibfnamefont{E.}~\bibnamefont{Shir}},
\bibinfo{year}{2006b},
\bibinfo{title}{``MEDUSA - New model of Internet topology using $k$-shell decomposition},''
\eprint{cond-mat/0601240}.


\bibitem[{\citenamefont{Carmi} \emph{et~al.}(2007)\citenamefont{Carmi, Havlin, Kirkpatrick, Shavitt, and Shir}}]
{Carmi:chk07}
\bibinfo{author}{\bibnamefont{Carmi},~\bibfnamefont{S.}},
\bibinfo{author}{\bibfnamefont{S.}~\bibnamefont{Havlin}},
\bibinfo{author}{\bibfnamefont{S.}~\bibnamefont{Kirkpatrick}},
\bibinfo{author}{\bibfnamefont{Y.}~\bibnamefont{Shavitt}},
and
\bibinfo{author}{\bibfnamefont{E.}~\bibnamefont{Shir}},
\bibinfo{year}{2007},
\bibinfo{title}{``New model of Internet topology using $k$-shell decomposition},''
\bibinfo{journal}{PNAS} \textbf{\bibinfo{volume}{104}},
\bibinfo{pages}{11150}.


\bibitem[{\citenamefont{Caruso} \emph{et~al.}(2006)\citenamefont{Caruso, Latora, Pluchino, Rapisarda, and Tadic}}]
{Caruso:clp06}
\bibinfo{author}{\bibnamefont{Caruso},~\bibfnamefont{F.}},
\bibinfo{author}{\bibfnamefont{V.}~\bibnamefont{Latora}},
\bibinfo{author}{\bibfnamefont{A.}~\bibnamefont{Pluchino}},
\bibinfo{author}{\bibfnamefont{A.}~\bibnamefont{Rapisarda}},
and
\bibinfo{author}{\bibfnamefont{B.}~\bibnamefont{Tadic}},
\bibinfo{year}{2006},
\bibinfo{title}{``Olami-Feder-Christensen model on different networks},''
\bibinfo{journal}{Eur. Phys. J. B} \textbf{\bibinfo{volume}{50}},
\bibinfo{pages}{243}.


\bibitem[{\citenamefont{Caruso} \emph{et~al.}(2007)\citenamefont{Caruso, Pluchino, Latora, Vinciguerra, and Rapisarda}}]
{Caruso:cpl07}
\bibinfo{author}{\bibnamefont{Caruso},~\bibfnamefont{F.}},
\bibinfo{author}{\bibfnamefont{A.}~\bibnamefont{Pluchino}},
\bibinfo{author}{\bibfnamefont{V.}~\bibnamefont{Latora}},
\bibinfo{author}{\bibfnamefont{S.}~\bibnamefont{Vinciguerra}},
and
\bibinfo{author}{\bibfnamefont{A.}~\bibnamefont{Rapisarda}},
\bibinfo{year}{2007},
\bibinfo{title}{``Analysis of self-organized criticality in the Olami-Feder-Christensen model and in real earthquakes},''
\bibinfo{journal}{Phys. Rev. E} \textbf{\bibinfo{volume}{75}},
\bibinfo{pages}{055101 (R)}.


\bibitem[{\citenamefont{Castellani} \emph{et~al.}(2005)\citenamefont{Castellani, Krz{\c a}ka{\l}a, and Ricci-Tersenghi}}]
{Castellani:ckr05}
\bibinfo{author}{\bibnamefont{Castellani},~\bibfnamefont{T.}},
\bibinfo{author}{\bibfnamefont{F.}~\bibnamefont{Krz{\c a}ka{\l}a}},
and
\bibinfo{author}{\bibfnamefont{F.}~\bibnamefont{Ricci-Tersenghi}},
\bibinfo{year}{2005},
\bibinfo{title}{``Spin glass models with ferromagnetically biased couplings on the Bethe lattice:
analytic solutions and numerical simulations},''
\bibinfo{journal}{Eur. Phys. J. B} \textbf{\bibinfo{volume}{47}},
\bibinfo{pages}{99}.



\bibitem[{\citenamefont{Castellano} \emph{et~al.}(2005)\citenamefont{Castellano, Loreto, Barrat, Cecconi, and Parisi}}]
{Castellano:clb05}
\bibinfo{author}{\bibnamefont{Castellano},~\bibfnamefont{C.}},
\bibinfo{author}{\bibfnamefont{V.}~\bibnamefont{Loreto}},
\bibinfo{author}{\bibfnamefont{A.}~\bibnamefont{Barrat}},
\bibinfo{author}{\bibfnamefont{F.}~\bibnamefont{Cecconi}},
and
\bibinfo{author}{\bibfnamefont{D.}~\bibnamefont{Parisi}},
\bibinfo{year}{2005},
\bibinfo{title}{``Comparison of voter and Glauber ordering dynamics on networks},''
\bibinfo{journal}{Phys. Rev. E} \textbf{\bibinfo{volume}{71}},
\bibinfo{pages}{066107}.


\bibitem[{\citenamefont{Castellano and Pastor-Satorras}(2006a)}]
{Castellano:cp06}
\bibinfo{author}{\bibnamefont{Castellano},~\bibfnamefont{C.}}
and
\bibinfo{author}{\bibfnamefont{R.}~\bibnamefont{Pastor-Satorras}},
\bibinfo{year}{2006a},
\bibinfo{title}{``Non mean-field behavior of the contact process on scale-free networks},''
\bibinfo{journal}{Phys. Rev. Lett.} \textbf{\bibinfo{volume}{96}},
\bibinfo{pages}{038701}.


\bibitem[{\citenamefont{Castellano and Pastor-Satorras}(2006b)}]
{Castellano:cp06b}
\bibinfo{author}{\bibnamefont{Castellano},~\bibfnamefont{C.}}
and
\bibinfo{author}{\bibfnamefont{R.}~\bibnamefont{Pastor-Satorras}},
\bibinfo{year}{2006b},
\bibinfo{title}{``Zero temperature Glauber dynamics on complex networks},''
\bibinfo{journal}{J. Stat. Mech.} 
\bibinfo{pages}{P05001}.


\bibitem[{\citenamefont{Castellano and Pastor-Satorras}(2007a)}]
{Castellano:cp07}
\bibinfo{author}{\bibnamefont{Castellano},~\bibfnamefont{C.}}
and
\bibinfo{author}{\bibfnamefont{R.}~\bibnamefont{Pastor-Satorras}},
\bibinfo{year}{2007a},
\bibinfo{title}{``Castellano and Pastor-Satorras reply},''
\bibinfo{journal}{Phys. Rev. Lett.} \textbf{\bibinfo{volume}{98}},
\bibinfo{pages}{029802}.


\bibitem[{\citenamefont{Castellano and Pastor-Satorras}(2007b)}]
{Castellano:cp07}
\bibinfo{author}{\bibnamefont{Castellano},~\bibfnamefont{C.}}
and
\bibinfo{author}{\bibfnamefont{R.}~\bibnamefont{Pastor-Satorras}},
\bibinfo{year}{2007b},
\bibinfo{title}{``Routes to thermodynamic limit on scale-free networks},''
\eprint{arXiv:0710.2784}.


\bibitem[{\citenamefont{Castellano} \emph{et~al.}(2003)\citenamefont{Castellano, Vilone, and Vespignani}}]
{Castellano:cvv03}
\bibinfo{author}{\bibnamefont{Castellano},~\bibfnamefont{C.}},
\bibinfo{author}{\bibfnamefont{D.}~\bibnamefont{Vilone}},
and
\bibinfo{author}{\bibfnamefont{A.}~\bibnamefont{Vespignani}},
\bibinfo{year}{2003},
\bibinfo{title}{``Incomplete ordering of the voter model on small-world networks},''
\bibinfo{journal}{Europhys. Lett.} \textbf{\bibinfo{volume}{63}},
\bibinfo{pages}{153}.


\bibitem[{\citenamefont{Catanzaro} \emph{et~al.}(2005)\citenamefont{Catanzaro, Bogu\~n\'a, and Pastor-Satorras}}]
{Catanzaro:cbp05}
\bibinfo{author}{\bibnamefont{Catanzaro},~\bibfnamefont{M.}},
\bibinfo{author}{\bibfnamefont{M.}~\bibnamefont{Bogu\~n\'a}},
and
\bibinfo{author}{\bibfnamefont{R.}~\bibnamefont{Pastor-Satorras}},
\bibinfo{year}{2005},
\bibinfo{title}{``Diffusion-annihilation processes in complex networks},''
\bibinfo{journal}{Phys. Rev. E} \textbf{\bibinfo{volume}{71}},
\bibinfo{pages}{056104}.




\bibitem[{\citenamefont{Chalupa} \emph{et~al.}(1979)\citenamefont{Chalupa, Leath, and Reich}}]
{Chalupa:clr79}
\bibinfo{author}{\bibnamefont{Chalupa},~\bibfnamefont{J.}},
\bibinfo{author}{\bibfnamefont{P.~L.}~\bibnamefont{Leath}},
and
\bibinfo{author}{\bibfnamefont{G.~R.}~\bibnamefont{Reich}},
\bibinfo{year}{1979},
\bibinfo{title}{``Bootstrap percolation on a Bethe lattice},''
\bibinfo{journal}{J. Phys. C} \textbf{\bibinfo{volume}{12}},
\bibinfo{pages}{L31}.



\bibitem[{\citenamefont{Chatterjee and Sen}(2006)}]
{Chatterjee:cs06}
\bibinfo{author}{\bibnamefont{Chatterjee},~\bibfnamefont{A.}}
and
\bibinfo{author}{\bibfnamefont{P.}~\bibnamefont{Sen}},
\bibinfo{year}{2006},
\bibinfo{title}{``Phase transitions in Ising model on a Euclidean network},''
\bibinfo{journal}{Phys. Rev. E} \textbf{\bibinfo{volume}{74}},
\bibinfo{pages}{036109}.




\bibitem[{\citenamefont{Chavez} \emph{et~al.}(2005)\citenamefont{Chavez, Hwang, Amann, Hentschel, and Boccaletti}}]
{Chavez:chahb05}
\bibinfo{author}{\bibnamefont{Chavez},~\bibfnamefont{M.}},
\bibinfo{author}{\bibfnamefont{D.-U.}~\bibnamefont{Hwang}},
\bibinfo{author}{\bibfnamefont{A.}~\bibnamefont{Amann}},
\bibinfo{author}{\bibfnamefont{H.~G.~E.}~\bibnamefont{Hentschel}},
and
\bibinfo{author}{\bibfnamefont{S.}~\bibnamefont{Boccaletti}},
\bibinfo{year}{2005},
\bibinfo{title}{``Synchronization is enhanced in weighted complex networks},''
\bibinfo{journal}{Phys. Rev. Lett.} \textbf{\bibinfo{volume}{94}},
\bibinfo{pages}{218701}.


\bibitem[{\citenamefont{Chavez} \emph{et~al.}(2006)\citenamefont{Chavez, Hwang, Martinerie, and Boccaletti}}]
{Chavez:chm06}
\bibinfo{author}{\bibnamefont{Chavez},~\bibfnamefont{M.}},
\bibinfo{author}{\bibfnamefont{D.-U.}~\bibnamefont{Hwang}},
\bibinfo{author}{\bibfnamefont{J.}~\bibnamefont{Martinerie}},
and
\bibinfo{author}{\bibfnamefont{S.}~\bibnamefont{Boccaletti}},
\bibinfo{year}{2006},
\bibinfo{title}{``Degree mixing and the enchancement of synchronization in complex weighted networks},''
\bibinfo{journal}{Phys. Rev. E} \textbf{\bibinfo{volume}{74}},
\bibinfo{pages}{066107}.


\bibitem[{\citenamefont{Chertkov and Chernyak}(2006a)}]
{Chertkov:cc06a}
\bibinfo{author}{\bibnamefont{Chertkov},~\bibfnamefont{M.}}
and
\bibinfo{author}{\bibfnamefont{V.~Y.}~\bibnamefont{Chernyak}},
\bibinfo{year}{2006a},
\bibinfo{title}{``Loop series for discrete statistical models on graphs},''
\bibinfo{journal}{J. Sta. Mech.} 
\bibinfo{pages}{P06009}.


\bibitem[{\citenamefont{Chertkov and Chernyak}(2006b)}]
{Chertkov:cc06b}
\bibinfo{author}{\bibnamefont{Chertkov},~\bibfnamefont{M.}}
and
\bibinfo{author}{\bibfnamefont{V. Y.}~\bibnamefont{Chernyak}},
\bibinfo{year}{2006b},
\bibinfo{title}{``Loop calculus helps to improve belief propagation and linear programming decodings of low-density-parity-check codes},''
\eprint{cs.IT/0609154}.


\bibitem[{\citenamefont{Chung}(1997)}]
{Chung:cbook97}
\bibinfo{author}{\bibnamefont{Chung}, \bibfnamefont{F.~R.~K.}},
\bibinfo{year}{1997},
\emph{\bibinfo{title}{Spectral graph theory}}
(\bibinfo{publisher}{American Mathematical Society, Providence, Rhode Island}).


\bibitem[{\citenamefont{Chung and Lu}(2002)}]
{Chung:cl02}
\bibinfo{author}{\bibnamefont{Chung},~\bibfnamefont{F.}}
and
\bibinfo{author}{\bibfnamefont{L.}~\bibnamefont{Lu}},
\bibinfo{year}{2002},
\bibinfo{title}{``The average distances in random graphs with given expected degrees},''
\bibinfo{journal}{PNAS} \textbf{\bibinfo{volume}{99}},
\bibinfo{pages}{15879}.


\bibitem[{\citenamefont{Chung} \emph{et~al.}(2003)\citenamefont{Chung, Lu, and Vu}}]
{Chung:clv03}
\bibinfo{author}{\bibnamefont{Chung},~\bibfnamefont{F.}},
\bibinfo{author}{\bibfnamefont{L.}~\bibnamefont{Lu}},
and
\bibinfo{author}{\bibfnamefont{V.}~\bibnamefont{Vu}},
\bibinfo{year}{2003},
\bibinfo{title}{``Spectra of random graphs with given expected
degrees},''
\bibinfo{journal}{PNAS} \textbf{\bibinfo{volume}{100}},
\bibinfo{pages}{6313}.


\bibitem[{\citenamefont{Clauset} \emph{et~al.}(2004)\citenamefont{Clauset, Newman, and Moore}}]
{Clauset:cnm04}
\bibinfo{author}{\bibnamefont{Clauset},~\bibfnamefont{A.}},
\bibinfo{author}{\bibfnamefont{M.~E.~J.}~\bibnamefont{Newman}},
and
\bibinfo{author}{\bibfnamefont{C.}~\bibnamefont{Moore}},
\bibinfo{year}{2004},
\bibinfo{title}{``Finding community structure in very large
networks},''
\bibinfo{journal}{Phys. Rev. E} \textbf{\bibinfo{volume}{70}},
\bibinfo{pages}{066111}.


\bibitem[{\citenamefont{Cohen} \emph{et~al.}(2002)\citenamefont{Cohen, ben-Avraham, and Havlin}}]
{Cohen:cbh02}
\bibinfo{author}{\bibnamefont{Cohen},~\bibfnamefont{R.}},
\bibinfo{author}{\bibfnamefont{D.}~\bibnamefont{ben-Avraham}},
and
\bibinfo{author}{\bibfnamefont{S.}~\bibnamefont{Havlin}},
\bibinfo{year}{2002},
\bibinfo{title}{``Percolation critical exponents in scale-free networks},''
\bibinfo{journal}{Phys. Rev. E} \textbf{\bibinfo{volume}{66}},
\bibinfo{pages}{036113}.


\bibitem[{\citenamefont{Cohen} \emph{et~al.}(2000)\citenamefont{Cohen, Erez, ben-Avraham, and Havlin}}]
{Cohen:ceb00}
\bibinfo{author}{\bibnamefont{Cohen},~\bibfnamefont{R.}},
\bibinfo{author}{\bibfnamefont{K.}~\bibnamefont{Erez}},
\bibinfo{author}{\bibfnamefont{D.}~\bibnamefont{ben-Avraham}},
and
\bibinfo{author}{\bibfnamefont{S.}~\bibnamefont{Havlin}},
\bibinfo{year}{2000},
\bibinfo{title}{``Resilience of the Internet to random breakdowns},''
\bibinfo{journal}{Phys. Rev. Lett.} \textbf{\bibinfo{volume}{85}},
\bibinfo{pages}{4626}.


\bibitem[{\citenamefont{Cohen} \emph{et~al.}(2001)\citenamefont{Cohen, Erez, ben-Avraham, and Havlin}}]
{Cohen:ceb01}
\bibinfo{author}{\bibnamefont{Cohen},~\bibfnamefont{R.}},
\bibinfo{author}{\bibfnamefont{K.}~\bibnamefont{Erez}},
\bibinfo{author}{\bibfnamefont{D.}~\bibnamefont{ben-Avraham}},
and
\bibinfo{author}{\bibfnamefont{S.}~\bibnamefont{Havlin}},
\bibinfo{year}{2001},
\bibinfo{title}{``Breakdown of the Internet under intentional attack},''
\bibinfo{journal}{Phys. Rev. Lett.} \textbf{\bibinfo{volume}{86}},
\bibinfo{pages}{3682}.


\bibitem[{\citenamefont{Cohen} \emph{et~al.}(2003a)\citenamefont{Cohen, Havlin, and ben-Avraham}}]
{Cohen:chb03a}
\bibinfo{author}{\bibnamefont{Cohen},~\bibfnamefont{R.}},
\bibinfo{author}{\bibfnamefont{S.}~\bibnamefont{Havlin}},
and
\bibinfo{author}{\bibfnamefont{D.}~\bibnamefont{ben-Avraham}},
\bibinfo{year}{2003a},
\bibinfo{title}{``Structural properties of scale free networks},''
in \emph{\bibinfo{booktitle}{Handbook of Graphs and Networks}},
edited by
\bibinfo{editor}{\bibfnamefont{S.}~\bibnamefont{Bornholdt}}
and
\bibinfo{editor}{\bibfnamefont{H.~G.}~\bibnamefont{Schuster}}
(\bibinfo{publisher}{Wiley-VCH GmbH \& Co., Weinheim}), p.
\bibinfo{pages}{85}.


\bibitem[{\citenamefont{Cohen} \emph{et~al.}(2003b)\citenamefont{Cohen, Havlin, and ben-Avraham}}]
{Cohen:chb03b}
\bibinfo{author}{\bibnamefont{Cohen},~\bibfnamefont{R.}},
\bibinfo{author}{\bibfnamefont{S.}~\bibnamefont{Havlin}},
and
\bibinfo{author}{\bibfnamefont{D.}~\bibnamefont{ben-Avraham}},
\bibinfo{year}{2003b},
\bibinfo{title}{``Efficient immunization strategies for computer networks and populations},''
\bibinfo{journal}{Phys. Rev. Lett.} \textbf{\bibinfo{volume}{91}},
\bibinfo{pages}{247901}.






\bibitem[{\citenamefont{Colizza} \emph{et~al.}(2007)\citenamefont{Colizza, Barrat, Barthelemy, Valleron, and Vespignani}}]
{Colizza:cbb07}
\bibinfo{author}{\bibnamefont{Colizza},~\bibfnamefont{V.}},
\bibinfo{author}{\bibfnamefont{A.}~\bibnamefont{Barrat}},
\bibinfo{author}{\bibfnamefont{M.}~\bibnamefont{Barthelemy}},
\bibinfo{author}{\bibfnamefont{A.-J.}~\bibnamefont{Valleron}},
and
\bibinfo{author}{\bibfnamefont{A.}~\bibnamefont{Vespignani}},
\bibinfo{year}{2007},
\bibinfo{title}{``Modeling the worldwide spread of pandemic influenza: Baseline case and containment interventions},''
\bibinfo{journal}{PLoS Med.} \textbf{\bibinfo{volume}{4}},
\bibinfo{pages}{e13}.


\bibitem[{\citenamefont{Colizza} \emph{et~al.}(2006)\citenamefont{Colizza, Flammini, Serrano and Vespignani}}]
{Colizza:cfsv06}
\bibinfo{author}{\bibnamefont{Colizza},~\bibfnamefont{V.}},
\bibinfo{author}{\bibfnamefont{A.}~\bibnamefont{Flammini}},
\bibinfo{author}{\bibfnamefont{M.~A.}~\bibnamefont{Serrano}},
and
\bibinfo{author}{\bibfnamefont{A.}~\bibnamefont{Vespignani}},
\bibinfo{year}{2006},
\bibinfo{title}{``Detecting rich-club ordering in complex networks},''
\bibinfo{journal}{Nature Physics} \textbf{\bibinfo{volume}{2}},
\bibinfo{pages}{110}.


\bibitem[{\citenamefont{Colizza} \emph{et~al.}(2007)\citenamefont{Colizza, Pastor-Satorras, and Vespignani}}]
{Colizza:gpv07}
\bibinfo{author}{\bibnamefont{Colizza},~\bibfnamefont{V.}},
\bibinfo{author}{\bibfnamefont{R.}~\bibnamefont{Pastor-Satorras}},
and
\bibinfo{author}{\bibfnamefont{A.}~\bibnamefont{Vespignani}},
\bibinfo{year}{2007},
\bibinfo{title}{``Reaction-diffusion processes and metapopulation models in heterogeneous networks},''
\bibinfo{journal}{Nature Physics} \textbf{\bibinfo{volume}{3}},
\bibinfo{pages}{276}.


\bibitem[{\citenamefont{Coolen} \emph{et~al.}(2005)\citenamefont{Coolen, Skantzos,
Castillo, Vicente, Hatchett, Wemmenhove, and Nikoletopoulos}}]
{Coolen:csc05}
\bibinfo{author}{\bibnamefont{Coolen},~\bibfnamefont{A.~C.~C.}},
\bibinfo{author}{\bibfnamefont{N.~S.}~\bibnamefont{Skantzos}},
\bibinfo{author}{\bibfnamefont{I.~P.}~\bibnamefont{Castillo}},
\bibinfo{author}{\bibfnamefont{C.~J.~P.}~\bibnamefont{Vicente}},
\bibinfo{author}{\bibfnamefont{J.~P.~L.}~\bibnamefont{Hatchett}},
\bibinfo{author}{\bibfnamefont{B.}~\bibnamefont{Wemmenhove}},
and
\bibinfo{author}{\bibfnamefont{T.}~\bibnamefont{Nikoletopoulos}},
\bibinfo{year}{2005},
\bibinfo{title}{Finitely connected vector spin systems with random matrix interactions},
\bibinfo{journal}{J. Phys. A} \textbf{\bibinfo{volume}{38}},
\bibinfo{pages}{8289}.


\bibitem[{\citenamefont{Copelli and Campos}(2007)}]
{Copelli:cc07}
\bibinfo{author}{\bibnamefont{Copelli},~\bibfnamefont{M.}}
and
\bibinfo{author}{\bibfnamefont{P.~R.~A.}~\bibnamefont{Campos}},
\bibinfo{year}{2007},
\bibinfo{title}{``Excitable scale free networks},''
\bibinfo{journal}{Eur. Phys. J. B} \textbf{\bibinfo{volume}{56}},
\bibinfo{pages}{273}.


\bibitem[{\citenamefont{Coppersmith} \emph{et~al.}(2004)\citenamefont{Coppersmith, Gamarnik, Hajiaghayi,
and Sorkin}}]
{Coppersmith:cghs04}
\bibinfo{author}{\bibnamefont{Coppersmith},~\bibfnamefont{D.}},
\bibinfo{author}{\bibfnamefont{D.}~\bibnamefont{Gamarnik}},
\bibinfo{author}{\bibfnamefont{M.~T.}~\bibnamefont{Hajiaghayi}},
and
\bibinfo{author}{\bibfnamefont{G.~B.}~\bibnamefont{Sorkin}},
\bibinfo{year}{2004},
\bibinfo{title}{``Random MAX SAT, random MAX CUT, and their phase transitions},''
\bibinfo{journal}{Random Structures and Algorithms} \textbf{\bibinfo{volume}{24}},
\bibinfo{pages}{502}.


\bibitem[{\citenamefont{Costin and Costin}(1991)}]
{Costin:cc91}
\bibinfo{author}{\bibnamefont{Costin},~\bibfnamefont{O.}}
and
\bibinfo{author}{\bibfnamefont{R.~D.}~\bibnamefont{Costin}},
\bibinfo{year}{1991},
\bibinfo{title}{``Limit probability distributions for an infinite-order phase transition model},''
\bibinfo{journal}{J. Stat. Phys.} \textbf{\bibinfo{volume}{64}},
\bibinfo{pages}{193}.


\bibitem[{\citenamefont{Costin} \emph{et~al.}(1990)\citenamefont{Costin, Costin, and Gr\"unfeld}}]
{Costin:ccd90}
\bibinfo{author}{\bibnamefont{Costin},~\bibfnamefont{O.}},
\bibinfo{author}{\bibfnamefont{R.~D.}~\bibnamefont{Costin}},
and
\bibinfo{author}{\bibfnamefont{C.~P.}~\bibnamefont{Gr\"unfeld}},
\bibinfo{year}{1990},
\bibinfo{title}{``Infinite-order phase transition in a classical spin system},''
\bibinfo{journal}{J. Stat. Phys.} \textbf{\bibinfo{volume}{59}},
\bibinfo{pages}{1531}.


\bibitem[{\citenamefont{Coulomb and Bauer}(2003)}]
{Coulomb:cb03}
\bibinfo{author}{\bibnamefont{Coulomb},~\bibfnamefont{M.}}
and
\bibinfo{author}{\bibfnamefont{S.}~\bibnamefont{Bauer}},
\bibinfo{year}{2003},
\bibinfo{title}{`` Asymmetric evolving random networks},''
\bibinfo{journal}{Eur. Phys. J. B} \textbf{\bibinfo{volume}{35}},
\bibinfo{pages}{377}.





\bibitem[{\citenamefont{Danila} \emph{et~al.}(2006)\citenamefont{Danila, Yu, Earl, Marsh, Toroczkai, and Bassler}}]
{Danila:dye06}
\bibinfo{author}{\bibnamefont{Danila},~\bibfnamefont{B.}},
\bibinfo{author}{\bibfnamefont{Y.}~\bibnamefont{Yu}},
\bibinfo{author}{\bibfnamefont{S.}~\bibnamefont{Earl}},
\bibinfo{author}{\bibfnamefont{J.~A.}~\bibnamefont{Marsh}},
\bibinfo{author}{\bibfnamefont{Z.}~\bibnamefont{Toroczkai}},
and
\bibinfo{author}{\bibfnamefont{K.~E.}~\bibnamefont{Bassler}},
\bibinfo{year}{2006},
\bibinfo{title}{``Congestion-gradient driven transport on complex networks},''
\bibinfo{journal}{Phys. Rev. E} \textbf{\bibinfo{volume}{74}},
\bibinfo{pages}{046114}.


\bibitem[{\citenamefont{Denker} \emph{et~al.}(2004)\citenamefont{Denker, Timme, Diesmann, Wolf, and Geisel}}]
{Denker:dtdwg04}
\bibinfo{author}{\bibnamefont{Denker},~\bibfnamefont{M.}},
\bibinfo{author}{\bibfnamefont{M.}~\bibnamefont{Timme}},
\bibinfo{author}{\bibfnamefont{M.}~\bibnamefont{Diesmann}},
\bibinfo{author}{\bibfnamefont{F.}~\bibnamefont{Wolf}},
and
\bibinfo{author}{\bibfnamefont{T.}~\bibnamefont{Geisel}},
\bibinfo{year}{2004},
\bibinfo{title}{``Breaking synchrony by heterogeneity in complex networks},''
\bibinfo{journal}{Phys. Rev. Lett.} \textbf{\bibinfo{volume}{92}},
\bibinfo{pages}{074103}.


\bibitem[{\citenamefont{Der\'enyi} \emph{et~al.}(2004)\citenamefont{Der\'enyi, Farkas, Palla, and Vicsek}}]
{Derenyi:dfp04}
\bibinfo{author}{\bibnamefont{Der\'enyi},~\bibfnamefont{I.}},
\bibinfo{author}{\bibfnamefont{I.}~\bibnamefont{Farkas}},
\bibinfo{author}{\bibfnamefont{G.}~\bibnamefont{Palla}},
and
\bibinfo{author}{\bibfnamefont{T.}~\bibnamefont{Vicsek}},
\bibinfo{year}{2004},
\bibinfo{title}{``Topological phase transitions of random networks},''
\bibinfo{journal}{Physica A} \textbf{\bibinfo{volume}{334}},
\bibinfo{pages}{583}.


\bibitem[{\citenamefont{Der\'{e}nyi} \emph{et~al.}(2005)\citenamefont{Der\'{e}nyi, Palla, and Vicsek}}]
{Derenyi:dpv05}
\bibinfo{author}{\bibnamefont{Der\'{e}nyi},~\bibfnamefont{I.}},
\bibinfo{author}{\bibfnamefont{G.}~\bibnamefont{Palla}},
and
\bibinfo{author}{\bibfnamefont{T.}~\bibnamefont{Vicsek}},
\bibinfo{year}{2005},
\bibinfo{title}{``Clique percolation in random networks},''
\bibinfo{journal}{Phys. Rev. Lett.} \textbf{\bibinfo{volume}{94}},
\bibinfo{pages}{160202}.


\bibitem[{\citenamefont{Derrida}(1981)}]
{Derrida:d81}
\bibinfo{author}{\bibnamefont{Derrida},~\bibfnamefont{B.}}
\bibinfo{year}{1981},
\bibinfo{title}{``Random-energy model: An exactly solvable model of disordered
systems},''
\bibinfo{journal}{Phys. Rev. B} \textbf{\bibinfo{volume}{24}},
\bibinfo{pages}{2613}.






\bibitem[{\citenamefont{Dezs\~o and Barab\'asi}(2002)}]
{Dezso:db02}
\bibinfo{author}{\bibnamefont{Dezs\~o},~\bibfnamefont{Z.}}
and
\bibinfo{author}{\bibfnamefont{A.-L.}~\bibnamefont{Barab\'asi}},
\bibinfo{year}{2002},
\bibinfo{title}{``Halting viruses in scale-free networks},''
\bibinfo{journal}{Phys. Rev. E} \textbf{\bibinfo{volume}{65}},
\bibinfo{pages}{055103}.


\bibitem[{\citenamefont{Dhar} \emph{et~al.}(1997)\citenamefont{Dhar, Shukla, and Sethnal}}]
{Dhar:dss97}
\bibinfo{author}{\bibnamefont{Dhar},~\bibfnamefont{D.}},
\bibinfo{author}{\bibfnamefont{P.}~\bibnamefont{Shukla}},
and
\bibinfo{author}{\bibfnamefont{J.~P.}~\bibnamefont{Sethnal}},
\bibinfo{year}{1997},
\bibinfo{title}{``Zero-temperature gysteresis in the random field Ising
model on a Bethe lattice},''
\bibinfo{journal}{J. Phys. A} \textbf{\bibinfo{volume}{30}},
\bibinfo{pages}{5259}.

\bibitem[{\citenamefont{Domb}(1960)}]
{Domb:domb60}
\bibinfo{author}{\bibnamefont{Domb}, \bibfnamefont{C.}},
\bibinfo{year}{1960},
\bibinfo{title}{``On the theory of cooperatve phenomena in crystals},''
\bibinfo{journal}{Adv. Phys.} \textbf{\bibinfo{volume}{9}},
\bibinfo{pages}{245}.





\bibitem[{\citenamefont{Dorogovtsev}(2003)}]
{Dorogovtsev:d03}
\bibinfo{author}{\bibnamefont{Dorogovtsev},~\bibfnamefont{S.~N.}},
\bibinfo{year}{2003},
\bibinfo{title}{``Renormalization group for evolving networks},''
\bibinfo{journal}{Phys. Rev. E} \textbf{\bibinfo{volume}{67}},
\bibinfo{pages}{045102 (R)}.


\bibitem[{\citenamefont{Dorogovtsev} \emph{et~al.}(2002a)\citenamefont{Dorogovtsev, Goltsev, and Mendes}}]
{Dorogovtsev:dgm02a}
\bibinfo{author}{\bibnamefont{Dorogovtsev},~\bibfnamefont{S.~N.}},
\bibinfo{author}{\bibfnamefont{A.~V.}~\bibnamefont{Goltsev}},
and
\bibinfo{author}{\bibfnamefont{J.~F.~F.}~\bibnamefont{Mendes}},
\bibinfo{year}{2002a},
\bibinfo{title}{``Pseudofractal scale-free web},''
\bibinfo{journal}{Phys. Rev. E} \textbf{\bibinfo{volume}{65}},
\bibinfo{pages}{066122}.


\bibitem[{\citenamefont{Dorogovtsev} \emph{et~al.}(2002b)\citenamefont{Dorogovtsev, Goltsev, and Mendes}}]
{Dorogovtsev:dgm02b}
\bibinfo{author}{\bibnamefont{Dorogovtsev},~\bibfnamefont{S.~N.}},
\bibinfo{author}{\bibfnamefont{A.~V.}~\bibnamefont{Goltsev}},
and
\bibinfo{author}{\bibfnamefont{J.~F.~F.}~\bibnamefont{Mendes}},
\bibinfo{year}{2002b},
\bibinfo{title}{``Ising model on networks with an arbitrary distribution of connections},''
\bibinfo{journal}{Phys. Rev. E} \textbf{\bibinfo{volume}{66}},
\bibinfo{pages}{016104}.


\bibitem[{\citenamefont{Dorogovtsev} \emph{et~al.}(2004)\citenamefont{Dorogovtsev, Goltsev, and Mendes}}]
{Dorogovtsev:dgm04}
\bibinfo{author}{\bibnamefont{Dorogovtsev},~\bibfnamefont{S.~N.}},
\bibinfo{author}{\bibfnamefont{A.~V.}~\bibnamefont{Goltsev}},
and
\bibinfo{author}{\bibfnamefont{J.~F.~F.}~\bibnamefont{Mendes}},
\bibinfo{year}{2004},
\bibinfo{title}{``Potts model on complex networks},''
\bibinfo{journal}{Eur. Phys. J. B} \textbf{\bibinfo{volume}{38}},
\bibinfo{pages}{177}.


\bibitem[{\citenamefont{Dorogovtsev} \emph{et~al.}(2005)\citenamefont{ Dorogovtsev, Goltsev, and Mendes}}]
{Dorogovtsev:dgm05}
\bibinfo{author}{\bibnamefont{Dorogovtsev},~\bibfnamefont{S.~N.}},
\bibinfo{author}{\bibfnamefont{A.~V.}~\bibnamefont{Goltsev}},
and
\bibinfo{author}{\bibfnamefont{J.~F.~F.}~\bibnamefont{Mendes}},
\bibinfo{year}{2005},
\bibinfo{title}{``Correlations in interacting systems with a network topology},''
\bibinfo{journal}{Phys. Rev. E} \textbf{\bibinfo{volume}{72}},
\bibinfo{pages}{066130}.


\bibitem[{\citenamefont{Dorogovtsev} \emph{et~al.}(2006a)\citenamefont{Dorogovtsev, Goltsev, and Mendes}}]
{Dorogovtsev:dgm06a}
\bibinfo{author}{\bibnamefont{Dorogovtsev},~\bibfnamefont{S.~N.}},
\bibinfo{author}{\bibfnamefont{A.~V.}~\bibnamefont{Goltsev}},
and
\bibinfo{author}{\bibfnamefont{J.~F.~F.}~\bibnamefont{Mendes}},
\bibinfo{year}{2006a},
\bibinfo{title}{``$k$-core organization of complex networks},''
\bibinfo{journal}{Phys. Rev. Lett.} \textbf{\bibinfo{volume}{96}},
\bibinfo{pages}{040601}.


\bibitem[{\citenamefont{Dorogovtsev} \emph{et~al.}(2006b)\citenamefont{Dorogovtsev, Goltsev, and Mendes}}]
{Dorogovtsev:dgm06b}
\bibinfo{author}{\bibnamefont{Dorogovtsev},~\bibfnamefont{S.~N.}},
\bibinfo{author}{\bibfnamefont{A.~V.}~\bibnamefont{Goltsev}},
and
\bibinfo{author}{\bibfnamefont{J.~F.~F.}~\bibnamefont{Mendes}},
\bibinfo{year}{2006b},
\bibinfo{title}{``$k$-core architecture and $k$-core percolation on complex networks},''
\bibinfo{journal}{Physica D} \textbf{\bibinfo{volume}{224}},
\bibinfo{pages}{7}.


\bibitem[{\citenamefont{Dorogovtsev} \emph{et~al.}(2003)\citenamefont{Dorogovtsev, Goltsev, Mendes, and Samukhin}}]
{Dorogovtsev:dgms03}
\bibinfo{author}{\bibnamefont{Dorogovtsev},~\bibfnamefont{S.~N.}},
\bibinfo{author}{\bibfnamefont{A.~V.}~\bibnamefont{Goltsev}},
\bibinfo{author}{\bibfnamefont{J.~F.~F.}~\bibnamefont{Mendes}},
and
\bibinfo{author}{\bibfnamefont{S.~N.}~\bibnamefont{Samukhin}},
\bibinfo{year}{2003},
\bibinfo{title}{``Spectra of complex networks},''
\bibinfo{journal}{Phys. Rev. E} \textbf{\bibinfo{volume}{68}},
\bibinfo{pages}{046109}.


\bibitem[{\citenamefont{Dorogovtsev}
\emph{et~al.}(2007)\citenamefont{Dorogovtsev, Krapivsky, and Mendes}}]
{Dorogovtsev:dkm07}
\bibinfo{author}{\bibnamefont{Dorogovtsev},~\bibfnamefont{S.~N.}},
\bibinfo{author}{\bibfnamefont{P.~L.}~\bibnamefont{Krapivsky}}, 
and
\bibinfo{author}{\bibfnamefont{J.~F.~F.}~\bibnamefont{Mendes}},
\bibinfo{year}{2007},
\bibinfo{title}{``Transition from a small to a large world in growing networks},'' 
\eprint{arXiv:0709.3094}. 


\bibitem[{\citenamefont{Dorogovtsev, and Mendes}(2001)}]
{Dorogovtsev:dm01}
\bibinfo{author}{\bibnamefont{Dorogovtsev},~\bibfnamefont{S.~N.}}
and
\bibinfo{author}{\bibfnamefont{J.~F.~F.}~\bibnamefont{Mendes}},
\bibinfo{year}{2001},
\bibinfo{title}{``Scaling properties of scale-free evolving networks: Continuous approach},''
\bibinfo{journal}{Phys. Rev. E} \textbf{\bibinfo{volume}{63}},
\bibinfo{pages}{056125}.


\bibitem[{\citenamefont{Dorogovtsev and Mendes}(2002)}]
{Dorogovtsev:dm02}
\bibinfo{author}{\bibnamefont{Dorogovtsev},~\bibfnamefont{S.~N.}}
and
\bibinfo{author}{\bibfnamefont{J.~F.~F.}~\bibnamefont{Mendes}},
\bibinfo{year}{2002},
\bibinfo{title}{``Evolution of networks},''
\bibinfo{journal}{Adv. Phys.} \textbf{\bibinfo{volume}{51}},
\bibinfo{pages}{1079}.


\bibitem[{\citenamefont{Dorogovtsev and Mendes}(2003)}]
{Dorogovtsev:dmbook03}
\bibinfo{author}{\bibnamefont{Dorogovtsev},~\bibfnamefont{S.~N.}},
and
\bibinfo{author}{\bibfnamefont{J.~F.~F.}~\bibnamefont{Mendes}},
\bibinfo{year}{2003},
\emph{\bibinfo{title}{Evolution of Networks: From Biological Nets to
the Internet and WWW}} (\bibinfo{publisher}{Oxford University Press,
Oxford}).


\bibitem[{\citenamefont{Dorogovtsev} \emph{et~al.}(2005)\citenamefont{Dorogovtsev, Mendes, Povolotsky, and Samukhin}}]
{Dorogovtsev:dmp05}
\bibinfo{author}{\bibnamefont{Dorogovtsev},~\bibfnamefont{S.~N.}},
\bibinfo{author}{\bibfnamefont{J.~F.~F.}~\bibnamefont{Mendes}},
\bibinfo{author}{\bibfnamefont{A.~M.}~\bibnamefont{Povolotsky}},
and
\bibinfo{author}{\bibfnamefont{A.~N.}~\bibnamefont{Samukhin}},
\bibinfo{year}{2005},
\bibinfo{title}{``Organization of complex networks without multiple connections},''
\bibinfo{journal}{Phys. Rev. Lett.} \textbf{\bibinfo{volume}{95}},
\bibinfo{pages}{195701}.


\bibitem[{\citenamefont{Dorogovtsev} \emph{et~al.}(2000)\citenamefont{Dorogovtsev, Mendes, and Samukhin}}]
{Dorogovtsev:dms00}
\bibinfo{author}{\bibnamefont{Dorogovtsev},~\bibfnamefont{S.~N.}},
\bibinfo{author}{\bibfnamefont{J.~F.~F.}~\bibnamefont{Mendes}},
and
\bibinfo{author}{\bibfnamefont{A.~N.}~\bibnamefont{Samukhin}},
\bibinfo{year}{2000},
\bibinfo{title}{``Exact solution of the Barab\'asi-Albert model},''
\bibinfo{journal}{Phys. Rev. Lett.} \textbf{\bibinfo{volume}{85}},
\bibinfo{pages}{4633}.


\bibitem[{\citenamefont{Dorogovtsev} \emph{et~al.}(2001a)\citenamefont{Dorogovtsev, Mendes, and Samukhin}}]
{Dorogovtsev:dms01a}
\bibinfo{author}{\bibnamefont{Dorogovtsev},~\bibfnamefont{S.~N.}},
\bibinfo{author}{\bibfnamefont{J.~F.~F.}~\bibnamefont{Mendes}},
and
\bibinfo{author}{\bibfnamefont{A.~N.}~\bibnamefont{Samukhin}},
\bibinfo{year}{2001a},
\bibinfo{title}{``Giant strongly connected component of directed networks},''
\bibinfo{journal}{Phys. Rev. E} \textbf{\bibinfo{volume}{64}},
\bibinfo{pages}{025101 (R)}.


\bibitem[{\citenamefont{Dorogovtsev} \emph{et~al.}(2001b)\citenamefont{Dorogovtsev, Mendes, and Samukhin}}]
{Dorogovtsev:dms01b}
\bibinfo{author}{\bibnamefont{Dorogovtsev},~\bibfnamefont{S.~N.}},
\bibinfo{author}{\bibfnamefont{J.~F.~F.}~\bibnamefont{Mendes}},
and
\bibinfo{author}{\bibfnamefont{A.~N.}~\bibnamefont{Samukhin}},
\bibinfo{year}{2001b},
\bibinfo{title}{``Anomalous percolation properties of growing networks},''
\bibinfo{journal}{Phys. Rev. E} \textbf{\bibinfo{volume}{64}},
\bibinfo{pages}{066110}.


\bibitem[{\citenamefont{Dorogovtsev} \emph{et~al.}(2001c)\citenamefont{Dorogovtsev, Mendes, and Samukhin}}]
{Dorogovtsev:dms01c}
\bibinfo{author}{\bibnamefont{Dorogovtsev},~\bibfnamefont{S.~N.}},
\bibinfo{author}{\bibfnamefont{J.~F.~F.}~\bibnamefont{Mendes}},
and
\bibinfo{author}{\bibfnamefont{A.~N.}~\bibnamefont{Samukhin}},
\bibinfo{year}{2001c},
\bibinfo{title}{``Size-dependent degree distribution of a scale-free growing network},''
\bibinfo{journal}{Phys. Rev. E} \textbf{\bibinfo{volume}{63}},
\bibinfo{pages}{062101}.


\bibitem[{\citenamefont{Dorogovtsev} \emph{et~al.}(2003a)\citenamefont{Dorogovtsev, Mendes, and Samukhin}}]
{Dorogovtsev:dms03a}
\bibinfo{author}{\bibnamefont{Dorogovtsev},~\bibfnamefont{S.~N.}},
\bibinfo{author}{\bibfnamefont{J.~F.~F.}~\bibnamefont{Mendes}},
and
\bibinfo{author}{\bibfnamefont{A.~N.}~\bibnamefont{Samukhin}},
\bibinfo{year}{2003a},
\bibinfo{title}{``Metric structure of random networks},''
\bibinfo{journal}{Nucl. Phys. B} \textbf{\bibinfo{volume}{653}},
\bibinfo{pages}{307}.


\bibitem[{\citenamefont{Dorogovtsev} \emph{et~al.}(2003b)\citenamefont{Dorogovtsev, Mendes, and Samukhin}}]
{Dorogovtsev:dms03b}
\bibinfo{author}{\bibnamefont{Dorogovtsev},~\bibfnamefont{S.~N.}},
\bibinfo{author}{\bibfnamefont{J.~F.~F.}~\bibnamefont{Mendes}},
and
\bibinfo{author}{\bibfnamefont{A. ~N.}~\bibnamefont{Samukhin}},
\bibinfo{year}{2003b},
\bibinfo{title}{``Principles of statistical mechanics of random networks},''
\bibinfo{journal}{Nucl. Phys. B} \textbf{\bibinfo{volume}{666}},
\bibinfo{pages}{396}.


\bibitem[{\citenamefont{Doye and Massen}(2005)}]
{Doye:dm05}
\bibinfo{author}{\bibnamefont{Doye},~\bibfnamefont{J.~P.~K.}}
and
\bibinfo{author}{\bibfnamefont{C.~P.}~\bibnamefont{Massen}}, \bibinfo{year}{2005},
\bibinfo{title}{``Self-similar disk packings as model spatial scale-free networks},''
\bibinfo{journal}{Phys. Rev. E} \textbf{\bibinfo{volume}{71}},
\bibinfo{pages}{016128}.


\bibitem[{\citenamefont{Durrett}(2006)}]
{Durrett:dbook06}
\bibinfo{author}{\bibnamefont{Durrett},~\bibfnamefont{R.}},
\bibinfo{year}{2006},
\emph{\bibinfo{title}{Random Graph Dynamics}}
(\bibinfo{publisher}{Cambridge University Press, Cambridge}).





\bibitem[{\citenamefont{Echenique} \emph{et~al.}(2005)\citenamefont{Echenique, G\'omez-Garde\~nes, and Moreno}}]
{Echenique:egm05}
\bibinfo{author}{\bibnamefont{Echenique},~\bibfnamefont{P.}},
\bibinfo{author}{\bibfnamefont{J.}~\bibnamefont{G\'omez-Garde\~nes}},
and
\bibinfo{author}{\bibfnamefont{Y.}~\bibnamefont{Moreno}},
\bibinfo{year}{2005},
\bibinfo{title}{``Dynamics of jamming transitions in complex networks},''
\bibinfo{journal}{Europhys. Lett.} \textbf{\bibinfo{volume}{71}},
\bibinfo{pages}{325}.


\bibitem[{\citenamefont{Eggarter}(1974)}]
{Eggarter:e74}
\bibinfo{author}{\bibnamefont{Eggarter},~\bibfnamefont{T. P.}},
\bibinfo{year}{1974},
\bibinfo{title}{``Cayley trees, the Ising problem, and the thermodynamic limit},''
\bibinfo{journal}{Phys. Rev. B} \textbf{\bibinfo{volume}{9}},
\bibinfo{pages}{2989}.


\bibitem[{\citenamefont{Ehrhardt and Marsili}(2005)}]
{Ehrhardt:em05}
\bibinfo{author}{\bibnamefont{Ehrhardt},~\bibfnamefont{G.~C.~M.~A.}}
and
\bibinfo{author}{\bibfnamefont{M.}~\bibnamefont{Marsili}},
\bibinfo{year}{2005},
\bibinfo{title}{``Potts model on random trees},''
\bibinfo{journal}{J. Stat. Mech.: Theor. Exp.}
\bibinfo{pages}{P02006}.


\bibitem[{\citenamefont{Ehrhardt} \emph{et~al.}(2006)\citenamefont{Ehrhardt, Marsili, and Vega-Redondo}}]
{Ehrhardt:emv06}
\bibinfo{author}{\bibnamefont{Ehrhardt},~\bibfnamefont{G.~C.~M.~A.}},
\bibinfo{author}{\bibfnamefont{M.}~\bibnamefont{Marsili}}, 
and
\bibinfo{author}{\bibfnamefont{F.}~\bibnamefont{Vega-Redondo}},
\bibinfo{year}{2006},
\bibinfo{title}{``Phenomenological models of socioeconomic network dynamics},''
\bibinfo{journal}{Phys. Rev. E} \textbf{\bibinfo{volume}{74}},
\bibinfo{pages}{036106}.




\bibitem[{\citenamefont{Erd\H os and R\'enyi}(1959)}]
{Erdos:er59}
\bibinfo{author}{\bibnamefont{Erd\H os},~\bibfnamefont{P.}}
and
\bibinfo{author}{\bibfnamefont{A.}~\bibnamefont{R\'enyi}},
\bibinfo{year}{1959},
\bibinfo{title}{``On random graphs},''
\bibinfo{journal}{Publicationes Mathematicae Debrencen} \textbf{\bibinfo{volume}{6}},
\bibinfo{pages}{290}.

\bibitem[{\citenamefont{Evans and Hanney}(2005)}]
{Evans:eh05}
\bibinfo{author}{\bibnamefont{Evans},~\bibfnamefont{M.~R.}}
and
\bibinfo{author}{\bibfnamefont{T.}~\bibnamefont{Hanney}},
\bibinfo{year}{2005},
\bibinfo{title}{``Nonequilibrium statistical mechanics of the zero-range process and related models},''
\bibinfo{journal}{J. Phys. A} \textbf{\bibinfo{volume}{38}},
\bibinfo{pages}{R195}.




\bibitem[{\citenamefont{Falk}(1975)}]
{Falk:f75}
\bibinfo{author}{\bibnamefont{Falk},~\bibfnamefont{H.}}
\bibinfo{year}{1975},
\bibinfo{title}{``Ising spin system on a Cayley tree: Correlation decomposition and phase transition},''
\bibinfo{journal}{Phys. Rev. B} \textbf{\bibinfo{volume}{12}},
\bibinfo{pages}{5184}.


\bibitem[{\citenamefont{Farkas} \emph{et~al.}(2004)\citenamefont{Farkas, Derenyi, Palla, and Vicsek}}]
{Farkas:fdp04}
\bibinfo{author}{\bibnamefont{Farkas},~\bibfnamefont{I.}},
\bibinfo{author}{\bibfnamefont{I.}~\bibnamefont{Derenyi}},
\bibinfo{author}{\bibfnamefont{G.}~\bibnamefont{Palla}},
and
\bibinfo{author}{\bibfnamefont{T.}~\bibnamefont{Vicsek}},
\bibinfo{year}{2004},
\bibinfo{title}{``Equilibrium statistical mechanics of network structures},''
\bibinfo{journal}{Lect. Notes Phys.} \textbf{\bibinfo{volume}{650}},
\bibinfo{pages}{163}.




\bibitem[{\citenamefont{Fernholz and Ramachandran}(2004)}]
{Fernholz:fr04}
\bibinfo{author}{\bibnamefont{Fernholz},~\bibfnamefont{D.}}
and
\bibinfo{author}{\bibfnamefont{V.}~\bibnamefont{Ramachandran}},
\bibinfo{year}{2004},
\bibinfo{title}{``Cores and connectivity in sparse random graphs},''
\bibinfo{journal}{UTCS Technical Report TR04-13}.


\bibitem[{\citenamefont{Fortuin and Kasteleyn}(1972)}]
{Fortuin:fk72}
\bibinfo{author}{\bibnamefont{Fortuin},~\bibfnamefont{C.~M.}}
and
\bibinfo{author}{\bibfnamefont{P.~W.}~\bibnamefont{Kasteleyn}},
\bibinfo{year}{1972},
\bibinfo{title}{``On the random-cluster model: I. Introduction and relation to other models},''
\bibinfo{journal}{Physica} \textbf{\bibinfo{volume}{57}},
\bibinfo{pages}{536}.


\bibitem[{\citenamefont{Freeman} \emph{et~al.}(2000)\citenamefont{Freeman, Pasztor, and Carmichael}}]
{Freeman:fpc00}
\bibinfo{author}{\bibnamefont{Freeman},~\bibfnamefont{W.~T.}},
\bibinfo{author}{\bibfnamefont{E.~C.}~\bibnamefont{Pasztor}},
and
\bibinfo{author}{\bibfnamefont{O.~T.}~\bibnamefont{Carmichael}},
\bibinfo{year}{2000},
\bibinfo{title}{``Learning low-level vision},''
\bibinfo{journal}{Int. J. Comput. Vis.} \textbf{\bibinfo{volume}{40}},
\bibinfo{pages}{25}.


\bibitem[{\citenamefont{Frey}(1998)}]
{Frey:98}
\bibinfo{author}{\bibnamefont{Frey}, \bibfnamefont{B.~J.}},
\bibinfo{year}{1998},
\emph{\bibinfo{title}{Graphical models for machine learning and
digital communication}}
(\bibinfo{publisher}{MIT Press, Cambridge}).


\bibitem[{\citenamefont{Frieze}(1990)}]
{Frieze:f90}
\bibinfo{author}{\bibnamefont{Frieze},~\bibfnamefont{A.~M.}}
\bibinfo{year}{1990},
\bibinfo{title}{``On the independence number of random graphs},''
\bibinfo{journal}{Discr. Math.} \textbf{\bibinfo{volume}{81}},
\bibinfo{pages}{171}. 


\bibitem[{\citenamefont{Fronczak and Fronczak}(2007)}]
{Fronczak:ff07}
\bibinfo{author}{\bibnamefont{Fronczak},~\bibfnamefont{A.}} 
and
\bibinfo{author}{\bibfnamefont{P.}~\bibnamefont{Fronczak}},
\bibinfo{year}{2007},
\bibinfo{title}{``Biased random walks on complex networks: the role of local navigation rules},''
\eprint{arXiv:0708.4404 [math]}





\bibitem[{\citenamefont{Gade and Hu}(2000)}]
{Gade:gh00}
\bibinfo{author}{\bibnamefont{Gade},~\bibfnamefont{P.~M.}}
and
\bibinfo{author}{\bibfnamefont{C.-K.}~\bibnamefont{Hu}},
\bibinfo{year}{2000},
\bibinfo{title}{``Synchronous chaos in coupled map lattices with small-world
interactions},''
\bibinfo{journal}{Phys. Rev. E} \textbf{\bibinfo{volume}{62}},
\bibinfo{pages}{6409}.






\bibitem[{\citenamefont{Galla}
\emph{et~al.}(2006)\citenamefont{Galla, Leone, Marsili, Sellitto, Weigt, and Zecchina}}]
{Galla:glm06}
\bibinfo{author}{\bibnamefont{Galla},~\bibfnamefont{T.}},
\bibinfo{author}{\bibfnamefont{M.}~\bibnamefont{Leone}},
\bibinfo{author}{\bibfnamefont{M.}~\bibnamefont{Marsili}},
\bibinfo{author}{\bibfnamefont{M.}~\bibnamefont{Sellitto}},
\bibinfo{author}{\bibfnamefont{M.}~\bibnamefont{Weigt}},
and
\bibinfo{author}{\bibfnamefont{R.}~\bibnamefont{Zecchina}},
\bibinfo{year}{2006},
\bibinfo{title}{``Statistical mechanics of combinatorial auctions},''
\bibinfo{journal}{Phys. Rev. Lett.} \textbf{\bibinfo{volume}{97}},
\bibinfo{pages}{128701}.




\bibitem[{\citenamefont{Gallos and Argyrakis}(2004)}]
{Gallos:ga04}
\bibinfo{author}{\bibnamefont{Gallos},~\bibfnamefont{L.~K.}}
and
\bibinfo{author}{\bibfnamefont{P.}~\bibnamefont{Argyrakis}},
\bibinfo{year}{2004},
\bibinfo{title}{``Absence of kinetic effects in reaction-diffusion processes in scale-free networks},''
\bibinfo{journal}{Phys. Rev. Lett.} \textbf{\bibinfo{volume}{92}},
\bibinfo{pages}{138301}.


\bibitem[{\citenamefont{Gallos} \emph{et~al.}(2005)\citenamefont{Gallos, Cohen, Argyrakis, Bunde, and Havlin}}]
{Gallos:gca05}
\bibinfo{author}{\bibnamefont{Gallos},~\bibfnamefont{L.~K.}},
\bibinfo{author}{\bibfnamefont{R.}~\bibnamefont{Cohen}},
\bibinfo{author}{\bibfnamefont{P.}~\bibnamefont{Argyrakis}},
\bibinfo{author}{\bibfnamefont{A.}~\bibnamefont{Bunde}},
and
\bibinfo{author}{\bibfnamefont{S.}~\bibnamefont{Havlin}},
\bibinfo{year}{2005},
\bibinfo{title}{``Stability and topology of scale-free networks under attack and defense strategies},''
\bibinfo{journal}{Phys. Rev. Lett.} \textbf{\bibinfo{volume}{94}},
\bibinfo{pages}{188701}.


\bibitem[{\citenamefont{Gallos} \emph{et~al.}(2007a)\citenamefont{Gallos, Liljeros, Argyrakis, Bunde, and Havlin}}]
{Gallos:gla07a}
\bibinfo{author}{\bibnamefont{Gallos},~\bibfnamefont{L.~K.}},
\bibinfo{author}{\bibfnamefont{F.}~\bibnamefont{Liljeros}},
\bibinfo{author}{\bibfnamefont{P.}~\bibnamefont{Argyrakis}},
\bibinfo{author}{\bibfnamefont{A.}~\bibnamefont{Bunde}},
and
\bibinfo{author}{\bibfnamefont{S.}~\bibnamefont{Havlin}},
\bibinfo{year}{2007a},
\bibinfo{title}{``Improving immunization strategies},''
\eprint{arXiv:0704.1589 [cond-mat]}.


\bibitem[{\citenamefont{Gallos} \emph{et~al.}(2007b)\citenamefont{Gallos, Song, Havlin, and Makse}}]
{Gallos:gsh07b}
\bibinfo{author}{\bibnamefont{Gallos},~\bibfnamefont{L.~K.}},
\bibinfo{author}{\bibfnamefont{C.}~\bibnamefont{Song}},
\bibinfo{author}{\bibfnamefont{S.}~\bibnamefont{Havlin}},
and
\bibinfo{author}{\bibfnamefont{H.~A.}~\bibnamefont{Makse}},
\bibinfo{year}{2007b},
\bibinfo{title}{``Scaling theory of transport in complex networks},''
\bibinfo{journal}{PNAS} \textbf{\bibinfo{volume}{104}},
\bibinfo{pages}{7746}.


\bibitem[{\citenamefont{Garey and Johnson}(1979)}]
{Garey:gj79}
\bibinfo{author}{\bibnamefont{Garey}, \bibfnamefont{M.~A.}} and
\bibinfo{author}{\bibfnamefont{D.~S.}~\bibnamefont{Johnson}},
\bibinfo{year}{1979}, \emph{\bibinfo{title}{Computers and
intractability}}
(\bibinfo{publisher}{Freeman, New York}).


\bibitem[{\citenamefont{Gil and Zanette}(2006)}]
{Gil:gz06}
\bibinfo{author}{\bibnamefont{Gil},~\bibfnamefont{S.}} 
and
\bibinfo{author}{\bibfnamefont{D.~H.}~\bibnamefont{Zanette}},
\bibinfo{year}{2006},
\bibinfo{title}{``Coevolution of agents and networks: Opinion spreading and community disconnection},''
\bibinfo{journal}{Phys. Lett. A} \textbf{\bibinfo{volume}{356}},
\bibinfo{pages}{89}.


\bibitem[{\citenamefont{Gilbert}(1959)}]
{Gilbert:g59}
\bibinfo{author}{\bibnamefont{Gilbert},~\bibfnamefont{E.~N.}},
\bibinfo{year}{1959},
\bibinfo{title}{``Random graphs},''
\bibinfo{journal}{Ann. Math. Stat.} \textbf{\bibinfo{volume}{30}},
\bibinfo{pages}{1141}.


\bibitem[{\citenamefont{Gitterman}(2000)}]
{Gitterman:g00}
\bibinfo{author}{\bibnamefont{Gitterman},~\bibfnamefont{M.}}
\bibinfo{year}{2000},
\bibinfo{title}{``Small-word phenomena in physics: the Ising \ model},''
\bibinfo{journal}{J. Phys. A} \textbf{\bibinfo{volume}{33}},
\bibinfo{pages}{8373}.


\bibitem[{\citenamefont{Giuraniuc} \emph{et~al.}(2005)\citenamefont{Giuraniuc, Hatchett, Indekeu, Leone,
P\'erez Castillo, Van Schaeybroeck, and Vanderzande}}]
{Giuraniuc:ghi05}
\bibinfo{author}{\bibnamefont{Giuraniuc},~\bibfnamefont{C.~V.}},
\bibinfo{author}{\bibfnamefont{J.~P.~L.}~\bibnamefont{Hatchett}},
\bibinfo{author}{\bibfnamefont{J.~O.}~\bibnamefont{Indekeu}},
\bibinfo{author}{\bibfnamefont{M.}~\bibnamefont{Leone}},
\bibinfo{author}{\bibfnamefont{I.}~\bibnamefont{P\'erez Castillo}},
\bibinfo{author}{\bibfnamefont{B.}~\bibnamefont{Van Schaeybroeck}},
and
\bibinfo{author}{\bibfnamefont{C.}~\bibnamefont{Vanderzande}},
\bibinfo{year}{2005},
\bibinfo{title}{``Trading interactions for topology in scale-free networks},''
\bibinfo{journal}{Phys. Rev. Lett.} \textbf{\bibinfo{volume}{95}},
\bibinfo{pages}{098701}.


\bibitem[{\citenamefont{Giuraniuc} \emph{et~al.}(2006)\citenamefont{Giuraniuc, Hatchett, Indekeu, Leone, Perez Castillo, Van Schaeybroeck, and Vanderzande}}]
{Giuraniuc:ghi06}
\bibinfo{author}{\bibnamefont{Giuraniuc},~\bibfnamefont{C. V.}},
\bibinfo{author}{\bibfnamefont{J. P. L.}~\bibnamefont{Hatchett}},
\bibinfo{author}{\bibfnamefont{J. O.}~\bibnamefont{Indekeu}},
\bibinfo{author}{\bibfnamefont{M.}~\bibnamefont{Leone}},
\bibinfo{author}{\bibfnamefont{I.}~\bibnamefont{Perez Castillo}},
\bibinfo{author}{\bibfnamefont{B.}~\bibnamefont{Van Schaeybroeck}},
and
\bibinfo{author}{\bibfnamefont{C.}~\bibnamefont{Vanderzande}},
\bibinfo{year}{2006},
\bibinfo{title}{``Criticality on networks with topology-dependent interactions},''
\bibinfo{journal}{Phys. Rev. E} \textbf{\bibinfo{volume}{74}},
\bibinfo{pages}{036108}.


\bibitem[{\citenamefont{Goh} \emph{et~al.}(2001)\citenamefont{Goh, Kahng, and Kim}}]
{Goh:gkk01}
\bibinfo{author}{\bibnamefont{Goh},~\bibfnamefont{K.-I.}},
\bibinfo{author}{\bibfnamefont{B.}~\bibnamefont{Kahng}},
and
\bibinfo{author}{\bibfnamefont{D.}~\bibnamefont{Kim}},
\bibinfo{year}{2001},
\bibinfo{title}{``Universal behavior of load distribution in scale-free networks},''
\bibinfo{journal}{Phys. Rev. Lett.} \textbf{\bibinfo{volume}{87}},
\bibinfo{pages}{278701}.


\bibitem[{\citenamefont{Goh} \emph{et~al.}(2003)\citenamefont{Goh, Lee, Kahng, and Kim}}]
{Goh:glk03}
\bibinfo{author}{\bibnamefont{Goh},~\bibfnamefont{K.-I.}},
\bibinfo{author}{\bibfnamefont{D.-S.}~\bibnamefont{Lee}},
\bibinfo{author}{\bibfnamefont{B.}~\bibnamefont{Kahng}},
and
\bibinfo{author}{\bibfnamefont{D.}~\bibnamefont{Kim}},
\bibinfo{year}{2003},
\bibinfo{title}{``Sandpile on scale-free networks},''
\bibinfo{journal}{Phys. Rev. Lett.} \textbf{\bibinfo{volume}{91}},
\bibinfo{pages}{148701}.


\bibitem[{\citenamefont{Goh} \emph{et~al.}(2005)\citenamefont{Goh, Lee, Kahng and Kim}}]
{Goh:glk05}
\bibinfo{author}{\bibnamefont{Goh},~\bibfnamefont{K.-I.}},
\bibinfo{author}{\bibfnamefont{D.-S.}~\bibnamefont{ Lee}},
\bibinfo{author}{\bibfnamefont{B.}~\bibnamefont{Kahng}},
and
\bibinfo{author}{\bibfnamefont{D.}~\bibnamefont{Kim}},
\bibinfo{year}{2005},
\bibinfo{title}{``Cascading toppling dynamics on scale-free networks},''
\bibinfo{journal}{Physica A} \textbf{\bibinfo{volume}{346}},
\bibinfo{pages}{011665}.


\bibitem[{\citenamefont{Goh} \emph{et~al.}(2006)\citenamefont{Goh, Salvi, Kahng, and Kim}}]
{Goh:gsk06}
\bibinfo{author}{\bibnamefont{Goh},~\bibfnamefont{K.-I.}},
\bibinfo{author}{\bibfnamefont{G.}~\bibnamefont{Salvi}},
\bibinfo{author}{\bibfnamefont{B.}~\bibnamefont{Kahng}},
and
\bibinfo{author}{\bibfnamefont{D.}~\bibnamefont{Kim}},
\bibinfo{year}{2006},
\bibinfo{title}{``Skeleton and fractal scaling in complex networks},''
\bibinfo{journal}{Phys. Rev. Lett.} \textbf{\bibinfo{volume}{96}},
\bibinfo{pages}{018701}.


\bibitem[{\citenamefont{Goldschmidt and Dominicis}(1990)}]
{Goldschmidt:gd90}
\bibinfo{author}{\bibnamefont{Goldschmidt}, \bibfnamefont{Y.~Y.}} and
\bibinfo{author}{\bibfnamefont{C.}~\bibnamefont{De Dominicis}},
\bibinfo{title}{``Replica symmetry breaking in the
spin-glass model on lattices with finite connectivity: Application
to graph partitioning},''
\bibinfo{year}{1990}, \bibinfo{journal}{Phys. Rev. B}
\textbf{\bibinfo{volume}{41}}, \bibinfo{pages}{2184}.


\bibitem[{\citenamefont{Goltsev} \emph{et~al.}(2003)\citenamefont{Goltsev, Dorogovtsev, and Mendes}}]
{Goltsev:gdm03}
\bibinfo{author}{\bibnamefont{Goltsev},~\bibfnamefont{A.~V.}},
\bibinfo{author}{\bibfnamefont{S.~N.}~\bibnamefont{Dorogovtsev}},
and
\bibinfo{author}{\bibfnamefont{J.~F.~F.}~\bibnamefont{Mendes}},
\bibinfo{year}{2003},
\bibinfo{title}{``Critical phenomena in networks},''
\bibinfo{journal}{Phys. Rev. E} \textbf{\bibinfo{volume}{67}},
\bibinfo{pages}{026123}.


\bibitem[{\citenamefont{Goltsev} \emph{et~al.}(2006)\citenamefont{Goltsev, Dorogovtsev, and Mendes}}]
{Goltsev:gdm06}
\bibinfo{author}{\bibnamefont{Goltsev},~\bibfnamefont{A.~V.}},
\bibinfo{author}{\bibfnamefont{S.~N.}~\bibnamefont{Dorogovtsev}},
and
\bibinfo{author}{\bibfnamefont{J.~F.~F.}~\bibnamefont{Mendes}},
\bibinfo{year}{2006},
\bibinfo{title}{``$k$-core (bootstrap) percolation on complex networks:
Critical phenomena and nonlocal effects},''
\bibinfo{journal}{Phys. Rev. E} \textbf{\bibinfo{volume}{73}},
\bibinfo{pages}{056101}.


\bibitem[{\citenamefont{G\'{o}mez-Garde\~{n}es} \emph{et~al.}(2007a)\citenamefont{G\'{o}mez-Garde\~{n}es, Moreno, and
Arenas}}]
{Gomez-Gardenes:g-gma07a}
\bibinfo{author}{\bibnamefont{G\'{o}mez-Garde\~{n}es},~\bibfnamefont{J.}},
\bibinfo{author}{\bibfnamefont{Y.}~\bibnamefont{Moreno}},
and
\bibinfo{author}{\bibfnamefont{A.}~\bibnamefont{Arenas}},
\bibinfo{year}{2007a},
\bibinfo{title}{``Paths to synchronization on complex networks},''
\bibinfo{journal}{Phys. Rev. Lett.} \textbf{\bibinfo{volume}{98}},
\bibinfo{pages}{034101}.


\bibitem[{\citenamefont{G\'{o}mez-Garde\~{n}es} \emph{et~al.}(2007b)\citenamefont{G\'{o}mez-Garde\~{n}es, Moreno, and
Arenas}}]
{Gomez-Gardenes:g-gma07b}
\bibinfo{author}{\bibnamefont{G\'{o}mez-Garde\~{n}es},~\bibfnamefont{J.}},
\bibinfo{author}{\bibfnamefont{Y.}~\bibnamefont{Moreno}},
and
\bibinfo{author}{\bibfnamefont{A.}~\bibnamefont{Arenas}},
\bibinfo{year}{2007b},
\bibinfo{title}{``Synchronizability determined by coupling strengths and topology on complex networks},''
\bibinfo{journal}{Phys. Rev. E} \textbf{\bibinfo{volume}{75}},
\bibinfo{pages}{066106}.




\bibitem[{\citenamefont{Gourley and Johnson}(2006)}]
{Gourley:gj06}
\bibinfo{author}{\bibnamefont{Gourley},~\bibfnamefont{S.}}
and
\bibinfo{author}{\bibfnamefont{N.~F.}~\bibnamefont{Johnson}},
\bibinfo{year}{2006},
\bibinfo{title}{``Effects of decision-making on the transport costs across complex networks},''
\bibinfo{journal}{Physica A} \textbf{\bibinfo{volume}{363}},
\bibinfo{pages}{82}.

\bibitem[{\citenamefont{Grassberger}(1983)}]
{Grassberger:g83}
\bibinfo{author}{\bibfnamefont{Grassberger}~\bibnamefont{P.}},
\bibinfo{year}{1983},
\bibinfo{title}{``Critical behavior of the general epidemic process and dynamical percolation},''
\bibinfo{journal}{Math. Biosci.} \textbf{\bibinfo{volume}{63}},
\bibinfo{pages}{157}.


\bibitem[{\citenamefont{Grassberger and de la Torre}(1979)}]
{Grassberger:gt79}
\bibinfo{author}{\bibnamefont{Grassberger},~\bibfnamefont{P.}}
and
\bibinfo{author}{\bibfnamefont{A.}~\bibnamefont{de la Torre}},
\bibinfo{year}{1979},
\bibinfo{title}{``Reggeon field theory (Schl\"{o}gl's first model) on a lattice: Monte Carlo calculations of critical behaviour},''
\bibinfo{journal}{Ann. Phys. (NY)} \textbf{\bibinfo{volume}{122}},
\bibinfo{pages}{373}.


\bibitem[{\citenamefont{Grinstein and Linsker}(2005)}]
{Grinstein:gl05}
\bibinfo{author}{\bibnamefont{Grinstein},~\bibfnamefont{G.}}
and
\bibinfo{author}{\bibfnamefont{R.}~\bibnamefont{Linsker}},
\bibinfo{year}{2005},
\bibinfo{title}{``Synchronous neural activity in scale-free
network models versus random network models},''
\bibinfo{journal}{Proc. Natl. Acad. Sci.} \textbf{\bibinfo{volume}{102}},
\bibinfo{pages}{9948}.


\bibitem[{\citenamefont{Gross and Blasius}(2007)}]
{Gross:gb07}
\bibinfo{author}{\bibnamefont{Gross},~\bibfnamefont{T.}}
and
\bibinfo{author}{\bibfnamefont{B.}~\bibnamefont{Blasius}},
\bibinfo{year}{2007},
\bibinfo{title}{``Adaptive coevolutionary networks --- a review},''
\eprint{arXive:0709.1858}.


\bibitem[{\citenamefont{Guardiola} \emph{et~al.}(2000)\citenamefont{Guardiola, D\'{i}az-Guilera, Llas,
and P\'{e}rez}}]
{Guardiola:gd-glp00}
\bibinfo{author}{\bibnamefont{Guardiola},~\bibfnamefont{X.}},
\bibinfo{author}{\bibfnamefont{A.}~\bibnamefont{D\'{i}az-Guilera}},
\bibinfo{author}{\bibfnamefont{M.}~\bibnamefont{Llas}},
and
\bibinfo{author}{\bibfnamefont{C.~J.}~\bibnamefont{P\'{e}rez}},
\bibinfo{year}{2000},
\bibinfo{title}{``Synchronization, diversity, and topology of networks of integrate
and fire oscillators},''
\bibinfo{journal}{Phys. Rev. E} \textbf{\bibinfo{volume}{62}},
\bibinfo{pages}{5565}.


\bibitem[{\citenamefont{Guimer\`a} \emph{et~al.}(2002)\citenamefont{Guimer\`a, D\'iaz-Guilera, Vega-Redondo, Cabrales, and Arenas}}]
{Guimera:gdv02}
\bibinfo{author}{\bibnamefont{Guimer\`a},~\bibfnamefont{R.}},
\bibinfo{author}{\bibfnamefont{A.}~\bibnamefont{D\'iaz-Guilera}},
\bibinfo{author}{\bibfnamefont{F.}~\bibnamefont{Vega-Redondo}},
\bibinfo{author}{\bibfnamefont{A.}~\bibnamefont{Cabrales}},
and
\bibinfo{author}{\bibfnamefont{A.}~\bibnamefont{Arenas}},
\bibinfo{year}{2002},
\bibinfo{title}{``Optimal network topologies for local search with congestion},''
\bibinfo{journal}{Phys. Rev. Lett.} \textbf{\bibinfo{volume}{89}},
\bibinfo{pages}{258701}.




\bibitem[{\citenamefont{Guimer\`a} \emph{et~al.}(2004)\citenamefont{Guimer\`a, Sales-Pardo, Amaral}}]
{Guimera:gsa04}
\bibinfo{author}{\bibnamefont{Guimer\`a},~\bibfnamefont{R.}},
\bibinfo{author}{\bibfnamefont{M.}~\bibnamefont{Sales-Pardo}},
and
\bibinfo{author}{\bibfnamefont{L.~A.~N.}~\bibnamefont{Amaral}},
\bibinfo{year}{2004},
\bibinfo{title}{``Modularity from fluctuations in random graphs and complex networks},''
\bibinfo{journal}{Phys. Rev. E} \textbf{\bibinfo{volume}{70}},
\bibinfo{pages}{025101}.



\bibitem[{\citenamefont{Ha} \emph{et~al.}(2007)\citenamefont{Ha, Hong, and Park}}]
{Ha:hhp06}
\bibinfo{author}{\bibnamefont{Ha},~\bibfnamefont{M.}},
\bibinfo{author}{\bibfnamefont{H.}~\bibnamefont{Hong}},
and
\bibinfo{author}{\bibfnamefont{H.}~\bibnamefont{Park}},
\bibinfo{year}{2007},
\bibinfo{title}{``Comment on `non-mean-field behavior of the contact process on scale-free networks'},''
\bibinfo{journal}{Phys. Rev. Lett.} \textbf{\bibinfo{volume}{98}},
\bibinfo{pages}{029801}.


\bibitem[{\citenamefont{H\"aggstr\"om}(2002)}]
{Haggstrom:h02}
\bibinfo{author}{\bibfnamefont{H\"aggstr\"om}~\bibnamefont{O.}},
\bibinfo{year}{2002},
\bibinfo{title}{``Zero-temperature dynamics for the ferromagnetic Ising model on random graphs},''
\bibinfo{journal}{Physica A} \textbf{\bibinfo{volume}{310}},
\bibinfo{pages}{275}.


\bibitem[{\citenamefont{Harris}(1974)}]
{Harris:h74}
\bibinfo{author}{\bibfnamefont{Harris}~\bibnamefont{T. E.}},
\bibinfo{year}{1974},
\bibinfo{title}{``Contact interactions on a lattice},''
\bibinfo{journal}{Ann. Probab.} \textbf{\bibinfo{volume}{2}},
\bibinfo{pages}{969}.


\bibitem[{\citenamefont{Harris}(1975)}]
{Harris:h75}
\bibinfo{author}{\bibnamefont{Harris},~\bibfnamefont{A.~B.}}
\bibinfo{year}{1975},
\bibinfo{title}{``Nature of the Griffiths singularity in dilute magnets},''
\bibinfo{journal}{Phys. Rev. B} \textbf{\bibinfo{volume}{12}},
\bibinfo{pages}{203}.


\bibitem[{\citenamefont{Harris}(1982)}]
{Harris:h82}
\bibinfo{author}{\bibfnamefont{Harris}~\bibnamefont{A. B.}},
\bibinfo{year}{1982},
\bibinfo{title}{``Exact solution of a model of localization},''
\bibinfo{journal}{Phys. Rev. Lett.} \textbf{\bibinfo{volume}{49}},
\bibinfo{pages}{296}.






\bibitem[{\citenamefont{Hartmann and Weigt}(2005)}]
{Hartmann:hw05}
\bibinfo{author}{\bibfnamefont{Hartmann}~\bibnamefont{A. K.}}
and
\bibinfo{author}{\bibfnamefont{M.}~\bibnamefont{Weigt}},
\bibinfo{year}{2005},
\emph{\bibinfo{title}{Phase Transitions in Combinatorial
Optimization Problems: Basics, Algorithms and Statistical
Mechanics}} (\bibinfo{publisher}{Wiley-VCH}).




\bibitem[{\citenamefont{Hase} \emph{et~al.}(2005)\citenamefont{Hase, de Almeida, and Salinas}}]
{Hase:has05}
\bibinfo{author}{\bibnamefont{Hase},~\bibfnamefont{M.~O.}},
\bibinfo{author}{\bibfnamefont{J.~R.~L.}~\bibnamefont{Almeida}},
and
\bibinfo{author}{\bibfnamefont{S.~R.}~\bibnamefont{Salinas}},
\bibinfo{year}{2005},
\bibinfo{title}{``Replica-symmetric solutions of a
dilute Ising ferromagnet in a random field},''
\bibinfo{journal}{Eur. Phys. J B} \textbf{\bibinfo{volume}{47}},
\bibinfo{pages}{245}.


\bibitem[{\citenamefont{Hase} \emph{et~al.}(2006)\citenamefont{Hase, de Almeida, and Salinas}}]
{Hase:has06}
\bibinfo{author}{\bibnamefont{Hase},~\bibfnamefont{M.~O.}},
\bibinfo{author}{\bibfnamefont{J.~R.~L.}~\bibnamefont{de Almeida}},
and
\bibinfo{author}{\bibfnamefont{S.~R.}~\bibnamefont{Salinas}},
\bibinfo{year}{2006},
\bibinfo{title}{``Spin glass behaviour on random lattices},''
\eprint{cond-mat/0604144}.

\bibitem[{\citenamefont{Hastings}(2003)}]
{Hastings:h03}
\bibinfo{author}{\bibnamefont{Hastings},~\bibfnamefont{M.~B.}},
\bibinfo{year}{2003},
\bibinfo{title}{``Mean-field and anomalous behavior on a small-world network},''
\bibinfo{journal}{Phys. Rev. Lett.} \textbf{\bibinfo{volume}{91}},
\bibinfo{pages}{098701}.

\bibitem[{\citenamefont{Hastings}(2006)}]
{Hastings:h06}
\bibinfo{author}{\bibfnamefont{Hastings}~\bibnamefont{M.~B.}},
\bibinfo{year}{2006},
\bibinfo{title}{``Systematic series expansions for processes on networks},''
\bibinfo{journal}{Phys. Rev. Lett.} \textbf{\bibinfo{volume}{96}},
\bibinfo{pages}{148701}.













\bibitem[{\citenamefont{Heimburg and Thomas}(1974)}]
{Heimburg:ht74}
\bibinfo{author}{\bibnamefont{von Heimburg},~\bibfnamefont{J.}}
and
\bibinfo{author}{\bibfnamefont{H.}~\bibnamefont{Thomas}},
\bibinfo{year}{1974},
\bibinfo{title}{``Phase transition of the Cayley tree with Ising interaction},''
\bibinfo{journal}{J. Phys. C} \textbf{\bibinfo{volume}{7}},
\bibinfo{pages}{3433}.


\bibitem[{\citenamefont{Helbing} \emph{et~al.}(2007)\citenamefont{Helbing, Siegmeier, and L\"ammer}}]
{Helbing:hsl07}
\bibinfo{author}{\bibnamefont{Helbing},~\bibfnamefont{D.}},
\bibinfo{author}{\bibfnamefont{J.}~\bibnamefont{Siegmeier}},
and
\bibinfo{author}{\bibfnamefont{S.}~\bibnamefont{L\"ammer}},
\bibinfo{year}{2007},
\bibinfo{title}{``Self-organized network flows},'' 
\bibinfo{journal}{Networks and Heterogeneous Media} \textbf{\bibinfo{volume}{2}},
\bibinfo{pages}{193}. 


\bibitem[{\citenamefont{Herrero}(2002)}]
{Herrero:h02}
\bibinfo{author}{\bibnamefont{Herrero},~\bibfnamefont{C.~P.}},
\bibinfo{year}{2002},
\bibinfo{title}{``Ising model in small-world networks},''
\bibinfo{journal}{Phys. Rev. E} \textbf{\bibinfo{volume}{65}},
\bibinfo{pages}{066110}.


\bibitem[{\citenamefont{Herrero}(2004)}]
{Herrero:h04}
\bibinfo{author}{\bibnamefont{Herrero},~\bibfnamefont{C.~P.}}
\bibinfo{year}{2004},
\bibinfo{title}{``Ising model in scale-free networks: A Monte Carlo simulation},''
\bibinfo{journal}{Phys. Rev. E} \textbf{\bibinfo{volume}{69}},
\bibinfo{pages}{067109}.

\bibitem[{\citenamefont{Hinczewski and Berker}(2006)}]
{Hinczewski:hb05}
\bibinfo{author}{\bibnamefont{Hinczewski},~\bibfnamefont{M.}}
and
\bibinfo{author}{\bibfnamefont{A.~N.}~\bibnamefont{Berker}},
\bibinfo{year}{2006},
\bibinfo{title}{``Inverted Berezinskii-Kosterlitz-Thouless singularity and high-temperature algebraic order in an Ising model on a scale-free hierarchical-lattice small-world network},''
\bibinfo{journal}{Phys. Rev. E} \textbf{\bibinfo{volume}{73}},
\bibinfo{pages}{066126}.


\bibitem[{\citenamefont{Hinrichsen}(2000)}]
{Hinrichsen:h00}
\bibinfo{author}{\bibnamefont{Hinrichsen},~\bibfnamefont{H.}},
\bibinfo{year}{2000},
\bibinfo{title}{``Non-equilibrium critical phenomena and phase transitions into absorbing states},''
\bibinfo{journal}{Adv. Phys.} \textbf{\bibinfo{volume}{49}},
\bibinfo{pages}{815}.


\bibitem[{\citenamefont{Holme}(2003)}]
{Holme:h03}
\bibinfo{author}{\bibnamefont{Holme},~\bibfnamefont{P.}},
\bibinfo{year}{2003},
\bibinfo{title}{``Congestion and centrality in traffic flow on complex networks},''
\bibinfo{journal}{Adv. Complex Systems} \textbf{\bibinfo{volume}{6}},
\bibinfo{pages}{163}.


\bibitem[{\citenamefont{Holme}(2007)}]
{Holme:h07}
\bibinfo{author}{\bibfnamefont{Holme}~\bibnamefont{P.}},
\bibinfo{year}{2007},
\bibinfo{title}{``Scale-free networks with a large- to hypersmall-world transition},''
\bibinfo{journal}{Physica A} \textbf{\bibinfo{volume}{377}},
\bibinfo{pages}{315}.




\bibitem[{\citenamefont{Holme} \emph{et~al.}(2003)\citenamefont{Holme, Liljeros, Edling, and Kim}}]
{Holme:hlek03}
\bibinfo{author}{\bibnamefont{Holme},~\bibfnamefont{P.}},
\bibinfo{author}{\bibfnamefont{F.}~\bibnamefont{Liljeros}},
\bibinfo{author}{\bibfnamefont{C.~R.}~\bibnamefont{Edling}},
and
\bibinfo{author}{\bibfnamefont{B.~J.}~\bibnamefont{Kim}},,
\bibinfo{year}{2003},
\bibinfo{title}{``Network bipartivity},''
\bibinfo{journal}{Phys. Rev. E} \textbf{\bibinfo{volume}{68}},
\bibinfo{pages}{056107}.


\bibitem[{\citenamefont{Holme and Newman}(2006)}]
{Holme:hn06}
\bibinfo{author}{\bibnamefont{Holme},~\bibfnamefont{P.}}
and
\bibinfo{author}{\bibfnamefont{M.~E.~J.}~\bibnamefont{Newman}},
\bibinfo{year}{2006},
\bibinfo{title}{``Nonequilibrium phase transition in the coevolution of networks and opinions},''
\bibinfo{journal}{Phys. Rev. E} \textbf{\bibinfo{volume}{74}},
\bibinfo{pages}{056108}.




\bibitem[{\citenamefont{Hong} \emph{et~al.}(2007b)\citenamefont{Hong, Chate, Park, and Tang}}]
{Hong:hcpt07}
\bibinfo{author}{\bibnamefont{Hong},~\bibfnamefont{H.}},
\bibinfo{author}{\bibfnamefont{H.}~\bibnamefont{Chat\'e}},
\bibinfo{author}{\bibfnamefont{H.}~\bibnamefont{Park}},
and
\bibinfo{author}{\bibfnamefont{L.-H.}~\bibnamefont{Tang}},
\bibinfo{year}{2007b},
\bibinfo{title}{``Entrainment transition in populations of random frequency oscillators},'' 
\bibinfo{journal}{Phys. Rev. Lett.} \textbf{\bibinfo{volume}{99}},
\bibinfo{pages}{184101}. 


\bibitem[{\citenamefont{Hong} \emph{et~al.}(2002a)\citenamefont{Hong, Choi, and Kim}}]
{Hong:hck02}
\bibinfo{author}{\bibnamefont{Hong},~\bibfnamefont{H.}},
\bibinfo{author}{\bibfnamefont{M.~Y.}~\bibnamefont{Choi}},
and
\bibinfo{author}{\bibfnamefont{B.~J.}~\bibnamefont{Kim}},
\bibinfo{year}{2002a},
\bibinfo{title}{``Synchronization on small-world networks},''
\bibinfo{journal}{Phys. Rev. E} \textbf{\bibinfo{volume}{65}},
\bibinfo{pages}{026139}.


\bibitem[{\citenamefont{Hong} \emph{et~al.}(2007a)\citenamefont{Hong, Ha, and Park}}]
{Hong:hhp07}
\bibinfo{author}{\bibnamefont{Hong},~\bibfnamefont{H.}},
\bibinfo{author}{\bibfnamefont{M.}~\bibnamefont{Ha}},
and
\bibinfo{author}{\bibfnamefont{H.}~\bibnamefont{Park}},
\bibinfo{year}{2007a},
\bibinfo{title}{``Finite-size scaling in complex networks},''
\bibinfo{journal}{Phys. Rev. Lett.} \textbf{\bibinfo{volume}{98}},
\bibinfo{pages}{258701}.


\bibitem[{\citenamefont{Hong} \emph{et~al.}(2002b)\citenamefont{Hong, Kim, and Choi}}]
{Hong:hkc02}
\bibinfo{author}{\bibnamefont{Hong},~\bibfnamefont{H.}},
\bibinfo{author}{\bibfnamefont{B.~J.}~\bibnamefont{Kim}},
and
\bibinfo{author}{\bibfnamefont{M.~Y.}~\bibnamefont{Choi}},
\bibinfo{year}{2002b},
\bibinfo{title}{``Comment on Ising model on a small world network},''
\bibinfo{journal}{Phys. Rev. E} \textbf{\bibinfo{volume}{66}},
\bibinfo{pages}{018101}.


\bibitem[{\citenamefont{Hong} \emph{et~al.}(2004a)\citenamefont{Hong, Kim, Choi and Park}}]
{Hong:hkcp04}
\bibinfo{author}{\bibnamefont{Hong},~\bibfnamefont{H.}},
\bibinfo{author}{\bibfnamefont{B.~J.}~\bibnamefont{Kim}},
\bibinfo{author}{\bibfnamefont{M.~Y.}~\bibnamefont{Choi}},
and
\bibinfo{author}{\bibfnamefont{H.}~\bibnamefont{Park}},
\bibinfo{year}{2004a},
\bibinfo{title}{``Factors that predict better synchronizability on complex networks},''
\bibinfo{journal}{Phys. Rev. E} \textbf{\bibinfo{volume}{69}},
\bibinfo{pages}{067105}.


\bibitem[{\citenamefont{Hong} \emph{et~al.}(2004b)\citenamefont{Hong, Park, and Choi}}]
{Hong:hpc04}
\bibinfo{author}{\bibnamefont{Hong},~\bibfnamefont{H.}},
\bibinfo{author}{\bibfnamefont{H.}~\bibnamefont{Park}}, 
and
\bibinfo{author}{\bibfnamefont{Choi}~\bibnamefont{M.~Y.}},
\bibinfo{year}{2004b}, 
\bibinfo{title}{``Collective phase synchronization in locally coupled limit-cycle oscillators},''
\bibinfo{journal}{Phys. Rev. E} \textbf{\bibinfo{volume}{70}},
\bibinfo{pages}{045204 (R)}.


\bibitem[{\citenamefont{Hong} \emph{et~al.}(2005)\citenamefont{Hong, Park, and Choi}}]
{Hong:hpc05}
\bibinfo{author}{\bibnamefont{Hong},~\bibfnamefont{H.}},
\bibinfo{author}{\bibfnamefont{H.}~\bibnamefont{Park}}, 
and
\bibinfo{author}{\bibfnamefont{Choi}~\bibnamefont{M.~Y.}},
\bibinfo{year}{2005},
\bibinfo{title}{``Collective synchronization in spatially extended systems of coupled oscillators with random frequencies},''
\bibinfo{journal}{Phys. Rev. E} \textbf{\bibinfo{volume}{72}},
\bibinfo{pages}{036217}.


\bibitem[{\citenamefont{Hong} \emph{et~al.}(2007c)\citenamefont{Hong, Park, and Tang}}]
{Hong:hpt07}
\bibinfo{author}{\bibnamefont{Hong},~\bibfnamefont{H.}},
\bibinfo{author}{\bibfnamefont{H.}~\bibnamefont{Park}},
and
\bibinfo{author}{\bibfnamefont{L.-H.}~\bibnamefont{Tang}},
\bibinfo{year}{2007c},
\bibinfo{title}{``Finite-size scaling of synchronized oscillation on complex networks},''
\eprint{arXiv:0710.1137}.




\bibitem[{\citenamefont{Hovorka and Friedman}(2007)}]
{Hovorka:hf07}
\bibinfo{author}{\bibnamefont{Hovorka},~\bibfnamefont{O.}}
and
\bibinfo{author}{\bibfnamefont{G.}~\bibnamefont{Frieman}},
\bibinfo{year}{2007},
\bibinfo{title}{``Non-converging hysteretic cycles in random spin networks},''
\eprint{cond-mat/0703525}.




\bibitem[{\citenamefont{Huang} \emph{et~al.}(2006)\citenamefont{Huang, Park, Lai, Yang, and Yang}}]
{Huang:hplyy06}
\bibinfo{author}{\bibnamefont{Huang},~\bibfnamefont{L.}},
\bibinfo{author}{\bibfnamefont{K.}~\bibnamefont{Park}},
\bibinfo{author}{\bibfnamefont{Y.-C.}~\bibnamefont{Lai}},
\bibinfo{author}{\bibfnamefont{L.}~\bibnamefont{Yang}},
and
\bibinfo{author}{\bibfnamefont{K.}~\bibnamefont{Yang}},
\bibinfo{year}{2006},
\bibinfo{title}{``Abnormal synchronization in complex clustered networks},''
\bibinfo{journal}{Phys. Rev. Lett.} \textbf{\bibinfo{volume}{97}},
\bibinfo{pages}{164101}.




\bibitem[{\citenamefont{Ichinomiya}(2004)}]
{Ichinomiya:i04}
\bibinfo{author}{\bibnamefont{Ichinomiya},~\bibfnamefont{T.}}
\bibinfo{year}{2004},
\bibinfo{title}{``Frequency synchronization in a random oscillator network},''
\bibinfo{journal}{Phys. Rev. E} \textbf{\bibinfo{volume}{70}},
\bibinfo{pages}{026116}.


\bibitem[{\citenamefont{Ichinomiya}(2005)}]
{Ichinomiya:i05}
\bibinfo{author}{\bibnamefont{Ichinomiya},~\bibfnamefont{T.}}
\bibinfo{year}{2005},
\bibinfo{title}{``Path-integral approach to the dynamics in sparse random network},''
\bibinfo{journal}{Phys. Rev. E} \textbf{\bibinfo{volume}{72}},
\bibinfo{pages}{016109}.


\bibitem[{\citenamefont{Igl\'{o}i and Turban}(2002)}]
{Igloi:it02}
\bibinfo{author}{\bibnamefont{Igl\'{o}i},~\bibfnamefont{F.}}
and
\bibinfo{author}{\bibfnamefont{L.}~\bibnamefont{Turban}},
\bibinfo{year}{2002},
\bibinfo{title}{``First- and second-order phase transitions in
scale-free networks},''
\bibinfo{journal}{Phys. Rev. E} \textbf{\bibinfo{volume}{66}},
\bibinfo{pages}{036140}.


\bibitem[{\citenamefont{Ihler} \emph{et~al.}(2005)\citenamefont{Ihler, Fischer, and Willsky}}]
{Ihler:ifw05}
\bibinfo{author}{\bibnamefont{Ihler},~\bibfnamefont{A.~T.}},
\bibinfo{author}{\bibfnamefont{J.~W.}~\bibnamefont{Fischer}},
and
\bibinfo{author}{\bibfnamefont{A.~S.}~\bibnamefont{Willsky}},
\bibinfo{year}{2005},
\bibinfo{title}{``Loopy belief propagation: convergence and effects of message errors},''
\bibinfo{journal}{J. Mach. Learning Res.} \textbf{\bibinfo{volume}{6}},
\bibinfo{pages}{905}.











\bibitem[{\citenamefont{Imry and Ma}(1975)}]
{Imry:im75}
\bibinfo{author}{\bibnamefont{Imry},~\bibfnamefont{Y.}}
and
\bibinfo{author}{\bibfnamefont{S.-K.}~\bibnamefont{Ma}},
\bibinfo{year}{1975},
\bibinfo{title}{``Random-field instability of the ordered state of continuous symmetry},''
\bibinfo{journal}{Phys. Rev. Lett.} \textbf{\bibinfo{volume}{35}},
\bibinfo{pages}{1399}.





\bibitem[{\citenamefont{Janson}(2007)}]
{Janson:j07}
\bibinfo{author}{\bibnamefont{Janson},~\bibfnamefont{S.}},
\bibinfo{year}{2007},
\bibinfo{title}{``The largest component in a subcritical random graph with a power law degree distribution},''
\eprint{arXiv:0708.4404 [math]}. 






\bibitem[{\citenamefont{Jeong} \emph{et~al.}(2003)\citenamefont{Jeong, Hong, Kim, and Choi}}]
{Jeong:jhk03}
\bibinfo{author}{\bibnamefont{Jeong},~\bibfnamefont{D.}},
\bibinfo{author}{\bibnamefont{H.},~\bibfnamefont{Hong}},
\bibinfo{author}{\bibfnamefont{B.~J.}~\bibnamefont{Kim}},
and
\bibinfo{author}{\bibfnamefont{M.~Y.}~\bibnamefont{Choi}},
\bibinfo{year}{2003},
\bibinfo{title}{``Phase transition in the Ising model on a small-world network with distance-dependent interactions},''
\bibinfo{journal}{Phys. Rev. E} \textbf{\bibinfo{volume}{68}},
\bibinfo{pages}{027101}.

\bibitem[{\citenamefont{Johnston and Plech\'a\v{c}}(1998)}]
{Johnston:jp98}
\bibinfo{author}{\bibnamefont{Johnston},~\bibfnamefont{D. A.}}
and
\bibinfo{author}{\bibfnamefont{P.}~\bibnamefont{Plech\'a\v{c}}},
\bibinfo{year}{1998},
\bibinfo{title}{``Equivalence of ferromagnetic spin models on
 trees and random graphs},''
\bibinfo{journal}{J. Phys. A} \textbf{\bibinfo{volume}{31}},
\bibinfo{pages}{475}.




\bibitem[{\citenamefont{Jost and Joy}(2001)}]
{Jost:jj01}
\bibinfo{author}{\bibnamefont{Jost},~\bibfnamefont{J.}}
and
\bibinfo{author}{\bibfnamefont{M.~P.}~\bibnamefont{Joy}},
\bibinfo{year}{2001},
\bibinfo{title}{``Spectral properties and synchronization in coupled map lattices},''
\bibinfo{journal}{Phys. Rev. E} \textbf{\bibinfo{volume}{65}},
\bibinfo{pages}{016201}.


\bibitem[{\citenamefont{Jung} \emph{et~al.}(2002)\citenamefont{Jung, Kim, and Kahng}}]
{Jung:jkk02}
\bibinfo{author}{\bibnamefont{Jung},~\bibfnamefont{S.}},
\bibinfo{author}{\bibfnamefont{S.}~\bibnamefont{Kim}},
and
\bibinfo{author}{\bibfnamefont{B.}~\bibnamefont{Kahng}},
\bibinfo{year}{2002},
\bibinfo{title}{``Geometric fractal growth model for scale-free networks},''
\bibinfo{journal}{Phys. Rev. E} \textbf{\bibinfo{volume}{65}},
\bibinfo{pages}{056101}.


\bibitem[{\citenamefont{Kalapala and Moore}(2002)}]
{Kalapala:km02}
\bibinfo{author}{\bibnamefont{Kalapala},~\bibfnamefont{V.}}
and
\bibinfo{author}{\bibfnamefont{C.}~\bibnamefont{Moore}},
\bibinfo{year}{2002},
\bibinfo{title}{``MAX-CUT on sparse random graphs},''
http://www.cs.unm.edu/\~{}treport/tr/02-08/maxcut.ps.gz.






\bibitem[{\citenamefont{Kalisky and Cohen}(2006)}]
{Kalisky:kc05}
\bibinfo{author}{\bibnamefont{Kalisky},~\bibfnamefont{T.}}
and
\bibinfo{author}{\bibfnamefont{R.}~\bibnamefont{Cohen}},
\bibinfo{year}{2006},
\bibinfo{title}{``Width of percolation transition in complex networks},''
\bibinfo{journal}{Phys. Rev. E} \textbf{\bibinfo{volume}{73}},
\bibinfo{pages}{035101}.


\bibitem[{\citenamefont{Kanter and Sompolinsky}(2000)}]
{Kanter:ks87}
\bibinfo{author}{\bibnamefont{Kanter},~\bibfnamefont{I.}}
and
\bibinfo{author}{\bibfnamefont{H.}~\bibnamefont{Sompolinsky}},
\bibinfo{year}{2000},
\bibinfo{title}{``Mean-field theory of spin-glasses with finite
coordination number},''
\bibinfo{journal}{Phys. Rev. Lett.} \textbf{\bibinfo{volume}{58}},
\bibinfo{pages}{164}.


\bibitem[{\citenamefont{Kappen}(2002)}]
{Kappen:k02}
\bibinfo{author}{\bibnamefont{Kappen}, \bibfnamefont{H.}},
\bibinfo{year}{2002},
in \emph{\bibinfo{booktitle}{Modelling Biomedical Signals}},
edited by \bibinfo{editor}{\bibfnamefont{G.}~\bibnamefont{Nardulli}} and
\bibinfo{editor}{\bibfnamefont{S.}~\bibnamefont{Stramaglia}}
(\bibinfo{publisher}{World Scientific, Singapore}), p. \bibinfo{pages}{3-27}.


\bibitem[{\citenamefont{Karp and Sipser}(1981)}]
{Karp:ks81}
\bibinfo{author}{\bibnamefont{Karp},~\bibfnamefont{R.~M.}}
and
\bibinfo{author}{\bibnamefont{M.},~\bibfnamefont{Sipser}},
\bibinfo{year}{1981},
\bibinfo{title}{``Maximum matchings is sparse random graphs},''
in
\emph{\bibinfo{booktitle}{Proceedings of the 22nd IEEE Symposium on Foundations of Computing}},
p. \bibinfo{pages}{364}.


\bibitem[{\citenamefont{Karsai} \emph{et~al.}(2007)\citenamefont{Karsai, d'Auriac, and Igl\'oi}}]
{Karsai:kdi07}
\bibinfo{author}{\bibnamefont{Karsai},~\bibfnamefont{M.}},
\bibinfo{author}{\bibfnamefont{J.-C. A.}~\bibnamefont{d'Auriac}},
and
\bibinfo{author}{\bibfnamefont{F.}~\bibnamefont{Igl\'oi}},
\bibinfo{year}{2007},
\bibinfo{title}{``Rounding of first-order phase transitions and optimal cooperation in scale-free networks},''
\eprint{arXiv:0704.1538 [cond-mat]}.


\bibitem[{\citenamefont{Karsai} \emph{et~al.}(2006)\citenamefont{Karsai, Juh\'asz, and Igl\'oi}}]
{Karsai:kji06}
\bibinfo{author}{\bibnamefont{Karsai},~\bibfnamefont{M.}},
\bibinfo{author}{\bibfnamefont{R.}~\bibnamefont{Juh\'asz}},
and
\bibinfo{author}{\bibfnamefont{F.}~\bibnamefont{Igl\'oi}},
\bibinfo{year}{2006},
\bibinfo{title}{``Nonequilibrium phase transitions and finite size scaling in weighted scale-free networks},''
\bibinfo{journal}{Phys. Rev. E} \textbf{\bibinfo{volume}{73}},
\bibinfo{pages}{036116}.



\bibitem[{\citenamefont{Kasteleyn and Fortuin}(1969)}]
{Kasteleyn:kf69}
\bibinfo{author}{\bibnamefont{Kasteleyn},~\bibfnamefont{P.~W.}}
and
\bibinfo{author}{\bibfnamefont{C.~M.}~\bibnamefont{Fortuin}},
\bibinfo{year}{1969},
\bibinfo{title}{``Phase transition in lattice systems with random local properties},''
\bibinfo{journal}{J. Phys. Soc. Jpn. Suppl.} \textbf{\bibinfo{volume}{26}},
\bibinfo{pages}{11}.


\bibitem[{\citenamefont{Kenah and Robins}(2006)}]
{Kenah:kr06}
\bibinfo{author}{\bibnamefont{Kenah},~\bibfnamefont{E.}}
and
\bibinfo{author}{\bibfnamefont{J.}~\bibnamefont{Robins}},
\bibinfo{year}{2006},
\bibinfo{title}{``Second look at the spread of epidemics on networks},''
\eprint{q-bio/0610057}.


\bibitem[{\citenamefont{Khajeh} \emph{et~al.}(2007)\citenamefont{Khajeh, Dorogovtsev, and Mendes}}]
{Khajeh:kdm07}
\bibinfo{author}{\bibnamefont{Khajeh},~\bibfnamefont{E.}},
\bibinfo{author}{\bibfnamefont{S.~N.}~\bibnamefont{Dorogovtsev}},
and
\bibinfo{author}{\bibfnamefont{J.~F.~F.}~\bibnamefont{Mendes}},
\bibinfo{year}{2007},
\bibinfo{title}{``BKT-like transition in the Potts model on an inhomogeneous annealed network},''
\bibinfo{journal}{Phys. Rev. E} \textbf{\bibinfo{volume}{75}},
\bibinfo{pages}{041112}.




\bibitem[{\citenamefont{Kikuchi}(1951)}]
{Kikuchi:k51}
\bibinfo{author}{\bibnamefont{Kikuchi}, \bibfnamefont{R.}}, \bibinfo{year}{1951},
\bibinfo{title}{``A theory of cooperative phenomena},''
\bibinfo{journal}{Phys. Rev.} \textbf{\bibinfo{volume}{81}},
\bibinfo{pages}{988}.


\bibitem[{\citenamefont{Kim}(2004)}]
{Kim:k04}
\bibinfo{author}{\bibnamefont{Kim}, \bibfnamefont{B.~J.}},
\bibinfo{year}{2004},
\bibinfo{title}{``Performance of networks of artificial neurons: the role of clustering},''
\bibinfo{journal}{Phys. Rev. E} \textbf{\bibinfo{volume}{69}},
\bibinfo{pages}{045101}.


\bibitem[{\citenamefont{Kim} \emph{et~al.}(2001)\citenamefont{Kim, Hong, Holme, Jeon, Minnhagen, and Choi}}]
{Kim:khh01}
\bibinfo{author}{\bibnamefont{Kim},~\bibfnamefont{B.~J.}},
\bibinfo{author}{\bibfnamefont{H.}~\bibnamefont{Hong}},
\bibinfo{author}{\bibfnamefont{P.}~\bibnamefont{Holme}},
\bibinfo{author}{\bibfnamefont{G.~S}~\bibnamefont{Jeon}},
\bibinfo{author}{\bibfnamefont{P.}~\bibnamefont{Minnhagen}},
and
\bibinfo{author}{\bibfnamefont{M.~Y.}~\bibnamefont{Choi}},
\bibinfo{year}{2001},
\bibinfo{title}{``XY model in small-world networks},''
\bibinfo{journal}{Phys. Rev. E} \textbf{\bibinfo{volume}{64}},
\bibinfo{pages}{056135}.


\bibitem[{\citenamefont{Kim} \emph{et~al.}(2002)\citenamefont{Kim, Krapivsky, Kahng, and Redner}}]
{Kim:kkk02}
\bibinfo{author}{\bibnamefont{Kim},~\bibfnamefont{J.}},
\bibinfo{author}{\bibfnamefont{P.~L.}~\bibnamefont{Krapivsky}},
\bibinfo{author}{\bibfnamefont{B.}~\bibnamefont{Kahng}},
and
\bibinfo{author}{\bibfnamefont{S.}~\bibnamefont{Redner}},
\bibinfo{year}{2002},
\bibinfo{title}{``Infinite-order percolation and giant fluctuations in a protein interaction network},''
\bibinfo{journal}{Phys. Rev. E} \textbf{\bibinfo{volume}{66}},
\bibinfo{pages}{055101}.


\bibitem[{\citenamefont{Kim and Motter}(2007)}]
{Kim:km07}
\bibinfo{author}{\bibnamefont{Kim},~\bibfnamefont{D.-H.}}
and
\bibinfo{author}{\bibfnamefont{A.~E.}~\bibnamefont{Motter}},
\bibinfo{year}{2007},
\bibinfo{title}{``Ensemble averageability in network spectra},'' 
\bibinfo{journal}{Phys. Rev. Lett.} \textbf{\bibinfo{volume}{98}},
\bibinfo{pages}{248701}.  



\bibitem[{\citenamefont{Kim} \emph{et~al.}(2005)\citenamefont{Kim, Rodgers, Kahng, and Kim}}]
{Kim:krkk05}
\bibinfo{author}{\bibnamefont{Kim},~\bibfnamefont{D.-H.}},
\bibinfo{author}{\bibfnamefont{G.~J.}~\bibnamefont{Rodgers}},
\bibinfo{author}{\bibfnamefont{B.}~\bibnamefont{Kahng}},
and
\bibinfo{author}{\bibfnamefont{D.}~\bibnamefont{Kim}},
\bibinfo{year}{2005},
\bibinfo{title}{``Spin glass transition on scale-free networks},''
\bibinfo{journal}{Phys. Rev. E} \textbf{\bibinfo{volume}{71}},
\bibinfo{pages}{056115}.


\bibitem[{\citenamefont{Kim} \emph{et~al.}(2005)\citenamefont{Kim, Trusina, Minnhagen, and Sneppen}}]
{Kim:ktm05}
\bibinfo{author}{\bibnamefont{Kim},~\bibfnamefont{B.~J.}},
\bibinfo{author}{\bibfnamefont{A.}~\bibnamefont{Trusina}},
\bibinfo{author}{\bibfnamefont{P.}~\bibnamefont{Minnhagen}},
and
\bibinfo{author}{\bibfnamefont{K.}~\bibnamefont{Sneppen}},
\bibinfo{year}{2005},
\bibinfo{title}{``Self organized scale-free networks from merging and regeneration},''
\bibinfo{journal}{Eur. Phys. J. B} \textbf{\bibinfo{volume}{43}},
\bibinfo{pages}{369}.


\bibitem[{\citenamefont{Kinouchi and Copelli}(2006)}]
{Kinouchi:kc06}
\bibinfo{author}{\bibnamefont{Kinouchi},~\bibfnamefont{O.}}
and
\bibinfo{author}{\bibfnamefont{M.}~\bibnamefont{Copelli}},
\bibinfo{year}{2006},
\bibinfo{title}{``Optimal dynamical range of excitable networks at criticality},''
\bibinfo{journal}{Nat. Phys.} \textbf{\bibinfo{volume}{2}},
\bibinfo{pages}{348}.


\bibitem[{\citenamefont{Kirkpatrick}(2005)}]
{Kirkpatrick:k05}
\bibinfo{author}{\bibnamefont{Kirkpatrick},~\bibfnamefont{S.}},
\bibinfo{year}{2005},
\bibinfo{title}{``Jellyfish and other interesting creatures of the Internet},''
http://www.cs.huji.ac.il/\symbol{126}kirk/Jellyfish\_Dimes.ppt.


\bibitem[{\citenamefont{Kleinberg}(1999)}]
{Kleinberg:k99}
\bibinfo{author}{\bibnamefont{Kleinberg},~\bibfnamefont{J.}},
\bibinfo{year}{1999},
\bibinfo{title}{``The small-world phenomenon: An algorithmic perspective.},''
\bibinfo{journal}{Cornell Computer Science Technical Report 99-1776 (October 1999).}


\bibitem[{\citenamefont{Kleinberg}(2000)}]
{Kleinberg:k00}
\bibinfo{author}{\bibfnamefont{Kleinberg}~\bibnamefont{J.}},
\bibinfo{year}{2000},
\bibinfo{title}{``Navigation in a small world},''
\bibinfo{journal}{Nature} \textbf{\bibinfo{volume}{406}},
\bibinfo{pages}{845}.


\bibitem[{\citenamefont{Kleinberg}(2006)}]
{Kleinberg:k06}
\bibinfo{author}{\bibfnamefont{Kleinberg}~\bibnamefont{J.}},
\bibinfo{year}{2006},
\bibinfo{title}{``Complex networks and decentralized search algorithms},''
http://www.cs.cornell.edu/home/kleinber/icm06-swn.pdf.


\bibitem[{\citenamefont{Klemm} \emph{et~al.}(2003)\citenamefont{Klemm, Egu\'iluz, Toral, and San Miguel}}]
{Klemm:ket03}
\bibinfo{author}{\bibnamefont{Klemm},~\bibfnamefont{K.}},
\bibinfo{author}{\bibfnamefont{V.~M.}~\bibnamefont{Egu\'iluz}},
\bibinfo{author}{\bibfnamefont{R.}~\bibnamefont{Toral}},
and
\bibinfo{author}{\bibfnamefont{M.}~\bibnamefont{San Miguel}},
\bibinfo{year}{2003},
\bibinfo{title}{``Nonequilibrium transitions in complex networks: A model of social interaction},''
\bibinfo{journal}{Phys. Rev. E} \textbf{\bibinfo{volume}{67}},
\bibinfo{pages}{026120}.



\bibitem[{\citenamefont{Kori and Mikhailov}(2004)}]
{Kori:km04}
\bibinfo{author}{\bibnamefont{Kori},~\bibfnamefont{H.}}
and
\bibinfo{author}{\bibfnamefont{A.~S.}~\bibnamefont{Mikhailov}},
\bibinfo{year}{2004},
\bibinfo{title}{``Entrainment of randomly coupled oscillator networks by a pacemaker},''
\bibinfo{journal}{Phys. Rev. Lett.} \textbf{\bibinfo{volume}{93}},
\bibinfo{pages}{254101}.


\bibitem[{\citenamefont{Kori and Mikhailov}(2006)}]
{Kori:km06}
\bibinfo{author}{\bibnamefont{Kori},~\bibfnamefont{H.}}
and
\bibinfo{author}{\bibfnamefont{A.~S.}~\bibnamefont{Mikhailov}},
\bibinfo{year}{2006},
\bibinfo{title}{``Strong effect of network architecture of coupled oscillator systems},''
\eprint{cond-mat/0608702}


\bibitem[{\citenamefont{Kosterlitz and Thouless}(1973)}]
{Kosterlitz:kt73}
\bibinfo{author}{\bibnamefont{Kosterlitz},~\bibfnamefont{J.~M.}}
and
\bibinfo{author}{\bibfnamefont{D.~J.}~\bibnamefont{Thouless}},
\bibinfo{year}{1973},
\bibinfo{title}{``Ordering, metastability and phase transitions in two-dimensional systems},''
\bibinfo{journal}{J. Phys. C} \textbf{\bibinfo{volume}{6}},
\bibinfo{pages}{1181}.


\bibitem[{\citenamefont{Kozma and Barrat}(2007)}]
{Kozma:kb07}
\bibinfo{author}{\bibnamefont{Kozma},~\bibfnamefont{B.}} 
and
\bibinfo{author}{\bibfnamefont{A.}~\bibnamefont{Barrat}},
\bibinfo{year}{2007},
\bibinfo{title}{``Consensus formation on adaptive networks},''
\eprint{arXiv:0707.4416 [physics]}.


\bibitem[{\citenamefont{Kozma} \emph{et~al.}(2004)\citenamefont{Kozma, Hastings, and Korniss}}]
{Kozma:khk04}
\bibinfo{author}{\bibnamefont{Kozma},~\bibfnamefont{B.}},
\bibinfo{author}{\bibfnamefont{M.~B.}~\bibnamefont{Hastings}},
and
\bibinfo{author}{\bibfnamefont{G.}~\bibnamefont{Korniss}},
\bibinfo{year}{2004},
\bibinfo{title}{``Roughness scaling for Edwards-Wilkinson
relaxation in small-world networks},''
\bibinfo{journal}{Phys. Rev. Lett.} \textbf{\bibinfo{volume}{92}},
\bibinfo{pages}{108701}.


\bibitem[{\citenamefont{Krapivsky}(1992)}]
{Krapivsky:k92}
\bibinfo{author}{\bibfnamefont{Krapivsky}~\bibnamefont{P.~L.}},
\bibinfo{year}{1992},
\bibinfo{title}{``Kinetics of monomer-monomer surface catalytic reactions},''
\bibinfo{journal}{Phys. Rev. A} \textbf{\bibinfo{volume}{45}},
\bibinfo{pages}{1067}.


\bibitem[{\citenamefont{Krapivsky and Derrida}(2004)}]
{Krapivsky:kd04}
\bibinfo{author}{\bibnamefont{Krapivsky},~\bibfnamefont{P.~L.}}
and
\bibinfo{author}{\bibfnamefont{B.}~\bibnamefont{Derrida}},
\bibinfo{year}{2004},
\bibinfo{title}{``Universal properties of growing networks},''
\bibinfo{journal}{Physica A} \textbf{\bibinfo{volume}{340}},
\bibinfo{pages}{714}.


\bibitem[{\citenamefont{Krapivsky and Redner}(2002)}]
{Krapivsky:kr02}
\bibinfo{author}{\bibnamefont{Krapivsky},~\bibfnamefont{P.~L.}}
and
\bibinfo{author}{\bibfnamefont{S.}~\bibnamefont{Redner}},
\bibinfo{year}{2002},
\bibinfo{title}{``Finiteness and fluctuations in growing networks},''
\bibinfo{journal}{J. Phys. A} \textbf{\bibinfo{volume}{35}},
\bibinfo{pages}{9517}.


\bibitem[{\citenamefont{Krapivsky} \emph{et~al.}(2000)\citenamefont{Krapivsky, Redner, and Leyvraz}}]
{Krapivsky:krl00}
\bibinfo{author}{\bibnamefont{Krapivsky},~\bibfnamefont{P.~L.}},
\bibinfo{author}{\bibfnamefont{S.}~\bibnamefont{Redner}},
and
\bibinfo{author}{\bibfnamefont{F.}~\bibnamefont{Leyvraz}},
\bibinfo{year}{2000},
\bibinfo{title}{``Connectivity of growing random networks},''
\bibinfo{journal}{Phys. Rev. Lett.} \textbf{\bibinfo{volume}{85}},
\bibinfo{pages}{4629}.


\bibitem[{\citenamefont{Krivelevich and Sudakov}(2003)}]
{Krivelevich:ks03}
\bibinfo{author}{\bibnamefont{Krivelevich},~\bibfnamefont{M.}}
and
\bibinfo{author}{\bibfnamefont{B.}~\bibnamefont{Sudakov}},
\bibinfo{year}{2003},
\bibinfo{title}{``The largest eigenvalue of sparse random graphs},''
\bibinfo{journal}{Combinatorics, Probab. Comput.} \textbf{\bibinfo{volume}{12}},
\bibinfo{pages}{61}.


\bibitem[{\citenamefont{Krz{\c a}ka{\l}a} \emph{et~al.}(2007)\citenamefont{Krz{\c a}ka{\l}a, Montanari, Ricci-Tersenghi, Semerjian and Zdeborova}}]
{Krzakala:kmr07}
\bibinfo{author}{\bibnamefont{Krz{\c a}ka{\l}a},~\bibfnamefont{F.}},
\bibinfo{author}{\bibfnamefont{A.}~\bibnamefont{Montanari}},
\bibinfo{author}{\bibfnamefont{F.}~\bibnamefont{Ricci-Tersenghi}},
\bibinfo{author}{\bibfnamefont{G.}~\bibnamefont{Semerjian}},
and
\bibinfo{author}{\bibfnamefont{L.}~\bibnamefont{Zdeborova}},
\bibinfo{year}{2007},
\bibinfo{title}{``Gibbs states and the set of solutions of random constraint satisfaction problems},''
\bibinfo{journal}{PNAS} \textbf{\bibinfo{volume}{104}},
\bibinfo{pages}{10318}.


\bibitem[{\citenamefont{Krz\c{a}ka{\l}a} \emph{et~al.}(2004)\citenamefont{Krz\c{a}ka{\l}a, Pagnani, and Weigt}}]
{Krzakala:kpw04}
\bibinfo{author}{\bibnamefont{Krz\c{a}ka{\l}a},~\bibfnamefont{F.}},
\bibinfo{author}{\bibfnamefont{A.}~\bibnamefont{Pagnani}},
and
\bibinfo{author}{\bibfnamefont{M.}~\bibnamefont{Weigt}},
\bibinfo{year}{2004},
\bibinfo{title}{``Threshold values, stability analysis, and
high-q asymptotics for the coloring problem on random graphs},''
\bibinfo{journal}{Phys. Rev. E} \textbf{\bibinfo{volume}{70}},
\bibinfo{pages}{046705}.



\bibitem[{\citenamefont{Kulkarni} \emph{et~al.}(1999)\citenamefont{Kulkarni, E. Almaas, and Stroud}}]
{Kulkarni:kas99}
\bibinfo{author}{\bibnamefont{Kulkarni},~\bibfnamefont{R.~V.}},
\bibinfo{author}{\bibnamefont{E.},~\bibfnamefont{Almaas}},
and
\bibinfo{author}{\bibfnamefont{D.}~\bibnamefont{Stroud}},
\bibinfo{year}{1999},
\bibinfo{title}{``Evolutionary dynamics in the Bak-Sneppen model on small-world networks},''
\eprint{cond-mat/9905066}.


\bibitem[{\citenamefont{Kumpula} \emph{et~al.}(2007)\citenamefont{Kumpula,Saramaki,
Kaski, and Kertesz}}]
{Kumpula:ksk07}
\bibinfo{author}{\bibnamefont{Kumpula},~\bibfnamefont{J.~M.}},
\bibinfo{author}{\bibfnamefont{J.}~\bibnamefont{Saramaki}},
\bibinfo{author}{\bibfnamefont{K.}~\bibnamefont{Kaski}},
and
\bibinfo{author}{\bibfnamefont{J.}~\bibnamefont{Kertesz}},
\bibinfo{year}{2007},
\bibinfo{title}{``Limited resolution
in complex network community detection with Potts model approach},''
\bibinfo{journal}{Eur. Phys. J. B} \textbf{\bibinfo{volume}{56}},
\bibinfo{pages}{41}.


\bibitem[{\citenamefont{Kuramoto}(1984)}]
{Kuramoto:kbook84}
\bibinfo{author}{\bibnamefont{Kuramoto}, \bibfnamefont{Y.}},
\bibinfo{year}{1984},
\emph{\bibinfo{title}{Chemical Oscillations, Waves and
Turbulence}}
(\bibinfo{publisher}{Springer, New York}).


\bibitem[{\citenamefont{Kwon and Moon}(2002)}]
{Kwon:km02}
\bibinfo{author}{\bibnamefont{Kwon},~\bibfnamefont{O.}}
and
\bibinfo{author}{\bibfnamefont{H.-T.}~\bibnamefont{Moon}},
\bibinfo{year}{2002},
\bibinfo{title}{``Coherence resonance in small-world networks of excitable cells},''
\bibinfo{journal}{Phys. Lett. A} \textbf{\bibinfo{volume}{298}},
\bibinfo{pages}{319}.


\bibitem[{\citenamefont{Kwon and Thouless}(1988)}]
{Kwon:kt88}
\bibinfo{author}{\bibnamefont{Kwon},~\bibfnamefont{C.}}
and
\bibinfo{author}{\bibfnamefont{D.~J.}~\bibnamefont{Thouless}},
\bibinfo{year}{1988},
\bibinfo{title}{``Ising spin glass at zero temperature on the Bethe lattice},''
\bibinfo{journal}{Phys. Rev. B} \textbf{\bibinfo{volume}{37}},
\bibinfo{pages}{7649}.



\bibitem[{\citenamefont{Lacour-Gayet and Toulouse}(1974)}]
{Lacour:lt74}
\bibinfo{author}{\bibnamefont{Lacour-Gayet},~\bibfnamefont{P.}}
and
\bibinfo{author}{\bibfnamefont{G.}~\bibnamefont{Toulouse}},
\bibinfo{year}{1974},
\bibinfo{title}{``Ideal Bose-Einstein condensation and disorder effects},''
\bibinfo{journal}{J. Phys. (Paris)} \textbf{\bibinfo{volume}{35}},
\bibinfo{pages}{425}.


\bibitem[{\citenamefont{Lago-Fern\'{a}ndez} \emph{et~al.}(2000)\citenamefont{Lago-Fern\'{a}ndez,
Huerta, Corbacho, and Sig\"{u}enza}}]
{Lago-Fernandez:lhcs00}
\bibinfo{author}{\bibnamefont{Lago-Fern\'{a}ndez},~\bibfnamefont{L.~F.}},
\bibinfo{author}{\bibfnamefont{R.}~\bibnamefont{Huerta}},
\bibinfo{author}{\bibfnamefont{F.}~\bibnamefont{Corbacho}},
and
\bibinfo{author}{\bibfnamefont{J.~A.}~\bibnamefont{Sig\"{u}enza}},
\bibinfo{year}{2000},
\bibinfo{title}{``Fast response and temporal coherent oscillations in
small-world networks},''
\bibinfo{journal}{Phys. Rev. Lett.} \textbf{\bibinfo{volume}{84}},
\bibinfo{pages}{2758}.




\bibitem[{\citenamefont{Lahtinen} \emph{et~al.}(2005)\citenamefont{Lahtinen, Kert\'esz, and Kaski}}]
{Lahtinen:lkk05}
\bibinfo{author}{\bibnamefont{Lahtinen},~\bibfnamefont{J.}},
\bibinfo{author}{\bibfnamefont{J.}~\bibnamefont{Kert\'esz}},
and
\bibinfo{author}{\bibfnamefont{K.}~\bibnamefont{Kaski}},
\bibinfo{year}{2005},
\bibinfo{title}{``Sandpiles on Watts-Strogatz type small-worlds},''
\bibinfo{journal}{Physica A} \textbf{\bibinfo{volume}{349}},
\bibinfo{pages}{535}.




\bibitem[{\citenamefont{Lambiotte and Ausloos}(2005)}]
{Lambiotte:la05}
\bibinfo{author}{\bibnamefont{Lambiotte},~\bibfnamefont{R.}}
and
\bibinfo{author}{\bibfnamefont{M.}~\bibnamefont{Ausloos}},
\bibinfo{year}{2005},
\bibinfo{title}{``Uncovering collective listening habits and music
genres in bipartite networks},''
\bibinfo{journal}{Phys. Rev. E} \textbf{\bibinfo{volume}{72}},
\bibinfo{pages}{066107}.


\bibitem[{\citenamefont{Lancaster}(2002)}]
{Lancaster:l02}
\bibinfo{author}{\bibnamefont{Lancaster},~\bibfnamefont{D.}},
\bibinfo{year}{2002},
\bibinfo{title}{``Cluster growth in two growing network models},''
\bibinfo{journal}{J. Phys. A} \textbf{\bibinfo{volume}{35}},
\bibinfo{pages}{1179}.


\bibitem[{\citenamefont{Lee}(2005)}]
{Lee:l05}
\bibinfo{author}{\bibnamefont{Lee},~\bibfnamefont{D.-S.}}
\bibinfo{year}{2005},
\bibinfo{title}{``Synchronization transition in scale-free networks: Clusters of
synchrony},''
\bibinfo{journal}{Phys. Rev. E} \textbf{\bibinfo{volume}{72}},
\bibinfo{pages}{026208}.


\bibitem[{\citenamefont{Lee} \emph{et~al.}(2004a)\citenamefont{Lee, Goh, Kahng, and Kim}}]
{Lee:lgk04a}
\bibinfo{author}{\bibnamefont{Lee},~\bibfnamefont{D.-S.}},
\bibinfo{author}{\bibfnamefont{K.-I.}~\bibnamefont{Goh}},
\bibinfo{author}{\bibfnamefont{B.}~\bibnamefont{Kahng}},
and
\bibinfo{author}{\bibfnamefont{D.}~\bibnamefont{Kim}},
\bibinfo{year}{2004a},
\bibinfo{title}{``Sandpile avalanche dynamics on scale-free networks,},''
\bibinfo{journal}{Physica A} \textbf{\bibinfo{volume}{338}},
\bibinfo{pages}{84}.


\bibitem[{\citenamefont{Lee} \emph{et~al.}(2004b)\citenamefont{Lee, Goh, Kahng, and Kim}}]
{Lee:lgk04b}
\bibinfo{author}{\bibnamefont{Lee},~\bibfnamefont{D.-S.}},
\bibinfo{author}{\bibfnamefont{K.-I.}~\bibnamefont{Goh}},
\bibinfo{author}{\bibfnamefont{B.}~\bibnamefont{Kahng}},
and
\bibinfo{author}{\bibfnamefont{D.}~\bibnamefont{Kim}},
\bibinfo{year}{2004b},
\bibinfo{title}{``Branching process approach to avalanche dynamics on complex networks},''
\bibinfo{journal}{J. Korean Phys. Soc.} \textbf{\bibinfo{volume}{44}},
\bibinfo{pages}{633}.


\bibitem[{\citenamefont{Lee} \emph{et~al.}(2004c)\citenamefont{Lee, Goh, Kahng and Kim}}]
{Lee:lgk04c}
\bibinfo{author}{\bibnamefont{Lee},~\bibfnamefont{D.-S.}},
\bibinfo{author}{\bibfnamefont{K.-I.}~\bibnamefont{Goh}},
\bibinfo{author}{\bibfnamefont{B.}~\bibnamefont{Kahng}},
and
\bibinfo{author}{\bibfnamefont{D.}~\bibnamefont{Kim}},
\bibinfo{year}{2004c},
\bibinfo{title}{``Evolution of scale-free random graphs: Potts model formulation},''
\bibinfo{journal}{Nucl. Phys. B} \textbf{\bibinfo{volume}{696}},
\bibinfo{pages}{351}.


\bibitem[{\citenamefont{Lee} \emph{et~al.}(2005a)\citenamefont{Lee, Goh, Kahng and Kim}}]
{Lee:lgk05}
\bibinfo{author}{\bibnamefont{ Lee},~\bibfnamefont{E.~J.}},
\bibinfo{author}{\bibfnamefont{K.-I.}~\bibnamefont{Goh}},
\bibinfo{author}{\bibfnamefont{B.}~\bibnamefont{Kahng}},
and
\bibinfo{author}{\bibfnamefont{D.}~\bibnamefont{Kim}},
\bibinfo{year}{2005a},
\bibinfo{title}{``Robustness of the avalanche dynamics in data packet transport on scale-free networks},''
\bibinfo{journal}{Phys. Rev. E} \textbf{\bibinfo{volume}{71}},
\bibinfo{pages}{056108}.


\bibitem[{\citenamefont{Lee} \emph{et~al.}(2005b)\citenamefont{Lee, Hong, and Lee}}]
{Lee:lhl05}
\bibinfo{author}{\bibnamefont{Lee},~\bibfnamefont{K.~E.}},
\bibinfo{author}{\bibfnamefont{B.~H.}~\bibnamefont{Hong}},
and
\bibinfo{author}{\bibfnamefont{J.~W.}~\bibnamefont{Lee}},
\bibinfo{year}{2005b},
\bibinfo{title}{``Universality class of Bak-Sneppen model on scale-free network},''
\eprint{cond-mat/0510067}.


\bibitem[{\citenamefont{Lee} \emph{et~al.}(2006a)\citenamefont{Lee, Goh, Kahng, and Kim}}]
{Lee:lgk06}
\bibinfo{author}{\bibnamefont{Lee},~\bibfnamefont{J.-S.}},
\bibinfo{author}{\bibfnamefont{K.-I.}~\bibnamefont{Goh}},
\bibinfo{author}{\bibfnamefont{B.}~\bibnamefont{ Kahng}},
and
\bibinfo{author}{\bibfnamefont{D.}~\bibnamefont{Kim}},
\bibinfo{year}{2006a},
\bibinfo{title}{``Intrinsic degree-correlations in static model of scale-free networks},''
\bibinfo{journal}{Eur. Phys. J. B} \textbf{\bibinfo{volume}{49}},
\bibinfo{pages}{231}. 


\bibitem[{\citenamefont{Lee} \emph{et~al.}(2006b)\citenamefont{Lee, Jeong, and Noh}}]
{Lee:ljn06}
\bibinfo{author}{\bibnamefont{Lee},~\bibfnamefont{S.~H.}},
\bibinfo{author}{\bibfnamefont{H.}~\bibnamefont{Jeong}},
and
\bibinfo{author}{\bibfnamefont{J.~D.}~\bibnamefont{Noh}},
\bibinfo{year}{2006b},
\bibinfo{title}{``Random field Ising model on networks with
inhomogeneous connections},''
\bibinfo{journal}{Phys. Rev. E} \textbf{\bibinfo{volume}{74}},
\bibinfo{pages}{031118}.


\bibitem[{\citenamefont{Leland} \emph{et~al.}(1994)\citenamefont{Leland, Taqqu, Willinger, and Wilson}}]
{Leland:ltw94}
\bibinfo{author}{\bibnamefont{Leland},~\bibfnamefont{W.~E.}},
\bibinfo{author}{\bibfnamefont{M.~S.}~\bibnamefont{Taqqu}},
\bibinfo{author}{\bibfnamefont{W.}~\bibnamefont{Willinger}},
and
\bibinfo{author}{\bibfnamefont{D.~W.}~\bibnamefont{Wilson}},
\bibinfo{year}{1994},
\bibinfo{title}{``On the self-similar nature of Ethernet traffic (extended version)},''
\bibinfo{journal}{IEEE/ACM Trans. Network.} \textbf{\bibinfo{volume}{2}},
\bibinfo{pages}{1}.




\bibitem[{\citenamefont{Leone} \emph{et~al.}(2002)\citenamefont{Leone, A. V\'azquez, A. Vespignani, and R. Zecchina}}]
{Leone:lvv02}
\bibinfo{author}{\bibnamefont{Leone},~\bibfnamefont{M.}},
\bibinfo{author}{\bibfnamefont{A.}~\bibnamefont{V\'azquez}},
\bibinfo{author}{\bibfnamefont{A.}~\bibnamefont{Vespignani}},
and
\bibinfo{author}{\bibfnamefont{R.}~\bibnamefont{Zecchina}},
\bibinfo{year}{2002},
\bibinfo{title}{``Ferromagnetic ordering in graphs with arbitrary degree distribution},''
\bibinfo{journal}{Eur. Phys. J. B} \textbf{\bibinfo{volume}{28}},
\bibinfo{pages}{191}.


\bibitem[{\citenamefont{Li} \emph{et~al.}(2007)\citenamefont{Li, Braunstein, Buldyrev, Havlin, and Stanley}}]
{Li:lbb07}
\bibinfo{author}{\bibnamefont{Li},~\bibfnamefont{G.}},
\bibinfo{author}{\bibfnamefont{L.~A.}~\bibnamefont{Braunstein}},
\bibinfo{author}{\bibfnamefont{S.~V.}~\bibnamefont{Buldyrev}},
\bibinfo{author}{\bibfnamefont{S.}~\bibnamefont{Havlin}},
and
\bibinfo{author}{\bibfnamefont{H.~E.}~\bibnamefont{Stanley}},
\bibinfo{year}{2007},
\bibinfo{title}{``Transport and percolation theory in weighted networks},''
\bibinfo{journal}{Phys. Rev. E} \textbf{\bibinfo{volume}{75}},
\bibinfo{pages}{045103}.


\bibitem[{\citenamefont{Liers} \emph{et~al.}(2003)\citenamefont{Liers, Palassini, Hartmann, and J\"{u}nger}}]
{Liers:lphj03}
\bibinfo{author}{\bibnamefont{Liers},~\bibfnamefont{F.}},
\bibinfo{author}{\bibfnamefont{M.}~\bibnamefont{Palassini}},
\bibinfo{author}{\bibfnamefont{A.~K.}~\bibnamefont{Hartmann}},
and
\bibinfo{author}{\bibfnamefont{M.}~\bibnamefont{J\"{u}nger}},
\bibinfo{year}{2003},
\bibinfo{title}{``Ground state of the Bethe lattice spin glass and running time of an exact optimization
algorithm},''
\bibinfo{journal}{Phys. Rev. B} \textbf{\bibinfo{volume}{68}},
\bibinfo{pages}{094406}.


\bibitem[{\citenamefont{Lima and Stauffer}(2006)}]
{Lima:ls06}
\bibinfo{author}{\bibnamefont{Lima},~\bibfnamefont{F.~W.~S.}}
and
\bibinfo{author}{\bibfnamefont{D.}~\bibnamefont{Stauffer}},
\bibinfo{year}{2006},
\bibinfo{title}{``Ising model simulations in directed lattices and networks},''
\bibinfo{journal}{Physica A} \textbf{\bibinfo{volume}{359}},
\bibinfo{pages}{423}.


\bibitem[{\citenamefont{Lind} \emph{et~al.}(2004)\citenamefont{Lind, Gallas, and Herrmann}}]
{Lind:lgh04}
\bibinfo{author}{\bibnamefont{Lind},~\bibfnamefont{P.~G.}},
\bibinfo{author}{\bibfnamefont{J.~A.~C.}~\bibnamefont{Gallas}},
and
\bibinfo{author}{\bibfnamefont{H.~J.}~\bibnamefont{Herrmann}},
\bibinfo{year}{2004},
\bibinfo{title}{``Coherence in scale-free networks of chaotic
maps},''
\bibinfo{journal}{Phys. Rev. E} \textbf{\bibinfo{volume}{70}},
\bibinfo{pages}{056207}.


\bibitem[{\citenamefont{Liu} \emph{et~al.}(2006)\citenamefont{Liu, Ma, Zhang, Sun, and Hui}}]
{Liu:lmz06}
\bibinfo{author}{\bibnamefont{Liu},~\bibfnamefont{Z.}},
\bibinfo{author}{\bibfnamefont{W.}~\bibnamefont{Ma}},
\bibinfo{author}{\bibfnamefont{H.}~\bibnamefont{Zhang}},
\bibinfo{author}{\bibfnamefont{Y.}~\bibnamefont{Sun}},
and
\bibinfo{author}{\bibfnamefont{P.~M.}~\bibnamefont{Hui}},
\bibinfo{year}{2006},
\bibinfo{title}{``An efficient approach of controlling traffic congestion in scale-free networks},''
\bibinfo{journal}{Physica A} \textbf{\bibinfo{volume}{370}},
\bibinfo{pages}{843}.


\bibitem[{\citenamefont{Lopes} \emph{et~al.}(2004)\citenamefont{Lopes, Pogorelov, dos Santos, and
Toral}}] {Lopes:lpd04}
\bibinfo{author}{\bibnamefont{Lopes},~\bibfnamefont{J.~V.}},
\bibinfo{author}{\bibfnamefont{Y.~G.}~\bibnamefont{Pogorelov}},
\bibinfo{author}{\bibfnamefont{J.~M.~B.~L.}~\bibnamefont{dos Santos}},
and
\bibinfo{author}{\bibfnamefont{R.}~\bibnamefont{Toral}},
\bibinfo{year}{2004},
\bibinfo{journal}{Phys. Rev. E} \textbf{\bibinfo{volume}{70}},
\bibinfo{pages}{026112}.




\bibitem[{\citenamefont{L\'opez} \emph{et~al.}(2007)\citenamefont{L\'opez, Parshani, Cohen, Carmi, and Havlin}}]
{Lopez:lpc07}
\bibinfo{author}{\bibnamefont{L\'opez},~\bibfnamefont{E.}},
\bibinfo{author}{\bibfnamefont{R.}~\bibnamefont{Parshani}},
\bibinfo{author}{\bibfnamefont{R.}~\bibnamefont{Cohen}},
\bibinfo{author}{\bibfnamefont{S.}~\bibnamefont{Carmi}},
and
\bibinfo{author}{\bibfnamefont{S.}~\bibnamefont{Havlin}},
\bibinfo{year}{2007},
\bibinfo{title}{``Limited path percolation in complex networks},''
\eprint{cond-mat/0702691}.

 
\bibitem[{\citenamefont{L\"ubeck and Janssen}(2005)}]
{Lubeck:lj05}
\bibinfo{author}{\bibnamefont{L\"ubeck},~\bibfnamefont{S.}} 
and
\bibinfo{author}{\bibfnamefont{H.-K.}~\bibnamefont{Janssen}},
\bibinfo{year}{2005},
\bibinfo{title}{``Finite-size scaling of directed percolation above the upper critical dimension},''
\bibinfo{journal}{Phys. Rev. E} \textbf{\bibinfo{volume}{72}},
\bibinfo{pages}{016119}.


\bibitem[{\citenamefont{Luczak}(1991)}]
{Luczak:l91}
\bibinfo{author}{\bibnamefont{Luczak},~\bibfnamefont{T.}}
\bibinfo{year}{1991},
\bibinfo{title}{``The chromatic number of random graphs},''
\bibinfo{journal}{Combinatorica} \textbf{\bibinfo{volume}{11}},
\bibinfo{pages}{45}.


\bibitem[{\citenamefont{Luijten and Bl\"ote}(1997)}]
{Luijten:lb97}
\bibinfo{author}{\bibnamefont{Luijten},~\bibfnamefont{E.}}
and
\bibinfo{author}{\bibfnamefont{H.~W.~J.}~\bibnamefont{Bl\"ote}},
\bibinfo{year}{1997},
\bibinfo{title}{``Classical critical behavior of spin models with long-range interactions},''
\bibinfo{journal}{Phys. Rev. B} \textbf{\bibinfo{volume}{56}},
\bibinfo{pages}{8945}.




\bibitem[{\citenamefont{Lyons}(1989)}]
{Lyons:l89}
\bibinfo{author}{\bibnamefont{Lyons},~\bibfnamefont{R.}},
\bibinfo{year}{1989},
\bibinfo{title}{``The Ising model and percolation on trees and tree-like graphs},''
\bibinfo{journal}{Commun. Math. Phys.} \textbf{\bibinfo{volume}{125}},
\bibinfo{pages}{337}.


\bibitem[{\citenamefont{Lyons}(1990)}]
{Lyons:l90}
\bibinfo{author}{\bibfnamefont{Lyons}~\bibnamefont{R.}},
\bibinfo{year}{1990},
\bibinfo{title}{``Random walks and percolation on trees},''
\bibinfo{journal}{Ann. Probab.} \textbf{\bibinfo{volume}{18}},
\bibinfo{pages}{931}.


\bibitem[{\citenamefont{Lyons} \emph{et~al.}(1996)\citenamefont{Lyons, Pemantle, and Peres}}]
{Lyons:lpp96}
\bibinfo{author}{\bibnamefont{Lyons},~\bibfnamefont{R.}},
\bibinfo{author}{\bibfnamefont{R.}~\bibnamefont{Pemantle}}, and
\bibinfo{author}{\bibfnamefont{Y.}~\bibnamefont{Peres}}, \bibinfo{year}{1996},
\bibinfo{title}{``Biased random walks on Galton-Watson trees},''
\bibinfo{journal}{Prob. Theor. Relat. Fields} \textbf{\bibinfo{volume}{106}},
\bibinfo{pages}{249}.





\bibitem[{\citenamefont{Malarz} \emph{et~al.}(2007)\citenamefont{Malarz, Antosiewicz, Karpinska, Kulakowski, and Tadic}}]
{Malarz:mak07}
\bibinfo{author}{\bibfnamefont{Malarz}~\bibnamefont{K.}},
\bibinfo{author}{\bibfnamefont{W.}~\bibnamefont{Antosiewicz}},
\bibinfo{author}{\bibfnamefont{J.}~\bibnamefont{Karpinska}},
\bibinfo{author}{\bibfnamefont{K.}~\bibnamefont{Kulakowski}},
and
\bibinfo{author}{\bibfnamefont{B.}~\bibnamefont{Tadi\'c}},
\bibinfo{year}{2007},
\bibinfo{title}{``Avalanches in complex spin networks},''
\bibinfo{journal}{Physica A} \textbf{\bibinfo{volume}{373}},
\bibinfo{pages}{785}.


\bibitem[{\citenamefont{Martel and Nguyen}(2004)}]
{Martel:mn04}
\bibinfo{author}{\bibnamefont{Martel},~\bibfnamefont{C.}}
and
\bibinfo{author}{\bibnamefont{V.},~\bibfnamefont{Nguyen}},
\bibinfo{year}{2004},
\bibinfo{title}{``Analyzing Kleinberg's (and other) small-world models},''
in \emph{\bibinfo{booktitle}{Proceedings of the twenty-third annual
ACM symposium on Principles of distributed computing, St. John's,
Newfoundland, Canada}},
p. \bibinfo{pages}{179}.



\bibitem[{\citenamefont{Maslov and Sneppen}(2002)}]
{Maslov:ms02}
\bibinfo{author}{\bibnamefont{Maslov},~\bibfnamefont{S.}}
and
\bibinfo{author}{\bibfnamefont{K.}~\bibnamefont{Sneppen}}, \bibinfo{year}{2002},
\bibinfo{title}{``Specificity and stability in topology of protein networks},''
\bibinfo{journal}{Science} \textbf{\bibinfo{volume}{296}},
\bibinfo{pages}{910}.


\bibitem[{\citenamefont{Masuda} \emph{et~al.}(2005)\citenamefont{Masuda, Goh, and Kahng}}]
{Masuda:mgk05}
\bibinfo{author}{\bibnamefont{Masuda},~\bibfnamefont{N.}},
\bibinfo{author}{\bibfnamefont{K.-I.}~\bibnamefont{Goh}},
and
\bibinfo{author}{\bibfnamefont{B.}~\bibnamefont{Kahng}},
\bibinfo{year}{2005},
\bibinfo{title}{``Extremal dynamics on complex networks: Analytic solutions},''
\bibinfo{journal}{Phys. Rev. E} \textbf{\bibinfo{volume}{72}},
\bibinfo{pages}{066106}.


\bibitem[{\citenamefont{Matsuda}(1974)}]
{Matsuda:74}
\bibinfo{author}{\bibnamefont{Matsuda},~\bibfnamefont{H.}}
\bibinfo{year}{1974},
\bibinfo{title}{``Infinite susceptibility without spontaneous magnetization: Exact
properties of the Ising model on the Cayley tree},''
\bibinfo{journal}{Prog. Theor. Phys.} \textbf{\bibinfo{volume}{51}},
\bibinfo{pages}{1053}.


\bibitem[{\citenamefont{May and Lloyd}(2001)}]
{May:ml01}
\bibinfo{author}{\bibfnamefont{May}~\bibnamefont{R. M.}}
and
\bibinfo{author}{\bibfnamefont{A. L.}~\bibnamefont{Lloyd}},
\bibinfo{year}{2001},
\bibinfo{title}{``Infection dynamics on scale-free networks},''
\bibinfo{journal}{Phys. Rev. E} \textbf{\bibinfo{volume}{64}},
\bibinfo{pages}{066112}.


\bibitem[{\citenamefont{McEliece} \emph{et~al.}(1998)\citenamefont{McEliece, MacKay, and Cheng}}]
{McEliece:mmc98}
\bibinfo{author}{\bibnamefont{McEliece},~\bibfnamefont{R.~J.}},
\bibinfo{author}{\bibfnamefont{D.~J.~C.}~\bibnamefont{MacKay}},
and
\bibinfo{author}{\bibfnamefont{J.~F.}~\bibnamefont{Cheng}},
\bibinfo{year}{1998},
\bibinfo{title}{``Turbo decoding as an instance of Pearl's "beliefpropagation" algorithm },''
\bibinfo{journal}{IEEE J. Select. Areas Commun.} \textbf{\bibinfo{volume}{16}},
\bibinfo{pages}{140}.


\bibitem[{\citenamefont{McGraw and Menzinger}(2006)}]
{McGraw:mm06}
\bibinfo{author}{\bibnamefont{McGraw},~\bibfnamefont{P.}}
and
\bibinfo{author}{\bibfnamefont{M.}~\bibnamefont{Menzinger}},
\bibinfo{year}{2006},
\bibinfo{title}{``Analysis of nonlinear synchronization dynamics of oscillator networks by Laplacian spectral methods},''
\eprint{cond-mat/0610522}.


\bibitem[{\citenamefont{Medvedyeva} \emph{et~al.}(2003)\citenamefont{Medvedyeva, Holme, Minnhagen, and Kim}}]
{Medvedyeva:mhm03}
\bibinfo{author}{\bibnamefont{Medvedyeva},~\bibfnamefont{K.}},
\bibinfo{author}{\bibfnamefont{P.}~\bibnamefont{Holme}},
\bibinfo{author}{\bibfnamefont{P.}~\bibnamefont{Minnhagen}},
and
\bibinfo{author}{\bibfnamefont{B.~J.}~\bibnamefont{Kim}},
\bibinfo{year}{2003},
\bibinfo{title}{``Dynamic critical behavior of the XY model in small-world networks},''
\bibinfo{journal}{Phys. Rev. E} \textbf{\bibinfo{volume}{67}},
\bibinfo{pages}{036118}.


\bibitem[{\citenamefont{Melin} \emph{et~al.}(1996)\citenamefont{Melin, Angles d'Auriac, Chandra, and Doucot}}]
{Melin:mac96}
\bibinfo{author}{\bibnamefont{Melin},~\bibfnamefont{R.}},
\bibinfo{author}{\bibfnamefont{J. C.}~\bibnamefont{Angles d'Auriac}},
\bibinfo{author}{\bibfnamefont{P.}~\bibnamefont{Chandra}},
and
\bibinfo{author}{\bibfnamefont{B.}~\bibnamefont{Doucot}},
\bibinfo{year}{1996},
\bibinfo{title}{``Glassy behavior in the ferromagnetic Ising model on a Cayley tree},''
\bibinfo{journal}{J. Phys. A} \textbf{\bibinfo{volume}{29}},
\bibinfo{pages}{5773}.




\bibitem[{\citenamefont{Mertens} \emph{et~al.}(2003)\citenamefont{Mertens, M\'ezard, and Zecchina,}}]
{Mertens:mmz03}
\bibinfo{author}{\bibnamefont{Mertens},~\bibfnamefont{S.}},
\bibinfo{author}{\bibfnamefont{M.}~\bibnamefont{M\'ezard}},
and
\bibinfo{author}{\bibfnamefont{R.}~\bibnamefont{Zecchina}},
\bibinfo{year}{2003},
\bibinfo{title}{``Threshold values of Random K-SAT from the cavity method},''
\bibinfo{journal}{Rand. Struct.and Alg.} \textbf{\bibinfo{volume}{28}},
\bibinfo{pages}{340}.




\bibitem[{\citenamefont{M\'ezard} \emph{et~al.}(2005)\citenamefont{M\'ezard, Mora, and Zecchina}}]
{Mezard:mmz05}
\bibinfo{author}{\bibnamefont{M\'ezard},~\bibfnamefont{M.}},
\bibinfo{author}{\bibfnamefont{T.}~\bibnamefont{Mora}},
and
\bibinfo{author}{\bibfnamefont{R.}~\bibnamefont{Zecchina}},
\bibinfo{year}{2005},
\bibinfo{title}{``Clustering of solutions in the random satisfiability problem},''
\bibinfo{journal}{Phys. Rev. Lett.} \textbf{\bibinfo{volume}{94}},
\bibinfo{pages}{197205}.


\bibitem[{\citenamefont{M\'ezard and Parisi}(2001)}]
{Mezard:mp01}
\bibinfo{author}{\bibnamefont{M\'ezard},~\bibfnamefont{M.}}
and
\bibinfo{author}{\bibfnamefont{G.}~\bibnamefont{Parisi}},
\bibinfo{year}{2001},
\bibinfo{title}{``The Bethe lattice spin glass revisited},''
\bibinfo{journal}{Eur. Phys. J. B} \textbf{\bibinfo{volume}{20}},
\bibinfo{pages}{217}.


\bibitem[{\citenamefont{M\'ezard and Parisi}(2003)}]
{Mezard:mp03}
\bibinfo{author}{\bibnamefont{M\'ezard},~\bibfnamefont{M.}}
and
\bibinfo{author}{\bibfnamefont{G.}~\bibnamefont{Parisi}},
\bibinfo{year}{2003},
\bibinfo{title}{``The cavity method at zero temperature},''
\bibinfo{journal}{J. Stat. Phys.} \textbf{\bibinfo{volume}{111}},
\bibinfo{pages}{1}.


\bibitem[{\citenamefont{M\'ezard}\emph{et~al.}(1987)\citenamefont{M\'ezard, Parisi, and Virasoro}}]
{Mezard:mpv87}
\bibinfo{author}{\bibnamefont{M\'ezard},~\bibfnamefont{M.}},
\bibinfo{author}{\bibfnamefont{G.}~\bibnamefont{Parisi}},
and
\bibinfo{author}{\bibfnamefont{M.~A.}~\bibnamefont{Virasoro}},
\bibinfo{year}{1987},
\emph{\bibinfo{title}{Spin glass theory and beoyond}}
(\bibinfo{publisher}{World Scientific, Singapur}).


\bibitem[{\citenamefont{M\'ezard} \emph{et~al.}(2002)\citenamefont{M\'ezard, Parisi, and Zecchina}}]
{Mezard:mpz02}
\bibinfo{author}{\bibnamefont{M\'ezard},~\bibfnamefont{M.}},
\bibinfo{author}{\bibfnamefont{G.}~\bibnamefont{Parisi}},
and
\bibinfo{author}{\bibfnamefont{R.}~\bibnamefont{Zecchina}},
\bibinfo{year}{2002},
\bibinfo{title}{``Analytic and algorithmic solution of random satisfiability problems},''
\bibinfo{journal}{Science} \textbf{\bibinfo{volume}{297}},
\bibinfo{pages}{812}.


\bibitem[{\citenamefont{M\'ezard and Tarzia}(2007)}]
{Mezard:mt07}
\bibinfo{author}{\bibnamefont{M\'ezard},~\bibfnamefont{M.}}
and
\bibinfo{author}{\bibfnamefont{M.}~\bibnamefont{Tarzia}},
\bibinfo{year}{2007},
\bibinfo{title}{``Statistical mechanics of the hyper vertex cover problem},''
\eprint{arXiv:0707.0189 [cond-mat]}.




\bibitem[{\citenamefont{M\'ezard and R. Zecchina}(2002)}]
{Mezard:mz02}
\bibinfo{author}{\bibnamefont{M\'ezard},~\bibfnamefont{M.}}
and
\bibinfo{author}{\bibfnamefont{R.}~\bibnamefont{Zecchina}},
\bibinfo{year}{2002},
\bibinfo{title}{``Random K-satisfiability problem: From an analytic solution to an efficient algorithm},''
\bibinfo{journal}{Phys. Rev. E} \textbf{\bibinfo{volume}{66}},
\bibinfo{pages}{056126}.




\bibitem[{\citenamefont{Miller}(2007)}]
{Miller:m07}
\bibinfo{author}{\bibnamefont{Miller},~\bibfnamefont{J.~C.}}, 
\bibinfo{year}{2007},
\bibinfo{title}{``Epidemic size and probability in populations with heterogeneous infectivity and susceptibility},''
\bibinfo{journal}{Phys. Rev. E} \textbf{\bibinfo{volume}{76}},
\bibinfo{pages}{010101 (R)}.


\bibitem[{\citenamefont{Minnhagen} \emph{et~al.}(2004)\citenamefont{Minnhagen, Rosvall, Sneppen, and Trusina}}]
{Minnhagen:mrs04}
\bibinfo{author}{\bibnamefont{Minnhagen},~\bibfnamefont{P.}},
\bibinfo{author}{\bibfnamefont{M.}~\bibnamefont{Rosvall}},
\bibinfo{author}{\bibfnamefont{K.}~\bibnamefont{Sneppen}},
and
\bibinfo{author}{\bibfnamefont{A.}~\bibnamefont{Trusina}},
\bibinfo{year}{2004},
\bibinfo{title}{``Self-organization of structures and networks from merging and small-scale fluctuations},''
\bibinfo{journal}{Physica A} \textbf{\bibinfo{volume}{340}},
\bibinfo{pages}{725}.





\bibitem[{\citenamefont{Molloy and Reed}(1995)}]
{Molloy:mr95}
\bibinfo{author}{\bibnamefont{Molloy},~\bibfnamefont{M.}}
and
\bibinfo{author}{\bibfnamefont{B.~A.}~\bibnamefont{Reed}},
\bibinfo{year}{1995},
\bibinfo{title}{``A critical point for random graphs with a given degree sequence},''
\bibinfo{journal}{Random Struct. Algor.} \textbf{\bibinfo{volume}{6}},
\bibinfo{pages}{161}.


\bibitem[{\citenamefont{Molloy and Reed}(1998)}]
{Molloy:mr98}
\bibinfo{author}{\bibnamefont{Molloy},~\bibfnamefont{M.}}
and
\bibinfo{author}{\bibfnamefont{B.~A.}~\bibnamefont{Reed}},
\bibinfo{year}{1998},
\bibinfo{title}{``The size of the giant component of a random graph with a given degree sequence},''
\bibinfo{journal}{Combin. Prob. Comp.} \textbf{\bibinfo{volume}{7}},
\bibinfo{pages}{295}.






\bibitem[{\citenamefont{Montanari}(2005)}]
{Montanari:m05}
\bibinfo{author}{\bibnamefont{Montanari},~\bibfnamefont{A.}},
\bibinfo{year}{2005},
\bibinfo{title}{``Two lectures on iterative coding and statistical mechanics},''
\eprint{cond-mat/0512296}.


\bibitem[{\citenamefont{Montanari and Rizzo}(2005)}]
{Montanari:mr05}
\bibinfo{author}{\bibnamefont{Montanari},~\bibfnamefont{A.}}
and
\bibinfo{author}{\bibfnamefont{T.}~\bibnamefont{Rizzo}},
\bibinfo{year}{2005},
\bibinfo{title}{``How to compute loop corrections to Bethe approximation},''
\bibinfo{journal}{J. Stat. Mech.} 
\bibinfo{pages}{P10011}.




\bibitem[{\citenamefont{Mooij and Kappen}(2005)}]
{Mooij:mk05}
\bibinfo{author}{\bibnamefont{Mooij},~\bibfnamefont{J.~M.}}
and
\bibinfo{author}{\bibfnamefont{H.~J.}~\bibnamefont{Kappen}},
\bibinfo{year}{2005},
\bibinfo{title}{``On the properties of the Bethe approximation and loopy belief propagation on binary networks},''
\bibinfo{journal}{J. Stat. Mech.}
\bibinfo{pages}{P11012}.


\bibitem[{\citenamefont{Moore and Newman}(2000a)}]
{Moore:mn00a}
\bibinfo{author}{\bibnamefont{Moore},~\bibfnamefont{C.}}
and
\bibinfo{author}{\bibfnamefont{M.~E.~J.}~\bibnamefont{Newman}},
\bibinfo{year}{2000a},
\bibinfo{title}{``Epidemics and percolation in small-world networks},''
\bibinfo{journal}{Phys. Rev. E} \textbf{\bibinfo{volume}{61}},
\bibinfo{pages}{5678}.


\bibitem[{\citenamefont{Moore and Newman}(2000b)}]
{Moore:mn00b}
\bibinfo{author}{\bibnamefont{Moore},~\bibfnamefont{C.}}
and
\bibinfo{author}{\bibfnamefont{M.~E.~J.}~\bibnamefont{Newman}},
\bibinfo{year}{2000b},
\bibinfo{title}{``Exact solution of site and bond percolation on small-world networks},''
\bibinfo{journal}{Phys. Rev. E} \textbf{\bibinfo{volume}{62}},
\bibinfo{pages}{7059}.


\bibitem[{\citenamefont{Moreno and Pacheco}(2004)}]
{Moreno:mp04}
\bibinfo{author}{\bibnamefont{Moreno},~\bibfnamefont{Y.}}
and
\bibinfo{author}{\bibfnamefont{A.~F.}~\bibnamefont{Pacheco}},
\bibinfo{year}{2004},
\bibinfo{title}{``Synchronization of Kuramoto oscillators in scale-free
networks},''
\bibinfo{journal}{Europhys. Lett.} \textbf{\bibinfo{volume}{68}},
\bibinfo{pages}{603}.


\bibitem[{\citenamefont{Moreno} \emph{et~al.}(2003)\citenamefont{Moreno, Pastor-Satorras, V\'azquez, and Vespignani}}]
{Moreno:mpv03}
\bibinfo{author}{\bibnamefont{Moreno},~\bibfnamefont{Y.}},
\bibinfo{author}{\bibfnamefont{R.}~\bibnamefont{Pastor-Satorras}},
\bibinfo{author}{\bibfnamefont{A.}~\bibnamefont{V\'azquez}},
and
\bibinfo{author}{\bibfnamefont{A.}~\bibnamefont{Vespignani}},
\bibinfo{year}{2003},
\bibinfo{title}{``Critical load and congestion instabilities in scale-free networks},''
\bibinfo{journal}{Europhys. Lett.} \textbf{\bibinfo{volume}{62}},
\bibinfo{pages}{292}.


\bibitem[{\citenamefont{Moreno} \emph{et~al.}(2002)\citenamefont{Moreno, Pastor-Satorras, and Vespignani}}]
{Moreno:mpv01}
\bibinfo{author}{\bibnamefont{Moreno},~\bibfnamefont{Y.}},
\bibinfo{author}{\bibfnamefont{R.}~\bibnamefont{Pastor-Satorras}},
and
\bibinfo{author}{\bibfnamefont{A.}~\bibnamefont{Vespignani}},
\bibinfo{year}{2002},
\bibinfo{title}{``Epidemic outbreaks in complex heterogeneous networks},''
\bibinfo{journal}{Eur. Phys. J. B} \textbf{\bibinfo{volume}{26}},
\bibinfo{pages}{521}.


\bibitem[{\citenamefont{Moreno and Vazquez}(2002)}]
{Moreno:mv01}
\bibinfo{author}{\bibnamefont{Moreno},~\bibfnamefont{Y.}}
and
\bibinfo{author}{\bibfnamefont{A.}~\bibnamefont{Vazquez}},
\bibinfo{year}{2002},
\bibinfo{title}{``The Bak-Sneppen model on scale-free networks},''
\bibinfo{journal}{Europhys. Lett.} \textbf{\bibinfo{volume}{57}},
\bibinfo{pages}{765}.


\bibitem[{\citenamefont{Moreno and V\'azquez}(2003)}]
{Moreno:mv02}
\bibinfo{author}{\bibnamefont{Moreno},~\bibfnamefont{Y.}}
and
\bibinfo{author}{\bibfnamefont{A.}~\bibnamefont{V\'azquez}},
\bibinfo{year}{2003},
\bibinfo{title}{``Disease spreading in structured scale-free networks},''
\bibinfo{journal}{Eur. Phys. J. B} \textbf{\bibinfo{volume}{31}},
\bibinfo{pages}{265}.


\bibitem[{\citenamefont{Moreno} \emph{et~al.}(2004)\citenamefont{Moreno, Vazquez-Prada, and Pacheco}}]
{Moreno:mvp04}
\bibinfo{author}{\bibnamefont{Moreno},~\bibfnamefont{Y.}},
\bibinfo{author}{\bibfnamefont{M.}~\bibnamefont{Vazquez-Prada}},
and
\bibinfo{author}{\bibfnamefont{A.~F.}~\bibnamefont{Pacheco}},
\bibinfo{year}{2004},
\bibinfo{title}{``Fitness for Synchronization of Network
Motifs},''
\bibinfo{journal}{Physica A} \textbf{\bibinfo{volume}{343}},
\bibinfo{pages}{279}.


\bibitem[{\citenamefont{Motter}(2004)}]
{Motter:m04}
\bibinfo{author}{\bibnamefont{Motter},~\bibfnamefont{A.~E.}},
\bibinfo{year}{2004},
\bibinfo{title}{``Cascade control and defense in complex networks},''
\bibinfo{journal}{Phys. Rev. Lett.} \textbf{\bibinfo{volume}{93}},
\bibinfo{pages}{098701}.


\bibitem[{\citenamefont{Motter}(2007)}]
{Motter:m07}
\bibinfo{author}{\bibfnamefont{Motter}~\bibnamefont{A.~E.}},
\bibinfo{year}{2007},
\bibinfo{title}{``Bounding network spectra for network design},'' 
\bibinfo{journal}{New J. Phys.} \textbf{\bibinfo{volume}{9}},
\bibinfo{pages}{182}. 


\bibitem[{\citenamefont{Motter and Lai}(2002)}]
{Motter:ml02}
\bibinfo{author}{\bibnamefont{Motter},~\bibfnamefont{A.~E.}}
and
\bibinfo{author}{\bibfnamefont{Y.-C.}~\bibnamefont{Lai}},
\bibinfo{year}{2002},
\bibinfo{title}{``Cascade-based attacks on complex networks},''
\bibinfo{journal}{Phys. Rev. E} \textbf{\bibinfo{volume}{66}},
\bibinfo{pages}{065102}.


\bibitem[{\citenamefont{Motter} \emph{et~al.}(2005a)\citenamefont{Motter, Zhou, and Kurths}}]
{Motter:mzk05a}
\bibinfo{author}{\bibnamefont{Motter},~\bibfnamefont{A.~E.}},
\bibinfo{author}{\bibfnamefont{C.}~\bibnamefont{Zhou}},
and
\bibinfo{author}{\bibfnamefont{J.}~\bibnamefont{Kurths}},
\bibinfo{year}{2005a},
\bibinfo{title}{``Network synchronization, diffusion, and the
paradox of heterogeneity},''
\bibinfo{journal}{Phys. Rev. E} \textbf{\bibinfo{volume}{71}},
\bibinfo{pages}{016116}.


\bibitem[{\citenamefont{Motter} \emph{et~al.}(2005b)\citenamefont{Motter, Zhou, and Kurths}}]
{Motter:mzk05b}
\bibinfo{author}{\bibnamefont{Motter},~\bibfnamefont{A.~E.}},
\bibinfo{author}{\bibfnamefont{C.}~\bibnamefont{Zhou}},
and
\bibinfo{author}{\bibfnamefont{J.}~\bibnamefont{Kurths}},
\bibinfo{year}{2005b},
\bibinfo{title}{``Enhancing complex-network synchronization},''
\bibinfo{journal}{Europhys. Lett.} \textbf{\bibinfo{volume}{69}},
\bibinfo{pages}{334}.


\bibitem[{\citenamefont{Motter} \emph{et~al.}(2005c)\citenamefont{Motter, Zhou, and Kurths}}]
{Motter:mzk05c}
\bibinfo{author}{\bibnamefont{Motter},~\bibfnamefont{A.~E.}},
\bibinfo{author}{\bibfnamefont{C.}~\bibnamefont{Zhou}},
and
\bibinfo{author}{\bibfnamefont{J.}~\bibnamefont{Kurths}},
\bibinfo{year}{2005c},
\bibinfo{title}{``Weighted networks are more synchronizable: How and why},''
\bibinfo{journal}{AIP Conf. Proceedings} \textbf{\bibinfo{volume}{776}},
\bibinfo{pages}{201}.



\bibitem[{\citenamefont{Mottishaw}(1987)}]
{Mottishaw:m87}
\bibinfo{author}{\bibnamefont{Mottishaw}, \bibfnamefont{P.}},
\bibinfo{year}{1987},
\bibinfo{title}{``Replica symmetry breaking and the spin-glass on a Bethe lattice},''
\bibinfo{journal}{Europhys. Lett.} \textbf{\bibinfo{volume}{4}},
\bibinfo{pages}{333}.


\bibitem[{\citenamefont{Moukarzel and de Menezes}(2002)}]
{Moukarzel:mm02}
\bibinfo{author}{\bibnamefont{Moukarzel},~\bibfnamefont{C.~F.}}
and
\bibinfo{author}{\bibfnamefont{M.~A.}~\bibnamefont{de Menezes}},
\bibinfo{year}{2002},
\bibinfo{title}{``Shortest paths on systems with power-law distributed long-range connections},''
\bibinfo{journal}{Phys. Rev. E} \textbf{\bibinfo{volume}{65}},
\bibinfo{pages}{056709}.


\bibitem[{\citenamefont{Mukamel}(1974)}]
{Mukamel:m74}
\bibinfo{author}{\bibnamefont{Mukamel},~\bibfnamefont{D.}}
\bibinfo{year}{1974},
\bibinfo{title}{``Two-spin correlation function of spin 1/2 Ising model on a Bethe lattice},''
\bibinfo{journal}{Phys. Lett. A} \textbf{\bibinfo{volume}{50}},
\bibinfo{pages}{339}.


\bibitem[{\citenamefont{Mulet} \emph{et~al.}(2002)\citenamefont{Mulet, Pagnani, Weigt, and Zecchina}}]
{Mulet:mpw02}
\bibinfo{author}{\bibnamefont{Mulet},~\bibfnamefont{R.}},
\bibinfo{author}{\bibfnamefont{A.}~\bibnamefont{Pagnani}},
\bibinfo{author}{\bibfnamefont{M.}~\bibnamefont{Weigt}},
and
\bibinfo{author}{\bibfnamefont{R.}~\bibnamefont{Zecchina}},
\bibinfo{year}{2002},
\bibinfo{title}{``Coloring random graphs},''
\bibinfo{journal}{Phys. Rev. Lett.} \textbf{\bibinfo{volume}{89}},
\bibinfo{pages}{268701}.


\bibitem[{\citenamefont{Muller-Hartmann and Zittartz}(1974)}]
{Muller-Hartmann:mz74}
\bibinfo{author}{\bibnamefont{Muller-Hartmann},~\bibfnamefont{E.}}
and
\bibinfo{author}{\bibfnamefont{J.}~\bibnamefont{Zittartz}},
\bibinfo{year}{1974},
\bibinfo{title}{``New type of phase transition},''
\bibinfo{journal}{Phys. Rev. Lett.} \textbf{\bibinfo{volume}{33}},
\bibinfo{pages}{893}.




\bibitem[{\citenamefont{N\^{a}sell}(2002)}]
{Nasell:n02}
\bibinfo{author}{\bibfnamefont{N\^{a}sell}~\bibnamefont{I.}},
\bibinfo{year}{2002},
\bibinfo{title}{``Stochastic models of some endemic infections},''
\bibinfo{journal}{Math. Biosci.} \textbf{\bibinfo{volume}{179}},
\bibinfo{pages}{1}.



\bibitem[{\citenamefont{Newman}(2000)}]
{Newman:n00}
\bibinfo{author}{\bibnamefont{Newman},~\bibfnamefont{M.~E.~J.}},
\bibinfo{year}{2000},
\bibinfo{title}{``Models of the small world},''
\bibinfo{journal}{J. Stat. Phys.} \textbf{\bibinfo{volume}{101}},
\bibinfo{pages}{819}.


\bibitem[{\citenamefont{Newman}(2002a)}]
{Newman:n02a}
\bibinfo{author}{\bibnamefont{Newman},~\bibfnamefont{M.~E.~J.}},
\bibinfo{year}{2002a},
\bibinfo{title}{``The spread of epidemic disease on networks},''
\bibinfo{journal}{Phys. Rev. E} \textbf{\bibinfo{volume}{66}},
\bibinfo{pages}{ 016128}.


\bibitem[{\citenamefont{Newman}(2002b)}]
{Newman:n02b}
\bibinfo{author}{\bibnamefont{Newman},~\bibfnamefont{M.~E.~J.}},
\bibinfo{year}{2002b},
\bibinfo{title}{``Assortative mixing in networks},''
\bibinfo{journal}{Phys. Rev. Lett.} \textbf{\bibinfo{volume}{89}},
\bibinfo{pages}{208701}.


\bibitem[{\citenamefont{Newman}(2003a)}]
{Newman:n03a}
\bibinfo{author}{\bibnamefont{Newman},~\bibfnamefont{M.~E.~J.}},
\bibinfo{year}{2003a},
\bibinfo{title}{``The structure and function of complex networks},''
\bibinfo{journal}{SIAM Review} \textbf{\bibinfo{volume}{45}},
\bibinfo{pages}{167}.


\bibitem[{\citenamefont{Newman}(2003b)}]
{Newman:n03b}
\bibinfo{author}{\bibfnamefont{Newman}~\bibnamefont{M.~E.~J.}},
\bibinfo{year}{2003b},
\bibinfo{title}{``Properties of highly clustered networks},''
\bibinfo{journal}{Phys. Rev. E} \textbf{\bibinfo{volume}{68}},
\bibinfo{pages}{026121}.


\bibitem[{\citenamefont{Newman}(2003c)}]
{Newman:n03c}
\bibinfo{author}{\bibnamefont{Newman},~\bibfnamefont{M.~E.~J.}},
\bibinfo{year}{2003c},
\bibinfo{title}{``Random graphs as models of networks},''
in \emph{\bibinfo{booktitle}{Handbook of Graphs and Networks: From
the Genome to the Internet}}, edited by
\bibinfo{editor}{\bibfnamefont{S.}~\bibnamefont{Bornholdt}}
and
\bibinfo{editor}{\bibfnamefont{H. G.}~\bibnamefont{Schuster},}
(\bibinfo{publisher}{Wiley-VCH, Berlin}), p. \bibinfo{pages}{35}.


\bibitem[{\citenamefont{Newman}(2003d)}]
{Newman:n03d}
\bibinfo{author}{\bibnamefont{Newman},~\bibfnamefont{M.~E.~J.}}
\bibinfo{year}{2003d},
\bibinfo{title}{``Mixing patterns in networks},''
\bibinfo{journal}{Phys. Rev. E} \textbf{\bibinfo{volume}{67}},
\bibinfo{pages}{026126}.


\bibitem[{\citenamefont{Newman}(2007)}]
{Newman:n07}
\bibinfo{author}{\bibfnamefont{Newman}~\bibnamefont{M.~E.~J.}},
\bibinfo{year}{2007},
\bibinfo{title}{``Component sizes in networks with arbitrary degree distributions},''
\eprint{arXiv:0707.0080 [cond-mat]}.




\bibitem[{\citenamefont{Newman and Girvan}(2004)}]
{Newman:ng04}
\bibinfo{author}{\bibnamefont{Newman},~\bibfnamefont{M.~E.~J.}}
and
\bibinfo{author}{\bibfnamefont{M.}~\bibnamefont{Girvan}},
\bibinfo{year}{2004},
\bibinfo{title}{``Finding and evaluating community structure in
networks},''
\bibinfo{journal}{Phys. Rev. E} \textbf{\bibinfo{volume}{69}},
\bibinfo{pages}{026113}.


\bibitem[{\citenamefont{Newman} \emph{et~al.}(2002)\citenamefont{Newman, Jensen, and Ziff}}]
{Newman:njz02}
\bibinfo{author}{\bibnamefont{Newman},~\bibfnamefont{M.~E.~J.}},
\bibinfo{author}{\bibfnamefont{I.}~\bibnamefont{Jensen}},
and
\bibinfo{author}{\bibfnamefont{R.~M.}~\bibnamefont{Ziff}},
\bibinfo{year}{2002},
\bibinfo{title}{``Percolation and epidemics in a two-dimensional small world},''
\bibinfo{journal}{Phys. Rev. E} \textbf{\bibinfo{volume}{65}},
\bibinfo{pages}{021904}.


\bibitem[{\citenamefont{Newman} \emph{et~al.}(2001)\citenamefont{Newman, Strogatz, and Watts}}]
{Newman:nsw01}
\bibinfo{author}{\bibnamefont{Newman},~\bibfnamefont{M.~E.~J.}},
\bibinfo{author}{\bibfnamefont{S.~H.}~\bibnamefont{Strogatz}},
and
\bibinfo{author}{\bibfnamefont{D.~J.}~\bibnamefont{Watts}},
\bibinfo{year}{2001},
\bibinfo{title}{``Random graphs with arbitrary degree distributions and their applications},''
\bibinfo{journal}{Phys. Rev. E} \textbf{\bibinfo{volume}{64}},
\bibinfo{pages}{026118}.


\bibitem[{\citenamefont{Newman and Watts}(1999a)}]
{Newman:nw99a}
\bibinfo{author}{\bibnamefont{Newman},~\bibfnamefont{M.~E.~J.}}
and
\bibinfo{author}{\bibfnamefont{D.~J.}~\bibnamefont{Watts}},
\bibinfo{year}{1999a},
\bibinfo{title}{``Scaling and percolation in the small-world network model},''
\bibinfo{journal}{Phys. Rev. E} \textbf{\bibinfo{volume}{60}},
\bibinfo{pages}{7332}.


\bibitem[{\citenamefont{Newman and Watts}(1999b)}]
{Newman:nw99b}
\bibinfo{author}{\bibnamefont{Newman},~\bibfnamefont{M.~E.~J.}}
and
\bibinfo{author}{\bibfnamefont{D.~J.}~\bibnamefont{Watts}},
\bibinfo{year}{1999b},
\bibinfo{title}{``Renormalization group analysis of the small-world network model},''
\bibinfo{journal}{Phys. Lett. A} \textbf{\bibinfo{volume}{263}},
\bibinfo{pages}{341}.


\bibitem[{\citenamefont{Nikoletopoulos}
\emph{et~al.}(2004)\citenamefont{Nikoletopoulos, Coolen,
P\'{e}rez Castillo, Skantzos, Hatchett and B Wemmenhove}}]
{Nikoletopoulos:nccshw04}
\bibinfo{author}{\bibnamefont{Nikoletopoulos},~\bibfnamefont{T.}},
\bibinfo{author}{\bibfnamefont{A.~C.~C.}~\bibnamefont{Coolen}},
\bibinfo{author}{\bibfnamefont{I.}~\bibnamefont{P\'{e}rez Castillo}},
\bibinfo{author}{\bibfnamefont{N.~S.}~\bibnamefont{Skantzos}},
\bibinfo{author}{\bibfnamefont{J.~P.~L.}~\bibnamefont{Hatchett}},
and
\bibinfo{author}{\bibfnamefont{B.}~\bibnamefont{Wemmenhove}},
\bibinfo{year}{2004},
\bibinfo{title}{``Replicated transfer matrix analysis of
Ising spin models on 'small world' lattices},''
\bibinfo{journal}{J. Phys. A} \textbf{\bibinfo{volume}{37}},
\bibinfo{pages}{6455}.


\bibitem[{\citenamefont{Nishikawa and Motter}(2006a)}]
{Nishikawa:nm06a}
\bibinfo{author}{\bibnamefont{Nishikawa},~\bibfnamefont{T.}}
and
\bibinfo{author}{\bibfnamefont{A.~E.}~\bibnamefont{Motter}},
\bibinfo{year}{2006a},
\bibinfo{title}{``Synchronization is optimal in non-diagonalizable
networks},''
\bibinfo{journal}{Phys. Rev. E} \textbf{\bibinfo{volume}{73}},
\bibinfo{pages}{065106}.


\bibitem[{\citenamefont{Nishikawa and Motter}(2006b)}]
{Nishikawa:nm06b}
\bibinfo{author}{\bibnamefont{Nishikawa},~\bibfnamefont{T.}}
and
\bibinfo{author}{\bibfnamefont{A.~E.}~\bibnamefont{Motter}},
\bibinfo{year}{2006b},
\bibinfo{title}{``Maximum performance at minimum cost in network synchronization},''
\bibinfo{journal}{Physica D} \textbf{\bibinfo{volume}{224}},
\bibinfo{pages}{77}.


\bibitem[{\citenamefont{Nishikawa} \emph{et~al.}(2003)\citenamefont{Nishikawa, Motter, Lai, and Hoppensteadt}}]
{Nishikawa:nmlh03}
\bibinfo{author}{\bibnamefont{Nishikawa},~\bibfnamefont{T.}},
\bibinfo{author}{\bibfnamefont{A.~E.}~\bibnamefont{Motter}},
\bibinfo{author}{\bibfnamefont{Y.-C.}~\bibnamefont{Lai}},
and
\bibinfo{author}{\bibfnamefont{F.~C.}~\bibnamefont{Hoppensteadt}},
\bibinfo{year}{2003},
\bibinfo{title}{``Heterogeneity in oscillator networks: Are smaller worlds easier to synchronize?},''
\bibinfo{journal}{Phys. Rev. Lett.} \textbf{\bibinfo{volume}{91}},
\bibinfo{pages}{014101}.




\bibitem[{\citenamefont{Noh}(2005)}]
{Noh:n05}
\bibinfo{author}{\bibnamefont{Noh},~\bibfnamefont{J.~D.}},
\bibinfo{year}{2005},
\bibinfo{title}{``Stationary and dynamical properties of a zero range process on scale-free networks},''
\bibinfo{journal}{Phys. Rev. E} \textbf{\bibinfo{volume}{72}},
\bibinfo{pages}{056123}.

 
\bibitem[{\citenamefont{Noh}(2007)}]
{Noh:n07}
\bibinfo{author}{\bibnamefont{Noh},~\bibfnamefont{J.~D.}},
\bibinfo{year}{2007},
\bibinfo{title}{``Percolation transition in networks with degree-degree correlation},''
\eprint{arXiv:0705.0087 [cond-mat]}.




\bibitem[{\citenamefont{Noh} \emph{et~al.}(2005)\citenamefont{Noh, Shim, and Lee}}]
{Noh:nsl05}
\bibinfo{author}{\bibnamefont{Noh},~\bibfnamefont{J.~D.}},
\bibinfo{author}{\bibfnamefont{G.~M.}~\bibnamefont{Shim}},
and
\bibinfo{author}{\bibfnamefont{H.}~\bibnamefont{Lee}},
\bibinfo{year}{2005},
\bibinfo{title}{``Complete condensation in a zero range process on scale-free networks},''
\bibinfo{journal}{Phys. Rev. Lett.} \textbf{\bibinfo{volume}{94}},
\bibinfo{pages}{198701}.






\bibitem[{\citenamefont{Oh} \emph{et~al.}(2007)\citenamefont{Oh, Lee, Kahng, and Kim}}]
{Oh:olkk07}
\bibinfo{author}{\bibnamefont{Oh},~\bibfnamefont{E.}},
\bibinfo{author}{\bibfnamefont{D.-S.}~\bibnamefont{Lee}},
\bibinfo{author}{\bibfnamefont{B.}~\bibnamefont{Kahng}},
and
\bibinfo{author}{\bibfnamefont{D.}~\bibnamefont{Kim}},
\bibinfo{year}{2007},
\bibinfo{title}{``Synchronization transition of heterogeneously
coupled oscillators on scale-free networks},''
\bibinfo{journal}{Phys. Rev. E} \textbf{\bibinfo{volume}{75}},
\bibinfo{pages}{011104}.


\bibitem[{\citenamefont{Oh} \emph{et~al.}(2005)\citenamefont{Oh, Rho, Hong, and Kahng}}]
{Oh:orh05}
\bibinfo{author}{\bibnamefont{Oh},~\bibfnamefont{E.}},
\bibinfo{author}{\bibfnamefont{K.}~\bibnamefont{Rho}},
\bibinfo{author}{\bibfnamefont{H.}~\bibnamefont{Hong}},
and
\bibinfo{author}{\bibfnamefont{B.}~\bibnamefont{Kahng}},
\bibinfo{year}{2005},
\bibinfo{title}{``Modular synchronization in complex networks},''
\bibinfo{journal}{Phys. Rev. E} \textbf{\bibinfo{volume}{72}},
\bibinfo{pages}{047101}.


\bibitem[{\citenamefont{Ohira and Sawatari}(1998)}]
{Ohira:os98}
\bibinfo{author}{\bibnamefont{Ohira},~\bibfnamefont{T.}}
and
\bibinfo{author}{\bibfnamefont{R.}~\bibnamefont{Sawatari}},
\bibinfo{year}{1998},
\bibinfo{title}{``Phase transition in a computer network traffic model},''
\bibinfo{journal}{Phys. Rev. E} \textbf{\bibinfo{volume}{58}},
\bibinfo{pages}{193}.


\bibitem[{\citenamefont{Ohkubo} \emph{et~al.}(2005)\citenamefont{Ohkubo, Yasuda, and Tanaka}}]
{Ohkubo:oyt05}
\bibinfo{author}{\bibnamefont{Ohkubo},~\bibfnamefont{J.}},
\bibinfo{author}{\bibfnamefont{M.}~\bibnamefont{Yasuda}},
and
\bibinfo{author}{\bibfnamefont{K.}~\bibnamefont{Tanaka}},
\bibinfo{year}{2005},
\bibinfo{title}{``Statistical-mechanical iterative algorithms on
complex networks},''
\bibinfo{journal}{Phys. Rev. E} \textbf{\bibinfo{volume}{72}},
\bibinfo{pages}{046135}.

\bibitem[{\citenamefont{Ostilli}(2006a)}]
{Ostilli:o06a}
\bibinfo{author}{\bibnamefont{Ostilli}, \bibfnamefont{M.}},
\bibinfo{year}{2006a},
\bibinfo{title}{``Ising spin glass models versus Ising models: an effective mapping at
high temperature I. General result},''
\bibinfo{journal}{J. Stat. Mech.} 
\bibinfo{pages}{P10004}.

\bibitem[{\citenamefont{Ostilli}(2006b)}]
{Ostilli:o06b}
\bibinfo{author}{\bibfnamefont{Ostilli}~\bibnamefont{M.}},
\bibinfo{year}{2006b},
\bibinfo{title}{``Ising spin glass models versus Ising models: an effective mapping at high temperature II. Applications to graphs and networks},''
\bibinfo{journal}{J. Stat. Mech.} 
\bibinfo{pages}{P10005}.

\bibitem[{\citenamefont{Ozana}(2001)}]
{Ozana:o01}
\bibinfo{author}{\bibfnamefont{Ozana}~\bibnamefont{M.}},
\bibinfo{year}{2001},
\bibinfo{title}{``Incipient spanning cluster on small-world networks},''
\bibinfo{journal}{Europhys. Lett.} \textbf{\bibinfo{volume}{55}},
\bibinfo{pages}{762}.


\bibitem[{\citenamefont{Pacheco} \emph{et~al.}(2006)\citenamefont{Pacheco, Traulsen, and Nowak}}]
{Pacheco:ptm07}
\bibinfo{author}{\bibnamefont{Pacheco},~\bibfnamefont{J.~M.}},
\bibinfo{author}{\bibfnamefont{A.}~\bibnamefont{Traulsen}},
and
\bibinfo{author}{\bibfnamefont{M.~A.}~\bibnamefont{Nowak}},
\bibinfo{year}{2006},
\bibinfo{title}{``Co-evolution of strategy and structure in complex networks with dynamical linking},''
\bibinfo{journal}{Phys. Rev. Lett.} \textbf{\bibinfo{volume}{97}},
\bibinfo{pages}{258103}.


\bibitem[{\citenamefont{Pagnani} \emph{et~al.}(2003)\citenamefont{Pagnani, Parisi, and Ratieville}}]
{Pagnani:ppr03}
\bibinfo{author}{\bibnamefont{Pagnani},~\bibfnamefont{A.}},
\bibinfo{author}{\bibfnamefont{G.}~\bibnamefont{Parisi}},
and
\bibinfo{author}{\bibfnamefont{M.}~\bibnamefont{Ratieville}},
\bibinfo{year}{2003},
\bibinfo{title}{``Metastable configurations on the Bethe
lattice},''
\bibinfo{journal}{Phys. Rev. E} \textbf{\bibinfo{volume}{67}},
\bibinfo{pages}{026116}.


\bibitem[{\citenamefont{Palla} \emph{et~al.}(2004)\citenamefont{Palla, Der\'enyi, Farkas, and Vicsek}}]
{Palla:pdf04}
\bibinfo{author}{\bibnamefont{Palla},~\bibfnamefont{G.}},
\bibinfo{author}{\bibfnamefont{I.}~\bibnamefont{Der\'enyi}},
\bibinfo{author}{\bibfnamefont{I.}~\bibnamefont{Farkas}},
and
\bibinfo{author}{\bibfnamefont{T.}~\bibnamefont{Vicsek}},
\bibinfo{year}{2004},
\bibinfo{title}{``Statistical mechanics of topological phase transitions in networks},''
\bibinfo{journal}{Phys. Rev. E} \textbf{\bibinfo{volume}{69}},
\bibinfo{pages}{046117}.


\bibitem[{\citenamefont{Palla} \emph{et~al.}(2007)\citenamefont{Palla, Der\'enyi, and Vicsek}}]
{Palla:pdv06}
\bibinfo{author}{\bibnamefont{Palla},~\bibfnamefont{G.}},
\bibinfo{author}{\bibfnamefont{I.}~\bibnamefont{Der\'enyi}},
and
\bibinfo{author}{\bibfnamefont{T.}~\bibnamefont{Vicsek}},
\bibinfo{year}{2007},
\bibinfo{title}{``The critical point of $k$-clique percolation in the Erd\H os-R\'enyi graph},''
\bibinfo{journal}{J.~Stat. Phys.} \textbf{\bibinfo{volume}{128}},
\bibinfo{pages}{219}. 




\bibitem[{\citenamefont{Parisi and Rizzo}(2006)}]
{Parisi:pr06}
\bibinfo{author}{\bibnamefont{Parisi},~\bibfnamefont{G.}}
and
\bibinfo{author}{\bibfnamefont{T.}~\bibnamefont{Rizzo}},
\bibinfo{year}{2006},
\bibinfo{title}{``On $k$-core percolation in four dimensions},''
\eprint{cond-mat/0609777}.


\bibitem[{\citenamefont{Parisi and Slanina}(2006)}]
{Parisi:ps06}
\bibinfo{author}{\bibnamefont{Parisi},~\bibfnamefont{G.}}
and
\bibinfo{author}{\bibfnamefont{F.}~\bibnamefont{Slanina}},
\bibinfo{year}{2006},
\bibinfo{title}{``Loop expansion around the Bethe-Peierls approximation for lattice models},''
\bibinfo{journal}{J. Stat. Mech.} \textbf{\bibinfo{volume}{0602}},
\bibinfo{pages}{L003}.


\bibitem[{\citenamefont{Park and Newman}(2004a)}]
{Park:pn04a}
\bibinfo{author}{\bibnamefont{Park},~\bibfnamefont{J.}}
and
\bibinfo{author}{\bibfnamefont{M.~E.~J.}~\bibnamefont{Newman}},
\bibinfo{year}{2004a},
\bibinfo{title}{``Solution of the 2-star model of a network},''
\bibinfo{journal}{Phys. Rev. E} \textbf{\bibinfo{volume}{70}},
\bibinfo{pages}{066146}.


\bibitem[{\citenamefont{Park and Newman}(2004b)}]
{Park:pn04b}
\bibinfo{author}{\bibnamefont{Park},~\bibfnamefont{J.}}
and
\bibinfo{author}{\bibfnamefont{M.~E.~J.}~\bibnamefont{Newman}},
\bibinfo{year}{2004b},
\bibinfo{title}{``The statistical mechanics of networks},''
\bibinfo{journal}{Phys. Rev. E} \textbf{\bibinfo{volume}{70}},
\bibinfo{pages}{066117}.




\bibitem[{\citenamefont{Pastor-Satorras and Vespignani}(2001)}]
{Pastor-Satorras:pv01}
\bibinfo{author}{\bibnamefont{Pastor-Satorras},~\bibfnamefont{R.}}
and
\bibinfo{author}{\bibfnamefont{A.}~\bibnamefont{Vespignani}},
\bibinfo{year}{2001},
\bibinfo{title}{``Epidemic spreading in scale-free networks},''
\bibinfo{journal}{Phys. Rev. Lett.} \textbf{\bibinfo{volume}{86}},
\bibinfo{pages}{3200}.


\bibitem[{\citenamefont{Pastor-Satorras and Vespignani}(2002a)}]
{Pastor-Satorras:pv02a}
\bibinfo{author}{\bibnamefont{Pastor-Satorras},~\bibfnamefont{R.}}
and
\bibinfo{author}{\bibfnamefont{A.}~\bibnamefont{Vespignani}},
\bibinfo{year}{2002a},
\bibinfo{title}{``Epidemic dynamics in finite size scale-free networks},''
\bibinfo{journal}{Phys. Rev. E} \textbf{\bibinfo{volume}{65}},
\bibinfo{pages}{035108}.


\bibitem[{\citenamefont{Pastor-Satorras and Vespignani}(2002b)}]
{Pastor-Satorras:pv02b}
\bibinfo{author}{\bibnamefont{Pastor-Satorras},~\bibfnamefont{R.}}
and
\bibinfo{author}{\bibfnamefont{A.}~\bibnamefont{Vespignani}},
\bibinfo{year}{2002b},
\bibinfo{title}{``Immunization of complex networks},''
\bibinfo{journal}{Phys. Rev. E} \textbf{\bibinfo{volume}{65}},
\bibinfo{pages}{036104}.




\bibitem[{\citenamefont{Pastor-Satorras and Vespignani}(2003)}]
{Pastor-Satorras:pv03}
\bibinfo{author}{\bibnamefont{Pastor-Satorras},~\bibfnamefont{R.}}
and
\bibinfo{author}{\bibnamefont{A.},~\bibfnamefont{Vespignani}},
\bibinfo{year}{2003},
\bibinfo{title}{``Epidemics and immunization in scale-free networks},''
in
\emph{\bibinfo{booktitle}{Handbook of Graphs and Networks: From the Genome to the Internet}},
edited by
\bibinfo{editor}{\bibfnamefont{S.}~\bibnamefont{Bornholdt}}
and
\bibinfo{editor}{\bibfnamefont{H. G.}~\bibnamefont{Schuster},}
(\bibinfo{publisher}{Wiley-VCH, Berlin}), p. \bibinfo{pages}{111}.


\bibitem[{\citenamefont{Pastor-Satorras and Vespignani}(2004)}]
{Pastor-Satorras:pvbook04}
\bibinfo{author}{\bibnamefont{Pastor-Satorras},~\bibfnamefont{R.}}
and
\bibinfo{author}{\bibfnamefont{A. }~\bibnamefont{Vespignani}},
\bibinfo{year}{2004},
\emph{\bibinfo{title}{Evolution and Structure of the Internet: A
Statistical Physics Approach}} (\bibinfo{publisher}{Cambridge
University Press, Cambridge}).


\bibitem[{\citenamefont{Paul} \emph{et~al.}(2007)\citenamefont{Paul, Cohen, Sreenivasan, Havlin, and Stanley}}]
{Paul:pcs07}
\bibinfo{author}{\bibnamefont{Paul},~\bibfnamefont{G.}},
\bibinfo{author}{\bibfnamefont{R.}~\bibnamefont{Cohen}},
\bibinfo{author}{\bibfnamefont{S.}~\bibnamefont{Sreenivasan}},
\bibinfo{author}{\bibfnamefont{S.}~\bibnamefont{Havlin}},
and
\bibinfo{author}{\bibfnamefont{H.~E.}~\bibnamefont{Stanley}},
\bibinfo{year}{2007},
\bibinfo{title}{``Graph partitioning induced phase transitions},''
\bibinfo{journal}{Phys. Rev. Lett.} \textbf{\bibinfo{volume}{99}},
\bibinfo{pages}{115701}.


\bibitem[{\citenamefont{Pearl}(1988)}]
{Pearl:p88}
\bibinfo{author}{\bibnamefont{Pearl}, \bibfnamefont{J.}},
\bibinfo{year}{1988},
\emph{\bibinfo{title}{Probabilistic Reasoning in Intelligent
Systems: Networks of Plausible Inference}}
(\bibinfo{publisher}{Morgan Kaufmann, San Francisco}).


\bibitem[{\citenamefont{Pecora and Carroll}(1998)}]
{Pecora:pc98}
\bibinfo{author}{\bibnamefont{Pecora},~\bibfnamefont{L. M.}}
and
\bibinfo{author}{\bibfnamefont{T. L.}~\bibnamefont{Carroll}},
\bibinfo{year}{1998},
\bibinfo{title}{``Master stability functions for synchronized coupled systems},''
\bibinfo{journal}{Phys. Rev. Lett.} \textbf{\bibinfo{volume}{80}},
\bibinfo{pages}{2109}.


\bibitem[{\citenamefont{Peierls}(1936)}]
{Peierls:p36}
\bibinfo{author}{\bibnamefont{Peierls},~\bibfnamefont{R.}},
\bibinfo{year}{1936},
\bibinfo{title}{``On Ising's ferromagnet model},''
\bibinfo{journal}{Proc. Cambridge Phil. Soc.} \textbf{\bibinfo{volume}{32}},
\bibinfo{pages}{477}.






\bibitem[{\citenamefont{P\c{e}kalski}(2001)}]
{Pekalski:p01}
\bibinfo{author}{\bibnamefont{P\c{e}kalski},~\bibfnamefont{A.}}
\bibinfo{year}{2001},
\bibinfo{title}{``Ising model on a small-world network},''
\bibinfo{journal}{Phys. Rev. E} \textbf{\bibinfo{volume}{64}},
\bibinfo{pages}{057104}.


\bibitem[{\citenamefont{Perkovic} \emph{et~al.}(1995)\citenamefont{Perkovic, Dahmen, and Sethna}}]
{Percovic:pds95}
\bibinfo{author}{\bibnamefont{Perkovic},~\bibfnamefont{O.}},
\bibinfo{author}{\bibfnamefont{K.}~\bibnamefont{Dahmen}},
and
\bibinfo{author}{\bibfnamefont{J.~P.}~\bibnamefont{Sethnal}},
\bibinfo{year}{1995},
\bibinfo{title}{``Avalanches, Barkhausen noise, and plain
old criticality},''
\bibinfo{journal}{Phys. Rev. Lett.} \textbf{\bibinfo{volume}{75}},
\bibinfo{pages}{4528}.


\bibitem[{\citenamefont{Petermann and De Los Rios}(2004)}]
{Petermann:pd04}
\bibinfo{author}{\bibnamefont{Petermann},~\bibfnamefont{T.}}
and
\bibinfo{author}{\bibfnamefont{P.}~\bibnamefont{De Los Rios}},
\bibinfo{year}{2004},
\bibinfo{title}{``Role of clustering and gridlike ordering in epidemic spreading},''
\bibinfo{journal}{Phys. Rev. E} \textbf{\bibinfo{volume}{69}},
\bibinfo{pages}{066116}.


\bibitem[{\citenamefont{Picard and Ratliff}(1975)}]
{Picard:pr75}
\bibinfo{author}{\bibnamefont{Picard},~\bibfnamefont{J.-C.}}
and
\bibinfo{author}{\bibfnamefont{H.}~\bibnamefont{Ratliff}},
\bibinfo{year}{1975},
\bibinfo{title}{``Minimum cuts and related problems},''
\bibinfo{journal}{Networks} \textbf{\bibinfo{volume}{5}},
\bibinfo{pages}{357}.


\bibitem[{\citenamefont{Pikovsky} \emph{et~al.}(2001)\citenamefont{Pikovsky, Rosenblum, and Kurths}}]
{Pikovsky:prk01}
\bibinfo{author}{\bibnamefont{Pikovsky},~\bibfnamefont{A.~S.}},
\bibinfo{author}{\bibfnamefont{M.~G.}~\bibnamefont{Rosenblum}},
and
\bibinfo{author}{\bibfnamefont{J.}~\bibnamefont{Kurths}},
\bibinfo{year}{2001}, \emph{\bibinfo{title}{Synchronization: A Universal Concept in Nonlinear Sciences}}
(\bibinfo{publisher}{Cambridge University Press, Cambridge}).


\bibitem[{\citenamefont{Pittel} \emph{et~al.}(1996)\citenamefont{Pittel, Spencer and Wormald}}]
{Pittel:p96}
\bibinfo{author}{\bibnamefont{Pittel},~\bibfnamefont{B.}},
\bibinfo{author}{\bibfnamefont{J.}~\bibnamefont{Spencer}},
and
\bibinfo{author}{\bibfnamefont{N.}~\bibnamefont{Wormald}},
\bibinfo{year}{1996},
\bibinfo{title}{``Sudden emergence of a giant $k$-core in a random graph},''
\bibinfo{journal}{J. Combin. Theor. B} \textbf{\bibinfo{volume}{67}},
\bibinfo{pages}{111}.


\bibitem[{\citenamefont{Pretti and Pelizzola}(2003)}]
{Pretti:pp03}
\bibinfo{author}{\bibnamefont{Pretti},~\bibfnamefont{M.}}
and
\bibinfo{author}{\bibfnamefont{A.}~\bibnamefont{Pelizzola}},
\bibinfo{year}{2003},
\bibinfo{title}{``Stable propagation algorithm for the minimization of the Bethe free energy},''
\bibinfo{journal}{J. Phys. A} \textbf{\bibinfo{volume}{36}},
\bibinfo{pages}{11201}.






\bibitem[{\citenamefont{Radicchi and Meyer-Ortmanns}(2006)}]
{Radicchi:rm06}
\bibinfo{author}{\bibnamefont{Radicchi},~\bibfnamefont{F.}}
and
\bibinfo{author}{\bibfnamefont{H.}~\bibnamefont{Meyer-Ortmanns}},
\bibinfo{year}{2006},
\bibinfo{title}{``Entrainment of coupled oscillator on regular networks by pacemakers},''
\bibinfo{journal}{Phys. Rev. E} \textbf{\bibinfo{volume}{73}},
\bibinfo{pages}{036218}.




\bibitem[{\citenamefont{Reichardt and Bornholdt}(2004)}]
{Reichardt:rb04}
\bibinfo{author}{\bibnamefont{Reichardt},~\bibfnamefont{J.}}
and
\bibinfo{author}{\bibfnamefont{S.}~\bibnamefont{Bornholdt}},
\bibinfo{year}{2004},
\bibinfo{title}{``Detecting fuzzy community structures in complex networks with a Potts model},''
\bibinfo{journal}{Phys. Rev. Lett.} \textbf{\bibinfo{volume}{93}},
\bibinfo{pages}{218701}.


\bibitem[{\citenamefont{Reichardt and Bornholdt}(2006)}]
{Reichardt:rb06}
\bibinfo{author}{\bibnamefont{Reichardt},~\bibfnamefont{J.}}
and
\bibinfo{author}{\bibfnamefont{S.}~\bibnamefont{Bornholdt}},
\bibinfo{year}{2006},
\bibinfo{title}{``Statistical mechanics of community detection},''
\bibinfo{journal}{Phys. Rev. E} \textbf{\bibinfo{volume}{74}},
\bibinfo{pages}{016110}.


\bibitem[{\citenamefont{Restrepo} \emph{et~al.}(2005)\citenamefont{Restrepo, Ott, and Hunt}}]
{Restrepo:roh05}
\bibinfo{author}{\bibnamefont{Restrepo},~\bibfnamefont{J.~G.}},
\bibinfo{author}{\bibfnamefont{E.}~\bibnamefont{Ott}},
and
\bibinfo{author}{\bibfnamefont{B.~R.}~\bibnamefont{Hunt}},
\bibinfo{year}{2005},
\bibinfo{title}{``Onset of synchronization in large networks of
coupled oscillators},''
\bibinfo{journal}{Phys. Rev. E} \textbf{\bibinfo{volume}{71}},
\bibinfo{pages}{036151}.




\bibitem[{\citenamefont{Rizzo} \emph{et~al.}(2006)\citenamefont{Rizzo, Wemmenhove, and Kappen}}]
{Rizzo:rwk06}
\bibinfo{author}{\bibnamefont{Rizzo},~\bibfnamefont{T.}},
\bibinfo{author}{\bibfnamefont{B.}~\bibnamefont{Wemmenhove}},
and
\bibinfo{author}{\bibfnamefont{H. J.}~\bibnamefont{Kappen}},
\bibinfo{year}{2006},
\bibinfo{title}{``On Cavity approximations for graphical models},''
\eprint{cond-mat/0608312}.






\bibitem[{\citenamefont{Rosvall} \emph{et~al.}(2004)\citenamefont{Rosvall, Minnhagen, and Sneppen}}]
{Rosvall:rms04}
\bibinfo{author}{\bibnamefont{Rosvall},~\bibfnamefont{M.}},
\bibinfo{author}{\bibfnamefont{P.}~\bibnamefont{Minnhagen}},
and
\bibinfo{author}{\bibfnamefont{K.}~\bibnamefont{Sneppen}},
\bibinfo{year}{2004},
\bibinfo{title}{``Navigating networks with limited information},''
\bibinfo{journal}{Phys. Rev. E} \textbf{\bibinfo{volume}{71}},
\bibinfo{pages}{066111}.  


\bibitem[{\citenamefont{Roy and Bhattacharjee}(2006)}]
{Roy:rb06}
\bibinfo{author}{\bibnamefont{Roy},~\bibfnamefont{S.}}
and
\bibinfo{author}{\bibfnamefont{S.~M.}~\bibnamefont{Bhattacharjee}},
\bibinfo{year}{2006},
\bibinfo{title}{``Is small-world network disordered?},''
\bibinfo{journal}{Phys. Letters A} \textbf{\bibinfo{volume}{352}},
\bibinfo{pages}{13}.


\bibitem[{\citenamefont{Rozenfeld and ben-Avraham}(2007)}]
{Rozenfeld:rb07}
\bibinfo{author}{\bibnamefont{Rozenfeld},~\bibfnamefont{H.~D.}}
and
\bibinfo{author}{\bibfnamefont{D.}~\bibnamefont{ben-Avraham}},
\bibinfo{year}{2007},
\bibinfo{title}{``Percolation in hierarchical scale-free nets},'' 
\bibinfo{journal}{Phys. Rev. E} \textbf{\bibinfo{volume}{75}},
\bibinfo{pages}{061102}. 








\bibitem[{\citenamefont{Sade} \emph{et~al.}(2005)\citenamefont{Sade, Kalisky, Havlin, and Berkovits}}]
{Sade:skh05}
\bibinfo{author}{\bibnamefont{Sade},~\bibfnamefont{M.}},
\bibinfo{author}{\bibfnamefont{T.}~\bibnamefont{Kalisky}},
\bibinfo{author}{\bibfnamefont{S.}~\bibnamefont{Havlin}},
and
\bibinfo{author}{\bibfnamefont{R.}~\bibnamefont{Berkovits}},
\bibinfo{year}{2005},
\bibinfo{title}{``Localization transition on complex networks via spectral statistics},''
\bibinfo{journal}{Phys. Rev. E} \textbf{\bibinfo{volume}{72}},
\bibinfo{pages}{066123}.

\bibitem[{\citenamefont{S\'anchez} \emph{et~al.}(2002)\citenamefont{S\'anchez, L\'opez, and Rodr\'iguez}}]
{Sanchez:slr02}
\bibinfo{author}{\bibnamefont{S\'anchez},~\bibfnamefont{A.~D.}},
\bibinfo{author}{\bibfnamefont{J.~M.}~\bibnamefont{L\'opez}},
and
\bibinfo{author}{\bibfnamefont{M.~A.}~\bibnamefont{Rodr\'iguez}},
\bibinfo{year}{2002},
\bibinfo{title}{``Nonequilibrium phase transitions in directed small-world networks},''
\bibinfo{journal}{Phys. Rev. Lett.} \textbf{\bibinfo{volume}{88}},
\bibinfo{pages}{048701}.




\bibitem[{\citenamefont{Scalettar}(1991)}]
{Scalettar:s91}
\bibinfo{author}{\bibfnamefont{Scalettar}~\bibnamefont{R.~T.}},
\bibinfo{year}{1991},
\bibinfo{title}{``Critical properties of an Ising model with dilute long range interactions},''
\bibinfo{journal}{Physica A} \textbf{\bibinfo{volume}{170}},
\bibinfo{pages}{282}.




\bibitem[{\citenamefont{Schneider and Pytte}(1977)}]
{Schneider:sp77}
\bibinfo{author}{\bibnamefont{Schneider},~\bibfnamefont{T.}}
and
\bibinfo{author}{\bibfnamefont{E.}~\bibnamefont{Pytte}},
\bibinfo{year}{1977},
\bibinfo{title}{``Random-field instability of the ferromagnetic state},''
\bibinfo{journal}{Phys. Rev. B} \textbf{\bibinfo{volume}{15}},
\bibinfo{pages}{1519}.


\bibitem[{\citenamefont{Schwartz} \emph{et~al.}(2002)\citenamefont{Schwartz, Cohen, ben-Avraham, Barab\'asi, and Havlin}}]
{Schwartz:scb02}
\bibinfo{author}{\bibnamefont{Schwartz},~\bibfnamefont{N.}},
\bibinfo{author}{\bibfnamefont{R.}~\bibnamefont{Cohen}},
\bibinfo{author}{\bibfnamefont{D.}~\bibnamefont{ben-Avraham}},
\bibinfo{author}{\bibfnamefont{A.-L.}~\bibnamefont{Barab\'asi}},
and
\bibinfo{author}{\bibfnamefont{S.}~\bibnamefont{Havlin}},
\bibinfo{year}{2002},
\bibinfo{title}{``Percolation in directed scale-free networks},''
\bibinfo{journal}{Phys. Rev. E} \textbf{\bibinfo{volume}{66}},
\bibinfo{pages}{015104}.


\bibitem[{\citenamefont{Schwartz} \emph{et~al.}(2006)\citenamefont{Schwartz, Liu, and Chayes}}]
{Schwartz:slc06}
\bibinfo{author}{\bibnamefont{Schwartz},~\bibfnamefont{J.~M.}},
\bibinfo{author}{\bibfnamefont{A.~J.}~\bibnamefont{Liu}},
and
\bibinfo{author}{\bibfnamefont{L.~Q.}~\bibnamefont{Chayes}},
\bibinfo{year}{2006},
\bibinfo{title}{``The onset of jamming as the sudden emergence of an infinite $k$-core cluster},''
\bibinfo{journal}{Europhys. Lett.} \textbf{\bibinfo{volume}{73}},
\bibinfo{pages}{560566}.







\bibitem[{\citenamefont{Serrano and Bogu\~n\'a}(2006a)}]
{Serrano:sb06a}
\bibinfo{author}{\bibnamefont{Serrano},~\bibfnamefont{M.~A.}}
and
\bibinfo{author}{\bibfnamefont{M.}~\bibnamefont{Bogu\~n\'a}},
\bibinfo{year}{2006a},
\bibinfo{title}{``Percolation and epidemic thresholds in clustered networks},''
\bibinfo{journal}{Phys. Rev. Lett.} \textbf{\bibinfo{volume}{97}},
\bibinfo{pages}{088701}.

\bibitem[{\citenamefont{Serrano and Bogu\~n\'a}(2006b}]
{Serrano:sb06b}
\bibinfo{author}{\bibnamefont{Serrano},~\bibfnamefont{M.~A.}}
and
\bibinfo{author}{\bibfnamefont{M.}~\bibnamefont{Boguna}},
\bibinfo{year}{2006b},
\bibinfo{title}{``Clustering in complex networks. I. General formalism},''
\bibinfo{journal}{Phys. Rev. E} \textbf{\bibinfo{volume}{74}},
\bibinfo{pages}{056114}.


\bibitem[{\citenamefont{Serrano and Bogu\~n\'a}(2006c)}]
{Serrano:sb06c}
\bibinfo{author}{\bibnamefont{Serrano},~\bibfnamefont{M.~A.}}
and
\bibinfo{author}{\bibfnamefont{M.}~\bibnamefont{Boguna}},
\bibinfo{year}{2006c},
\bibinfo{title}{``Clustering in complex networks. II. Percolation properties},''
\bibinfo{journal}{Phys. Rev. E} \textbf{\bibinfo{volume}{74}},
\bibinfo{pages}{056115}.


\bibitem[{\citenamefont{Serrano and De Los Rios}(2007)}]
{Serrano:sd07}
\bibinfo{author}{\bibnamefont{Serrano},~\bibfnamefont{M.~A.}}
and
\bibinfo{author}{\bibfnamefont{P.}~\bibnamefont{De Los Rios}},
\bibinfo{year}{2007},
\bibinfo{title}{``Interfaces and the edge percolation map of random directed networks},''
\eprint{arXiv:0706.3156 [cond-mat]}.


\bibitem[{\citenamefont{Sethna} \emph{et~al.}(1993)\citenamefont{Sethna, Dahmen, Kartha,
Krumhansl, Roberts, and Shore}}]
{Sethna:sdkkrs93}
\bibinfo{author}{\bibnamefont{Sethna},~\bibfnamefont{J.~P.}},
\bibinfo{author}{\bibfnamefont{K.}~\bibnamefont{Dahmen}},
\bibinfo{author}{\bibfnamefont{S.}~\bibnamefont{Kartha}},
\bibinfo{author}{\bibfnamefont{J.~A.}~\bibnamefont{Krumhansl}},
\bibinfo{author}{\bibfnamefont{B.~W.}~\bibnamefont{Roberts}},
and
\bibinfo{author}{\bibfnamefont{J.~D.}~\bibnamefont{Shore}},
\bibinfo{year}{1993},
\bibinfo{title}{``Hysteresis and hierarchies: Dynamics of disorder-driven
first-order phase transformations},''
\bibinfo{journal}{Phys. Rev. Lett.} \textbf{\bibinfo{volume}{70}},
\bibinfo{pages}{3347}.


\bibitem[{\citenamefont{Sethna} \emph{et~al.}(2001)\citenamefont{Sethna, Dahmen, and Myers}}]
{Sethna:sdm01}
\bibinfo{author}{\bibnamefont{Sethna},~\bibfnamefont{J.~P.}},
\bibinfo{author}{\bibfnamefont{K.~A.}~\bibnamefont{Dahmen}},
and
\bibinfo{author}{\bibfnamefont{C.~R.}~\bibnamefont{Myers}},
\bibinfo{year}{2001},
\bibinfo{title}{``Crackling noise},''
\bibinfo{journal}{Nature} \textbf{\bibinfo{volume}{410}},
\bibinfo{pages}{242}.


\bibitem[{\citenamefont{Seyed-allaei} \emph{et~al.}(2006)\citenamefont{Seyed-allaei, Bianconi, and Marsili}}]
{Seyed-allaei:sbm06}
\bibinfo{author}{\bibnamefont{Seyed-allaei},~\bibfnamefont{H.}},
\bibinfo{author}{\bibfnamefont{G.}~\bibnamefont{Bianconi}},
and
\bibinfo{author}{\bibfnamefont{M.}~\bibnamefont{Marsili}},
\bibinfo{year}{2006},
\bibinfo{title}{``Scale-free networks with an exponent less than two},''
\bibinfo{journal}{Phys. Rev. E} \textbf{\bibinfo{volume}{73}},
\bibinfo{pages}{046113}.



\bibitem[{\citenamefont{Skantzos} \emph{et~al.}(2005)\citenamefont{Skantzos, Castillo, and Hatchett}}]
{Skantzos:sch05}
\bibinfo{author}{\bibfnamefont{Skantzos}~\bibnamefont{N.~S.}},
\bibinfo{author}{\bibfnamefont{I.~P.}~\bibnamefont{Castillo}},
and
\bibinfo{author}{\bibfnamefont{J.~P.~L.}~\bibnamefont{Hatchett}},
\bibinfo{year}{2005},
\bibinfo{title}{Cavity approach for real variables on diluted graphs and application
to synchronization in small-world lattices},
\bibinfo{journal}{Phys. Rev. E} \textbf{\bibinfo{volume}{72}},
\bibinfo{pages}{066127}.


\bibitem[{\citenamefont{Skantzos and Hatchett}(2007)}]
{Skantzos:sh07}
\bibinfo{author}{\bibnamefont{Skantzos},~\bibfnamefont{N.~S.}}
and
\bibinfo{author}{\bibfnamefont{J.~P.~L.}~\bibnamefont{Hatchett}},
\bibinfo{year}{2007},
\bibinfo{title}{``Statics and dynamics of the Lebwohl-Lasher model in the Bethe
approximation},''
\bibinfo{journal}{Physica A} \textbf{\bibinfo{volume}{381}},
\bibinfo{pages}{202}.





\bibitem[{\citenamefont{Soderberg}(2002)}]
{Soderberg:s02}
\bibinfo{author}{\bibnamefont{Soderberg},~\bibfnamefont{B.}},
\bibinfo{year}{2002},
\bibinfo{title}{``A general formalism for inhomogeneous random graphs},''
\bibinfo{journal}{Phys. Rev. E} \textbf{\bibinfo{volume}{66}},
\bibinfo{pages}{066121}.


\bibitem[{\citenamefont{Sol\'e and Valverde}(2001)}]
{Sole:sv01}
\bibinfo{author}{\bibnamefont{Sol\'e},~\bibfnamefont{R.~V.}}
and
\bibinfo{author}{\bibfnamefont{S.}~\bibnamefont{Valverde}},
\bibinfo{year}{2001},
\bibinfo{title}{``Information transfer and phase transitions in a model of internet traffic},''
\bibinfo{journal}{Physica A} \textbf{\bibinfo{volume}{289}},
\bibinfo{pages}{595}.


\bibitem[{\citenamefont{Solomonoff and Rapoport}(1951)}]
{Solomonoff:sr51}
\bibinfo{author}{\bibnamefont{Solomonoff},~\bibfnamefont{R.}}
and
\bibinfo{author}{\bibfnamefont{A.}~\bibnamefont{Rapoport}},
\bibinfo{year}{1951},
\bibinfo{title}{``Connectivity of random nets},''
\bibinfo{journal}{Bull. Math. Biophys.} \textbf{\bibinfo{volume}{13}},
\bibinfo{pages}{107}.


\bibitem[{\citenamefont{Son} \emph{et~al.}(2006)\citenamefont{Son, Jeong, and Noh}}]
{Son:sjn06}
\bibinfo{author}{\bibnamefont{Son},~\bibfnamefont{S.-W.}},
\bibinfo{author}{\bibfnamefont{H.}~\bibnamefont{Jeong}},
and
\bibinfo{author}{\bibfnamefont{J.~D.}~\bibnamefont{Noh}},
\bibinfo{year}{2006},
\bibinfo{title}{``Random field Ising model and community structure in complex
networks},''
\bibinfo{journal}{Eur. Phys. J. B} \textbf{\bibinfo{volume}{50}},
\bibinfo{pages}{431}.


\bibitem[{\citenamefont{Song} \emph{et~al.}(2005)\citenamefont{Song, Havlin, and Makse}}]
{Song:shm05}
\bibinfo{author}{\bibnamefont{Song},~\bibfnamefont{C.}},
\bibinfo{author}{\bibfnamefont{S.}~\bibnamefont{Havlin}}, 
and
\bibinfo{author}{\bibfnamefont{H.~A.}~\bibnamefont{Makse}},
\bibinfo{year}{2005},
\bibinfo{title}{``Self-similarity of complex networks},''
\bibinfo{journal}{Nature} \textbf{\bibinfo{volume}{433}},
\bibinfo{pages}{392}. 


\bibitem[{\citenamefont{Song} \emph{et~al.}(2006)\citenamefont{Song, Havlin, and Makse}}]
{Song:shm06}
\bibinfo{author}{\bibnamefont{Song},~\bibfnamefont{C.}},
\bibinfo{author}{\bibfnamefont{S.}~\bibnamefont{Havlin}}, 
and
\bibinfo{author}{\bibfnamefont{H.~A.}~\bibnamefont{Makse}},
\bibinfo{year}{2006},
\bibinfo{title}{``Origins of fractality in the growth of complex networks},''
\bibinfo{journal}{Nature Physics} \textbf{\bibinfo{volume}{2}},
\bibinfo{pages}{275}.


\bibitem[{\citenamefont{Song} \emph{et~al.}(2007)\citenamefont{Song, Gallos, Havlin, and Makse}}]
{Song:sghm07}
\bibinfo{author}{\bibnamefont{Song},~\bibfnamefont{C.}},
\bibinfo{author}{\bibfnamefont{L.~K.}~\bibnamefont{Gallos}},
\bibinfo{author}{\bibfnamefont{S.}~\bibnamefont{Havlin}}, 
and
\bibinfo{author}{\bibfnamefont{H.~A.}~\bibnamefont{Makse}},
\bibinfo{year}{2007},
\bibinfo{title}{``How to calculate the fractal dimension of a complex network: the box covering algorithm},''
\bibinfo{journal}{J.~Stat. Mech.} 
\bibinfo{pages}{P03006}.


\bibitem[{\citenamefont{Sood and Grassberger}(2007)}]
{Sood:sg07}
\bibinfo{author}{\bibnamefont{Sood},~\bibfnamefont{V.}}
and
\bibinfo{author}{\bibfnamefont{P.}~\bibnamefont{Grassberger}},
\bibinfo{year}{2007},
\bibinfo{title}{``Localization transition of biased random walks on random networks},''
\bibinfo{journal}{Phys. Rev. Lett.} \textbf{\bibinfo{volume}{99}},
\bibinfo{pages}{098701}. 


\bibitem[{\citenamefont{Sood and Redner}(2005)}]
{Sood:sr05}
\bibinfo{author}{\bibnamefont{Sood},~\bibfnamefont{V.}}
and
\bibinfo{author}{\bibfnamefont{S.}~\bibnamefont{Redner}},
\bibinfo{year}{2005},
\bibinfo{title}{``Voter model on heterogeneous graphs},''
\bibinfo{journal}{Phys. Rev. Lett.} \textbf{\bibinfo{volume}{94}},
\bibinfo{pages}{178701}.







\bibitem[{\citenamefont{Stosic} \emph{et~al.}(1998)\citenamefont{Stosic, Stosic, and Fittipaldi}}]
{Stosic:ssf98}
\bibinfo{author}{\bibnamefont{Stosic},~\bibfnamefont{T.}},
\bibinfo{author}{\bibfnamefont{B. D.}~\bibnamefont{Stosic}},
and
\bibinfo{author}{\bibfnamefont{I. P.}~\bibnamefont{Fittipaldi}},
\bibinfo{year}{1998},
\bibinfo{title}{``Exact zero-field susceptibility of the Ising model on a Cayley tree},''
\bibinfo{journal}{J. Magn. Magn. Mater.} \textbf{\bibinfo{volume}{177}},
\bibinfo{pages}{185}.


\bibitem[{\citenamefont{Strauss}(1986)}]
{Strauss:s86}
\bibinfo{author}{\bibfnamefont{Strauss}~\bibnamefont{D.}},
\bibinfo{year}{1986},
\bibinfo{title}{``On a general class of models for interaction},''
\bibinfo{journal}{SIAM Review} \textbf{\bibinfo{volume}{28}},
\bibinfo{pages}{513}.


\bibitem[{\citenamefont{Strogatz}(2000)}]
{Strogatz:s00}
\bibinfo{author}{\bibnamefont{Strogatz},~\bibfnamefont{S.~H.}}
\bibinfo{year}{2000},
\bibinfo{title}{``From Kuramoto to Crawford: Exploring the onset of synchronization in
populations of coupled oscillators},''
\bibinfo{journal}{Physica D} \textbf{\bibinfo{volume}{143}},
\bibinfo{pages}{1}.

\bibitem[{\citenamefont{Strogatz}(2003)}]
{Strogatz:sbook03}
\bibinfo{author}{\bibnamefont{Strogatz}, \bibfnamefont{S.~H.}},
\bibinfo{year}{2003},
\emph{\bibinfo{title}{Sync: The Emerging Science of
Spontaneous Order}}
(\bibinfo{publisher}{Hyperion, New York}).


\bibitem[{\citenamefont{Suchecki} \emph{et~al.}(2005a)\citenamefont{Suchecki, Egu\'iluz, and San Miguel}}]
{Suchecki:sem05a}
\bibinfo{author}{\bibnamefont{Suchecki},~\bibfnamefont{K.}},
\bibinfo{author}{\bibfnamefont{V.~M.}~\bibnamefont{Egu\'iluz}},
and
\bibinfo{author}{\bibfnamefont{M.}~\bibnamefont{San Miguel}},
\bibinfo{year}{2005a},
\bibinfo{title}{``Conservation laws for the voter model in complex networks},''
\bibinfo{journal}{Europhys. Lett.} \textbf{\bibinfo{volume}{69}},
\bibinfo{pages}{228}.


\bibitem[{\citenamefont{Suchecki} \emph{et~al.}(2005b)\citenamefont{Suchecki, Egu\'iluz, and San Miguel}}]
{Suchecki:sem05b}
\bibinfo{author}{\bibnamefont{Suchecki},~\bibfnamefont{K.}},
\bibinfo{author}{\bibfnamefont{V.~M.}~\bibnamefont{Egu\'iluz}},
and
\bibinfo{author}{\bibfnamefont{M.}~\bibnamefont{San Miguel}},
\bibinfo{year}{2005b},
\bibinfo{title}{``Voter model dynamics in complex networks: Role of dimensionality, disorder, and degree distribution},''
\bibinfo{journal}{Phys. Rev. E} \textbf{\bibinfo{volume}{72}},
\bibinfo{pages}{036132}.


\bibitem[{\citenamefont{Svenson}(2001)}]
{Svenson:s01}
\bibinfo{author}{\bibfnamefont{Svenson}~\bibnamefont{P.}},
\bibinfo{year}{2001},
\bibinfo{title}{``Freezing in random graph ferromagnets},''
\bibinfo{journal}{Phys. Rev. E} \textbf{\bibinfo{volume}{64}},
\bibinfo{pages}{036122}.


\bibitem[{\citenamefont{Svenson and Nordahl}(1999)}]
{Svenson:sn99}
\bibinfo{author}{\bibnamefont{Svenson},~\bibfnamefont{P.}}
and
\bibinfo{author}{\bibfnamefont{M.~G.}~\bibnamefont{Nordahl}},
\bibinfo{year}{1999},
\bibinfo{title}{``Relaxation in graph coloring and satisfiability problem},''
\bibinfo{journal}{Phys. Rev. E} \textbf{\bibinfo{volume}{59}},
\bibinfo{pages}{3983}.








\bibitem[{\citenamefont{Szab\'o} \emph{et~al.}(2003)\citenamefont{Szab\'o, Alava and Kert\'esz}}]
{Szabo:sak03}
\bibinfo{author}{\bibnamefont{Szab\'o},~\bibfnamefont{G.}},
\bibinfo{author}{\bibfnamefont{M.~J.}~\bibnamefont{Alava}},
and
\bibinfo{author}{\bibfnamefont{J.}~\bibnamefont{Kert\'esz}},
\bibinfo{year}{2003},
\bibinfo{title}{``Structural transitions in scale free networks},''
\bibinfo{journal}{Phys. Rev. E 67} \textbf{\bibinfo{volume}{67}},
\bibinfo{pages}{056102}.








\bibitem[{\citenamefont{Tadi\'c} \emph{et~al.}(2005)\citenamefont{Tadi\'c, Malarz, and Kulakowski}}]
{Tadic:tmk05}
\bibinfo{author}{\bibfnamefont{Tadi\'c}~\bibnamefont{B.}},
\bibinfo{author}{\bibfnamefont{K.}~\bibnamefont{Malarz}},
and
\bibinfo{author}{\bibfnamefont{K.}~\bibnamefont{Kulakowski}},
\bibinfo{year}{2005},
\bibinfo{title}{``Magnetization reversal in spin patterns with complex geometry},''
\bibinfo{journal}{Phys. Rev. Lett.} \textbf{\bibinfo{volume}{94}},
\bibinfo{pages}{137204}.


\bibitem[{\citenamefont{Tadi\'c} \emph{et~al.}(2007)\citenamefont{Tadi\'c, Rodgers, and Thurner}}]
{Tadic:trt06}
\bibinfo{author}{\bibnamefont{Tadi\'c},~\bibfnamefont{B.}},
\bibinfo{author}{\bibfnamefont{G.~J.}~\bibnamefont{ Rodgers}},
and
\bibinfo{author}{\bibfnamefont{S.}~\bibnamefont{Thurner}},
\bibinfo{year}{2007},
\bibinfo{title}{``Transport on complex networks: Flow, jamming and optimization},''
\bibinfo{journal}{Int. J. Bifurcation and Chaos} \textbf{\bibinfo{volume}{17}},
\bibinfo{pages}{2363}. 


\bibitem[{\citenamefont{Tadi\'c and Thurner}(2004)}]
{Tadic:tt04}
\bibinfo{author}{\bibnamefont{Tadi\'c},~\bibfnamefont{B.}}
and
\bibinfo{author}{\bibfnamefont{S.}~\bibnamefont{Thurner}},
\bibinfo{year}{2004},
\bibinfo{title}{``Information super-diffusion on structured networks},''
\bibinfo{journal}{Physica A} \textbf{\bibinfo{volume}{332}},
\bibinfo{pages}{566}.


\bibitem[{\citenamefont{Tadi\'c} \emph{et~al.}(2004)\citenamefont{Tadi\'c, Thurner, and Rodgers}}]
{Tadic:ttr04}
\bibinfo{author}{\bibnamefont{Tadi\'c},~\bibfnamefont{B.}},
\bibinfo{author}{\bibfnamefont{S.}~\bibnamefont{Thurner}},
and
\bibinfo{author}{\bibfnamefont{G.~J.}~\bibnamefont{Rodgers}},
\bibinfo{year}{2004},
\bibinfo{title}{``Traffic on complex networks: Towards understanding global statistical properties from microscopic density fluctuations},''
\bibinfo{journal}{Phys. Rev. E} \textbf{\bibinfo{volume}{69}},
\bibinfo{pages}{036102}.


\bibitem[{\citenamefont{Tanaka}(2002)}]
{Tanaka:t02}
\bibinfo{author}{\bibnamefont{Tanaka},~\bibfnamefont{K.}}
\bibinfo{year}{2002},
\bibinfo{title}{``Statistical-mechanical approach to image processing},''
\bibinfo{journal}{J. Phys. A} \textbf{\bibinfo{volume}{35}},
\bibinfo{pages}{R81}.


\bibitem[{\citenamefont{Tanaka} \emph{et~al.}(1997)\citenamefont{Tanaka, Lichtenberg, and Oishi}}]
{Tanaka:tlo97}
\bibinfo{author}{\bibnamefont{Tanaka},~\bibfnamefont{H.-A.}},
\bibinfo{author}{\bibfnamefont{A.~J.}~\bibnamefont{Lichtenberg}},
and
\bibinfo{author}{\bibfnamefont{S.}~\bibnamefont{Oishi}},
\bibinfo{year}{1997},
\bibinfo{title}{``First order phase transition resulting
from finite inertia in coupled oscillator systems},''
\bibinfo{journal}{Phys. Rev. Lett.} \textbf{\bibinfo{volume}{78}},
\bibinfo{pages}{2104}.


\bibitem[{\citenamefont{Tang} \emph{et~al.}(2006)\citenamefont{Tang, Liu, and Zhou}}]
{Tang:tlz06}
\bibinfo{author}{\bibnamefont{Tang},~\bibfnamefont{M.}},
\bibinfo{author}{\bibfnamefont{Z.}~\bibnamefont{Liu}},
and
\bibinfo{author}{\bibfnamefont{J.}~\bibnamefont{Zhou}},
\bibinfo{year}{2006},
\bibinfo{title}{``Condensation in a zero range process on weighted scale-free networks},''
\bibinfo{journal}{Phys. Rev. E} \textbf{\bibinfo{volume}{74}},
\bibinfo{pages}{036101}.


\bibitem[{\citenamefont{Thouless}(1986)}]
{Thouless:t86}
\bibinfo{author}{\bibnamefont{Thouless}, \bibfnamefont{D.~J.}}, \bibinfo{year}{1986},
\bibinfo{title}{``Spin-Glass on a Bethe Lattice},''
\bibinfo{journal}{Phys. Rev. Lett.} \textbf{\bibinfo{volume}{56}},
\bibinfo{pages}{1082}.

\bibitem[{\citenamefont{Thouless} \emph{et~al.}(1977)\citenamefont{Thouless, Anderson, and Palmer}}]
{Thouless:tap77}
\bibinfo{author}{\bibnamefont{Thouless},~\bibfnamefont{D.~J.}},
\bibinfo{author}{\bibfnamefont{P.~W.}~\bibnamefont{Anderson}},
and
\bibinfo{author}{\bibfnamefont{R.~G.}~\bibnamefont{Palmer}},
\bibinfo{year}{1977},
\bibinfo{title}{``Solution of solvable model of a spin glass},''
\bibinfo{journal}{Philos. Mag.} \textbf{\bibinfo{volume}{35}},
\bibinfo{pages}{593}.


\bibitem[{\citenamefont{Timme}(2006)}]
{Timme:t06}
\bibinfo{author}{\bibfnamefont{Timme}~\bibnamefont{M.}},
\bibinfo{year}{2006},
\bibinfo{title}{``Does dynamics reflect topology in directed networks?},''
\bibinfo{journal}{Europhys. Lett.} \textbf{\bibinfo{volume}{76}},
\bibinfo{pages}{367}.


\bibitem[{\citenamefont{Timme} \emph{et~al.}(2004)\citenamefont{Timme, Wolf, and Geisel}}]
{Timme:twg04}
\bibinfo{author}{\bibnamefont{Timme},~\bibfnamefont{M.}},
\bibinfo{author}{\bibfnamefont{F.}~\bibnamefont{Wolf}},
and
\bibinfo{author}{\bibfnamefont{T.}~\bibnamefont{Geisel}},
\bibinfo{year}{2004},
\bibinfo{title}{``Topological speed limits to network synchronization},''
\bibinfo{journal}{Phys. Rev. Lett.} \textbf{\bibinfo{volume}{92}},
\bibinfo{pages}{074101}.


\bibitem[{\citenamefont{Toroczkai and Bassler}(2004)}]
{Toroczkai:tb04}
\bibinfo{author}{\bibnamefont{Toroczkai},~\bibfnamefont{Z.}}
and
\bibinfo{author}{\bibfnamefont{K.~E.}~\bibnamefont{Bassler}},
\bibinfo{year}{2004},
\bibinfo{title}{``Jamming is limited in scale-free systems},''
\bibinfo{journal}{Nature} \textbf{\bibinfo{volume}{428}},
\bibinfo{pages}{716}.








\bibitem[{\citenamefont{Vazquez}(2006a)}]
{Vazquez:v06a}
\bibinfo{author}{\bibfnamefont{Vazquez}~\bibnamefont{A.}},
\bibinfo{year}{2006a},
\bibinfo{title}{``Polynomial growth in branching processes with diverging reproductive number},''
\bibinfo{journal}{Phys. Rev. Lett.} \textbf{\bibinfo{volume}{96}},
\bibinfo{pages}{038702}.


\bibitem[{\citenamefont{Vazquez}(2006b)}]
{Vazquez:v06b}
\bibinfo{author}{\bibfnamefont{Vazquez}~\bibnamefont{A.}},
\bibinfo{year}{2006b},
\bibinfo{title}{``Spreading dynamics on small-world networks with connectivity fluctuations and correlations},''
\eprint{q-bio/0603010}.


\bibitem[{\citenamefont{V\'azquez and Moreno}(2003)}]
{Vazquez:vm03}
\bibinfo{author}{\bibnamefont{V\'azquez},~\bibfnamefont{A.}}
and
\bibinfo{author}{\bibfnamefont{Y.}~\bibnamefont{Moreno}},
\bibinfo{year}{2003},
\bibinfo{title}{``Resilience to damage of graphs with degree correlations},''
\bibinfo{journal}{Phys. Rev. E} \textbf{\bibinfo{volume}{67}},
\bibinfo{pages}{015101 (R)}.


\bibitem[{\citenamefont{V\'azquez and Weigt}(2003)}]
{Vazquez:vw03}
\bibinfo{author}{\bibnamefont{V\'azquez},~\bibfnamefont{A.}}
and
\bibinfo{author}{\bibfnamefont{M.}~\bibnamefont{Weigt}},
\bibinfo{year}{2003},
\bibinfo{title}{``Computational complexity arising from degree correlations in networks},''
\bibinfo{journal}{Phys. Rev. E} \textbf{\bibinfo{volume}{67}},
\bibinfo{pages}{027101}.


\bibitem[{\citenamefont{Viana and Bray}(1985)}]
{Viana:vb85}
\bibinfo{author}{\bibnamefont{Viana},~\bibfnamefont{L.}}
and
\bibinfo{author}{\bibfnamefont{A.~J.}~\bibnamefont{Bray}},
\bibinfo{year}{1985},
\bibinfo{title}{``Phase diagrams for dilute spin glasses},''
\bibinfo{journal}{J. Phys. C} \textbf{\bibinfo{volume}{18}},
\bibinfo{pages}{3037}.


\bibitem[{\citenamefont{Vilone and Castellano}(2004)}]
{Vilone:vc04}
\bibinfo{author}{\bibnamefont{Vilone},~\bibfnamefont{D.}}
and
\bibinfo{author}{\bibfnamefont{C.}~\bibnamefont{Castellano}},
\bibinfo{year}{2004},
\bibinfo{title}{``Solution of voter model dynamics on annealed small-world networks},''
\bibinfo{journal}{Phys. Rev. E} \textbf{\bibinfo{volume}{69}},
\bibinfo{pages}{016109}.




\bibitem[{\citenamefont{Waclaw} \emph{et~al.}(2007)\citenamefont{Waclaw, Bogacz, Burda, and Janke}}]
{Waclaw:wbb07}
\bibinfo{author}{\bibnamefont{Waclaw},~\bibfnamefont{B.}},
\bibinfo{author}{\bibfnamefont{L.}~\bibnamefont{Bogacz}},
\bibinfo{author}{\bibfnamefont{Z.}~\bibnamefont{Burda}},
and
\bibinfo{author}{\bibfnamefont{W.}~\bibnamefont{Janke}},
\bibinfo{year}{2007},
\bibinfo{title}{``Condensation in zero-range processes on inhomogeneous networks},''
\bibinfo{journal}{Phys. Rev. E} \textbf{\bibinfo{volume}{76}},
\bibinfo{pages}{046114}. 


\bibitem[{\citenamefont{Waclaw and Sokolov}(2007)}]
{Waclaw:ws07}
\bibinfo{author}{\bibnamefont{Waclaw},~\bibfnamefont{B.}}
and
\bibinfo{author}{\bibfnamefont{I.~M.}~\bibnamefont{Sokolov}},
\bibinfo{year}{2007},
\bibinfo{title}{``Finite size effects in Barab\'asi-Albert growing networks},''
\bibinfo{journal}{Phys. Rev. E} \textbf{\bibinfo{volume}{75}},
\bibinfo{pages}{056114}.


\bibitem[{\citenamefont{Walsh}(1999)}]
{Walsh:99}
\bibinfo{author}{\bibfnamefont{Walsh}~\bibnamefont{T.}},
\bibinfo{year}{1999},
\bibinfo{title}{``Search in a small world},''
in \emph{\bibinfo{booktitle}{Proceedings of the 16the International Joint Conference on Artificial Intelligence}},
edited by
\bibinfo{editor}{\bibfnamefont{T.}~\bibnamefont{Dean}}
(\bibinfo{publisher}{Morgan Kaufmann, San Francisco, CA}), p.
\bibinfo{pages}{1172}.



\bibitem[{\citenamefont{Wang and Chen}(2002)}]
{Wang:wc02}
\bibinfo{author}{\bibnamefont{Wang},~\bibfnamefont{X.~F.}}
and
\bibinfo{author}{\bibfnamefont{G.}~\bibnamefont{Chen}},
\bibinfo{year}{2002},
\bibinfo{title}{``Synchronization in small-world dynamical networks},''
\bibinfo{journal}{Int. J. Bifurcation Chaos Appl. Sci. Eng.} \textbf{\bibinfo{volume}{12}},
\bibinfo{pages}{187}. 


\bibitem[{\citenamefont{Wang} \emph{et~al.}(2006)\citenamefont{Wang, Wang, Yin, Xie, and Zhou}}]
{Wang:wwy06}
\bibinfo{author}{\bibnamefont{Wang},~\bibfnamefont{W.-X.}},
\bibinfo{author}{\bibfnamefont{B.-H.}~\bibnamefont{Wang}},
\bibinfo{author}{\bibfnamefont{C.-Y.}~\bibnamefont{Yin}},
\bibinfo{author}{\bibfnamefont{Y.-B.}~\bibnamefont{Xie}},
and
\bibinfo{author}{\bibfnamefont{T.}~\bibnamefont{Zhou}},
\bibinfo{year}{2006},
\bibinfo{title}{``Traffic dynamics based on local routing protocol on a scale-free network},''
\bibinfo{journal}{Phys. Rev. E} \textbf{\bibinfo{volume}{73}},
\bibinfo{pages}{026111}.


\bibitem[{\citenamefont{Warren} \emph{et~al.}(2003)\citenamefont{Warren, Sander, and Sokolov}}]
{Warren:wss01}
\bibinfo{author}{\bibnamefont{Warren},~\bibfnamefont{C.~P.}},
\bibinfo{author}{\bibfnamefont{L.~M.}~\bibnamefont{Sander}},
and
\bibinfo{author}{\bibfnamefont{I.~M.}~\bibnamefont{Sokolov}},
\bibinfo{year}{2003},
\bibinfo{title}{``Epidemics, disorder, and percolation},''
\bibinfo{journal}{Physica A} \textbf{\bibinfo{volume}{325}},
\bibinfo{pages}{1}.


\bibitem[{\citenamefont{Watts}(1999)}]
{Watts:wbook99}
\bibinfo{author}{\bibnamefont{Watts},~\bibfnamefont{D.~J.}},
\bibinfo{year}{1999},
\emph{\bibinfo{title}{Small Worlds: The Dynamics of Networks between
Order and Randomness}} (\bibinfo{publisher}{Princeton University
Press, Princeton, NJ}).


\bibitem[{\citenamefont{Watts}(2002)}]
{Watts:w02}
\bibinfo{author}{\bibnamefont{Watts},~\bibfnamefont{D.~J.}},
\bibinfo{year}{2002},
\bibinfo{title}{``A simple model of global cascades on random networks},''
\bibinfo{journal}{PNAS} \textbf{\bibinfo{volume}{99}},
\bibinfo{pages}{5766}.


\bibitem[{\citenamefont{Watts} \emph{et~al.}(2002)\citenamefont{Watts, Dodds, and Newman}}]
{Watts:wdn02}
\bibinfo{author}{\bibnamefont{Watts},~\bibfnamefont{D.~J.}},
\bibinfo{author}{\bibfnamefont{P.~S.}~\bibnamefont{Dodds}},
and
\bibinfo{author}{\bibfnamefont{M.~E.~J.}~\bibnamefont{Newman}},
\bibinfo{year}{2002},
\bibinfo{title}{``Identity and search in social networks},''
\bibinfo{journal}{Science} \textbf{\bibinfo{volume}{296}},
\bibinfo{pages}{1302}.


\bibitem[{\citenamefont{Watts and Strogatz}(1998)}]
{Watts:ws98}
\bibinfo{author}{\bibnamefont{Watts},~\bibfnamefont{D.~J.}}
and
\bibinfo{author}{\bibfnamefont{S.~H.}~\bibnamefont{Strogatz}},
\bibinfo{year}{1998},
\bibinfo{title}{``Collective dynamics of `small-world' networks},''
\bibinfo{journal}{Nature} \textbf{\bibinfo{volume}{393}},
\bibinfo{pages}{440}.


\bibitem[{\citenamefont{Weigt and Hartmann}(2000)}]
{Weigt:wh00}
\bibinfo{author}{\bibnamefont{Weigt},~\bibfnamefont{M.}}
and
\bibinfo{author}{\bibfnamefont{A.~K.}~\bibnamefont{Hartmann}},
\bibinfo{year}{2000},
\bibinfo{title}{``Number of guards needed by a museum: A phase transition in vertex covering
of random graphs},''
\bibinfo{journal}{Phys. Rev. Lett.} \textbf{\bibinfo{volume}{84}},
\bibinfo{pages}{6118}.


\bibitem[{\citenamefont{Weigt and Hartmann}(2001)}]
{Weigt:wh01}
\bibinfo{author}{\bibnamefont{Weigt},~\bibfnamefont{M.}}
and
\bibinfo{author}{\bibfnamefont{A.~K.}~\bibnamefont{Hartmann}},
\bibinfo{year}{2001},
\bibinfo{title}{``Minimal vertex covers on finite-connectivity random graphs: A
hard-sphere lattice-gas picture},''
\bibinfo{journal}{Phys. Rev. E} \textbf{\bibinfo{volume}{63}},
\bibinfo{pages}{056127}.




\bibitem[{\citenamefont{Weigt and Zhou}(2006)}]
{Weigt:wz06}
\bibinfo{author}{\bibnamefont{Weigt},~\bibfnamefont{M.}}
and
\bibinfo{author}{\bibfnamefont{H.}~\bibnamefont{Zhou}},
\bibinfo{year}{2006},
\bibinfo{title}{``Message passing for vertex covers},''
\bibinfo{journal}{Phys. Rev. E} \textbf{\bibinfo{volume}{74}},
\bibinfo{pages}{046110}.




\bibitem[{\citenamefont{Willinger} \emph{et~al.}(2002)\citenamefont{Willinger, Govindan, Jamin, Paxson, and Shenker}}]
{Willinger:wgj02}
\bibinfo{author}{\bibnamefont{Willinger},~\bibfnamefont{W.}},
\bibinfo{author}{\bibfnamefont{R.}~\bibnamefont{Govindan}},
\bibinfo{author}{\bibfnamefont{S.}~\bibnamefont{Jamin}},
\bibinfo{author}{\bibfnamefont{V.}~\bibnamefont{Paxson}},
and
\bibinfo{author}{\bibfnamefont{S.}~\bibnamefont{Shenker}},
\bibinfo{year}{2002},
\bibinfo{title}{``Scaling phenomena in the Internet: Critically examining criticality},''
\bibinfo{journal}{PNAS} \textbf{\bibinfo{volume}{99}},
\bibinfo{pages}{2573}.




\bibitem[{\citenamefont{Wo{\l}oszyn} \emph{et~al.}(2007)\citenamefont{Wo{\l}oszyn, Stauffer, and Kulakowski}}]
{Woloszyn:wsk06}
\bibinfo{author}{\bibnamefont{Wo{\l}oszyn},~\bibfnamefont{M.}},
\bibinfo{author}{\bibfnamefont{D.}~\bibnamefont{Stauffer}},
and
\bibinfo{author}{\bibfnamefont{K.}~\bibnamefont{Kulakowski}},
\bibinfo{year}{2007},
\bibinfo{title}{``Order-disorder phase transition in a cliquey social network},'' 
\bibinfo{journal}{Eur. Phys. J. B} \textbf{\bibinfo{volume}{57}},
\bibinfo{pages}{331}. 






\bibitem[{\citenamefont{Wu}(1982)}]
{Wu:w82}
\bibinfo{author}{\bibnamefont{Wu},~\bibfnamefont{F.~Y.}},
\bibinfo{year}{1982},
\bibinfo{title}{``The Potts model},''
\bibinfo{journal}{Rev. Mod. Phys.} \textbf{\bibinfo{volume}{54}},
\bibinfo{pages}{235}.


\bibitem[{\citenamefont{Wu} \emph{et~al.}(2006)\citenamefont{Wu, Braunstein, Vittoria Colizza, Reuven Cohen, Shlomo Havlin, and Stanley}}]
{Wu:wbc06}
\bibinfo{author}{\bibnamefont{Wu},~\bibfnamefont{Z.}},
\bibinfo{author}{\bibfnamefont{L.~A.}~\bibnamefont{Braunstein}},
\bibinfo{author}{\bibfnamefont{V.}~\bibnamefont{Colizza}},
\bibinfo{author}{\bibfnamefont{R.}~\bibnamefont{Cohen}},
\bibinfo{author}{\bibfnamefont{S.}~\bibnamefont{Havlin}},
and
\bibinfo{author}{\bibfnamefont{H.~E.}~\bibnamefont{Stanley}},
\bibinfo{year}{2006},
\bibinfo{title}{``Optimal paths in complex networks with correlated weights: The worldwide airport network},''
\bibinfo{journal}{Phys. Rev. E} \textbf{\bibinfo{volume}{74}},
\bibinfo{pages}{056104}.


\bibitem[{\citenamefont{Wu and Huberman}(2004)}]
{Wu:wh04}
\bibinfo{author}{\bibnamefont{Wu},~\bibfnamefont{F.}}
and
\bibinfo{author}{\bibfnamefont{B.~A.}~\bibnamefont{Huberman}},
\bibinfo{year}{2004},
\bibinfo{title}{``Social structure and opinion formation},''
\eprint{cond-mat/0407252}.


\bibitem[{\citenamefont{Wu} \emph{et~al.}(2007a)\citenamefont{Wu, Lagorio, Braunstein, Cohen, Havlin, and Stanley}}]
{Wu:wlb07}
\bibinfo{author}{\bibnamefont{Wu},~\bibfnamefont{Z.}},
\bibinfo{author}{\bibfnamefont{C.}~\bibnamefont{Lagorio}},
\bibinfo{author}{\bibfnamefont{L.~A.}~\bibnamefont{Braunstein}},
\bibinfo{author}{\bibfnamefont{R.}~\bibnamefont{Cohen}},
\bibinfo{author}{\bibfnamefont{S.}~\bibnamefont{Havlin}},
and
\bibinfo{author}{\bibfnamefont{H.~E.}~\bibnamefont{Stanley}},
\bibinfo{year}{2007a},
\bibinfo{title}{``Numerical evaluation of the upper critical dimension of percolation in scale-free networks},''
\bibinfo{journal}{Phys. Rev. E} \textbf{\bibinfo{volume}{75}},
\bibinfo{pages}{066110}.


\bibitem[{\citenamefont{Wu} \emph{et~al.}(2007b)\citenamefont{Wu, Xu, and Wang}}]
{Wu:wxw07}
\bibinfo{author}{\bibnamefont{Wu},~\bibfnamefont{A.-C.}},
\bibinfo{author}{\bibfnamefont{X.-J.}~\bibnamefont{Xu}},
and
\bibinfo{author}{\bibfnamefont{Y.-H.}~\bibnamefont{Wang}},
\bibinfo{year}{2007b},
\bibinfo{title}{``Excitable Greenberg-Hastings cellular automaton model on scale-free networks},''
\bibinfo{journal}{Phys. Rev. E} \textbf{\bibinfo{volume}{75}},
\bibinfo{pages}{032901}.




\bibitem[{\citenamefont{Yamada}(2002)}]
{Yamada:y02}
\bibinfo{author}{\bibnamefont{Yamada},~\bibfnamefont{H.}}
\bibinfo{year}{2002},
\bibinfo{title}{``Phase-locked and phase drift solutions of phase oscillators with asymmetric coupling strengths},''
\bibinfo{journal}{Progr. Theor. Phys.} \textbf{\bibinfo{volume}{108}},
\bibinfo{pages}{13}.


\bibitem[{\citenamefont{Yedidia} \emph{et~al.}(2001)\citenamefont{Yedidia, Freeman, and Weiss}}]
{Yedidia:yfw01}
\bibinfo{author}{\bibnamefont{Yedidia}, \bibfnamefont{J.~S.}},
\bibinfo{author}{\bibfnamefont{W.~T.} \bibnamefont{Freeman}}, and
\bibinfo{author}{\bibfnamefont{Y.}~\bibnamefont{Weiss}},
\bibinfo{year}{2001}, in \emph{\bibinfo{booktitle}{Advances in Neural
Information Processing Systems}}, edited by
\bibinfo{editor}{\bibfnamefont{T.~K.}~\bibnamefont{Leen}},
\bibinfo{editor}{\bibfnamefont{T.~G.}~\bibnamefont{Dietterich}}, and
\bibinfo{editor}{\bibfnamefont{V.}~\bibnamefont{Tresp}},
(\bibinfo{publisher}{MA: MIT Press, Cambridge}),
\bibinfo{title}{``Generalized belief propagation},''
\textbf{\bibinfo{volume}{13}}, pp. \bibinfo{pages}{689-695}.






\bibitem[{\citenamefont{Zanette}(2007)}]
{Zanette:z07}
\bibinfo{author}{\bibnamefont{Zanette},~\bibfnamefont{D.~H.}},
\bibinfo{year}{2007},
\bibinfo{title}{``Coevolution of agents and networks in an epidemiological model},''
\eprint{0707.1249 [cond-mat]}.


\bibitem[{\citenamefont{Zdeborov\'a and Krz\c{a}ka{\l}a}(2007)}]
{Zdeborova:zk07}
\bibinfo{author}{\bibnamefont{Zdeborov\'a},~\bibfnamefont{L.}}
and
\bibinfo{author}{\bibfnamefont{F.}~\bibnamefont{Krz\c{a}ka{\l}a}},
\bibinfo{year}{2007},
\bibinfo{title}{``Phase transitions in the coloring of random graphs},''
\bibinfo{journal}{Phys. Rev. E} \textbf{\bibinfo{volume}{76}},
\bibinfo{pages}{031131}.




\bibitem[{\citenamefont{Zhang} \emph{et~al.}(2007)\citenamefont{Zhang, Liu, Tang, and Hui}}]
{Zhang:hlt07}
\bibinfo{author}{\bibnamefont{Zhang},~\bibfnamefont{H.}},
\bibinfo{author}{\bibfnamefont{Z.}~\bibnamefont{Liu}},
\bibinfo{author}{\bibfnamefont{M.}~\bibnamefont{Tang}},
and
\bibinfo{author}{\bibfnamefont{P.~M.}~\bibnamefont{Hui}},
\bibinfo{year}{2007},
\bibinfo{title}{``An adaptive routing strategy for packet delivery in complex networks},''
\bibinfo{journal}{Phys. Lett. A} \textbf{\bibinfo{volume}{364}},
\bibinfo{pages}{177}.


\bibitem[{\citenamefont{Zhang and Novotny}(2006)}]
{Zhang:zn06}
\bibinfo{author}{\bibnamefont{Zhang},~\bibfnamefont{X.}}
and
\bibinfo{author}{\bibfnamefont{M.~A.}~\bibnamefont{Novotny}},
\bibinfo{year}{2006},
\bibinfo{title}{``Critical behavior of Ising models with random long-range (small-world) interactions},''
\bibinfo{journal}{Brazilian J. Phys.} \textbf{\bibinfo{volume}{36}},
\bibinfo{pages}{664}. 


\bibitem[{\citenamefont{Zhou}(2003)}]
{Zhou:z03}
\bibinfo{author}{\bibnamefont{Zhou},~\bibfnamefont{H.}}
\bibinfo{year}{2003},
\bibinfo{title}{``Vertex cover problem studied by cavity method: Analytics and population dynamics},''
\bibinfo{journal}{Eur. Phys. J. B} \textbf{\bibinfo{volume}{32}},
\bibinfo{pages}{265}.


\bibitem[{\citenamefont{Zhou}(2005)}]
{Zhou:z05}
\bibinfo{author}{\bibnamefont{Zhou},~\bibfnamefont{H.}}
\bibinfo{year}{2005},
\bibinfo{title}{``Long range frustrations in a spin glass model of the vertex cover problem },''
\bibinfo{journal}{Phys. Rev. Lett.} \textbf{\bibinfo{volume}{94}},
\bibinfo{pages}{217203}.


\bibitem[{\citenamefont{Zhou and Lipowsky}(2005)}]
{Zhou:zl05}
\bibinfo{author}{\bibnamefont{Zhou},~\bibfnamefont{H.}}
and
\bibinfo{author}{\bibfnamefont{R.}~\bibnamefont{Lipowsky}},
\bibinfo{year}{2005},
\bibinfo{title}{``Dynamic pattern evolution on scale-free networks},''
\bibinfo{journal}{PNAS} \textbf{\bibinfo{volume}{102}},
\bibinfo{pages}{10052}.


\bibitem[{\citenamefont{Zhou and Mondrag\'{o}n}(2004)}]
{Zhou:zm04}
\bibinfo{author}{\bibnamefont{Zhou},~\bibfnamefont{S.}}
and
\bibinfo{author}{\bibfnamefont{R.~J.}~\bibnamefont{Mondrag\'{o}n}},
\bibinfo{year}{2004},
\bibinfo{title}{``The rich-club phenomenon in the Internet topology},''
\bibinfo{journal}{IEEE Commun. Lett.} \textbf{\bibinfo{volume}{8}},
\bibinfo{pages}{180}.


\bibitem[{\citenamefont{Zhou} \emph{et~al.}(2006)\citenamefont{Zhou, Motter, and Kurths}}]
{Zhou:zmk06}
\bibinfo{author}{\bibnamefont{Zhou},~\bibfnamefont{C.}},
\bibinfo{author}{\bibfnamefont{A.~E.}~\bibnamefont{Motter}},
and
\bibinfo{author}{\bibfnamefont{J.}~\bibnamefont{Kurths}},
\bibinfo{year}{2006},
\bibinfo{title}{``Universality in the synchronization of weighted random networks},''
\bibinfo{journal}{Phys. Rev. Lett.} \textbf{\bibinfo{volume}{96}},
\bibinfo{pages}{034101}.


\bibitem[{\citenamefont{Zimmermann} \emph{et~al.}(2004)\citenamefont{Zimmermann, Egu\'iluz, and San Miguel}}]
{Zimmermann:zes04}
\bibinfo{author}{\bibnamefont{Zimmermann},~\bibfnamefont{M.~G.}},
\bibinfo{author}{\bibfnamefont{Egu\'iluz}~\bibnamefont{V.~M.}},
and
\bibinfo{author}{\bibfnamefont{M.}~\bibnamefont{San Miguel}},
\bibinfo{year}{2004},
\bibinfo{title}{``Coevolution of dynamical states and interactions in dynamic networks},''
\bibinfo{journal}{Phys. Rev. E} \textbf{\bibinfo{volume}{69}},
\bibinfo{pages}{065102}.






















































































































\end{thebibliography}




\end{document}
\end







\bibitem[{\citenamefont{}(200)}]
{}
\bibinfo{author}{\bibnamefont{},~\bibfnamefont{}}
and
\bibinfo{author}{\bibfnamefont{}~\bibnamefont{}},
\bibinfo{year}{200},
\bibinfo{title}{``},''
\eprint{}.


\bibitem[{\citenamefont{} \emph{et~al.}(200)\citenamefont{}}]
{}
\bibinfo{author}{\bibnamefont{},~\bibfnamefont{}},
\bibinfo{author}{\bibfnamefont{}~\bibnamefont{}},
and
\bibinfo{author}{\bibfnamefont{}~\bibnamefont{}},
\bibinfo{year}{200},
\bibinfo{title}{``},''
\eprint{}.


\bibitem[{\citenamefont{}(200)}]
{}
\bibinfo{author}{\bibnamefont{},~\bibfnamefont{}}
and
\bibinfo{author}{\bibfnamefont{}~\bibnamefont{}},
\bibinfo{year}{200},
\bibinfo{title}{``},''
\bibinfo{journal}{} \textbf{\bibinfo{volume}{}},
\bibinfo{pages}{}.


\bibitem[{\citenamefont{} \emph{et~al.}(200)\citenamefont{}}]
{}
\bibinfo{author}{\bibnamefont{},~\bibfnamefont{}},
\bibinfo{author}{\bibfnamefont{}~\bibnamefont{}},
\bibinfo{author}{\bibfnamefont{}~\bibnamefont{}},
\bibinfo{author}{\bibfnamefont{}~\bibnamefont{}},
and
\bibinfo{author}{\bibfnamefont{}~\bibnamefont{}},
\bibinfo{year}{200},
\bibinfo{title}{``},''
\bibinfo{journal}{} \textbf{\bibinfo{volume}{}},
\bibinfo{pages}{}.


\bibitem[{\citenamefont{}\emph{et~al.}(200)\citenamefont{}}]
{}
\bibinfo{author}{\bibnamefont{},~\bibfnamefont{}},
\bibinfo{author}{\bibfnamefont{}~\bibnamefont{}},
and
\bibinfo{author}{\bibfnamefont{}~\bibnamefont{}},
\bibinfo{year}{200},
\emph{\bibinfo{title}{}} (\bibinfo{publisher}{}).


\bibitem[{\citenamefont{}(196)}]
{}
\bibinfo{author}{\bibnamefont{},~\bibfnamefont{}},
\bibinfo{year}{196},
\bibinfo{title}{``},''
in \emph{\bibinfo{booktitle}{}}, edited by
\bibinfo{editor}{\bibfnamefont{}~\bibnamefont{}}
and
\bibinfo{editor}{\bibfnamefont{}~\bibnamefont{}}
(\bibinfo{publisher}{}), p. \bibinfo{pages}{}.











\end{document}